\documentclass[11pt,a4paper]{book}
\pdfoutput=1

\RequirePackage{ifpdf}

\usepackage{amsmath,amsfonts,amssymb,amsthm,mathtools} 
\usepackage{pstricks}
\usepackage[final]{pdfpages} 
\usepackage{slashed}
\usepackage{color,xcolor} 
\usepackage{fixmath}
\usepackage{caption} 
\usepackage{subcaption} 
\usepackage{cite}
\usepackage{float}
\usepackage{afterpage}
\allowdisplaybreaks[4]

\usepackage{graphicx}
\usepackage{setspace}
\usepackage[T1]{fontenc}
\usepackage{txfonts}
\usepackage[margin=3.0cm]{geometry}
\usepackage[toc,page]{appendix}

\usepackage{tikz}
\usetikzlibrary{positioning,arrows}
\usetikzlibrary{decorations.pathmorphing}
\usetikzlibrary{decorations.markings}
\usetikzlibrary{shapes.geometric}
\usepackage{pgflibraryarrows}
\usepackage{pgflibrarysnakes}
\tikzset{
    vector/.style={decorate, decoration={snake}, draw},
    provector/.style={decorate, decoration={snake,amplitude=2.5pt}, draw},
    antivector/.style={decorate, decoration={snake,amplitude=-2.5pt}, draw},
    fermion/.style={draw=black, postaction={decorate},decoration={markings,mark=at position .55 with {\arrow[draw=black]{>}}}},
    fermionbar/.style={draw=black, postaction={decorate},
                       decoration={markings,mark=at position .55 with {\arrow[draw=black]{<}}}},
    fermionnoarrow/.style={draw=black},
    gluon/.style={decorate, draw=black,decoration={coil,amplitude=4pt, segment length=5pt}},
    scalar/.style={dashed,draw=black, postaction={decorate},decoration={markings,mark=at position .55 with {\arrow[draw=black]{>}}}},
    scalarbar/.style={dashed,draw=black, postaction={decorate},decoration={markings,mark=at position .55 with {\arrow[draw=black]{<}}}},
    scalarnoarrow/.style={dashed,draw=black},
    electron/.style={draw=black, postaction={decorate},decoration={markings,mark=at position .55 with {\arrow[draw=black]{>}}}},
    bigvector/.style={decorate, decoration={snake,amplitude=4pt}, draw},
}

\usepackage[font=small,labelfont=bf,margin=3mm,labelsep=period,tableposition=top]{caption}
\usepackage{booktabs}

\newcommand{\mytitle}{Higher Order Corrections in pQCD}
\newcommand{\myauthor}{Taushif Ahmed}

\definecolor{urlblue}{rgb}{0.2,0.4,0.7}
\definecolor{citegreen}{rgb}{0,0.6,0.2}
\definecolor{linkred}{rgb}{0.9,0.2,0.1}
\usepackage{hyperref}\hypersetup{
colorlinks,citecolor=citegreen,linkcolor=blue,urlcolor=urlblue,linktocpage,linktoc=all%
pdfauthor={\myauthor},pdftitle={\mytitle}}

\definecolor{headercolor}{gray}{0.65} 
\usepackage{fancyhdr}
\fancypagestyle{plain}{%
\fancyhf{} 
\fancyfoot[C]{\fontfamily{ppl}\selectfont \small\bfseries \thepage} 
 
}
\usepackage{emptypage}

\definecolor{halfgray}{gray}{0.55} 
\newfont{\chapNumFont}{eurb10 scaled 7000}
\newfont{\chapTitFont}{pplr9d}
\usepackage[explicit]{titlesec}
\titleformat{\section}[hang]{\bfseries\Large}{\fontfamily{ppl}\selectfont \thesection}{15pt}{\fontfamily{ppl}\selectfont #1}
\titleformat{\subsection}[hang]{\bfseries\large}{\fontfamily{ppl}\selectfont \thesubsection}{15pt}{\fontfamily{ppl}\selectfont #1}
\titleformat{\subsubsection}[hang]{\bfseries}{\fontfamily{ppl}\selectfont \thesubsubsection}{15pt}{\fontfamily{ppl}\selectfont #1}
\titleformat{\chapter}[block]%
{\Huge}{\raggedleft{\color{halfgray}\chapNumFont\thechapter}}{20pt}%
{\raggedright{\fontfamily{ppl}\selectfont #1}}

\usepackage{titletoc}
\titlecontents{chapter}[1.4em]{\vspace{8pt}}{\fontfamily{ppl}\selectfont\bfseries \contentslabel{1.4em}}%
{\fontfamily{ppl}\selectfont\bfseries \hspace*{-1.4em}}{\titlerule*[1pc]{.}\fontfamily{ppl}\selectfont\bfseries\contentspage}
\titlecontents{section}[4em]{}{\fontfamily{ppl}\selectfont\small \contentslabel{2.2em}}{}{\titlerule*[1pc]{.}\fontfamily{ppl}\selectfont\small\contentspage}
\titlecontents{subsection}[7em]{}{\fontfamily{ppl}\selectfont \contentslabel{2.6em}}{}{\hfill\contentspage}

\usepackage[notoccite,undotted]{minitoc}
\let\minitocORIG\minitoc
\renewcommand{\minitoc}{\minitocORIG \vspace{1.5em}}

\numberwithin{equation}{section}

\begin{document}
\unitlength1cm


\def\zo{\overline{z}_1}
\def\zt{\overline{z}_2}
\def\C{\overline{C}}
\def\D{{\cal D}}
\def\DD{\overline{\cal D}}
\def\g{\overline{\cal G}}
\def\gm{\gamma}
\def\M{{\cal M}}
\def\ep{\epsilon}
\def\epm1{\frac{1}{\epsilon}}
\def\epm2{\frac{1}{\epsilon^{2}}}
\def\epm3{\frac{1}{\epsilon^{3}}}
\def\epm4{\frac{1}{\epsilon^{4}}}
\def\unM{\hat{\cal M}}
\def\ashat{\hat{a}_{s}}
\def\asmur{a_{s}^{2}(\mu_{R}^{2})}
\def\sigbar{{{\overline {\sigma}}}\left(a_{s}(\mu_{R}^{2}), L\left(\mu_{R}^{2}, m_{H}^{2}\right)\right)}
\def\sigbarn{{{{\overline \sigma}}_{n}\left(a_{s}(\mu_{R}^{2}) L\left(\mu_{R}^{2}, m_{H}^{2}\right)\right)}}
\def\unas{ \left( \frac{\hat{a}_s}{\mu_0^{\epsilon}} S_{\epsilon} \right) }
\def\rnM{{\cal M}}
\def\bt{\beta}
\def\cD{{\cal D}}
\def\cC{{\cal C}}
\def\ca{\text{\tiny C}_\text{\tiny A}}
\def\cf{\text{\tiny C}_\text{\tiny F}}
\def\ct{{\red []}}
\def\sv{\text{SV}}
\def\murOmu{\left( \frac{\mu_{R}^{2}}{\mu^{2}} \right)}
\def\bb{b{\bar{b}}}
\def\bt0{\beta_{0}}
\def\bt1{\beta_{1}}
\def\bt2{\beta_{2}}
\def\bt3{\beta_{3}}
\def\gm0{\gamma_{0}}
\def\gm1{\gamma_{1}}
\def\gm2{\gamma_{2}}
\def\gm3{\gamma_{3}}
\def\nn{\nonumber}
\def\l{\left}
\def\r{\right}
\def\CA{\mathbf{C_A}}
\def\F{{\cal F}}
\newcommand{\dis}{}
\newcommand{\overbar}[1]{mkern-1.5mu\overline{\mkern-1.5mu#1\mkern-1.5mu}\mkern
1.5mu}
\newcommand{\iu}{{i\mkern1mu}}
\newcommand{\ConferenceEntry}[7]{
		\noindent #1 \href{#3}{\textbf{\textit{#2}}}, #4,
                \href{#6}{#5} #7 }
\newcommand{\TalkEntry}[4]{
		\noindent #1 \textbf{\textit{#2}}, #3 #4}

\thispagestyle{empty}

\newlength{\centeroffset}
\setlength{\centeroffset}{0cm}
\setlength{\centeroffset}{-0.5\oddsidemargin}
\addtolength{\centeroffset}{0.5\evensidemargin}

\noindent\vspace*{-10ex}\hspace*{\centeroffset}\makebox[\textwidth]{%
\begin{minipage}{\textwidth}
\begin{center}
\noindent\textsc{\fontfamily{ppl}\selectfont Ph.D. thesis}\\[1ex]
\noindent\textsc{\footnotesize\fontfamily{ppl}\selectfont in Theoretical Physics}\\
\end{center}
\end{minipage}}

\noindent\hspace*{\centeroffset}\makebox[\textwidth]{%
\begin{minipage}{\textwidth}
\includegraphics[height=0.15\textwidth]{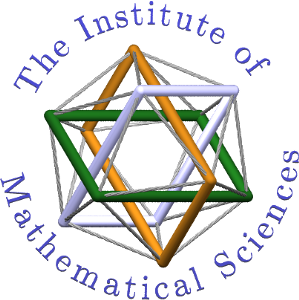}
\hspace{\stretch{1}}
\includegraphics[height=0.15\textwidth]{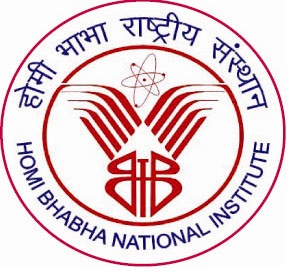}
\end{minipage}}

\vskip 5cm

\noindent\hspace*{\centeroffset}\makebox[0pt][l]{%
\begin{minipage}{\textwidth}
\begin{center}
{\setstretch{3}
\noindent{\fontfamily{ppl}\fontsize{24.88pt}{0pt}\selectfont QCD Radiative Corrections to Higgs Physics}\\[2cm]
}
\noindent{\fontfamily{ppl}\selectfont by}\\[3ex]
\noindent{\Large\bfseries\fontfamily{ppl}\selectfont Taushif Ahmed}
\end{center}
\vspace*{0.5cm}
\end{minipage}}

\vspace{3cm}

\noindent\hspace*{\centeroffset}\makebox[0pt][l]{%
\begin{minipage}{\textwidth}
\begin{flushright}
\begin{tabular}{rr}
{\fontfamily{ppl}\selectfont Supervisor:}
&{\bfseries\fontfamily{ppl}\selectfont V. Ravindran}\\[2ex]
{\fontfamily{ppl}\selectfont Examiners:}
&{\bfseries\fontfamily{ppl}\selectfont Asmita Mukherjee}\\
&{\bfseries\fontfamily{ppl}\selectfont Matthias Steinhauser}\\[2ex]
{\fontfamily{ppl}\selectfont Exam date:}
&{\bfseries\fontfamily{ppl}\selectfont February 21, 2017}\\
\end{tabular}
\end{flushright}
\end{minipage}}

\vspace{\stretch{1}}

\noindent\hspace*{\centeroffset}\makebox[0pt][l]{%
\begin{minipage}{\textwidth}
\begin{center}
\textsc{\fontfamily{ppl}\selectfont \small Homi Bhaba National Institute\\The Institute of Mathematical Sciences\\ IV Cross Road, CIT Campus, Taramani, Chennai 600113, India} \\ {\href{mailto:taushif.ahmed@kit.edu}{taushif.ahmed@kit.edu}}

\end{center}
\begin{center}
{\color{white}\tiny Last modified: \today}
\end{center}
\end{minipage}}

\newpage\null\thispagestyle{empty}\newpage

\newpage
\thispagestyle{empty}
\vspace*{\fill}
\centerline{{\bf {\large Dedicated to}}}
\vskip 0.5cm

\centerline{{\bf {\large My Parents and Sister}}}
\vspace*{\fill}

\newpage\null\thispagestyle{empty}\newpage

 \newpage
\thispagestyle{empty}

\centerline{{\bf{\large ABSTRACT}}}
\vskip 0.3cm

The central part of this thesis deals with the quantum chromodynamics (QCD) radiative corrections to some important observables associated with the Drell-Yan, scalar and pseudo-scalar Higgs boson productions at next-to-next-to-next-to-leading order (N$^{3}$LO) aiming to uplift the accuracy of theoretical results. The Higgs bosons are produced dominantly at the Large Hadron Collider (LHC) via gluon fusion through top quark loop, while one of the subdominant ones takes place through bottom quark annihilation whose contribution is equally important and must be included in precision studies. Here, we have computed analytically the inclusive cross section of the Higgs boson produced through this channel under the soft-virtual (SV) approximation at N$^{3}$LO QCD following an elegant formalism. Moreover, the differential rapidity distribution is another most important observable, which is expected to be measured in upcoming days at the LHC. This immediately calls for very precise theoretical predictions. The analytical expressions of the SV corrections to this observable at N$^{3}$LO for the Higgs boson, produced through gluon fusion, and leptonic pair in Drell-Yan (DY) production are computed and the numerical impacts of these results are demonstrated. In addition, the CP-odd/pseudo-scalar Higgs boson, which is one of the most prime candidates in BSM scenarios, is studied in great details, taking into account the QCD radiative corrections. We have computed the analytical results of the three loop QCD corrections to the pseudo-scalar Higgs boson production and consequently, obtained the inclusive production cross section at N$^{3}$LO under SV approximation. These indeed help to reduce the theoretical uncertainties arising from the renormalisation and factorisation scales and undoubtedly, improve the reliabilities of the theoretical results.

\newpage\null\thispagestyle{empty}\newpage

\newpage
\thispagestyle{empty}
\hrulefill
\centerline{{\bf{\large LIST OF PUBLICATIONS (Included in this thesis)}}}

\begin{enumerate}

\item \textbf{Higgs boson production through ${b{\bar b}}$
    annihilation at threshold in N$^3$LO QCD}\\ Taushif Ahmed, Narayan
  Rana and V. Ravindran\\
  \href{http://dx.doi.org/10.1007/JHEP10(2014)139}{\textit{JHEP 1410,
      139 (2014)}}

\item \textbf{Rapidity Distributions in Drell-Yan and Higgs
    Productions at Threshold to Third Order in QCD} \\Taushif Ahmed,
  M. K. Mandal, Narayan Rana and V. Ravindran\\
  \href{http://dx.doi.org/10.1103/PhysRevLett.113.212003}{\textit{Phys.Rev.Lett. 113,
      212003 (2014)}}

\item \textbf{Pseudo-scalar Form Factors at Three Loops in QCD}\\
Taushif Ahmed, Thomas Gehrmann, Prakash Mathews, Narayan Rana and
V. Ravindran\\
\href{http://dx.doi.org/10.1007/JHEP11(2015)169}{\textit{JHEP 1511,
    169 (2015)}}

\item \textbf{Pseudo-scalar Higgs Boson Production at Threshold N$^3$LO
    and N$^3$LL QCD} \\ Taushif Ahmed, M. C. Kumar, Prakash Mathews,
  Narayan Rana and V. Ravindran \\
  \href{http://link.springer.com/article/10.1140/epjc/s10052-016-4199-1?wt_mc=Internal.Event.1.SEM.ArticleAuthorIncrementalIssue}{\textit{Eur. Phys. J. C (2016) 76:355 }}  
  \end{enumerate}

\vspace{8mm}

\centerline{{\bf{\large LIST OF PUBLICATIONS (Not included in this thesis)}}}

\begin{enumerate}

\item \textbf{Two-loop QCD corrections to Higgs $\rightarrow b +
    \bar{b} + g$ amplitude} \\ Taushif Ahmed, Maguni Mahakhud, Prakash
  Mathews, Narayan Rana and V. Ravindran \\
\href{http://dx.doi.org/10.1007/JHEP08(2014)075}{\textit{JHEP 1408,
    075 (2014)}}

\item \textbf{Drell-Yan Production at Threshold to Third Order in QCD}
  \\ Taushif Ahmed, Maguni Mahakhud, Narayan Rana and V. Ravindran \\
\href{http://dx.doi.org/10.1103/PhysRevLett.113.112002}{\textit{Phys.Rev.Lett. 113,
    112002 (2014)}} 

\item \textbf{Two-Loop QCD Correction to massive spin-2 resonance
    $\rightarrow$ 3-gluons} \\ Taushif Ahmed, Maguni Mahakhud, Prakash
  Mathews, Narayan Rana and V. Ravindran \\
  \href{http://dx.doi.org/10.1007/JHEP05(2014)107}{\textit{JHEP 1405,
      107 (2014)}}

\item \textbf{Spin-2 form factors at three loop in QCD} 
\\ Taushif Ahmed, Goutam Das, Prakash Mathews, Narayan
  Rana and V. Ravindran \\
  \href{http://dx.doi.org/10.1007/JHEP12(2015)084}{\textit{JHEP 1512,
      084 (2015)}}

\item \textbf{Higgs Rapidity Distribution in $b \bar{b}$ Annihilation
    at Threshold in N$^{3}$LO QCD}\\ Taushif Ahmed, M. K. Mandal,
    Narayan Rana and V. Ravindran \\
    \href{http://dx.doi.org/10.1007/JHEP02(2015)131}{\textit{JHEP
        1502, 131 (2015)}}

\item \textbf{Pseudo-scalar Higgs boson production at N$^3$LO$_{A}$
    +N$^3$LL'} \\ Taushif Ahmed, Marco Bonvini, M.C. Kumar, Prakash Mathews, Narayan
Rana, V. Ravindran, Luca Rottoli  \\
  \href{http://dx.doi.org/10.1140/epjc/s10052-016-4510-1}{\textit{Eur.Phys.J. C76 (2016) no.12, 663}} 

\item \textbf{The two-loop QCD correction to massive spin-2 resonance $\rightarrow q {\bar q} g$ } \\ Taushif Ahmed, Goutam Das, Prakash Mathews, Narayan Rana and V. Ravindran \\
  \href{http://dx.doi.org/10.1140/epjc/s10052-016-4478-x}{\textit{Eur.Phys.J. C76 (2016) no.12, 667}} 

\item \textbf{NNLO QCD Corrections to the Drell-Yan Cross Section in
    Models of TeV-Scale Gravity} \\ Taushif Ahmed, Pulak Banerjee, Prasanna K. Dhani, M.C. Kumar,
Prakash Mathews, Narayan Rana and V. Ravindran \\
  \href{http://dx.doi.org/10.1140/epjc/s10052-016-4587-6}{\textit{Eur.Phys.J. C77 (2017) no.1, 22}} 

\item \textbf{Three loop form factors of a massive spin-2 with non-universal coupling} \\ Taushif Ahmed, Pulak Banerjee, Prasanna K. Dhani, 
Prakash Mathews, Narayan Rana and V. Ravindran \\
  \href{http://dx.doi.org/10.1103/PhysRevD.95.034035}{\textit{Phys.Rev. D95 (2017) no.3, 034035 }}

\thispagestyle{empty}

\end{enumerate}

\vspace{8mm}

\centerline{{\bf{\large LIST OF PREPRINTS (Not included in this thesis)}}}

\begin{enumerate}

\item \textbf{RG improved Higgs boson production to N$^{3}$LO in QCD}
  \\ Taushif Ahmed, Goutam Das, M. C. Kumar, Narayan Rana and
  V. Ravindran \\
  \href{https://inspirehep.net/record/1373312}{\textit{arXiv:1505.07422
      [hep-ph]}} 

\item \textbf{Konishi Form Factor at Three Loop in ${\cal N}=4$ SYM } \\ Taushif Ahmed, Pulak Banerjee, Prasanna K. Dhani, Narayan Rana, V. Ravindran and Satyajit Seth \\
  \href{https://inspirehep.net/record/1492546}{\textit{arXiv:1610.05317 [hep-th]}} (Under review in PRL)

\end{enumerate}

\vspace{8mm}

\centerline{{\bf{\large LIST OF CONFERENCE PROCEEDINGS}}}

\begin{enumerate}

\item \textbf{Pseudo-scalar Higgs boson form factors at 3 loops in QCD }
  \\ Taushif Ahmed, Thomas Gehrmann, Prakash Mathews, Narayan Rana and
  V. Ravindran \\
  \href{http://inspirehep.net/record/1491483/files/PoS(LL2016)026.pdf}{\textit{PoS LL2016 (2016) 026 }} 

\end{enumerate}

\vspace{8mm}

\centerline{{\bf{\large CONFERENCE ATTENDED}}}

\begin{enumerate}

\item \ConferenceEntry{16 - 19 March 2016 :}{MHV@30: Amplitudes and
    Modern Applications}{}{Fermi National Accelerator
    Laboratory}{}{}{Chicago, USA.} 

\item\ConferenceEntry{23 - 27 February 2016 :}{Multiloop and Multiloop
    Processes for Precision Physics at the LHC}{}{Saha Institute of
    Nuclear Physics}{}{}{Kolkata, India.} 

\item\ConferenceEntry{11 - 12 October 2015 :}{Recent Trends in
    AstroParticle and Particle Physics}{}{Indian Institute of
    Science}{}{}{Bangalore, India.} 

\item\ConferenceEntry{15 - 26 June 2015 :}{ICTP Summer School on
    Particle Physics}{}{International Centre for Theoretical
    Physics}{}{}{Trieste, Italy.} 

\item\ConferenceEntry{23 - 28 February 2015 :}{Workshop on LHC and
    Dark Matter}{}{Indian Association for the Cultivation of
    Science}{}{}{Kolkata, India.} 

\item\ConferenceEntry{2 - 5 December 2014}{International Workshop on
    Frontiers of QCD}{}{Indian Institute of Technology
    Bombay}{}{}{Mumbai, India.} 

\item\ConferenceEntry{5 - 10 March 2014}{Discussion Meeting on
    Radiative Correction}{}{Institute of Physics}{}{}{Bhubaneswar,
    India.} 

\item\ConferenceEntry{7 - 17 July 2013}{CTEQ School on QCD and
    Electroweak Phenomenology}{}{University of
    Pittsburgh}{}{}{Pennsylvania, USA.} 

\item\ConferenceEntry{7 - 10 January 2013}{Lecture Workshop in High
    Energy Physics}{}{Indian Institute of Technology
    Bombay}{}{}{Mumbai, India.} 

\item\ConferenceEntry{10 - 13 December 2012}{Frontiers of High Energy
    Physics IMSc Golden Jubilee Symposium}{}{The Institute of
    Mathematical Sciences}{}{}{Chennai, India.} 

\thispagestyle{empty}

\end{enumerate}

\vspace{8mm}

\centerline{{\bf{\large SEMINARS PRESENTED}}}

\begin{enumerate}

\item\TalkEntry{March 2016}{Pseudo-Scalar Form Factors at 3-Loop in
    QCD}{University of Buffalo, USA.}{} 

\item\TalkEntry{July 2015}{Threshold Corrections to DY and Higgs at
    N$^{3}$LO QCD}{Wuppertal University, Germany.}{}

\item\TalkEntry{July 2015}{Threshold Corrections to DY and Higgs at
    N$^{3}$LO QCD}{DESY Hamburg, Germany.}{}

\item\TalkEntry{July 2015}{Threshold Corrections to DY and Higgs at
    N$^{3}$LO QCD}{INFN Milan, Italy.}{}

\item\TalkEntry{July 2015}{Threshold Corrections to DY and Higgs at
    N$^{3}$LO QCD}{INFN Sezione Di Torino, Italy.}{}

\item\TalkEntry{July 2015}{Threshold Corrections to DY and Higgs at
    N$^{3}$LO QCD}{Johannes Gutenberg Universitat Mainz, Germany.}{} 

\item\TalkEntry{July 2014}{Two-loop QCD Correction to Massive Spin-2
    Resonance $\rightarrow$ 3-gluons}{Harish-Chandra Research
    Institute, India.}{}

\end{enumerate}

\begingroup
\hypersetup{linkcolor=blue}
\dominitoc
\tableofcontents
\adjustmtc
\endgroup

\chapter*{SYNOPSIS}

%

\addcontentsline{toc}{chapter}{Synopsis}

The Standard Model (SM) of particle physics is one of the most
remarkably successful fundamental theories of all time which got its
finishing touch on the eve of July 2012 through the discovery of the
long-awaited particle, ``the Higgs boson'', at the biggest underground
particle research amphitheater, the Large Hadron Collider (LHC). It
would take a while to make the conclusive remarks about the true
identity of the newly-discovered particle. However,  
after the discovery of this SM-like-Higgs boson, the high
energy physics community is standing on the verge of a very crucial era
where the new physics may show up as tiny deviations from the
predictions of the SM. To exploit this possibility, it is a crying need to make the
theoretical predictions, along with the revolutionary experimental
progress, to a spectacularly high accuracy within the SM and beyond
(BSM). 

The most successful and celebrated methodology to perform the theoretical
calculations within the SM and BSM are based on the perturbation
theory, due to our inability to solve the theory exactly. Under the
prescriptions of perturbation
theory, all the observables are expanded in powers of the coupling
constants present in the underlying Lagrangian. The result obtained
from the first term of
perturbative series is called the leading order (LO), the next one
is called next-to-leading order (NLO) and so on. In most of the cases,
the LO results fail miserably to deliver a reliable theoretical prediction of the
associated observables, one must go beyond the wall of
LO result to achieve a higher accuracy.

Due to the presence of three fundamental forces within the SM, any
observable can be expanded in powers of the coupling constants
associated with the corresponding forces, namely, electromagnetic
($\alpha_{\rm EM}$), weak ($\alpha_{\rm EW}$) and strong ($\alpha_s$)
ones and consequently, perturbative calculations can be
performed with respect to each of these constants. However, at typical
energy scales, at which the hadron colliders undergo operations, the
contributions arising from the $\alpha_s$ expansion dominate over the
others due to comparatively large values of $\alpha_s$. Hence, to
catch the dominant contributions to any observables, 
we must concentrate on
the $\alpha_s$ expansion and evaluate the terms beyond LO. These are
called Quantum Chromo-dynamics (QCD) radiative or perturbative QCD (pQCD) corrections. 
%
In addition, the pQCD predictions depend on two
unphysical scales, the renormalisation ($\mu_R$) and factorisation ($\mu_F$) scales, which
are required to introduced in the process of renormalising the
theory. The $\mu_R$ arises from the ultraviolet (UV) renormalisation,
whereas the mass factorisation (removes collinear
singularities) introduces the $\mu_F$. Any fixed order results do
depend on these unphysical scales which happens due to the truncation of
the perturbative expansion at any finite order. As we include the
contributions from higher and higher orders, the dependence of any
physical observable on these unphysical scales gradually goes down. 
Hence, to make a reliable theoretical prediction, it is absolutely
necessary to take into account the contributions arising from the
higher order QCD corrections to any observable at the hadron
colliders. 

\textit{This thesis arises exactly in this context. The central
part of this thesis deals with the QCD radiative corrections to some
important observables associated with the Drell-Yan, scalar and pseudo
scalar Higgs boson production at three loop or N$^3$LO order}. In the
subsequent discussions, we will concentrate only on these three
processes.

\section{Soft-Virtual QCD Corrections to Cross Section at N$^3$LO}
\label{sec:Synop-SV-CS}

The Higgs bosons are produced dominantly at the LHC 
via gluon fusion through top quark loop, while  
one of the sub-dominant ones take place through bottom quark
annihilation. In the SM, the interaction between the Higgs boson and
bottom quarks is controlled through the Yukawa coupling which is
reasonably small at typical energy scales. However, in
the minimal super symmetric SM (MSSM), this 
channel can contribute substantially due to enhanced coupling between the
Higgs boson and bottom quarks in the large $\tan\beta$ region, where
 $\tan\beta$ is the ratio of vacuum expectation values of the up
and down type Higgs fields. In the present run of LHC, the measurements
of the various coupling constants including this one are underway which can
shed light on the properties of the newly discovered Higgs boson.
Most importantly, for the precision studies we
must take into account all the contributions, does not matter how tiny
those are, arising from sub-dominant
channels along with the dominant ones to reduce the dependence on the
unphysical scales and make a reliable prediction.

The computations of the higher order QCD corrections beyond leading
order often becomes quite challenging because of the large number of Feynman
diagrams and, presence of the complicated loop and phase space
integrals. Under this circumstance, when we fail to compute the
complete result at certain order, it is quite natural to try an
alternative approach to capture 
the dominant contributions from the missing higher order corrections.
It has been observed for many processes that the dominant contributions to an observable
often comes from the soft gluon emission diagrams. The contributions
arising from the
associated soft gluon emission along with the virtual Feynman 
diagrams are known as
the soft-virtual (SV) corrections. \textit{The goal of this section is to
discuss the SV
QCD corrections to the production cross section of the Higgs boson,
produced through bottom quark annihilation.}

The NNLO QCD corrections to this channel are already present in the
literature. In addition, the partial result for the N$^3$LO
corrections under the SV approximation were also computed long
back. \textit{In this work, we have computed the missing part and 
completed the full SV corrections to the cross section at N$^3$LO}. 

The infrared safe contributions from the soft gluons are obtained by
adding the soft part of the cross section with the UV
renormalized virtual part and performing mass factorisation using
appropriate counter terms. The main ingredients are the
form factors, overall operator UV renormalization constant, 
soft-collinear distribution arising from the real radiations in the partonic 
subprocesses and mass factorization kernels. The computations of
SV cross section at N$^3$LO QCD require all of these above quantities
up to 3-loop order. The relevant form factor becomes available very
recently. The soft-collinear distribution at N$^3$LO was computed by
us around the same time. This was calculated from the recent result of
N$^3$LO SV cross section of the Higgs boson productions in gluon
fusion by employing a symmetry (maximally 
non-Abelian property). Prior to this, this symmetry was verified
explicitly up to NNLO order. However, neither there was any clear
reason to believe that the symmetry would fail
nor there was any transparent indication of holding it beyond this
order. Nevertheless, we postulate that the relation would hold true
even at N$^3$LO order! This is inspired by the universal properties of
the soft gluons which are the  
underlying reasons behind the existence of this remarkable
symmetry. Later, this conjecture is verified by explicit computations
performed by two different groups on Drell-Yan process. This symmetry plays the most important role in
achieving our goal. With these, along with the existing results of the
remaining required ingredients, we obtain the complete analytical
expressions of N$^3$LO SV
cross section of the Higgs boson production through bottom quark
annihilation. It reduces the scale dependence and
provides a more precise result. We demonstrate the impact of this result numerically at
the Large Hadron Collider (LHC) briefly. \textit{This is the most accurate
result for this channel which exists in the literature till date and
it is expected to play an important role in coming days at the LHC.}

\section{Soft-Virtual QCD Corrections to Rapidity at N$^3$LO}

The productions of the Higgs boson in gluon fusion and leptonic pair
in DY are among the most important processes at the LHC which are studied not
only to test the SM to an unprecedented accuracy but also to explore
the new physics under BSM. During the present run at the LHC, in addition to the inclusive production cross section, the differential rapidity distribution is among the most important observables,
which is expected to be measured in upcoming days. This immediately
calls for very precise theoretical predictions.

In the same spirit of the SV corrections to
the inclusive production cross section, the dominant contributions to
the differential rapidity distributions often arise from the soft
gluon emission diagrams. Hence, in the absence of complete fixed
order result, the rapidity distribution under SV approximation is the
best available alternative in order to capture the dominant contributions from
the missing higher orders and stabilise the dependence on unphysical
scales. For the Higgs boson production through gluon fusion, we work in the effective
theory where the top quark is integrated out.
 \textit{This section is devoted to demonstrate the SV corrections
to this observable at N$^3$LO for the Higgs boson, produced through
gluon fusion, and leptonic pair in Drell-Yan (DY) production.}

For the processes under considerations, the NNLO QCD
corrections are present, computed long back, and in addition, the partial N$^3$LO SV
results are also available. However, due to reasonably large scale
uncertainties and crying demand of uplifting the accuracy of theoretical
predictions, we must push the boundaries of existing results. \textit{In this work,
  we have computed the missing part and completed the SV corrections
  to the rapidity distributions at N$^3$LO QCD}.

The prescription which has been employed to calculate the SV QCD
corrections is similar to that of the inclusive cross section, more
specifically, it is a generalisation of the other one. The infrared
safe contributions under SV approximation
can be computed by adding the soft part of the rapidity distribution
with the UV renormalised virtual part and performing the mass
factorisation using appropriate counter terms. Similar to the
inclusive case, the main ingredients to perform this computation are
the form factors, overall UV operator renormalisation constant,
soft-collinear distribution for rapidity and mass factorisation
kernels. These quantities are required up to N$^3$LO to calculate the
rapidity at this order. The three loop quark and gluon form factors were
calculated long back. The operator renormalisation constants are also
present. For DY, this constant is not required or equivalently equals
to unity. The mass factorisation kernels are also available in the
literature to the required order. The only missing part was the
soft-collinear distribution for rapidity at N$^3$LO. This was
not possible to compute until very recently. Because of the universal
behaviour of the soft gluons, the soft-collinear distributions for
rapidity and inclusive cross section can be related to all orders in
perturbation theory. Employing this beautiful relation, we obtain this
quantity at N$^3$LO
from the results of soft-collinear distribution of the inclusive cross
section. Using this, along with the existing results of the other
relevant quantities, we compute the complete analytical expressions of
N$^3$LO SV correction to the rapidity distributions for the Higgs
boson in gluon fusion and leptonic pair in DY. We demonstrate the
numerical impact of this correction for the case of Higgs boson at the
LHC. This indeed reduces the scale dependence significantly and
provides a more reliable theoretical predictions. \textit{These are the
  most accurate results for the rapidity distributions of the Higgs
  boson and DY pair which exist in the literature and undoubtedly, 
  expected to play very important role in the upcoming run at the LHC.}

\section{Pseudo-Scalar Form Factors at Three Loops in QCD}

One of the most popular extensions of the SM, namely, the MSSM and two
Higgs doublet model have richer Higgs sector containing more than one
Higgs boson and there have been intense search strategies to observe
them at the LHC. In particular, the production of CP-odd Higgs
boson/pseudo-scalar at the LHC has been studied in detail, taking into
account higher order QCD radiative corrections, due to similarities
with its CP-even counter part. Very recently, the N$^{3}$LO QCD corrections
to the inclusive production cross section of the CP-even Higgs boson is
computed. So, it is very natural to extend the theoretical accuracy for the CP-odd Higgs boson
to the same order of N$^{3}$LO. This requires the 3-loop quark and gluon form factors for
the pseudo-scalar which are the only missing ingredients to achieve this goal.

Multiloop and multileg computations play a crucial role to achieve the golden
task of making precise theoretical predictions. However, the
complexity of these computations grows very rapidly 
with the increase of number of loops and/or external
particles. Nevertheless, it has become a reality due to several
remarkable developments in due course of time. \textit{This section is
devoted to demonstrate the computations of the 3-loop quark and gluon
form factors for the pseudo-scalar operators in QCD}.

The coupling of a pseudo-scalar Higgs boson to gluons is mediated through a heavy quark loop. In the limit of large quark mass, it is described by an effective Lagrangian that only admits light degrees of freedom. 
In this effective theory, we compute the 3-loop massless QCD corrections to the form factor that describes the coupling of a pseudo-scalar Higgs boson to gluons.
The evaluation of this 3-loop
form factors is truly a non-trivial task not only because of the
involvement of a large number of Feynman diagrams but also due to
the presence of the axial vector coupling. We work in dimensional regularisation and use the 't Hooft-Veltman  prescription for the axial vector current,
The state-of-the-art techniques including
integration-by-parts (IBP) and Lorentz invariant (LI) identities
have been employed to accomplish this task. The UV
renormalisation is quite involved since the two operators, present in
the Lagrangian, mix under UV renormalization due to the axial anomaly and
additionally, a finite renormalisation constant needs to be introduced
in order to fulfill the chiral Ward identities. 
Using the universal infrared (IR) factorization properties, we independently derive the three-loop operator mixing and finite operator renormalisation from the renormalisation group equation for the form factors, thereby confirming recent results, which were computed following a completely different methodology, in the operator product expansion.  
This form factor is an important ingredient to the precise
prediction of the pseudo-scalar Higgs boson production cross section
at hadron colliders. We derive the hard matching coefficient in soft-collinear effective theory (SCET). We 
also study the form factors in the context of leading transcendentality principle and we find that the 
diagonal form factors become identical to those of ${\cal N}=4$ upon imposing some identification
on the quadratic Casimirs.  Later, these form factors are used to
calculate the SV corrections to the pseudo-scalar production cross
section at N$^3$LO and N$^3$LL QCD.

\graphicspath{ {figures/} }
\phantomsection
\addcontentsline{toc}{chapter}{\listfigurename}
\listoffigures
\adjustmtc

\graphicspath{ {tables/} }
\phantomsection
\addcontentsline{toc}{chapter}{\listtablename}
\listoftables
\adjustmtc

\chapter{\label{chap:Intro}Introduction}

 \begingroup
 \hypersetup{linkcolor=blue}
 \minitoc
 \endgroup



The Standard Model (SM) of Particle Physics is one of most remarkably successful
fundamental theories which encapsulates the governing principles of
elementary constituents of matter and their interactions. Its
development throughout the latter half of the 20th century resulting
from an unprecedented collaborative effort of the brightest minds
around the world is undoubtedly one of the greatest achievements in
human history. Over the duration of many decades around 1970s, the
theoretical predictions of the SM were verified one after
another with a spectacular accuracy and it got the ultimate credence
through the announcement, 
made on a fine morning of 4th July 2012 at CERN in Geneva: 
\begin{quote}
  \textit{``If we combine $ZZ$ and $\gamma\gamma$, this is what we get: they line
up extremely well in a region of 125 GeV with the combine significance
of 5 standard deviation!''}
\end{quote}

The SM relies on the mathematical framework of quantum field theory
(QFT), in which a Lagrangian controls the dynamics
and kinematics of the theory. Each kind of particle is described in
terms of a dynamical field that pervades space-time. The construction
of the SM proceeds through the modern methodology of constructing a QFT,
it happens through postulating a set of underlying symmetries of
the system and writing down the most general renormalisable Lagrangian
from its field content. 

The underlying symmetries of the QFT can be largely categorized into
global and local ones. The global Poincar\'e symmetry is postulated for
all the relativistic QFT. It consists of the familiar
translational symmetry, rotational symmetry and the inertial reference
frame invariance central to the special theory of relativity. Being
global, its operations must be simultaneously applied to all points of space-time
On the
other hand, the local gauge symmetry is an internal symmetry that
plays the most crucial role in determining the predictions of the
underlying QFT. These are the symmetries that act independently at each point in space-time.
The SM relies on the local
SU(3)$\times$SU(2)$_{\rm L}\times$U(1)$_{\rm Y}$ gauge symmetry. Each
gauge symmetry manifestly 
gives rise to a fundamental interaction: the
electromagnetic interactions are characterized by an U(1), the
weak interactions by an SU(2) and the strong interactions by an SU(3)
symmetry. 

In its current formulation of the SM, it includes three different families
of elementary particles. The first ones are called fermions arising
from the quantisation of the fermionic fields. These constitute the
matter content of the theory. The quanta of the bosonic fields, which
form the second family, are the force carriers i.e. the mediators of
the strong, weak, and electromagnetic fundamental interactions. In
addition to the these, there is a third boson, the Higgs boson
resulting from the quantum excitation of the Higgs field. This is the only
known scalar particle that was postulated long ago and
observed very recently at the Large Hadron Collider
(LHC)~\cite{Aad:2012tfa, Chatrchyan:2012ufa}. The presence of this
field, now believed to be confirmed, explains the
mechanisms of acquiring mass of some of the fundamental
particles when, based on the underlying gauge symmetries controlling
their interactions, they should be massless. This mechanism, which is
believed to be one of the most revolutionary ideas of the last
century, is known as Brout-Englert-Higgs-Kibble (BEHK) mechanism. 

Two of the four known fundamental forces, electromagnetism and weak forces
which appear very different at low energies, 
are actually unified to so called electro-weak force in high
energy. The structure of this unified picture is accomplished under
the gauge group SU(2)$_{\rm L} \times$ U(1)$_{\rm Y}$. The
corresponding gauge bosons are the three $W$ bosons of weak isospin from SU(2) and the
$B$ boson of weak hypercharge from U(1), all of which are
massless. Upon spontaneous symmetry breaking from SU(2)$_{\rm L}
\times$ U(1)$_{\rm Y}$ to U(1)$_{\rm EM}$, caused by the BEHK
mechanism, the three mediators of the electro-weak force, the
$W^{\pm},Z$ bosons acquire mass, leaving the mediator of the
electromagnetic force, the photon, as massless. Finally, the theory of
strong interactions, Quantum Chromo-Dynamics (QCD) is governed by the
unbroken SU(3) gauge group, whose force carriers, the gluons remain
massless. 

Although the SM is believed to be theoretically
self-consistent with a spectacular accuracy and has demonstrated huge and continued successes
in providing experimental predictions, it indeed does leave some phenomena
unexplained and it falls short of being a complete theory of
fundamental interactions. It fails to incorporate the full theory of
gravitation as described by general relativity, or account for the
accelerating expansion of the universe (as possibly described by dark
energy). The model does not contain any viable dark matter particle
that possesses all of the required properties deduced from
observational cosmology. It also does not incorporate neutrino
oscillations (and their non-zero masses).

Currently, the high energy physics community is standing on the verge of a
crucial era where the new physics may show up as tiny deviations from
the prediction of the SM! To exploit this possibility it is absolutely
necessary to make the theoretical predictions, along with the
revolutionary experimental progress, to a very high accuracy within the SM and beyond. 
The relevance of this thesis arises exactly in this context.

The most crucial quantity in the process of accomplishing the job of
making any prediction based on QFT is undoubtedly the scattering
amplitude. This is the fundamental building block of any observable in
QFT. In the upcoming section, we will elaborate on the idea of
scattering amplitude which will be followed by a brief description of
QCD. We will close the chapter of introduction by introducing the
concept of computing the observables under certain approximation,
known as soft-virtual approximation.

\section{Scattering Amplitudes}
\label{sec:Intro-ScattAmp}

The fundamental quantity of any QFT which encodes all the underlying
symetries of the theory is called the action. This is constructed out
of Lagrangian density, which is a functional of the fields present in
the theory, and integrating over all space time points:
\begin{align}
\label{eq:Intro-action}
S &= \int d^4x \,{\cal L}\l[\phi_i(x)\r]\,.
\end{align} 

By construction the QFT is a probabilistic theory and all the
observables calculated based on this theory always carry a
probabilistic interpretation. For example, an important observable is
the total cross section which measures the total probability of any
event to happen in colliders. The computation of the cross section, and in fact, almost
all the observables in QFT requires the evaluation of scattering matrix ($S$-matrix)
elements which describe the evolution of the system from asymptotic initial to final states 
due to presence of the interaction. The $S$-matrix elements are defined as
\begin{align}
\langle f|S| i \rangle = \delta_{fi} + i (2\pi)^{4} \delta^{(4)}\l(p_{f}-p_{i}\r) {\cal M}_{i \rightarrow f}
\end{align}
where, the $\delta_{fi}$ represents the unscattered forward scattering states, while the other part
${\cal M}_{i \rightarrow f}$ encapsulates the ``actual'' interaction (For simplicity, we
will call ${\cal M}_{i \rightarrow f}$ as scattering matrix element.). So, the calculation of all those
observables essentially boils down to the computation of the second quantity. However, the exact computation
of this quantity is
incredibly difficult in any general 
field theory. The only viable methodology is provided under the
framework of perturbation theory where the matrix elements as well as
the observables are expanded in powers of coupling constants, $c$, present
in the theory:
\begin{align}
\label{eq:Intro-expand-coupling}
{\cal M}_{i \rightarrow f} = \sum\limits_{n=0}^{\infty} c^n {\cal M}_{i \rightarrow f}^{(n)}\,.
\end{align}
If the coupling constant is small enough, the evaluation of only the first term of
the perturbative expansion often turns out to be a very good
approximation that provides a 
reliable prediction to any observable. However, in QFT, it is a
well-known fact that the coupling constants are truly not `constants', their
strength depends on the energy scale at which the interaction takes
place. This evolution of the coupling constant may make it
comparatively large at some energy scale. In case of Quantum
Electro-Dynamics (QED), quantum field theory of electromagnetism, the
magnitude of the coupling constant, $c=\alpha_{\rm EM}$, increases with
the increase of momentum transfer:
\begin{align}
\label{eq:Intro-QED-flow}
\alpha_{\rm EM}(Q^2 \approx 0) \approx \frac{1}{137}\,, \qquad {\rm
  and} \qquad \alpha_{\rm EM}(Q^2 \approx m_W^2) \approx \frac{1}{128}
\end{align}
where, $m_W \approx 80$ GeV is the invariant mass of the $W$
boson. The smallness of $\alpha_{\rm EM}$ at all typical energy scales
which can be probed in all collider experiments guarantees very fast
convergence of the perturbation series to what we expect to be real
non-perturbative result. However, this picture no longer holds true in
case of QCD where the coupling constant, $c=\alpha_s$, may become quite large at
certain energy scales:
\begin{align}
\label{eq:Intro-QCD-flow}
\alpha_{s} (m_p^2) \approx 0.55, \qquad {\rm and} \qquad \alpha_{s} (m_Z^2) \approx 0.1
\end{align}
where, $m_p \approx 938 MeV$ and $m_Z \approx 90 GeV$ are the masses
of the proton and $Z$ boson. Clearly the magnitude $0.55$ is far from
being small! Hence, computation of only the leading term in
perturbative series often turns out to be a very crude approximation
which fails to deliver a reliable prediction. We must take into
account the contributions arising beyond leading term. 

In perturbation theory, the most acceptable and well known
prescription to compute the terms in a perturbative series is
provided by Feynman diagrams. Every term of a series is represented
through a set of Feynman diagrams and each diagram corresponds to a
mathematical expression. Hence, evaluation of a term in any perturbative
series boils down to the computation of all the corresponding Feynman
diagrams. Given an action of a QFT, one first requires to derive a set
of rules, called Feynman rules, which essentially establish the correspondence
between the Feynman diagrams and mathematical expressions. With the
rules in hand, we just need to draw all the possible Feynman diagrams
contributing to the order of our interest and eventually evaluate those using the
rules. Needless to say, as the perturbative order increases, the
number of Feynman diagrams to be drawn grows so rapidly that after certain order
it becomes almost prohibitively large to draw.

In this thesis, we will concentrate only on the aspects of
perturbative QCD. We will start our discussion of QCD by introducing
the basic aspects of this QFT which will be followed by the writing
down the quantum action and corresponding Feynman rules. Then we will
discuss how to compute amplitudes beyond leading order in QCD and
eventually get reliable numerical predictions at hadron colliders for
any process. 

\section{Quantum Chromo-Dynamics}
\label{sec:QCD}

Quantum Chromo-dynamics, familiarly called QCD, is the modern theory
of strong interactions, a fundamental force describing the
interactions between quarks and gluons which make up hadrons such as
the protons, neutrons and pions. QCD is a type of QFT called non-Abelian
gauge theory that has underlying SU(3) gauge symmetry. It appears as an 
expanded version of QED. Whereas in QED there is just
one kind of charge, namely electric charge, QCD has three different kinds of
charge, labeled by ``colour''.  Avoiding chauvinism, those are 
chosen as red, green, and blue. But, of course, the colour
charges of QCD have nothing to do with optical colours.
Rather, they have properties analogous to electric charges in QED.
In particular, the colour charges are conserved in all physical
processes. There are also photon-like massless gauge bosons,
called gluons, that act as the mediators of the strong interactions between
spin-1/2 quarks. Unlike the photons, which mediate the electromagnetic
interaction but lacks an electric charge, the gluons themselves carry
color charges. Gluons, as a consequence, participate in the strong
interactions in addition to mediating it, making QCD substantially
harder to analyse than QED.

In sharp contrast to other gauge theories, QCD enjoys two salient
features: confinement and asymptotic freedom. The force among
quarks/gluons fields does not diminish as they are separated from each
others. With the increase in mutual distance between them, the
mediating gluon fields gather enough energy to create a 
pair of quarks/gluons which forbids them to be found as free
particles; they are thus forever bound into hadrons such as the
protons, neutrons, pions or kaons. Although literature lacks the
satisfactory theoretical explanation, confinement is believed to be
true as it explains the consistent failure of finding free quarks or
gluons. The other interesting property, the asymptotic
freedom~\cite{Gross:1973id, Gross:1973ju, Politzer:1973fx, Gross:1974cs,
  Politzer:1974fr}, causes 
bonds between quarks/gluons become asymptotically weaker as energy
increases or distance decreases which allows us to perform the
calculation in QCD using the technique of perturbation theory. The Nobel
prize was awarded for this remarkable discovery of last century. 

In perturbative QCD, the basic building blocks of performing any
calculation are the Feynman rules, which will be discussed in next
subsection.

\subsection{The QCD Lagrangian and Feynman Rules}
\label{ss:QCD-Feynman-Rules}

The first step in performing perturbative calculations in a
QFT is to work out the Feynman rules. The SU(N) gauge invariant classical
Lagrangian density encapsulating the interaction between fermions and non-Abelian gauge
fields is 
\begin{align}
\label{eq:Intro-Lag-QCD-1}
{\cal L}_{classical} = - \frac{1}{4} F^a_{\mu\nu} F^{a,\mu\nu} +
 \sum\limits_{f=1}^{n_f}  {\overline \psi}_{\alpha,i}^{(f)} \left( \iu
  {\slashed D}_{\alpha\beta,ij} - 
  m_f \delta_{\alpha\beta} \delta_{ij} \right) \psi_{\beta,j}^{(f)}\,.
\end{align}
In the above expression, 
\begin{align}
\label{eq:Intro-Lag-QCD-2}
&F^a_{\mu\nu} = \partial_{\mu}A_{\nu}^a - \partial_{\nu}A_{\mu}^a +
  g_s f^{abc} A_{\mu}^b A_{\nu}^c\,,
\nonumber\\
&{\slashed D}_{\alpha\beta,ij} \equiv \gamma^{\mu}_{\alpha\beta} D_{\mu,ij} = \gamma^{\mu}
  \left( \delta_{ij} \partial_{\mu} - \iu g_s T^a_{ij} A^a_{\mu} \right)
\end{align}
where, $A^a_{\mu}$ and $\psi_{\alpha,i}^{(f)}$ are the guage and
fermionic quark fields, respectively. The indices represent the following things:
\begin{align}
\label{eq:Intro-indices}
&a, b, \cdots: \quad \text{color indices in the adjoint
  representation} \Rightarrow  [1, N^2-1]\,,
\nonumber\\
&i,j, \cdots: \quad \text{~color indices in the fundamental 
  representation} \Rightarrow  [1, N]\,,
\nonumber\\
&\alpha,\beta, \cdots: \quad \text{Dirac spinor indices} \Rightarrow
  [1, d]\,,
\nonumber\\
&\mu, \nu, \cdots: \quad \text{\,Lorentz indices} \Rightarrow [1,d]\,.
\end{align}
Numbers within the `[]' signifies the range of the corresponding
indices. $d$ is the space-time dimensions. $f$ is the quark flavour index which runs from 1 to $n_f$. $m_f$ and 
$g_s$ are the mass of the quark corresponding to $\psi^{(f)}$ and
strong coupling constant, 
respectively. $f^{abc}$ are the structure constants of SU(N)
group. These are related to the Gellmann matrices $T^a$, generators of the
fundamental representations of SU(N), through
\begin{align}
\label{eq:Intro-Gellman-fabc}
\l[T^a,T^b\r]=i f^{abc} T^{c}\,.
\end{align}
The $T^a$ are traceless, Hermitian matrices and these are
normalised with 
\begin{align}
\label{eq:Intro-Gellman-Norm}
Tr\l(T^{a}T^{b}\r) = T_{F} \delta^{ab}
\end{align}
where, $T_F=\frac{1}{2}$.
They satisfy the following completeness relation
\begin{align}
\label{eq:Intro-Gellman-Complete}
\sum\limits_{a} T^a_{ij} T^a_{kl} = \frac{1}{2} \left( \delta_{il} \delta_{kj} -
  \frac{1}{N} \delta_{ij} \delta_{kl} \right)\,.
\end{align}
In addition to the above three parent identities expressed through the
Eq.~(\ref{eq:Intro-Gellman-fabc}, \ref{eq:Intro-Gellman-Norm},
\ref{eq:Intro-Gellman-Complete}), we can have some auxiliary ones
which are often useful in simplifying colour algebra:
\begin{align}
\label{eq:Intro-aux-identities}
&\sum\limits_a (T^a T^a)_{ij} = C_F \delta_{ij}\,,
\nonumber\\
&f^{acd}f^{bcd} = C_A \delta^{ab}\,.
\end{align}  
The $C_{A}=N$ and $C_F=\frac{N^2-1}{2N}$ are the quadratic Casimirs of
the SU(N) group in the adjoint and fundamental representations,
respectively. For QCD, the SU(N) group index, $N=3$ and the flavor number
$n_f=6$. 

The quantisation of the non-Abelian gauge theory or the Yang-Mills (YM) theory
faces an immediate problem, namely, the propagator of gauge fields
cannot be obtained unambiguously. This is directly related to the
presence of gauge degrees of freedom inherent into the ${\cal
  L}_{classical}$. We need to perform the gauge fixing in order to get
rid of this problem. The gauge fixing in a covariant way, when done
through the path integral formalism, generates new particles called
Faddeev-Popov (FP) ghosts having spin-0 but obeying fermionic
statistics. The absolute necessity of introducing
the ghosts in the process of quantising the YM theory is a horrible
consequence of the Lagrangian formulation of QFT. There is no
observable consequence of these particles, we just need them in order
to describe an interacting theory of a massless spin-1 particle using
a local manifestly Lorentz invariant Lagrangian. These particles never
appear as physical external states 
but must be included in internal lines to cancel the unphysical
degrees of freedom of the gauge fields. Some alternative formulations of
non-Abelian gauge theory (such as the lattice) also do not
require ghosts. Perturbative gauge theories in certain non-covariant
gauges, such as light-cone or axial gauges, are also ghost free. However, to
maintain manifest Lorentz invariance in a perturbative gauge theory,
it seems ghosts are unavoidable and in this thesis we will be remained
within the regime of covariant gauge and consequently will include
ghost fields consistently into our computations. 

Upon applying this technique to quantise the YM theory, we end up with
getting the following full quantum Lagrangian density:
\begin{align}
\label{eq:Intro-Lag-Quant}
{\cal L}_{YM} &= {\cal L}_{classical} + {\cal L}_{gauge-fix} + {\cal L}_{ghost}
\end{align}
where, the second and third terms on the right hand side correspond to
the gauge fixing and FP contributions, respectively. These are
obtained as
\begin{align}
\label{eq:Intro-Lag-GF-FP}
&{\cal L}_{gauge-fix} = - \frac{1}{2 \xi}
  \left( \partial^{\mu}A^a_{\mu} \right)^2\,,
\nonumber\\
&{\cal L}_{ghost} = \left( \partial^{\mu} \chi^{a*} \right) D_{\mu,ab} \,\chi^b
\end{align}
with
\begin{align}
\label{eq:Intro-Lag-Dmuab}
D_{\mu,ab} \equiv \delta_{ab} \partial_{\mu} - g_s f_{abc} A^c_{\mu}\,.
\end{align}
The gauge parameter $\xi$ is arbitrary and it is introduced in order
to specify the gauge in a covariant way. This prescription of fixing
gauge in a covariant way is known as $R_{\xi}$ gauge. A typical choice
which is often used is $\xi=1$, known as Feynman gauge. We will be
working in this Feynman gauge throughout this thesis, unless otherwise
mentioned specifically. However, we emphasize that the physical
results are independent of the choice of the gauges. The field
$\chi^a$ and $\chi^{a*}$ are ghost and anti-ghost fields,
respectively.  

All the Feynman rules can be read off from the quantized Lagrangian
${\cal L}_{YM}$ in Eq.~\ref{eq:Intro-Lag-Quant}. We will denote the
quarks through straight lines, gluons through curly and ghosts through
dotted lines. We provide the rules in $R_{\xi}$ gauge.

\begin{itemize}
\item The propagators for quarks, gluons and ghosts are obtained as
  respectively:

\begin{figure}[H]
\qquad
  \begin{minipage}{2in}
\begin{tikzpicture}[line width=0.6 pt, scale=0.7]
\draw[fermion] (-2,0) -- (2,0);
\node at (-2.2,0.8) {$j,\beta$};
\node at (2.2,0.8) {$i,\alpha$};
\draw[->] (-1.5,-0.5) -- (-2,-0.5);
\node at (-1.7,-1) {$p_2$};
\draw[->] (1.5,-0.5) -- (2,-0.5);
\node at (1.7,-1) {$p_1$};
\end{tikzpicture}
  \end{minipage}
\hspace{-0.5cm}
  \begin{minipage}{3in}
\begin{align*}
\iu \left( 2\pi \right)^4 \delta^{(4)}\l(p_1+p_2\r) \delta_{ij} \left(
  \frac{1}{{\slashed p_1}-m_f+\iu \varepsilon} \right)_{\alpha\beta}
\end{align*}
  \end{minipage}

\vspace{1cm}

\qquad
  \begin{minipage}{2in}
\begin{tikzpicture}[line width=0.6 pt, scale=0.7]
\draw[gluon] (-2,0) -- (2,0);
\node at (-2.2,0.8) {$b,\nu$};
\node at (2.2,0.8) {$a,\mu$};
\draw[->] (-1.5,-0.5) -- (-2,-0.5);
\node at (-1.7,-1) {$p_2$};
\draw[->] (1.5,-0.5) -- (2,-0.5);
\node at (1.7,-1) {$p_1$};
\end{tikzpicture}
  \end{minipage}
  \begin{minipage}{3in}
\begin{align*}
\iu \left( 2\pi \right)^4 \delta^{(4)}\l(p_1+p_2\r) \delta_{ab}
  \frac{1}{p_1^2} \left[ -g_{\mu\nu} + (1-\xi) \frac{p_{1\mu} p_{1\nu}}{p_1^{2}} \right]
\end{align*}
  \end{minipage}

\vspace{1cm}

\qquad
  \begin{minipage}{2in}
\begin{tikzpicture}[line width=0.6 pt, scale=0.7]
\draw[scalar] (-2,0) -- (2,0);
\node at (-2.2,0.8) {$b$};
\node at (2.2,0.8) {$a$};
\draw[->] (-1.5,-0.5) -- (-2,-0.5);
\node at (-1.7,-1) {$p_2$};
\draw[->] (1.5,-0.5) -- (2,-0.5);
\node at (1.7,-1) {$p_1$};
\end{tikzpicture}
  \end{minipage}
\hspace{-1.8cm}
  \begin{minipage}{3in}
\begin{align*}
\iu \left( 2\pi \right)^4 \delta^{(4)}\l(p_1+p_2\r) \delta_{ab} \frac{1}{p_1^2}
\end{align*}
  \end{minipage}
\label{fig:Intro-QCD-Prop}
\end{figure}


\item The interacting vertices are given by:

\begin{figure}[H]
\qquad
  \begin{minipage}{2in}
\begin{tikzpicture}[line width=0.6 pt, scale=0.7]
\draw[fermion] (-2,-1) -- (0,1);
\draw[fermion] (0,1) -- (2,-1);
\draw[gluon] (0,1) -- (0,3);
\draw[->] (0.5,3) -- (0.5,2.5);
\draw[->] (-2.6,-0.9) -- (-2.2,-0.5);
\draw[->] (2.6,-0.9) -- (2.2,-0.5);
\node at (2.9,-0.6) {$p_1$};
\node at (-2.9,-0.6) {$p_2$};
\node at (1,2.8) {$p_3$};
\node at (2,-1.5) {$i,\alpha$};
\node at (-2,-1.5) {$j,\beta$};
\node at (0,3.5) {$a,\mu$};
\end{tikzpicture}
  \end{minipage}
\hspace{-0.5cm}
  \begin{minipage}{3in}
\begin{align*}
\iu g_s \left( 2\pi \right)^4 \delta^{(4)}\l(p_1+p_2+p_3\r) T^{a}_{ij}
  \,\l(\gamma^{\mu}\r)_{\alpha\beta} 
\end{align*}
  \end{minipage}
\end{figure}


\begin{figure}[H]
\qquad
  \begin{minipage}{2in}
\begin{tikzpicture}[line width=0.6 pt, scale=0.7]
\draw[gluon] (-2,-1) -- (0,1);
\draw[gluon] (0,1) -- (2,-1);
\draw[gluon] (0,1) -- (0,3);
\draw[->] (0.5,3) -- (0.5,2.5);
\draw[->] (-2.6,-0.9) -- (-2.2,-0.5);
\draw[->] (2.6,-0.9) -- (2.2,-0.5);
\node at (2.9,-0.6) {$p_1$};
\node at (-2.9,-0.6) {$p_2$};
\node at (1,2.8) {$p_3$};
\node at (2,-1.5) {$a,\mu$};
\node at (-2,-1.5) {$b,\nu$};
\node at (0,3.5) {$c,\rho$};
\end{tikzpicture}
  \end{minipage}
\quad
  \begin{minipage}{3in}
\begin{align*}
&\frac{g_s}{3!} \left( 2\pi \right)^4 \delta^{(4)}\l(p_1+p_2+p_3\r) f^{abc}
\\& \times \left[ g^{\mu\nu}(p_{1}-p_2)^{\rho} + g^{\nu\rho}(p_{2}-p_3)^{\mu} +
  g^{\rho\mu}(p_{3}-p_1)^{\nu} \right] 
\end{align*}
  \end{minipage}
\end{figure}


\begin{figure}[H]
\qquad
  \begin{minipage}{2in}
\begin{tikzpicture}[line width=0.6 pt, scale=0.7]
\draw[scalar] (-2,-1) -- (0,1);
\draw[scalar] (0,1) -- (2,-1);
\draw[gluon] (0,1) -- (0,3);
\draw[->] (0.5,3) -- (0.5,2.5);
\draw[->] (-2.6,-0.9) -- (-2.2,-0.5);
\draw[->] (2.6,-0.9) -- (2.2,-0.5);
\node at (2.9,-0.6) {$p_1$};
\node at (-2.9,-0.6) {$p_2$};
\node at (1,2.8) {$p_3$};
\node at (2,-1.5) {$b$};
\node at (-2,-1.5) {$c$};
\node at (0,3.5) {$a,\mu$};
\end{tikzpicture}
  \end{minipage}
\hspace{-0.5cm}
  \begin{minipage}{3in}
\begin{align*}
- g_s\left( 2\pi \right)^4 \delta^{(4)}\l(p_1+p_2+p_3\r) f^{abc} p_1^{\mu}
\end{align*}
  \end{minipage}
\end{figure}


\begin{figure}[H]
\qquad
  \begin{minipage}{2in}
\begin{tikzpicture}[line width=0.6 pt, scale=0.7]
\draw[gluon] (-2,-1) -- (0,1);
\draw[gluon] (0,1) -- (2,-1);
\draw[gluon] (-2,3) -- (0,1);
\draw[gluon] (0,1) -- (2,3);
\draw[->] (-2.6,-0.9) -- (-2.2,-0.5);
\draw[->] (2.6,-0.9) -- (2.2,-0.5);
\draw[->] (-2.6,3) -- (-2.2,2.6);
\draw[->] (2.6,3) -- (2.2,2.6);
\node at (2.9,-0.6) {$p_1$};
\node at (-2.9,-0.6) {$p_2$};
\node at (-2.9,2.6) {$p_3$};
\node at (2.9,2.6) {$p_4$};
\node at (2,-1.5) {$a,\mu$};
\node at (-2,-1.5) {$b,\nu$};
\node at (-2,3.5) {$c,\rho$};
\node at (2,3.5) {$d,\sigma$};
\end{tikzpicture}
  \end{minipage}
\quad
  \begin{minipage}{3in}
\begin{align*}
&-\frac{g_s^2}{4!} \left( 2\pi \right)^4
  \delta^{(4)}\l(p_1+p_2+p_3+p_4\r) 
\nonumber\\
&\Bigg\{ \l( f^{ac,bd} -f^{ad,cb} \r) g^{\mu\nu} g^{\rho\sigma} 
+ \l( f^{ab,cd} -f^{ad,bc} \r) g^{\mu\rho} g^{\nu\sigma} 
\nonumber\\
&+ \l( f^{ac,db} -f^{ab,cd} \r) g^{\mu\sigma} g^{\nu\rho} \Bigg\}
\intertext{with}
&f^{ab,cd} \equiv f^{abx} f^{cdx}
\end{align*}
  \end{minipage}
\end{figure}

\end{itemize}

In addition to these rules, we have keep in mind the following points:
\begin{itemize}
\item For any Feynman diagram, the symmetry factor needs to be
  multiplied appropriately. The symmetry factor is defined as the
  number of ways one can obtain the topological configuration of the Feynman diagram
  under consideration.
\item For each loop momenta, the integration over the loop momenta,
  $k$, needs to be performed with the integration measure
  $d^dk/(2\pi)^d$ in d-dimensions (in dimensional regularisation).
\item For each quark/ghost loop, one has to multiply a factor of (-1).
\end{itemize}

\section{Perturbative Calculations in QCD}
\label{sec:Intro-pQCD}

The asymptotic freedom of the QCD allows us to perform the calculations in
high energy regime using the techniques of perturbative QCD (pQCD). In
pQCD, we make the theoretical predictions through the computations of
the scattering matrix (S-matrix) elements. The S-matrix elements are
directly related to the scattering amplitude which is formally
expanded, within the framework of perturbation theory, in powers of
coupling constants. This expansion is represented through the set of Feynman
diagrams and the Feynman rules encapsulate the connection between
these these two. Hence, the theoretical predictions boil down to
evaluate the set of Feynman diagrams. Using the Feynman rules
presented in the previous Sec.~\ref{ss:QCD-Feynman-Rules}, we can
evaluate all the Feynman diagrams. 

Achieving precise theoretical predictions demand to go beyond leading
order which consists of evaluating the virtual/loop as well as real emission
diagrams. However, the contribution arising from the individual one is
not finite. The resulting expressions from the evaluation of loop
diagrams contain the ultraviolet (UV), soft and collinear
divergences. For simplicity, together we call the soft and collinear as
infrared (IR) divergence. 

The UV divergences arise from the region of large
momentum or very high energy (approaching infinity) of the Feynman
integrals, or, equivalently, because of the physical phenomena at very
short distances. We get rid of this through UV renormalisation. Before
performing the UV renormalisation, we need to regulate the Feynman
integrals which is essentially required to identify the true nature of
divergences. There are several ways to regulate the integrals. The
most consistent and beautiful way is the framework of dimensional
regularisation~\cite{'tHooft:1972fi,Cicuta:1972jf,Bollini:1972ui}. Within
this, we need to perform the integrals in general 
$d$-dimensions which is taken as $4+\epsilon$ in this thesis. Upon
performing the integrals, all the UV singularities arise as poles in
$\epsilon$. The UV renormalisation, which is performed through redefining
all the quantities present in the Lagrangian, absorbs these poles and
gives rise a UV finite result. The UV renormalisation is done at
certain energy scale, known as renormalisation scale, $\mu_{R}$. On
the other hand, the soft divergences 
arise from the low momentum  
limit (approaching zero) of the loop integrals and the collinear ones
arise when any loop momentum becomes collinear to any of
the external massless particles. The collinear divergence is a property of
theories with massless particles. Hence, even after performing the UV
renormalisation, the resulting expressions obtained through the
evaluation the loop integrals are not finite, they contain poles
arising from soft and collinear regions of the loop integrals.

To remove the residual IR divergences, we need to add the
contributions arising from the real emission diagrams. The latter
contains soft as well as collinear divergences which have the same
form as that of loop integrals. Once we add the virtual and real
emission diagrams and evaluate the phase space integrals, the
resulting expressions are guaranteed to be freed from UV, soft and
final state collinear singularities, thanks to the
Kinoshita-Lee-Nauenberg (KLN) theorem. An analogous result for quantum electrodynamics alone is known as Bloch?Nordsieck cancellation. However, the collinear
singularities arising from the collinear configurations involving
initial state particles remain. Those are removed at the hadronic level through the
techniques, known as mass factorisation, where the residual
singularities are absorbed into the bare parton distribution
functions (PDF). So, the observables at the hadronic level are finite
which are compared with the experimental outcomes at the hadron
colliders. Just like UV renormalisation, mass factorisation is
done at some energy scale, called factorisation scale, $\mu_F$. The
$\mu_R$ as well as $\mu_F$ are unphysical scales. The dependence of
the fixed order results on these scale is an artifact of
the truncation of the perturbative series to a finite order. If we can
capture the results to all order, then the dependence goes away.

The core part of this thesis deals with the higher order QCD
corrections employing the methodology of perturbation theory to some
of the very important processes within the SM and 
beyond. More specifically, the thesis contains 
\begin{itemize}
\item the soft-virtual QCD corrections to the inclusive cross section of the Higgs
  boson production through bottom quark annihilation at
  next-to-next-to-next-to-leading order
  (N$^3$LO)~\cite{Ahmed:2014cha},
\item the soft-virtual QCD corrections at N$^{3}$LO to the differential rapidity
  distributions of the productions of the Higgs boson in gluon fusion
  and of the leptonic pair in Drell-Yan~\cite{Ahmed:2014uya},
\item the three loop QCD corrections to the pseudo-scalar form
  factors~\cite{Ahmed:2015qpa,Ahmed:2015pSSV}.
\end{itemize}
In the subsequent subsections, we will discuss the above things in
brief.

\subsection{Soft-Virtual Corrections To Cross Section at N$^3$LO QCD}
\label{ss:Intro-SV}

The Higgs bosons are produced dominantly at the LHC 
via gluon fusion through top quark loop, while  
one of the sub-dominant ones take place through bottom quark
annihilation. In the SM, the interaction between the Higgs boson and
bottom quarks is controlled through the Yukawa coupling which is
reasonably small at typical energy scales. However, in
the minimal super symmetric SM (MSSM), this 
channel can contribute substantially due to enhanced coupling between the
Higgs boson and bottom quarks in the large $\tan\beta$ region, where
 $\tan\beta$ is the ratio of vacuum expectation values of the up
and down type Higgs fields. In the present run of LHC, the measurements
of the various coupling constants including this one are underway which can
shed light on the properties of the newly discovered Higgs boson~\cite{Aad:2012tfa,
  Chatrchyan:2012xdj}. 
Most importantly, for the precision studies we
must take into account all the contributions, does not matter how tiny
those are, arising from sub-dominant
channels along with the dominant ones to reduce the dependence on the
unphysical scales and make a reliable prediction.

The computations of the higher order QCD corrections beyond leading
order often become quite challenging because of the large number of Feynman
diagrams and, presence of the complicated loop and phase space
integrals. Under this circumstance, when we fail to compute the
complete result at certain order, it is quite natural to try an
alternative approach to capture 
the dominant contributions from the missing higher order corrections.
It has been observed for many processes that the dominant contributions to an observable
often comes from the soft gluon emission diagrams. The contributions
arising from the
associated soft gluon emission along with the virtual Feynman 
diagrams are known as
the soft-virtual (SV) corrections. \textit{The goal of the works published in the
  article~\cite{Ahmed:2014cha} is to 
discuss the SV
QCD corrections to the production cross section of the Higgs boson,
produced through bottom quark annihilation.}

The next-to-next-to-leading order (NNLO) QCD corrections to this channel are already present
in the variable flavour scheme (VFS) \cite{Dicus:1988cx, Dicus:1998hs,
  Maltoni:2003pn, Olness:1987ep, Gunion:1986pe, Harlander:2003ai},
while it is known to NLO in the fixed 
flavour scheme (FFS) \cite{Reina:2001sf, Beenakker:2001rj,
  Dawson:2002tg, Beenakker:2002nc, Raitio:1978pt, Kunszt:1984ri}.    
In addition, the partial result for the N$^3$LO
corrections~\cite{Ravindran:2005vv,Ravindran:2006cg,Kidonakis:2007ww} under the SV
approximation were also computed long 
back. In both~\cite{Ravindran:2005vv,Ravindran:2006cg} and~\cite{Kidonakis:2007ww},
it was not possible to determine the complete contribution at
N$^{3}$LO due to the lack of information on three loop finite part of
bottom anti-bottom Higgs form factor in QCD and the soft gluon
radiation at N$^{3}$LO level. \textit{In this
  work~\cite{Ahmed:2014cha}, we have computed the missing part and  
completed the full SV corrections to the cross section at N$^3$LO}. 

The infrared safe contributions from the soft gluons are obtained by
adding the soft part of the cross section with the UV
renormalized virtual part and performing mass factorisation using
appropriate counter terms. The main ingredients are the
form factors, overall operator UV renormalization constant, 
soft-collinear distribution arising from the real radiations in the partonic 
subprocesses and mass factorization kernels. The computations of
SV cross section at N$^3$LO QCD require all of these above quantities
up to 3-loop order. The relevant form factor becomes available very
recently in~\cite{Gehrmann:2014vha}. The soft-collinear distribution at
N$^3$LO was computed by 
us around the same time in~\cite{Ahmed:2014cla}. This was calculated
from the recent result of 
N$^3$LO SV cross section of the Higgs boson productions in gluon
fusion~\cite{Anastasiou:2014vaa} by employing a symmetry (maximally 
non-Abelian property). Prior to this, this symmetry was verified
explicitly up to NNLO order. However, neither there was any clear
reason to believe that the symmetry would fail
nor there was any transparent indication of holding it beyond this
order. Nevertheless, we conjecture~\cite{Ahmed:2014cla} that the relation would hold true
even at N$^3$LO order! This is inspired by the universal properties of
the soft gluons which are the  
underlying reasons behind the existence of this remarkable
symmetry. Later, this conjecture is verified by explicit computations
performed by two different groups on Drell-Yan
process~\cite{Catani:2014uta, Li:2014bfa}. This symmetry
plays the most important role in 
achieving our goal. With these, along with the existing results of the
remaining required ingredients, we obtain the complete analytical
expressions of N$^3$LO SV
cross section of the Higgs boson production through bottom quark
annihilation~\cite{Ahmed:2014cha} employing the methodology prescribed
in~\cite{Ravindran:2005vv,Ravindran:2006cg}. It reduces the scale dependence and
provides a more precise result. We demonstrate the impact of this result numerically at
the LHC briefly. \textit{This is the most accurate
result for this channel which exists in the literature till date and
it is expected to play an important role in coming days at the LHC.}

\subsection{Soft-Virtual QCD Corrections to Rapidity at N$^3$LO}
\label{ss:Intro-SV-Rap}

The productions of the Higgs boson in gluon fusion and leptonic pair
in Drell-Yan (DY) are among the most important processes at the LHC which are studied not
only to test the SM to an unprecedented accuracy but also to explore
the physics beyond Standard Model (BSM). During the present
run at the LHC, in addition to the inclusive production cross section,
the differential rapidity distribution is among the most important
observables, 
which is expected to be measured in upcoming days. This immediately
calls for very precise theoretical predictions.

In the same spirit of the SV corrections to
the inclusive production cross section, the dominant contributions to
the differential rapidity distributions often arise from the soft
gluon emission diagrams. Hence, in the absence of complete fixed
order result, the rapidity distribution under SV approximation is the
best available alternative in order to capture the dominant contributions from
the missing higher orders and stabilise the dependence on unphysical
scales. For the Higgs boson production through gluon fusion, we work in the effective
theory where the top quark is integrated out.
 \textit{This work published in the article~\cite{Ahmed:2014uya} is devoted to demonstrate the SV corrections
to this observable at N$^3$LO for the Higgs boson, produced through
gluon fusion, and leptonic pair in DY production.}

For the processes under considerations, the NNLO QCD
corrections are present~\cite{Anastasiou:2003yy, Anastasiou:2004xq,
  Anastasiou:2005qj}, computed long back, and in addition, the partial
N$^3$LO SV 
results~\cite{Ravindran:2006bu} are also available. However, due to reasonably large scale
uncertainties and crying demand of uplifting the accuracy of theoretical
predictions, we must push the boundaries of existing
results. \textit{In this work~\cite{Ahmed:2014uya}, 
  we have computed the missing part and completed the SV corrections
  to the rapidity distributions at N$^3$LO QCD}.

The prescription~\cite{Ravindran:2006bu} which has been employed to calculate the SV QCD
corrections is similar to that of the inclusive cross section, more
specifically, it is a generalisation of the other one. The infrared
safe contributions under SV approximation
can be computed by adding the soft part of the rapidity distribution
with the UV renormalised virtual part and performing the mass
factorisation using appropriate counter terms. Similar to the
inclusive case, the main ingredients to perform this computation are
the form factors, overall UV operator renormalisation constant,
soft-collinear distribution for rapidity and mass factorisation
kernels. These quantities are required up to N$^3$LO to calculate the
rapidity at this order. The three loop quark and gluon form
factors~\cite{Moch:2005tm, Baikov:2009bg, Gehrmann:2010ue, Gehrmann:2010tu} were
calculated long back. The operator renormalisation constants are also
present. For DY, this constant is not required or equivalently equals
to unity. The mass factorisation kernels are also available in the
literature to the required order. The only missing part was the
soft-collinear distribution for rapidity at N$^3$LO. This was
not possible to compute until very recently. Because of the universal
behaviour of the soft gluons, the soft-collinear distributions for
rapidity and inclusive cross section can be related to all orders in
perturbation theory~\cite{Ravindran:2006bu}. Employing this beautiful
relation, we obtain this 
quantity at N$^3$LO
from the results of soft-collinear distribution of the inclusive cross
section~\cite{Ahmed:2014cla}. Using this, along with the existing results of the other
relevant quantities, we compute the complete analytical expressions of
N$^3$LO SV correction to the rapidity distributions for the Higgs
boson in gluon fusion and leptonic pair in DY~\cite{Ahmed:2014uya}. We demonstrate the
numerical impact of this correction for the case of Higgs boson at the
LHC. This indeed reduces the scale dependence significantly and
provides a more reliable theoretical predictions. \textit{These are the
  most accurate results for the rapidity distributions of the Higgs
  boson and DY pair which exist in the literature and undoubtedly, 
  expected to play very important role in the upcoming run at the LHC.}

\subsection{Pseudo-Scalar Form Factors at Three Loops in QCD}
\label{ss:Intro-PS}

One of the most popular extensions of the SM, namely, the MSSM and two
Higgs doublet model have richer Higgs sector containing more than one
Higgs boson and there have been intense search strategies to observe
them at the LHC. In particular, the production of CP-odd Higgs
boson/pseudo-scalar at the LHC has been studied in detail, taking into
account higher order QCD radiative corrections, due to similarities
with its CP-even counter part. Very recently, the N$^{3}$LO QCD corrections
to the inclusive production cross section of the CP-even Higgs boson
becomes available~\cite{Anastasiou:2015ema}. So, it is very natural to
extend the theoretical accuracy 
for the CP-odd Higgs boson 
to the same order of N$^{3}$LO. This requires the 3-loop quark and gluon form factors for
the pseudo-scalar which are the only missing ingredients to achieve this goal.

Multiloop and multileg computations play a crucial role to achieve the golden
task of making precise theoretical predictions. However, the
complexity of these computations grows very rapidly 
with the increase of number of loops and/or external
particles. Nevertheless, it has become a reality due to several
remarkable developments in due course of time. \textit{These
  articles~\cite{Ahmed:2015qpa, Ahmed:2015pSSV} are
devoted to demonstrate the computations of the 3-loop quark and gluon
form factors for the pseudo-scalar operators in QCD}.

The coupling of a pseudo-scalar Higgs boson to gluons is mediated
through a heavy quark loop. In the limit of large quark mass, it is
described by an effective Lagrangian~\cite{Chetyrkin:1998mw} that only
admits light degrees of freedom.  
In this effective theory, we compute the 3-loop massless QCD
corrections to the form factor that describes the coupling of a
pseudo-scalar Higgs boson to gluons. 
The evaluation of this 3-loop 
form factors is truly a non-trivial task not only because of the
involvement of a large number of Feynman diagrams but also due to
the presence of the axial vector coupling. We work in dimensional
regularisation and use the 't Hooft-Veltman
prescription~\cite{'tHooft:1972fi} for the axial vector current, 
The state-of-the-art techniques including
integration-by-parts~\cite{Tkachov:1981wb,Chetyrkin:1981qh} and
Lorentz invariant~\cite{Gehrmann:1999as} identities 
have been employed to accomplish this task. The UV
renormalisation is quite involved since the two operators, present in
the Lagrangian, mix under UV renormalization due to the axial anomaly and
additionally, a finite renormalisation constant needs to be introduced
in order to fulfill the chiral Ward identities.  
Using the universal infrared factorization properties, we
independently derive~\cite{Ahmed:2015qpa} the three-loop operator
mixing and finite operator renormalisation from the renormalisation
group equation for the form factors, thereby confirming recent
results~\cite{Larin:1993tq, Zoller:2013ixa}, which were computed
following a completely different methodology, in the operator product
expansion.   
This form factor~\cite{Ahmed:2015qpa,Ahmed:2015pSSV} is an important
ingredient to the precise 
prediction of the pseudo-scalar Higgs boson production cross section
at hadron colliders. We derive the hard matching coefficient in
soft-collinear effective theory (SCET). We  
also study the form factors in the context of leading
transcendentality principle and we find that the  
diagonal form factors become identical to those of ${\cal N}=4$ upon
imposing some identification 
on the quadratic Casimirs.  Later, these form factors are used to
calculate the SV corrections~\cite{Ahmed:2015pSSV} to the pseudo-scalar production cross
section at N$^3$LO and next-to-next-to-next-to-leading logarithm (N$^3$LL) QCD.

\chapter{\label{chap:bBCS}Higgs boson production through $b \bar b$ annihilation at threshold in N$^3$LO QCD}

\textit{\textbf{The materials presented in this chapter are the result of an original research done in collaboration with Narayan Rana and V. Ravindran, and these are based on the published article~\cite{Ahmed:2014cha}}}.
\\
\begingroup
\hypersetup{linkcolor=blue}
\minitoc
\endgroup



\section{Prologue}
\setcounter{equation}{0}
\label{sec:intro}

The discovery of Higgs boson by ATLAS \cite{Aad:2012tfa} and CMS \cite{Chatrchyan:2012ufa}
collaborations of the LHC
at CERN has not only shed the light on the dynamics behind the electroweak symmetry breaking
but also put the SM of particle physics on a firmer ground.  In the SM, the 
elementary particles such as quarks, leptons and gauge bosons, $Z,W^\pm$ acquire 
their masses through spontaneous symmetry breaking (SSB).  
The Higgs mechanism provides the framework for 
SSB. The SM predicts the existence of a Higgs boson whose mass is a 
parameter of the model. 
The recent discovery of the SM Higgs boson like particle provides a valuable information on this, namely on
its mass which is about 125.5 GeV.   
The searches for the Higgs boson have been going on for several decades
in various experiments.  
Earlier experiments such as LEP \cite{Barate:2003sz} and Tevatron \cite{Aaltonen:2010yv} played
an important role in the discovery by the LHC collaborations 
through narrowing down its possible mass range.
LEP excluded Higgs boson of mass below 114.4 GeV 
and their precision electroweak measurements \cite{ALEPH:2010aa} 
hinted the mass less than 152 GeV at $95\%$ confidence level (CL), while
Tevatron excluded Higgs boson of mass in the range $162-166$ GeV at $95\%$ CL.   

Higgs bosons are produced dominantly at the LHC 
via gluon gluon fusion through top quark loop, while  
the sub-dominant ones 
are vector boson fusion, associated production of
Higgs boson with vector bosons, with top anti-top pairs and also in bottom anti-bottom 
annihilation.  The inclusive productions of Higgs boson in gluon gluon \cite{ALEPH:2010aa, Djouadi:1991tka, Dawson:1990zj, Spira:1995rr, Catani:2001ic, Harlander:2001is, Harlander:2002wh, Anastasiou:2002yz, Ravindran:2003um}, 
vector boson fusion processes \cite{Bolzoni:2010xr} and associated production with vector
bosons \cite{Han:1991ia} are known to NNLO
accuracy in QCD.     
Higgs production in bottom anti-bottom annihilation is also known to NNLO accuracy
in the variable flavour scheme (VFS) \cite{Dicus:1988cx, Dicus:1998hs, Maltoni:2003pn, Olness:1987ep, Gunion:1986pe, Harlander:2003ai}, while it is known to NLO in the fixed
flavour scheme (FFS) \cite{Reina:2001sf, Beenakker:2001rj, Dawson:2002tg, Beenakker:2002nc, Raitio:1978pt, Kunszt:1984ri}.   
In the MSSM, 
the coupling of bottom quarks to Higgs 
becomes large in the large $\tan\beta$ region, where
$\tan\beta$ is the ratio of vacuum expectation values of up and down type Higgs fields.
This can enhance contributions from bottom anti-bottom annihilation subprocesses.  

While the theoretical predictions of NNLO \cite{ALEPH:2010aa, Djouadi:1991tka, Dawson:1990zj, Spira:1995rr, Catani:2001ic, Harlander:2001is, Harlander:2002wh, Anastasiou:2002yz, Ravindran:2003um} and next to next to 
leading log (NNLL) \cite{Catani:2003zt} QCD corrections and of two loop 
electroweak effects \cite{Aglietti:2004nj, Actis:2008ug} played an important role in the Higgs discovery,
the theoretical uncertainties resulting from 
factorization and renormalization scales are not fully under control.  
Hence, the efforts to go beyond NNLO are going on intensively.
Some of the ingredients to obtain N$^3$LO QCD corrections
are already available.  For example, quark and gluon form
factors \cite{Moch:2005id, Moch:2005tm, Gehrmann:2005pd, Baikov:2009bg, Gehrmann:2010ue}, 
the mass factorization kernels \cite{Moch:2004pa}
and the renormalization constant \cite{Chetyrkin:1997un} for 
the effective operator describing the coupling of
Higgs boson with the SM fields in the infinite top quark mass 
limit up to three loop level in dimensional regularization are known for some time.  
In addition, NNLO soft contributions are known \cite{deFlorian:2012za} to all
orders in $\epsilon$ for both DY and Higgs productions using dimensional regularization
with space time dimension being $d=4+\epsilon$.
They were used to obtain the partial N$^3$LO
threshold effects \cite{Moch:2005ky, Laenen:2005uz, Idilbi:2005ni, Ravindran:2005vv, Ravindran:2006cg} to 
Drell-Yan production of di-leptons and inclusive productions of Higgs boson through gluon gluon 
fusion and in bottom anti-bottom annihilation.  
Threshold contribution to the inclusive production
cross section is expanded in terms of $\delta(1-z)$ and ${\cal D}_i(z)$ where
\begin{eqnarray}
{\cal D}_i(z) = \left(\frac{\ln^i (1-z)}{1-z}\right)_+ 
\end{eqnarray}
with the scaling parameter 
$z=m_H^2/\hat s$ for Higgs 
and $z=m_{l^+l^-}^2/\hat s$ for DY.
Here $m_H$, $m_{l^+l^-}$ and $\hat s$ are mass of the Higgs boson,
invariant mass of the di-leptons and square of the center of mass energy of the partonic reaction responsible for production mechanism respectively.
The missing $\delta(1-z)$ terms for the complete N$^3$LO threshold contributions 
to the Higgs production through gluon gluon fusion are now available 
due to the seminal work by Anastasiou et al \cite{Anastasiou:2014vaa} 
where the relevant soft contributions were obtained from the real radiations at 
N${}^3$LO level.  
The resummation of threshold effects \cite{Sterman:1986aj, Catani:1989ne} 
to infra-red safe observables 
resulting from their factorization properties  
as well as Sudakov resummation of soft gluons  
provides an elegant framework to obtain threshold enhanced 
contributions to inclusive
and semi inclusive observables order by order in perturbation theory.
In \cite{Ahmed:2014cla}, using this framework, we exploited the universal
structure of the soft radiations to obtain the corresponding soft
gluon contributions to DY production, which led to the evaluation of missing  
$\delta(1-z)$ part of the N$^3$LO threshold corrections. 
In \cite{Li:2014bfa},  relevant one loop double real emissions from light-like Wilson lines
were computed to obtain threshold corrections to Higgs as well as Drell-Yan productions
up to N$^3$LO level providing an independent approach.
In \cite{Catani:2014uta} the universality of soft gluon contributions near threshold and
the results of \cite{Anastasiou:2014vaa} were used to obtain general expression 
of the hard-virtual coefficient which contributes 
to N$^3$LO threshold as well as 
threshold resummation at next-to-next-to-next-to-leading-logarithmic (N$^3$LL) 
accuracy for the production cross section of a 
colourless heavy particle at hadron colliders.
For the Higgs production through $b \bar{b}$ annihilation,
till date, only partial N${}^3$LO threshold corrections are known \cite{Ravindran:2005vv, Ravindran:2006cg, Kidonakis:2007ww} where again the framework of threshold resummation was used.
In both \cite{Ravindran:2005vv, Ravindran:2006cg} and \cite{Kidonakis:2007ww}, it was not possible 
to determine the $\delta(1-z)$ at N$^3$LO
due to the lack of information on 
three loop finite part of bottom anti-bottom higgs form factor in QCD and the 
soft gluon radiation at N$^3$LO level.  In \cite{Kidonakis:2007ww}, subleading corrections
were also obtained through the method of Mellin moments.
The recent results on Higgs form factor with bottom anti-bottom by Gehrmann and Kara 
\cite{Gehrmann:2014vha} and on the universal soft distribution obtained
for the Drell-Yan production \cite{Ahmed:2014cla} can now be used   
to obtain $\delta(1-z)$ part of the threshold N$^3$LO contribution.
For the soft gluon radiations in the $b \bar{b}$ annihilation, 
the results from \cite{Ahmed:2014cla} can be 
used as they do not depend on the flavour of the incoming quark states.
We have set bottom quark mass to be zero throughout except in the Yukawa coupling. 

We begin by writing down the relevant interacting Lagrangian in Sec.~\ref{sec:bBH-Lag}. 
In the Sec.~\ref{sec:bBH-ThreResu}, we present the formalism of computing threshold QCD corrections to the cross-section and in Sec.~\ref{sec:bBH-Res}, we present 
our results for threshold N$^3$LO QCD contributions to Higgs production through $b\bar{b}$ annihilation at hadron colliders and their numerical impact .
The numerical impact of threshold enhanced  N$^3$LO contributions
is demonstrated for the LHC energy $\sqrt{s} = 14$ TeV by studying 
the stability of the perturbation theory under factorization and renormalization scales.
Finally we give a brief summary of our findings in Sec.~\ref{sec:bBH-Summary}.

\section{The Effective Lagrangian}
\label{sec:bBH-Lag}

The interaction of bottom quarks and the scalar Higgs
boson is given by the action 
\begin{align}
\label{eq:bBH-Lag}
{\cal L}^{b} = \phi(x) {O}^b(x) \equiv - {\lambda \over \sqrt{2}} \phi(x) \overline \psi_b(x) \psi_b(x)
\end{align}
where, $\psi_b(x)$ and $\phi(x)$ denote the bottom quark and scalar
Higgs field, respectively.
$\lambda$ is the Yukawa coupling given by $\sqrt{2} m_b/\nu$, with 
the bottom quark mass $m_b$ and 
the vacuum expectation value $\nu\approx 246$ GeV.  In MSSM,
for the pseudo-scalar Higgs boson, we need to replace $\lambda \phi(x) \overline \psi_b(x) \psi_b(x)$ by 
$\tilde \lambda \tilde \phi(x) \overline \psi_b(x) \gamma_5 \psi_b(x)$ in the above
equation.  The MSSM couplings are given by
\[
 \tilde{\lambda} = \left\{
  \begin{array}{ll}
    -  \frac{\sqrt{2} m_b \sin\alpha}{\nu \cos\beta}  \,,& \qquad \tilde{\phi} = h\,,\\
    \phantom{-}  \frac{\sqrt{2} m_b \cos\alpha}{\nu \cos\beta}  \,,& \qquad \tilde{\phi}=H\,,\\
    \phantom{-}  \frac{\sqrt{2} m_b \tan\beta}{\nu } \,, & \qquad \tilde{\phi}=A\,
  \end{array}
  \right.
\]
respectively. The angle $\alpha$ measures the mixing of weak and mass eigenstates
of neutral Higgs bosons.  We use VFS scheme throughout, hence except in the Yukawa coupling,
$m_b$ is taken to be zero like other light quarks in the theory. 

\section{Theoretical Framework for Threshold Corrections}
\label{sec:bBH-ThreResu}

\begin{figure}[htb]
\begin{center}
\begin{tikzpicture}[line width=1.5 pt, scale=1]
\draw[fermion] (-2.5,0) -- (0,-1.3);
\draw[fermion] (0,-1.3) -- (-2.5,-2.6);
\draw[vector] (0,-1.3) -- (2,-1.3);
\draw[fermion] (2,-1.3) -- (4.5,0);
\draw[fermionbar] (2,-1.3) -- (4.5,-2.6);
\node at (-2.8,0) {$q$};
\node at (-2.8,-2.6) {${\bar q}$};
\node at (1.1,-0.7) {$\gamma^{*}/Z$};
\node at (4.8,0) {$l^{+}$};
\node at (4.8,-2.6) {$l^{-}$};
\end{tikzpicture}
\caption{Drell-Yan pair production}
\label{fig:1-qQll}
\end{center}
\end{figure}
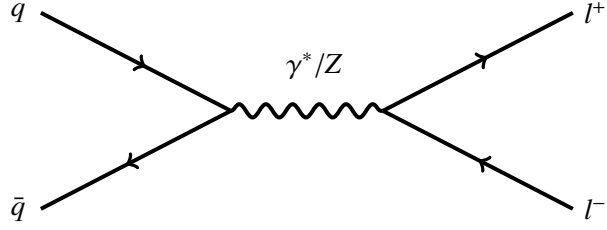

The inclusive cross-section for the production of a colorless
particle, namely, a Higgs boson through gluon fusion/bottom quark
annihilation or a pair of leptons in the Drell-Yan at the hadron
colliders is computed using  
\begin{align}
\label{eq:bBH-1}
\sigma^{I}(\tau, q^{2}) = \sigma^{I,(0)}(\mu_R^2) \sum\limits_{i,j=q,{\bar q},g}
  \int\limits_{\tau}^{1} dx \;\Phi_{ij}(x,\mu_F^{2})\;
  \Delta^I_{ij}\left(\frac{\tau}{x}, q^{2}, \mu_R^2, \mu_F^2\right)
\end{align}
with the partonic flux
\begin{align}
\label{eq:bBH-2}
\Phi_{ij}(x, \mu_{F}^2) = \int\limits_x^1 \frac{dy}{y} f_i(y, \mu_F^2)
  \;f_j\left(\frac{x}{y}, \mu_F^2 \right)\,.
\end{align}
In the above expressions, $f_i (y,\mu_F^2)$ and $f_j \left(\frac{x}{y},\mu_F^2\right)$
are the parton distribution functions (PDFs) of the initial state
partons $i$ and $j$ with momentum fractions $y$ and $x/y$,
respectively. These are renormalized at the factorization scale
$\mu_{F}$. The dimensionless quantity $\Delta^{I}_{ij}\left(\frac{\tau}{x}, q^2,
  \mu_{R}^{2}, \mu_F^2\right)$ is called the 
coefficient function of the partonic cross section for the
production of a colorless particle from partons $i$ and $j$, computed
after performing the UV renormalization at scale $\mu_{R}$ and mass
factorization at $\mu_{F}$. The quantity $\sigma^{I,(0)}$ is a pre-factor of the born level cross section. The variable $\tau$ is defined as
$q^{2}/s$, where  
\begin{equation}
\label{eq:q2}
    q^{2} =  
\begin{cases}
    ~m_{H}^{2}& ~\text{for}~ I=H\, ,\\
    ~m_{l^+l^-}^{2}& ~\text{for}~ I=DY\, .
\end{cases}
\end{equation}
$m_{H}$ is the mass of the Higgs boson and $m_{l^+l^-}$is the invariant
mass of the final state dilepton pair ($l^{+}l^{-}$), which can be
$e^{+}e^{-},\mu^{+}\mu^{-}, \tau^{+}\tau^{-}$, in the DY production. 
$\sqrt{s}$ and $\sqrt{\hat{s}}$ stand for the hadronic and partonic
center of mass energy, respectively. Throughout this chapter, we
denote $I=H$ for the 
productions of the Higgs boson 
through gluon ($gg$) fusion (Fig.~\ref{fig:1-ggH}) and bottom quark ($b{\bar b}$)
annihilation (Fig.~\ref{fig:1-bBH}), whereas we write $I=DY$ for the production of a pair of
leptons in the Drell-Yan (Fig.~\ref{fig:1-qQll}).
\begin{figure}[htb]
\begin{center}
\begin{tikzpicture}[line width=1.5 pt, scale=1]
\draw[gluon] (-2.5,0) -- (0,0);
\draw[gluon] (-2.5,-2) -- (0,-2);
\draw[fermion] (0,0) -- (2,-1);
\draw[fermion] (2,-1) -- (0,-2);
\draw[fermion] (0,-2) -- (0,0);
\draw[scalarnoarrow] (2,-1) -- (4,-1);
\node at (-2.8,0) {$g$};
\node at (-2.8,-2) {$g$};
\node at (4.3,-1) {$H$};
\end{tikzpicture}
\caption{Higgs boson production in gluon fusion}
\label{fig:1-ggH}
\end{center}
\end{figure}
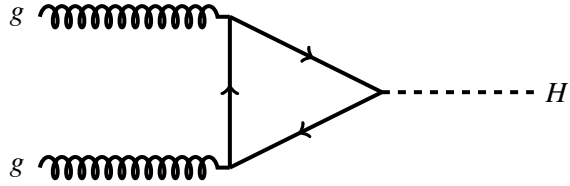
\begin{figure}[htb]
\begin{center}
\begin{tikzpicture}[line width=1.5 pt, scale=1]
\draw[fermion] (-2.5,0) -- (0,-1.3);
\draw[fermion] (0,-1.3) -- (-2.5,-2.6);
\draw[scalarnoarrow] (0,-1.3) -- (2,-1.3);
\node at (-2.8,0) {$b$};
\node at (-2.8,-2.6) {${\bar b}$};
\node at (2.3,-1.3) {$H$};
\end{tikzpicture}
\caption{Higgs boson production through bottom quark annihilation}
\label{fig:1-bBH}
\end{center}
\end{figure}

One of the \textit{goals} of this chapter is to study the impact of
the contributions arising from the soft gluons to the cross section of
a colorless particle production at Hadron colliders. The infrared safe
contributions from the soft gluons is obtained by adding the soft part
of the cross section with the UV renormalized virtual
part and performing mass factorisation using appropriate counter
terms. This combination is often called the soft-plus-virtual
cross section whereas the remaining portion is known as hard
part. Hence, we write the partonic cross section by decomposing into
two parts as 
\begin{align}
\label{eq:bBH-PartsOfDelta}
 &{\Delta}^{I}_{ij} (z, q^{2}, \mu_{R}^{2}, \mu_F^2) = {\Delta}^{I,
   \text{SV}}_{ij} (z, q^{2}, \mu_{R}^{2}, \mu_F^2) + {\Delta}^{I,
   \text{hard}}_{ij} (z, q^{2}, \mu_{R}^{2}, \mu_F^2) \,  
\end{align}
with $z \equiv q^{2}/\hat{s}$. The SV contributions
${\Delta}^{I, \text{SV}}_{ij} (z, q^{2}, \mu_{R}^{2}, \mu_F^2)$
contains only the distributions of kind $\delta(1-z)$ and
${\cal{D}}_{i}$, where the latter one is defined through 
\begin{align}
\label{eq:bBH-calD}
{\cal{D}}_{i} \equiv \left[ \frac{\ln^{i}(1-z)}{(1-z)} \right]_{+}\, .
\end{align}
This is also known as the threshold contributions. On the other hand,
the hard part ${\Delta}^{I, \text{hard}}_{ij}$ contains all the
regular terms in $z$. The SV cross section in $z$-space is
computed in $d=4+\ep$ dimensions, as formulated in
\cite{Ravindran:2005vv, Ravindran:2006cg}, using  
\begin{align}
\label{eq:bBH-sigma}
\Delta^{I, \sv}_{ij} (z, q^2, \mu_{R}^{2}, \mu_F^2) = 
{\cal C} \exp \Big( \Psi^I_{ij} \left(z, q^2, \mu_R^2, \mu_F^2, \epsilon
  \right)  \Big) \Big|_{\epsilon = 0} 
\end{align}
where, $\Psi^I_{ij} \left(z, q^2, \mu_R^2, \mu_F^2, \epsilon \right)$ is a
finite distribution and ${\cal C}$ is the convolution defined
as  
\begin{equation}
\label{eq:bBH-conv}
 {\cal C} e^{f(z)} = \delta(1-z) + \frac{1}{1!} f(z) + \frac{1}{2!}
 f(z) \otimes f(z) + \cdots  \,.
\end{equation}
Here $\otimes$ represents Mellin convolution and $f(z)$ is a
distribution of the kind $\delta(1-z)$ and ${\cal D}_i$. The
equivalent formalism of the SV approximation in the Mellin (or
$N$-moment) space, where instead of distributions in $z$, the dominant
contributions come from the meromorphic functions of the variable $N$
(see~\cite{Sterman:1986aj, Catani:1989ne}) and the threshold limit of
$z \rightarrow 1$ is translated to $N \rightarrow \infty$. 
The $\Psi^I_{ij} \left(z, q^2, \mu_R^2, \mu_F^2, \epsilon \right)$ is
constructed from the form factors ${\cal F}^I_{ij} (\hat{a}_s, Q^2, \mu^2,
\epsilon)$ with $Q^{2}=-q^{2}$, the overall operator UV
renormalization constant $Z^I_{ij}(\hat{a}_s, 
\mu_R^2, \mu^2, \epsilon)$, 
the soft-collinear distribution $\Phi^I_{ij}(\hat{a}_s, q^2, \mu^2, z, \epsilon)$
arising from the real radiations in the partonic subprocesses and the
mass factorization kernels $\Gamma^{I}_{ij} (\hat{a}_s, \mu^2, \mu_F^2, z,
\epsilon)$. In 
terms of the above-mentioned quantities it takes the following form,
as presented in \cite{Ravindran:2006cg, Ahmed:2014cla, Ahmed:2014cha} 
\begin{align}
\label{eq:bBH-psi}
\Psi^{I}_{ij} \left(z, q^2, \mu_R^2, \mu_F^2, \epsilon \right) = &\left( \ln \Big[ Z^I_{ij} (\hat{a}_s, \mu_R^2, \mu^2, \epsilon) \Big]^2 
+ \ln \Big|  {\cal F}^I_{ij} (\hat{a}_s, Q^2, \mu^2, \epsilon)  \Big|^2
                                                              \right)
                                                              \delta(1-z)  
\nonumber\\
& + 2 \Phi^I_{ij} (\hat{a}_s, q^2, \mu^2, z, \epsilon) - 2 {\cal C} \ln
  \Gamma^{I}_{ij} (\hat{a}_s, \mu^2, \mu_F^2, z, \epsilon) \, . 
\end{align}
In this expression, ${\hat{a}_s} \equiv {\hat{g}}_{s}^{2}/16\pi^{2}$
is the unrenormalized strong coupling constant which is related to the
renormalized one $a_{s}(\mu_{R}^{2})\equiv a_{s}$ through the renormalization
constant $Z_{a_{s}}(\mu_{R}^{2}) \equiv Z_{a_{s}}$ as  
\begin{align}
\label{eq:bBH-ashatANDas}
&{\hat a_s} S_{\epsilon}  = \left( \frac{\mu^{2}}{\mu_R^{2}}
  \right)^{\epsilon/2}  Z_{a_{s}} a_s\,,
\end{align}
where, $S_{\ep} = \exp\left[(\gamma_{E}-\text{ln} 4\pi)\ep/2)\right]$
and  $\mu$ is the mass scale introduced to keep the ${\hat{a}_s}$
dimensionless in $d$-dimensions. ${\hat g}_{s}$ is the coupling
constant appearing in the bare Lagrangian of QCD. $Z_{a_{s}}$ can be obtained by
solving the underlying renormalisation group equation (RGE)
\begin{align}
\label{eq:bBH-RGEZ}
\mu_{R}^{2} \frac{d}{d\mu_{R}^{2}}\ln Z_{a_{s}} = \frac{1}{a_{s}}
  \sum_{k=0}^{\infty} a_{s}^{k+2}\beta_{k} 
\end{align}
where, $\beta_{k}$'s are the coefficients of the QCD
$\beta$-function. The solution of the above RGE in terms of the
$\beta_{k}$'s and $\epsilon$ up to ${\cal O}(a_s^4)$comes out to be 
\begin{align}
\label{eq:bBH-Zas}
Z_{a_{s}} &= 1+ a_s \l[\frac{2}{\ep} \beta_0\right]
           + a_s^2 \left[\frac{4}{\ep^2 } \beta_0^2  + \frac{1}{\ep} \beta_1 \right]
           + a_s^3 \left[\frac{8}{ \ep^3} \beta_0^3 +\frac{14}{3
            \ep^2} \beta_0 \beta_1 + \frac{2}{3 \ep}  \beta_2 \right] 
\nn\\
          &+ a_s^4 \left[\frac{16}{\ep^{4}}\beta_{0}^{4} +
            \frac{46}{3\ep^{3}}\beta_{0}^{2}\beta_{1} +
            \frac{1}{\ep^{2}}\l(\frac{3}{2}\beta_{1}^{2} +
            \frac{10}{3}\beta_{0}\beta_{2}\r) +
            \frac{1}{2\ep}\beta_{3} \right]\,. 
\end{align}
Results beyond this order involve $\beta_{4}$ and higher order
$\beta_{k}$'s which are not available yet in the literature. The
$\beta_{k}$ up to $k=3$ are given 
by~\cite{Tarasov:1980au} 
\begin{align}
\label{eq:bBH-beta}
  \beta_0&={11 \over 3 } C_A - {2 \over 3 } n_f \, ,
           \nonumber \\[0.5ex]
  \beta_1&={34 \over 3 } C_A^2- 2 n_f C_F -{10 \over 3} n_f C_A \, ,
           \nonumber \\[0.5ex]
  \beta_2&={2857 \over 54} C_A^3 
           -{1415 \over 54} C_A^2 n_f
           +{79 \over 54} C_A n_f^2
           +{11 \over 9} C_F n_f^2
           -{205 \over 18} C_F C_A n_f
           + C_F^2 n_f\,,
\nonumber\\[0.5ex]
    \beta_3 &=
       N^2   \;\Bigg(  - \frac{40}{3} + 352 \zeta_3 \Bigg)
       + N^4   \;\Bigg(  - \frac{10}{27} + \frac{88}{9} \zeta_3 \Bigg)
       + n_f N   \;\Bigg( \frac{64}{9} - \frac{208}{3} \zeta_3 \Bigg)
\nonumber\\
&+ n_f N^3   \;\Bigg( \frac{32}{27} - \frac{104}{9} \zeta_3 \Bigg)
       + n_f^2 N^{-2}   \;\Bigg(  - \frac{44}{3} + 32 \zeta_3 \Bigg)
       + n_f^2   \;\Bigg( \frac{44}{9} - \frac{32}{3} \zeta_3 \Bigg)
\nonumber\\
&      + n_f^2 N^2   \;\Bigg(  - \frac{22}{27} + \frac{16}{9} \zeta_3 \Bigg)
       + C_F n_f^3   \;\Bigg( \frac{154}{243} \Bigg)
       + C_F^2 n_f^2   \;\Bigg( \frac{338}{27} - \frac{176}{9} \zeta_3 \Bigg)
      + C_F^3 n_f   \;\Bigg( 23 \Bigg)
\nonumber\\
&       + C_A n_f^3   \;\Bigg( \frac{53}{243} \Bigg)
       + C_A C_F n_f^2   \;\Bigg( \frac{4288}{243} + \frac{112}{9} \zeta_3 \Bigg)
       + C_A C_F^2 n_f   \;\Bigg(  - \frac{2102}{27} + \frac{176}{9} \zeta_3 \Bigg)
\nonumber\\
&       + C_A^2 n_f^2   \;\Bigg( \frac{3965}{162} + \frac{56}{9} \zeta_3 \Bigg)
       + C_A^2 C_F n_f   \;\Bigg( \frac{7073}{486} - \frac{328}{9} \zeta_3 \Bigg)
       + C_A^3 n_f   \;\Bigg(  - \frac{39143}{162} + \frac{68}{3} \zeta_3 \Bigg)
\nonumber\\
&       + C_A^4   \;\Bigg( \frac{150653}{486} - \frac{44}{9} \zeta_3 \Bigg)
\end{align}
with the SU(N) quadratic casimirs
\begin{equation}
  C_A=N,\quad \quad \quad C_F={N^2-1 \over 2 N}\,.
\end{equation}
$n_f$ is the number of active light quark flavors.

In this chapter, we will confine our discussion on the threshold
corrections to the Higgs boson production through bottom quark
annihilation and more precisely our main goal is to compute the SV
cross section of this process at N$^{3}$LO QCD. In the subsequent
sections, we will demonstrate the methodology to obtain the ingredients, Eq.~(\ref{eq:bBH-psi}) 
for computing the SV cross section of scalar Higgs boson production at
N$^3$LO QCD.

\subsection{The Form Factor}
\label{ss:bBH-FF}

The quark and gluon form factors represent the QCD loop corrections to
the transition matrix element from an on-shell quark-antiquark pair or
two gluons to a color-neutral operator $O$.  For the scalar Higgs boson
production through $b {\bar b}$ annihilation, we consider Yukawa
interaction, encapsulated through the operator ${O}^{b}$ present
in the interacting Lagrangian~\ref{eq:bBH-Lag}. For the process under
consideration, we need to consider bottom quark form factors. The
unrenormalised quark form factors at ${\cal O}({\hat a}_{s}^{n})$ 
are defined through
\begin{align}
  \label{eq:bBH-DefFb}
  {\hat{\cal F}}^{H,(n)}_{b{\bar b}} \equiv 
\frac{\langle{\hat{\cal
 M}}^{H,(0)}_{b{\bar b}}|{\hat{\cal M}}^{H,(n)}_{b{\bar b}}\rangle}{\langle{\hat{\cal
 M}}^{H,(0)}_{b{\bar b}}|{\hat{\cal M}}^{H,(0)}_{b{\bar b}}\rangle}\, ,  
\end{align}
where, $n=0, 1, 2, 3, \ldots$\,\,. In the above expressions
$|{\hat{\cal M}}^{H,(n)}_{b{\bar b}}\rangle$ is the
${\cal O}({\hat a}_{s}^{n})$ contribution to the unrenormalised matrix
element for the production of the Higgs boson from on-shell $b{\bar
  b}$ annihilation. In terms of these quantities, the full matrix element and the full
form factors can be written as a series expansion in ${\hat a}_{s}$ as
\begin{align}
  \label{eq:bBH-DefFlambda}
  |{\cal M}^{H}_{b{\bar b}}\rangle \equiv \sum_{n=0}^{\infty} {\hat
  a}^{n}_{s} S^{n}_{\epsilon}
  \left( \frac{Q^{2}}{\mu^{2}} \right)^{n\frac{\epsilon}{2}}
  |{\hat{\cal M}}^{H,(n)}_{b{\bar b}} \rangle \, , 
  \qquad \qquad 
  {\cal F}^{H}_{b{\bar b}} \equiv
  \sum_{n=0}^{\infty} \left[ {\hat a}_{s}^{n} S_{\epsilon}^{n}
  \left( \frac{Q^{2}}{\mu^{2}} \right)^{n\frac{\epsilon}{2}}
    {\hat{\cal F}}^{H,(n)}_{b{\bar b}}\right]\, ,
\end{align}
where $Q^{2}=-2\, p_{1}.p_{2}=-q^{2}$ and $p_i$ ($p_{i}^{2}=0$) are
the momenta of the external on-shell bottom quarks. The results of the
form factors up to two loop were present for a long time in~\cite{Harlander:2003ai,
  Anastasiou:2011qx} and the three loop one was
computed recently in~\cite{Gehrmann:2014vha}.

The form factor ${\cal F}^{H}_{b{\bar b}}(\hat{a}_{s}, Q^{2}, \mu^{2}, \epsilon)$
satisfies the $KG$-differential equation \cite{Sudakov:1954sw,
  Mueller:1979ih, Collins:1980ih, Sen:1981sd, Magnea:1990zb} which is
a direct consequence of the facts that QCD amplitudes exhibit
factorisation property, gauge and renormalisation group (RG)
invariances:
\begin{equation}
  \label{eq:bBH-KG}
  Q^2 \frac{d}{dQ^2} \ln {\cal F}^{H}_{b{\bar b}} (\hat{a}_s, Q^2, \mu^2, \epsilon)
  = \frac{1}{2} \left[ K^{H}_{b{\bar b}} \left(\hat{a}_s, \frac{\mu_R^2}{\mu^2}, \epsilon
    \right)  + G^{H}_{b{\bar b}} \left(\hat{a}_s, \frac{Q^2}{\mu_R^2},
      \frac{\mu_R^2}{\mu^2}, \epsilon \right) \right]\,. 
\end{equation}
In the above expression, all the poles in dimensional regularisation
parameter $\ep$ are captured in the $Q^{2}$ independent function
$K^{H}_{b{\bar b}}$ and the quantities which are finite as
$\epsilon \rightarrow 0$ are encapsulated in $G^{H}_{b{\bar b}}$. The
solution of the above $KG$ equation can be obtained as~\cite{Ravindran:2005vv} (see also
\cite{Ahmed:2014cla,Ahmed:2014cha})
\begin{align}
  \label{eq:bBH-lnFSoln}
  \ln {\cal F}^{H}_{b{\bar b}}(\hat{a}_s, Q^2, \mu^2, \epsilon) =
  \sum_{k=1}^{\infty} {\hat a}_{s}^{k}S_{\epsilon}^{k} \left(\frac{Q^{2}}{\mu^{2}}\right)^{k
  \frac{\epsilon}{2}}  \hat {\cal L}_{b{\bar b}, k}^{H}(\epsilon)
\end{align}
with
\begin{align}
  \label{eq:bBH-lnFitoCalLF}
  \hat {\cal L}_{b{\bar b},1}^{H}(\ep) =& { \frac{1}{\ep^2} } \Bigg\{-2 A^{H}_{{b{\bar b}},1}\Bigg\}
                                  + { \frac{1}{\ep}
                                  }
                                  \Bigg\{G^{H}_{{b{\bar b}},1}
                                  (\ep)\Bigg\}\, ,
                                  \nonumber\\
  \hat {\cal L}_{{b{\bar b}},2}^{H}(\ep) =& { \frac{1}{\ep^3} } \Bigg\{\beta_0 A^{H}_{{b{\bar b}},1}\Bigg\}
                                  + {
                                  \frac{1}{\ep^2} }
                                  \Bigg\{-  {
                                  \frac{1}{2} }  A^{H}_{{b{\bar b}},2}
                                  - \beta_0   G^{H}_{{b{\bar b}},1}(\ep)\Bigg\}
                                  + { \frac{1}{\ep}
                                  } \Bigg\{ {
                                  \frac{1}{2} }  G^{H}_{{b{\bar b}},2}(\ep)\Bigg\}\, ,
                                  \nonumber\\
  \hat {\cal L}_{{b{\bar b}},3}^{H}(\ep) =& { \frac{1}{\ep^4}
                                  } \Bigg\{- {
                                  \frac{8}{9} }  \beta_0^2 A^{H}_{{b{\bar b}},1}\Bigg\}
                                  + {
                                  \frac{1}{\ep^3} }
                                  \Bigg\{ { \frac{2}{9} } \beta_1 A^{H}_{{b{\bar b}},1}
                                  + { \frac{8}{9} }
                                  \beta_0 A^{H}_{{b{\bar b}},2}  + { \frac{4}{3} }
                                  \beta_0^2 G^{H}_{{b{\bar b}},1}(\ep)\Bigg\}
                                  \nonumber\\
                                &
                                  + { \frac{1}{\ep^2} } \Bigg\{- { \frac{2}{9} } A^{H}_{{b{\bar b}},3}
                                  - { \frac{1}{3} } \beta_1 G^{H}_{{b{\bar b}},1}(\ep)
                                  - { \frac{4}{3} } \beta_0 G^{H}_{{b{\bar b}},2}(\ep)\Bigg\}
                                  + { \frac{1}{\ep}
                                  } \Bigg\{  { \frac{1}{3} }
                                  G^{H}_{b{\bar b},3}(\ep)\Bigg\}\, .
\end{align}
In Appendix~\ref{chpt:App-KGSoln}, the derivation of the above
solution is discussed in great details.
$A^{H}_{b{\bar b}}$'s are called the cusp anomalous
dimensions. The constants $G^{H}_{{b{\bar b}},i}$'s are the coefficients of $a_{s}^{i}$ in the following
expansions: 
\begin{align}
  \label{eq:bBH-GandAExp}
  G^{H}_{b{\bar b}}\left(\hat{a}_s, \frac{Q^2}{\mu_R^2}, \frac{\mu_R^2}{\mu^2},
  \epsilon \right) &= G^{H}_{b{\bar b}}\left(a_{s}(Q^{2}), 1, \epsilon \right)
                     + \int_{\frac{Q^{2}}{\mu_{R}^{2}}}^{1}
                     \frac{dx}{x} A^{H}_{b{\bar b}}(a_{s}(x\mu_{R}^{2}))
                     \nonumber\\
                   &= \sum_{i=1}^{\infty}a_{s}^{i}(Q^{2})
                     G^{H}_{{b{\bar b}},i}(\epsilon) +
                     \int_{\frac{Q^{2}}{\mu_{R}^{2}}}^{1} 
                     \frac{dx}{x} A^{H}_{b{\bar b}}(a_{s}(x\mu_{R}^{2}))\,.
\end{align}
However, the solutions of the logarithm of the form factor involves
the unknown functions $G^{H}_{{b{\bar b}},i}$ which are observed to fulfill
\cite{Ravindran:2004mb, Moch:2005tm} the following decomposition in
terms of collinear ($B^{H}_{b{\bar b}}$), soft ($f^{H}_{b{\bar b}}$) and UV
($\gamma^{H}_{b{\bar b}}$) anomalous dimensions:
\begin{align}
  \label{eq:bBH-GIi}
  G^{H}_{{b{\bar b}},i} (\ep) = 2 \left(B^{H}_{{b{\bar b}},i} -
  \gamma^{H}_{{b{\bar b}},i}\right)  + f^{H}_{{b{\bar b}},i} +
  C^{H}_{{b{\bar b}},i}  + \sum_{k=1}^{\infty} \epsilon^k g^{H,k}_{{b{\bar b}},i} \, ,
\end{align}
where, the constants $C^{H}_{{b{\bar b}},i}$ are given by
\cite{Ravindran:2006cg}
\begin{align}
  \label{eq:bBH-Cg}
  C^{H}_{{b{\bar b}},1} &= 0\, ,
                \nonumber\\
  C^{H}_{{b{\bar b}},2} &= - 2 \beta_{0} g^{H,1}_{{b{\bar b}},1}\, ,
                \nonumber\\
  C^{H}_{{b{\bar b}},3} &= - 2 \beta_{1} g^{H,1}_{{b{\bar b}},1} - 2
                \beta_{0} \left(g^{H,1}_{{b{\bar b}},2}  + 2 \beta_{0} g^{H,2}_{{b{\bar b}},1}\right)\, .
\end{align}
In the above expressions, $X^{H}_{{b{\bar b}},i}$ with $X=A,B,f$ and
$\gamma^{H}_{{b{\bar b}}, i}$ are defined through the series expansion in powers
of $a_{s}$:
\begin{align}
  \label{eq:bBH-ABfgmExp}
  X^{H}_{b{\bar b}} &\equiv \sum_{i=1}^{\infty} a_{s}^{i}
              X^{H}_{{b{\bar b}},i}\,,
              \qquad \text{and} \qquad
              \gamma^{H}_{b{\bar b}} \equiv \sum_{i=1}^{\infty} a_{s}^{i} \gamma^{H}_{{b{\bar b}},i}\,\,.
\end{align}
$f_{i~{\bar i}}^{I}$ are introduced for the first time in the
article~\cite{Ravindran:2004mb} where it is shown to fulfill the
maximally non-Abelian property up to two loop level whose validity is
reconfirmed in~\cite{Moch:2005tm} at three loop level:
\begin{align}
\label{eq:bBH-MaxNAf}
f^{H}_{b{\bar b}} = \frac{C_F}{C_A} f^{H}_{gg}\,.
\end{align}
This identity implies the soft anomalous dimensions for the Higgs
boson production in bottom quark annihilation are related to the same
appearing in the Higgs boson production in gluon fusion through a
simple ratio of the quadratic casimirs of SU(N) gauge group. The same property is also obeyed by the
cusp anomalous dimensions up to three loop level:
\begin{align}
\label{eq:bBH-MaxNAA}
A^{H}_{b{\bar b}} = \frac{C_F}{C_A} A^{H}_{gg}\,.
\end{align}
It is not clear whether this nice property holds true beyond this
order of perturbation theory. Moreover, due
to universality of the quantities denoted by $X$, these are
independent of the operators insertion. These are only dependent on the
initial state partons of any process. Moreover, these are independent of
the quark flavors. Hence, being a process of 
quark annihilation, we can make use of the existing results up to
three loop which are employed in case of DY pair productions:
\begin{align}
  \label{eq:1}
  X^{H}_{b{\bar b}} = X^{DY}_{q{\bar q}} = X^{I}_{q{\bar q}}=X_{q{\bar q}}\,.
\end{align}
Here, $q$ denotes the independence of the quantities on the quark flavors and absence of $I$ represents the independence of the quantities on the nature of colorless particles.
$f^{H}_{b{\bar b}}$ can be found in \cite{Ravindran:2004mb, Moch:2005tm},
$A^{H}_{{b{\bar b}}}$ in~\cite{Moch:2004pa, Moch:2005tm, Vogt:2004mw, Vogt:2000ci}
and $B^{H}_{{b{\bar b}}}$ in \cite{Vogt:2004mw, Moch:2005tm} up to three loop
level. For readers' convenience we list them all up to three loop
level in the
Appendix~\ref{chpt:App-AnoDim}.
Utilising the results of these known quantities and comparing
the above expansions of $G^{H}_{{b{\bar b}},i}(\ep)$, Eq.~(\ref{eq:bBH-GIi}), with
the results of the logarithm of the form factors, we extract the
relevant $g_{{b{\bar b}},i}^{H,k}$ and $\gamma^{H}_{{b{\bar b}},i}$'s up to three
loop. For soft-virtual cross section at N$^{3}$LO we need
$g^{H,1}_{b{\bar b},3}$ in addition to the quantities arising from one and two
loop. The form factors for the Higgs boson production in $b{\bar b}$
annihilation up to two loop
can be found in~\cite{Harlander:2003ai,
  Ravindran:2005vv, Ravindran:2006cg} and the three loop one is
calculated very recently in the article~\cite{Gehrmann:2014vha}. 
These results are employed to extract the
required $g^{H,k}_{{b{\bar b}},i}$'s using Eq.~(\ref{eq:bBH-lnFSoln}),
(\ref{eq:bBH-lnFitoCalLF}) and (\ref{eq:bBH-GIi}). 
The relevant one loop terms are found to be 
\begin{align}
\label{eq:bBH-gk1}
 g_{b{\bar b},1}^{H,1} = C_F \Bigg\{ - 2 + \zeta_2 \Bigg\},\, \quad \quad 
 g_{b{\bar b},1}^{H,2} = C_F \Bigg\{ 2 - \frac{7}{3} \zeta_3 \Bigg\} , \,\quad \quad 
 g_{b{\bar b},1}^{H,3} = C_F \Bigg\{ - 2 + \frac{1}{4} \zeta_2 + \frac{47}{80} \zeta_2^2 \Bigg\}\,,
\end{align}
the relevant two loop terms are
\begin{align}
\label{eq:bBH-gk2}
g_{b{\bar b},2}^{H,1} &= C_F n_f \Bigg\{ \frac{616}{81} + \frac{10}{9} \zeta_2 - \frac{8}{3} \zeta_3 \Bigg\}
          + C_F C_A \Bigg\{ - \frac{2122}{81} - \frac{103}{9} \zeta_2 +
                        \frac{88}{5} {\zeta_2}^2 + \frac{152}{3}
                        \zeta_3 \Bigg\} 
\nonumber \\
          & + C_F^2 \Bigg\{ 8 + 32 \zeta_2 - \frac{88}{5} {\zeta_2}^2 - 60 \zeta_3 \Bigg\}  \,,
 \nonumber \\
g_{b{\bar b},2}^{H,2} &=
C_F n_f \Bigg\{ \frac{7}{12} {\zeta_2}^2 - \frac{55}{27} \zeta_2 +
                        \frac{130}{27} \zeta_3 - \frac{3100}{243}
                        \Bigg\} 
+ C_A C_F  \Bigg\{ - \frac{365}{24} {\zeta_2}^2 + \frac{89}{3} \zeta_2 \zeta_3 + \frac{1079}{54} \zeta_2 
\nonumber \\
& - \frac{2923}{27} \zeta_3 - 51 \zeta_5 + \frac{9142}{243} \Bigg\} 
+ C_F^2 \Bigg\{ \frac{ 96}{5} {\zeta_2}^2 - 28 \zeta_2 \zeta_3 
 - 44 \zeta_2 + 116 \zeta_3 + 12 \zeta_5 - 24 \Bigg\}
\nonumber
\end{align}
and the required three loop term is  
\begin{align}
g_{b{\bar b},3}^{H,1} &= 
C_A^2 C_F   \Bigg\{ - \frac{6152}{63} {\zeta_2}^3 + \frac{2738}{9} {\zeta_2}^2
 + \frac{976}{9} \zeta_2 \zeta_3 - \frac{342263}{486} \zeta_2
 - \frac{1136}{3} {\zeta_3}^2 + \frac{19582}{9} \zeta_3 
\nonumber \\
&
 + \frac{1228}{3} \zeta_5 
 + \frac{4095263}{8748} \Bigg\}
+ C_A C_F^2  \Bigg\{ - \frac{15448}{105} {\zeta_2}^3 - \frac{3634}{45} {\zeta_2}^2
 - \frac{2584}{3} \zeta_2 \zeta_3 + \frac{13357}{9} \zeta_2 
\nonumber \\
&
 + 296 \zeta_3^2
 - \frac{11570}{9} \zeta_3 - \frac{1940}{3} \zeta_5 - \frac{613}{3} \Bigg\}
+ C_A C_F n_f  \Bigg\{ - \frac{1064}{45} {\zeta_2}^2 + \frac{392}{9} \zeta_2 \zeta_3 
 + \frac{44551}{243} \zeta_2 
\nonumber \\
&
 - \frac{41552}{81} \zeta_3 
 - 72 \zeta_5 - \frac{6119}{4374} \Bigg\}
+ C_F^2 n_f  \Bigg\{ \frac{772}{45} {\zeta_2}^2 - \frac{152}{3} \zeta_2 \zeta_3
  - \frac{3173}{18} \zeta_2 + \frac{15956}{27} \zeta_3 
  \nonumber\\
  &-\frac{368}{3} \zeta_5
  + \frac{32899}{324}\Bigg\}
+ C_F n_f^2  \Bigg\{ - \frac{40}{9} {\zeta_2}^2 - \frac{892}{81} \zeta_2 
  + \frac{320}{81} \zeta_3 - \frac{27352}{2187} \Bigg\}
\nonumber \\
&
+ C_F^3 \Bigg\{ \frac{21584}{105} {\zeta_2}^3 - \frac{1644}{5} {\zeta_2}^2
  + 624 \zeta_2 \zeta_3 
  - 275 \zeta_2 + 48 \zeta_3^2 
  - 2142 \zeta_3 + 1272 \zeta_5 + 603 \Bigg\} \,.
\end{align}
The results up to two loop were present in the literature~\cite{Ravindran:2005vv, Ravindran:2006cg}, however the
three loop result is the new one which is computed in this thesis for
the first time. The other constants $\gamma^{H}_{b{\bar b},i}$,
appearing in the Eq.~(\ref{eq:bBH-GIi}), up to three loop
($i=3$) are obtained as 
\begin{align}
\label{eq:bBH-gamma}
\gamma^H_{b{\bar b}, 1}&= 3 C_F \, ,
\nonumber \\
\gamma^H_{b{\bar b}, 2}&= \frac{3}{2} C_F^2
           + \frac{97}{6} C_F C_A
           - \frac{5}{3} C_F n_f \, ,
\nonumber \\
\gamma^H_{b{\bar b}, 3}&= \frac{129}{2} C_F^3 
           - \frac{129}{4} C_F^2 C_A
           + \frac{11413}{108} C_F C_A^2
           +\Big(-23+24 \zeta_3\Big) C_F^2 n_f
\nonumber \\
&           +\left(-\frac{278}{27} -24 \zeta_3\right) C_F C_A n_f
           - \frac{35}{27} C_F n_f^2 \, .
\end{align}
These will be utilised in the next subsection to determine overall
operator renormalisation constant.

\subsection{Operator Renormalisation Constant}
\label{ss:bBH-OOR}

The strong coupling constant renormalisation through $Z_{a_{s}}$ is
not sufficient to make the form factor ${\cal F}^{H}_{b{\bar b}}$ completely UV
finite, one needs to perform additional renormalisation to remove the
residual UV divergences which is reflected through the presence of
non-zero $\gamma^{H}_{b{\bar b}}$ in Eq.~(\ref{eq:bBH-GIi}). This additional
renormalisation is called the overall operator renormalisation which
is performed through the constant $Z^{H}_{b{\bar b}}$. This is determined by
solving the underlying RG equation:
\begin{align}
  \label{eq:bBH-ZRGE}
  \mu_{R}^{2} \frac{d}{d\mu_{R}^{2}} \ln Z^{H}_{b{\bar b}} \left( {\hat a}_{s},
  \mu_{R}^{2}, \mu^{2}, \epsilon \right) = \sum_{i=1}^{\infty}
  a_{s}^{i}(\mu_R^2) \gamma^{H}_{b{\bar b},i}\,. 
\end{align}
Using the results of $\gamma^{H}_{b{\bar b},i}$ from Eq.~(\ref{eq:bBH-gamma}) and
solving the above RG equation following the methodology described in
the Appendix~\ref{chpt:App-SolRGEZas}, we obtain the following overall renormalisation
constant up to three loop level:
\begin{align}
\label{eq:bBH-OOR-Soln}
Z^H_{b{\bar b}} &= 1+ \sum\limits_{k=1}^{\infty} {\hat a}_s^k S_{\epsilon}^k \left( \frac{\mu_R^2}{\mu^2}
            \right)^{k\frac{\epsilon}{2}} {\hat Z}_{b{\bar b}}^{H,(k)}
\end{align}
where,
\begin{align}
\label{eq:bBH-OOR-Soln-1}
{\hat Z}_{b{\bar b}}^{H,(1)} &= \frac{1}{\epsilon}   \Bigg\{
           6 C_F  \Bigg\}\,,
\nonumber\\
{\hat Z}_{b{\bar b}}^{H,(2)} &= \frac{1}{\epsilon^2}   \Bigg\{
          - 22 C_F C_A
          + 18 C_F^2
          + 4 n_f C_F
          \Bigg\}
         + \frac{1}{\epsilon}   \Bigg\{
           \frac{97}{6} C_F C_A
          + \frac{3}{2} C_F^2
          - \frac{5}{3} n_f C_F 
          \Bigg\}\,,
\nonumber\\
{\hat Z}_{b{\bar b}}^{H,(3)} &=
\frac{1}{\epsilon^3}   \Bigg\{
           \frac{968}{9} C_F C_A^2
          - 132 C_F^2 C_A
          + 36 C_F^3
          - \frac{352}{9} n_f C_F C_A
          + 24 n_f C_F^2
          + \frac{32}{9} n_f^2 C_F
          \Bigg\}
 \nonumber\\
& 
       + \frac{1}{\epsilon^2}   \Bigg\{
          - \frac{4880}{27} C_F C_A^2
          + \frac{247}{3} C_F^2 C_A
          + 9 C_F^3
          + \frac{1396}{27} n_f C_F C_A
          - \frac{10}{3} n_f C_F^2
          - \frac{80}{27} n_f^2 C_F
          \Bigg\}
\nonumber\\
&
       + \frac{1}{\epsilon}   \Bigg\{
           \frac{11413}{162} C_F C_A^2
          - \frac{43}{2} C_F^2 C_A
          + 43 C_F^3
          - \frac{556}{81} n_f C_F C_A
          - \frac{46}{3} n_f C_F^2
          - \frac{70}{81} n_f^2 C_F
\nonumber\\
&
          - 16 \zeta_3 n_f C_F C_A
          + 16 \zeta_3 n_f C_F^2
          \Bigg\}\,.
\end{align}

\subsection{Mass Factorisation Kernel}
\label{ss:bBH-MFK}

The UV finite form factor contains additional divergences arising from
the soft and collinear regions of the loop momenta. In this section,
we address the issue of collinear divergences and describe a
prescription to remove them. The collinear singularities that arise in
the massless limit of partons are removed by absorbing the divergences
in the bare PDF through
renormalisation of the PDF. This prescription is called the mass
factorisation (MF) and is performed at the factorisation scale
$\mu_F$. In the process of performing this, one needs to introduce mass factorisation kernels
$\Gamma^I_{ij}(\hat{a}_s, \mu^2, \mu_F^2, z, \epsilon)$ which
essentially absorb the collinear singularities. More specifically, MF
removes the collinear singularities arising from the collinear
configuration associated with the initial state partons. The final
state collinear singularities are guaranteed to go away once the phase
space integrals are performed after summing over the contributions from
virtual and real emission diagrams, thanks to
Kinoshita-Lee-Nauenberg (KLN) theorem.
The kernels satisfy
the following RG equation :
\begin{align}
  \label{eq:bBH-kernelRGE}
  \mu_F^2 \frac{d}{d\mu_F^2} \Gamma^I_{ij}(z,\mu_F^2,\epsilon) =
  \frac{1}{2} \sum\limits_{k} P^I_{ik} \left(z,\mu_F^2\right) \otimes \Gamma^I_{kj} \left(z,\mu_F^2,\epsilon \right) 
\end{align}
where, $P^I\left(z,\mu_{F}^{2}\right)$ are Altarelli-Parisi splitting
functions (matrix valued). Expanding $P^{I}\left(z,\mu_{F}^{2}\right)$ and
$\Gamma^{I}(z,\mu_F^2,\epsilon)$ in powers of the strong coupling constant
we get
\begin{align}
  \label{eq:bBH-APexpand}
  &P^{I}_{ij}(z,\mu_{F}^{2}) = \sum_{k=1}^{\infty} a_{s}^{k}(\mu_{F}^{2})P^{I,(k-1)}_{ij}(z)\, 
    \intertext{and}
  &\Gamma^I_{ij}(z,\mu_F^2,\epsilon) = \delta_{ij}\delta(1-z) + \sum_{k=1}^{\infty}
    {\hat a}_{s}^{k}  S_{\ep}^{k} \l(\frac{\mu_{F}^{2}}{\mu^{2}}\r)^{k
    \frac{\ep}{2}}  {\hat \Gamma}^{I,(k)}_{ij}(z,\ep)\, .
\end{align}
The RG equation of $\Gamma^{I}(z,\mu_F^2,\epsilon)$,
Eq.~(\ref{eq:bBH-kernelRGE}), can be solved in dimensional regularisation
in powers of ${\hat a}_{s}$.  In the $\overline{MS}$ scheme, the
kernel contains only the poles in $\ep$. The solutions~\cite{Ravindran:2005vv} up to the
required order $\Gamma^{I,(3)}(z,\epsilon)$ in terms of $P^{I,(k)}(z)$
are presented in the Appendix~(\ref{eq:App-Gamma-GenSoln}). 
The relevant ones up
to three loop, $P^{I,(0)}(z), P^{I,(1)}(z) ~\text{and}~ P^{I,(2)}(z)$ are
computed in the articles~\cite{Moch:2004pa, Vogt:2004mw}. For the SV
cross section only the diagonal parts of the splitting functions
$P^{I,(k)}_{ij}(z)$ and kernels $\Gamma^{I,(k)}_{ij}(z,\ep)$
contribute since the diagonal elements of $P^{I,(k)}_{ij}(z)$ contain
$\delta(1-z)$ and ${\cal D}_{0}$ whereas the off-diagonal elements are
regular in the limit $z \rightarrow 1$.
The most remarkable fact is that these quantities are universal, independent of
the operators insertion. Hence, for the process under consideration,
we make use of the existing process independent results of kernels and splitting
functions:
\begin{align}
\label{eq:bBH-Gamma-P-ProcessInd}
\Gamma^H_{ij} = \Gamma^I_{ij} = \Gamma_{ij} \qquad \text{and} \qquad P^H_{ij} = P^I_{ij}=P_{ij}\,.
\end{align}
The absence of $I$ represents the independence of these quantities on
$I$. In the next subsection, we discuss the only remaining ingredient,
namely, the
soft-collinear distribution.

\subsection{Soft-Collinear Distribution}
\label{ss:bBH-SCD}

The resulting expression from form factor along with the operator
renormalisation constant and mass factorisation kernel is not
completely finite, it contains some residual divergences which get
cancelled against the contribution arising from soft gluon
emissions. Hence, the finiteness of $\Delta_{b{\bar b}}^{H, \sv}$ in
Eq.~(\ref{eq:bBH-sigma}) in the limit 
$\ep \rightarrow 0$ demands that the soft-collinear distribution,
$\Phi^H_{b{\bar b}} (\hat{a}_s, q^2, \mu^2, z, \epsilon)$, has pole structure
in $\ep$ similar to that of residual divergences. In
articles~\cite{Ravindran:2005vv} and \cite{Ravindran:2006cg}, it was
shown that $\Phi^{H}_{b{\bar b}}$ must obey $KG$ type integro-differential
equation, which we call ${\overline{KG}}$ equation, to remove that
residual divergences:
\begin{align}
  \label{eq:bBH-KGbarEqn}
  q^2 \frac{d}{dq^2} \Phi^H_{b{\bar b}}\left(\hat{a}_s, q^2, \mu^2, z,
    \ep\right)   = \frac{1}{2} \left[ \overline K^H_{b{\bar b}}
  \left(\hat{a}_s, \frac{\mu_R^2}{\mu^2}, z, 
      \ep \right)  + \overline G^H_{b{\bar b}} \left(\hat{a}_s,
      \frac{q^2}{\mu_R^2},  \frac{\mu_R^2}{\mu^2}, z, \ep \right) \right] \, .
\end{align}

${\overline K}^{H}_{b{\bar b}}$ and ${\overline G}^{H}_{b{\bar b}}$ play similar roles as those of $K^{H}_{b{\bar b}}$ 
and $G^{H}_{b{\bar b}}$ in Eq.~(\ref{eq:bBH-KG}), respectively. Also,
$\Phi^H_{b{\bar b}} (\hat{a}_s, q^2, \mu^2, z, \ep)$ being independent of
$\mu_{R}^{2}$ satisfy the RG equation
\begin{align}
  \label{eq:bBH-RGEphi}
  \mu_{R}^{2}\frac{d}{d\mu_{R}^{2}}\Phi^H_{b{\bar b}} (\hat{a}_s, q^2, \mu^2, z, \epsilon) = 0\, .
\end{align}
This RG invariance and the demand of cancellation of all the residual
divergences arising from ${\cal F}^H_{b{\bar b}}, Z^H_{b{\bar b}}$ and $\Gamma^H_{b{\bar b}}$
against $\Phi^{H}_{b{\bar b}}$ implies the solution of the ${\overline {KG}}$
equation as~\cite{Ravindran:2005vv, Ravindran:2006cg}
\begin{align}
  \label{eq:bBH-PhiSoln}
  \Phi^H_{b{\bar b}} (\hat{a}_s, q^2, \mu^2, z, \epsilon) &=
                                                            \Phi^H_{b{\bar
                                                            b}}
                                                            (\hat{a}_s,
                                                            q^2(1-z)^{2},
                                                            \mu^2,
                                                            \epsilon) 
                                                    \nonumber\\
                                                  &= \sum_{i=1}^{\infty}
                                                    {\hat a}_{s}^{i}  
                                                    \l(\frac{q^{2}(1-z)^{2}}{\mu^{2}}\r)^{i  
                                                    \frac{\ep}{2}}
                                                    S_{\ep}^{i}   
                                                    \l(\frac{i\ep}{1-z}\r)
                                                    {\hat
                                                    \Phi}^{H}_{b{\bar b},i}(\ep)
\end{align} 
with
\begin{align}
  \label{eq:bBH-phiHatIi}
  {\hat \Phi}^H_{b{\bar b},i}(\ep) = {\hat {\cal L}}_{b{\bar b},i}^{H}(\ep)\l(A^H_{b{\bar b},j}
  \rightarrow - A^H_{b{\bar b},j},  G^H_{b{\bar b},j}(\ep) \rightarrow
  {\overline{\cal G}}^H_{b{\bar b},j}(\ep)\r) 
\end{align}
where, ${\hat {\cal L}}_{b{\bar b},i}^{H}(\ep)$ are defined in
Eq.~(\ref{eq:bBH-lnFitoCalLF}). In
Appendix~\ref{chpt:App-Soft-Col-Dist}, the derivation of this
solution is depicted in great details.
The z-independent constants ${\overline{\cal G}}^{H}_{b{\bar b},i}(\ep)$ can
be obtained by comparing the poles as well as non-pole terms in $\ep$
of ${\hat \Phi}^{H}_{b{\bar b},i}(\ep)$ with those arising from form factor,
overall renormalisation constant and splitting functions. We find
\begin{align}
  \label{eq:bBH-calGexpans}
  \overline {\cal G}^{H}_{b{\bar b},i}(\ep)&= - f_{b{\bar b},i}^H + {\overline
                                     C}_{b{\bar b},i}^{H}  +
                                             \sum_{k=1}^\infty \ep^k
                                             {\overline {\cal
                                             G}}^{H,k}_{b{\bar b},i} \, ,
\end{align}
where,
\begin{align}
  \label{eq:bBH-overlineCiI}
  &{\overline C}_{b{\bar b},1}^{H} = 0\, ,
    \nonumber\\
  &{\overline C}_{b{\bar b},2}^{H} = -2\beta_{0}{\g}_{b{\bar b},1}^{H,1}\, ,
    \nonumber\\
  &{\overline C}_{b{\bar b},3}^{H} = -2\beta_{1}{\g}_{b{\bar b},1}^{H,1} -
    2\beta_{0}\left({\g}_{b{\bar b},2}^{H,1}  + 2\beta_{0}{\g}_{b{\bar
    b},1}^{H,2} \right)\, .
\end{align}
However, due to the universality of the soft gluon contribution,
$\Phi^{H}_{b{\bar b}}$ must be the same as that of the DY pair
production in quark annihilation since this quantity only depends on
the initial state partons, it does not depend on the final state
colorless particle:
\begin{align}
  \label{eq:PhiAgPhiHg}
  \Phi^H_{b{\bar b}} &= \Phi^{DY}_{q{\bar q}} = \Phi^I_{q{\bar q}}
                 \nonumber\\
  \text{i.e.}~~{\g}^{H,k}_{b{\bar b},i} &= {\g}^{DY,k}_{q{\bar q},i} =
                                          {\g}^{I,k}_{q{\bar q},i}\,. 
\end{align}
In the above expression, $\Phi^{I}_{q{\bar q}}$ and ${\g}^{I,k}_{q{\bar q},i}$ are written
in order to emphasise the universality of these quantities i.e.
$\Phi^{I}_{q{\bar q}}$ and ${\g}^{I,k}_{q{\bar q},i}$ can be used for
any quark annihilation process, these are independent of the operators
insertion. In the beginning, it was observed
in~\cite{Ravindran:2006cg,Ravindran:2005vv} that these quantities satisfy the maximally
non-Abelian property up to ${\cal O}(a_s^2)$: 
\begin{align}
\label{eq:bBH-Phi-MaxNA}
\Phi^{I}_{q{\bar q}} = \frac{C_F}{C_A} \Phi^I_{gg} \qquad \text{and}
  \qquad {\g}^{I,k}_{q{\bar q}, i} = \frac{C_F}{C_A} {\g}^{I,k}_{gg,i}\,.
\end{align}
Some of the relevant
constants, namely, ${\g}_{q{\bar q},1}^{I,1},{\g}_{q{\bar
    q},1}^{I,2},{\g}_{q{\bar q},2}^{I,1}$ are
computed~\cite{Ravindran:2006cg,Ravindran:2005vv} from the 
results of the explicit computations of soft gluon emissions to the DY
productions. However, to calculate the SV cross section at N$^3$LO, we
need to have the results of ${\g}_{q{\bar q},1}^{I,3},{\g}_{q{\bar
    q},2}^{I,2}$. These are obtained by employing the above
symmetry~(\ref{eq:bBH-Phi-MaxNA}). In~\cite{deFlorian:2012za}, the
soft corrections to the production
cross section of the Higgs boson through gluon fusion to ${\cal
  O}(a_s^2)$ was computed to
all orders in dimensional regularisation parameter
$\epsilon$. Utilising this all order result, we
extract ${\g}_{gg,1}^{H,3}$ and ${\g}_{gg,2}^{H,2}$. These essentially
lead us to obtain the corresponding quantities for DY production by
means of the maximally non-Abelian symmetry.  The
third order constant ${\g}_{gg,3}^{H,1}$ is extracted from the result of
SV cross section for the production of the Higgs boson at
N$^{3}$LO~\cite{Anastasiou:2014vaa}. We conjecture that the symmetry
relation~(\ref{eq:bBH-Phi-MaxNA}) holds true even at the three loop
level! Therefore, utilising that property we obtain the corresponding
quantity for the DY production, ${\g}_{q{\bar q},3}^{DY,1}$ which was
presented for the first time in the 
article~\cite{Ahmed:2014cla}. Later the result was reconfirmed through threshold
resummation in~\cite{Catani:2014uta} and
explicit computations in~\cite{Li:2014bfa}. This, in turn, establishes
our conjecture of maximally non-Abelian property at N$^3$LO. Being
flavor independent, we can employ 
all these constants to the problem under consideration. Below, we list the
relevant ones that contribute up to N$^3$LO level:
\begin{align}
\label{eq:bBH-calGres}
  {\overline {\cal G}}^{H,1}_{b{\bar b},1} &= C_F \Bigg\{ - 3 \zeta_2 \Bigg\} \,, 
\nonumber\\
  {\overline {\cal G}}^{H,2}_{b{\bar b},1} &= C_F \Bigg\{ \frac{7}{3}
                                             \zeta_3 \Bigg\} \,, 
\nonumber\\
  {\overline {\cal G}}^{H,3}_{b{\bar b},1} &=  C_F \Bigg\{ - \frac{3}{16}
                                             {\zeta_2}^2 \Bigg\} \,, 
\nonumber\\
  {\overline {\cal G}}^{H,1}_{b{\bar b},2} &=  C_F n_f  \Bigg\{ -
                                             \frac{328}{81} +
                                             \frac{70}{9} \zeta_2 +
                                             \frac{32}{3} \zeta_3
                                             \Bigg\} 
             + C_A C_F  \Bigg\{ \frac{2428}{81} - \frac{469}{9} \zeta_2 
                       + 4 {\zeta_2}^2 - \frac{176}{3} \zeta_3 \Bigg\}
                                             \,, 
\nonumber\\
  {\overline {\cal G}}^{H,2}_{b{\bar b},2} &=  C_A C_F \Bigg\{
                                             \frac{11}{40} {\zeta_2}^2
                                             - \frac{203}{3} {\zeta_2}
                                             {\zeta_3} 
             + \frac{1414}{27} {\zeta_2} + \frac{2077}{27} {\zeta_3} +
                                             43 {\zeta_5} -
                                             \frac{7288}{243}  \Bigg\}
\nonumber\\
           & + C_F n_f \Bigg\{ -\frac{1}{20} {\zeta_2}^2 -
             \frac{196}{27} {\zeta_2} - \frac{310}{27} {\zeta_3} +
             \frac{976}{243} \Bigg\} 
\nonumber\\
{\overline {\cal G}}^{H,1}_{b{\bar b},3} &= 
C_F  {C_A}^2 \Bigg\{\frac{152}{63} \;{\zeta_2}^3 + \frac{1964}{9} \;{\zeta_2}^2
+ \frac{11000}{9} \;{\zeta_2} {\zeta_3} - \frac{765127}{486} \;{\zeta_2}
+\frac{536}{3} \;{\zeta_3}^2 - \frac{59648}{27} \;{\zeta_3} 
\nonumber\\
&
- \frac{1430}{3} \;{\zeta_5}
+\frac{7135981}{8748}\Bigg\}
+ C_F {C_A} {n_f} \
\Bigg\{-\frac{532}{9} \;{\zeta_2}^2 - \frac{1208}{9} \;{\zeta_2} {\zeta_3}
+\frac{105059}{243} \;{\zeta_2} 
\nonumber\\
&+ \frac{45956}{81} \;{\zeta_3} 
+\frac{148}{3} \;{\zeta_5} - \frac{716509}{4374} \Bigg\}
+ {C_F^2} {n_f} \
\Bigg\{\frac{152}{15} \;{\zeta_2}^2 
- 88 \;{\zeta_2} {\zeta_3} 
+\frac{605}{6} \;{\zeta_2} + \frac{2536}{27} \;{\zeta_3}
\nonumber\\
&+\frac{112}{3} \;{\zeta_5} 
- \frac{42727}{324}\Bigg\}
+ C_F {n_f}^2 
\Bigg\{ \frac{32}{9} \;{\zeta_2}^2 - \frac{1996}{81} \;{\zeta_2}
-\frac{2720}{81} \;{\zeta_3} + \frac{11584}{2187}\Bigg\}  \,.
\end{align}
The above ${\g}^{H,k}_{b{\bar b},i}$ enable us to get the
$\Phi^{H}_{b{\bar b}}$
up to three loop level. This completes all the ingredients required to
compute the SV cross section up to N$^{3}$LO that are presented in the
next section.

\section{Results of the SV Cross Sections}
\label{sec:bBH-Res}

In this section, we present our findings of the SV cross section at
N$^{3}$LO along with the results of previous orders. Expanding the SV
cross section $\Delta^{H, {\rm SV}}_{b{\bar b}}$, Eq.~(\ref{eq:bBH-sigma}), in
powers of $a_{s}(\mu_F^2)$, we obtain
\begin{align}
  \label{eq:bBH-SVRenExp}
  \Delta_{b{\bar b}}^{H, {\rm SV}}(z, q^{2}, \mu_{F}^{2}) =
 \sum_{i=1}^\infty a_s^i(\mu_F^2) \Delta_{b{\bar b},i}^{H, {\rm SV}} (z, q^{2}, \mu_{F}^{2}) 
\end{align}
where,
\begin{align}
  \nn
  \Delta_{b{\bar b},i}^{H, {\rm SV}} =
  \Delta_{b{\bar b},i}^{H, {\rm SV}}|_\delta
  \delta(1-z) 
  + \sum_{j=0}^{2i-1} 
  \Delta_{b{\bar b},i}^{H, {\rm SV}}|_{{\cal D}_j}
  {\cal D}_j \, .
\end{align}
Before presenting the final result, we present the general
results of the SV cross section 
in terms of the anomalous 
dimensions $A^H_{b{\bar b}}$, $B^H_{b{\bar b}}$, $f^H_{b{\bar
    b}}$, $\gamma^{H}_{b{\bar b}}$ and other quantities arising from form
factor and soft-collinear distribution below:

Upon substituting the values of all the anomalous dimensions, beta
functions and
$g^{H,k}_{b{\bar b},i}$, $\overline{\cal G}^{H,k}_{b{\bar b},i}$,
we obtain the results of the scalar Higgs boson production cross
section at threshold in
$b{\bar b}$ annihilation  up to
N$^{3}$LO for the choices of the scale
$\mu_{R}^{2}=\mu_{F}^{2}$:
\begin{align}
\label{eq:bBH-SV-Res}
\Delta_{b{\bar b},1}^{H, {\rm SV}} &= \delta(1-z) \Bigg[ \dis{C_{F}}  \Bigg\{
          - 4
          + 8 \zeta_2
          \Bigg\} \Bigg]
       + {\cal D}_0 \Bigg[ \dis{C_F}     \Bigg\{
          8 \log\l(\frac{q^2}{\mu_F^2}\r)
          \Bigg\}
       + {\cal D}_1 \Bigg[ \dis{C_F}   \Bigg\{
          16
          \Bigg\} \Bigg]\,,
\nonumber\\
\Delta_{b{\bar b},2}^{H, {\rm SV}} &=   
       \delta(1-z) \Bigg[ \dis{C_F} \dis{C_A}   \Bigg\{
           \frac{166}{9}
          - 8 \zeta_3
          + \frac{232}{9} \zeta_2
          - \frac{12}{5} \zeta_2^2
          \Bigg\}
       +  \dis{C_F^2}   \Bigg\{
           16
          - 60 \zeta_3
          + \frac{8}{5} \zeta_2^2
          \Bigg\}
\nonumber\\
&       +  \dis{n_f} \dis{C_F}   \Bigg\{
           \frac{8}{9}
          + 8 \zeta_3
          - \frac{40}{9} \zeta_2
          \Bigg\}
       +  \log\l(\frac{q^2}{\mu_F^2}\r) \dis{C_F} \dis{C_A}   \Bigg\{
          - 12
          - 24 \zeta_3
          \Bigg\}
\nonumber\\
&       +  \log\l(\frac{q^2}{\mu_F^2}\r) \dis{C_F^2}   \Bigg\{
           176 \zeta_3
          - 24 \zeta_2
          \Bigg\}
       +  \log^2\l(\frac{q^2}{\mu_F^2}\r) \dis{C_F^2}   \Bigg\{
          - 32 \zeta_2
          \Bigg\} \Bigg]
\nonumber\\
&       + {\cal D}_0 \Bigg[ \dis{C_F} \dis{C_A}   \Bigg\{
          - \frac{1616}{27}
          + 56 \zeta_3
          + \frac{176}{3} \zeta_2
          \Bigg\}
       +  \dis{C_F^2}   \Bigg\{
           256 \zeta_3
          \Bigg\}
\nonumber\\
&       +  \dis{n_f} \dis{C_F}   \Bigg\{
           \frac{224}{27}
          - \frac{32}{3} \zeta_2
          \Bigg\}
       +  \log\l(\frac{q^2}{\mu_F^2}\r) \dis{C_F} \dis{C_A}   \Bigg\{
           \frac{536}{9}
          - 16 \zeta_2
          \Bigg\}
\nonumber\\
&       +  \log\l(\frac{q^2}{\mu_F^2}\r) \dis{C_F^2}   \Bigg\{
          - 32
          - 64 \zeta_2
          \Bigg\}
       +  \log\l(\frac{q^2}{\mu_F^2}\r) \dis{n_f} \dis{C_F}   \Bigg\{
          - \frac{80}{9}
          \Bigg\}
 \nonumber\\
&      +  \log^2\l(\frac{q^2}{\mu_F^2}\r) \dis{C_F} \dis{C_A}   \Bigg\{
          - \frac{44}{3}
          \Bigg\}
       +  \log^2\l(\frac{q^2}{\mu_F^2}\r) \dis{n_f} \dis{C_F}   \Bigg\{
           \frac{8}{3}
          \Bigg\} \Bigg]
\nonumber\\
&       + {\cal D}_1 \Bigg[ \dis{C_F} \dis{C_A}   \Bigg\{
           \frac{1072}{9}
          - 32 \zeta_2
          \Bigg\}
       +  \dis{C_F^2}   \Bigg\{
          - 64
          - 128 \zeta_2
          \Bigg\}
       +  \dis{n_f} \dis{C_F}   \Bigg\{
          - \frac{160}{9}
          \Bigg\}
\nonumber\\
&       +  \log\l(\frac{q^2}{\mu_F^2}\r) \dis{C_F} \dis{C_A}   \Bigg\{
          - \frac{176}{3}
          \Bigg\}
       +  \log\l(\frac{q^2}{\mu_F^2}\r) \dis{n_f} \dis{C_F}   \Bigg\{
          + \frac{32}{3}
          \Bigg\}
       +  \log^2\l(\frac{q^2}{\mu_F^2}\r) \dis{C_F^2}   \Bigg\{
           64
          \Bigg\} \Bigg]
\nonumber\\
&       + {\cal D}_2 \Bigg[ \dis{C_F} \dis{C_A}   \Bigg\{
          - \frac{176}{3}
          \Bigg\}
       +  \dis{n_f} \dis{C_F}   \Bigg\{
           \frac{32}{3}
          \Bigg\}
       + \log\l(\frac{q^2}{\mu_F^2}\r) \dis{C_F^2}   \Bigg\{
           192
          \Bigg\} \Bigg]
\nonumber\\
&       + {\cal D}_3 \Bigg[ \dis{C_F^2}   \Bigg\{
           128
          \Bigg\} \Bigg]\,,
\nonumber\\
\Delta_{b{\bar b},3}^{H, {\rm SV}} &=   
        \delta(1-z) \Bigg[ \dis{C_F} \dis{C_A^2}   \Bigg\{
           \frac{68990}{81}
          - 84 \zeta_5
          - \frac{14212}{81} \zeta_3
          - \frac{400}{3} \zeta_3^2
          - 272 \zeta_2
          - \frac{1064}{3} \zeta_2 \zeta_3
\nonumber\\
&          + \frac{2528}{27} \zeta_2^2
          + \frac{13264}{315} \zeta_2^3
          \Bigg\}
       +  \dis{C_F^2} \dis{C_A}   \Bigg\{
          - \frac{982}{3}
          - \frac{37144}{9} \zeta_5
          - \frac{10940}{9} \zeta_3
          + \frac{3280}{3} \zeta_3^2
\nonumber\\
&          + \frac{22106}{27} \zeta_2
          + \frac{27872}{9} \zeta_2 \zeta_3
          - \frac{62468}{135} \zeta_2^2
          - \frac{20816}{315} \zeta_2^3
          \Bigg\}
       +  \dis{C_F^3}   \Bigg\{
            \frac{1078}{3}
          + 848 \zeta_5
\nonumber\\
&          - 1188 \zeta_3
          + \frac{10336}{3} \zeta_3^2
          - \frac{550}{3} \zeta_2
          - 64 \zeta_2 \zeta_3
          + \frac{152}{5} \zeta_2^2
          - \frac{184736}{315} \zeta_2^3
          \Bigg\}
\nonumber\\
&       +  \dis{n_f} \dis{C_F} \dis{C_A}   \Bigg\{
          - \frac{11540}{81}
          - 8 \zeta_5
          + \frac{2552}{81} \zeta_3
          + \frac{3368}{81} \zeta_2
          + \frac{208}{3} \zeta_2 \zeta_3
          - \frac{6728}{135} \zeta_2^2
          \Bigg\}
\nonumber\\
&       +  \dis{n_f} \dis{C_F^2}   \Bigg\{
          - \frac{70}{9}
          + \frac{5536}{9} \zeta_5
          + \frac{4088}{9} \zeta_3
          - \frac{2600}{27} \zeta_2
          - \frac{5504}{9} \zeta_2 \zeta_3
          + \frac{12152}{135} \zeta_2^2
          \Bigg\}
\nonumber\\
&       +  \dis{n_f^2} \dis{C_F}   \Bigg\{
            \frac{16}{27}
          - \frac{1120}{81} \zeta_3
          - \frac{32}{81} \zeta_2
          + \frac{128}{27} \zeta_2^2
          \Bigg\}
       +  \log\l(\frac{q^2}{\mu_F^2}\r) \dis{C_F} \dis{C_A^2}   \Bigg\{
          - \frac{1180}{3}
          + 80 \zeta_5
\nonumber\\
&          - \frac{2576}{9} \zeta_3
          + \frac{160}{3} \zeta_2
          + \frac{68}{5} \zeta_2^2
          \Bigg\}
       +  \log\l(\frac{q^2}{\mu_F^2}\r) \dis{C_F^2} \dis{C_A}   \Bigg\{
           \frac{388}{3}
          + 240 \zeta_5
          + \frac{27040}{9} \zeta_3
\nonumber\\
&          - \frac{3380}{27} \zeta_2
          - 1120 \zeta_2 \zeta_3
          - \frac{8}{3} \zeta_2^2
          \Bigg\}
       +  \log\l(\frac{q^2}{\mu_F^2}\r) \dis{C_F^3}   \Bigg\{
          - 100
          + 5664 \zeta_5
          - 568 \zeta_3
          + 132 \zeta_2
\nonumber\\
&          - 2752 \zeta_2 \zeta_3
          - \frac{384}{5} \zeta_2^2
          \Bigg\}
       +  \log\l(\frac{q^2}{\mu_F^2}\r) \dis{n_f} \dis{C_F} \dis{C_A}   \Bigg\{
           \frac{196}{3}
          + \frac{208}{9} \zeta_3
          - \frac{16}{3} \zeta_2
          - \frac{8}{5} \zeta_2^2
          \Bigg\}
\nonumber\\
&       +  \log\l(\frac{q^2}{\mu_F^2}\r) \dis{n_f} \dis{C_F^2}   \Bigg\{
           \frac{8}{3}
          - \frac{4528}{9} \zeta_3
          + \frac{152}{27} \zeta_2
          + \frac{112}{15} \zeta_2^2
          \Bigg\}
\nonumber\\
&       +  \log\l(\frac{q^2}{\mu_F^2}\r) \dis{n_f^2} \dis{C_F}   \Bigg\{
            \frac{64}{9} \zeta_3
          \Bigg\}
       +  \log^2\l(\frac{q^2}{\mu_F^2}\r) \dis{C_F} \dis{C_A^2}   \Bigg\{
           44
          + 88 \zeta_3
          \Bigg\}
\nonumber\\
&       +  \log^2\l(\frac{q^2}{\mu_F^2}\r) \dis{C_F^2} \dis{C_A}   \Bigg\{
          - 880 \zeta_3
          - \frac{3496}{9} \zeta_2
          + 128 \zeta_2^2
          \Bigg\}
       +  \log^2\l(\frac{q^2}{\mu_F^2}\r) \dis{C_F^3}   \Bigg\{
           128 \zeta_2
\nonumber\\
&          - \frac{1792}{5} \zeta_2^2
          \Bigg\}
       +  \log^2\l(\frac{q^2}{\mu_F^2}\r) \dis{n_f} \dis{C_F} \dis{C_A}   \Bigg\{
          - 8
          - 16 \zeta_3
          \Bigg\}
       +  \log^2\l(\frac{q^2}{\mu_F^2}\r) \dis{n_f} \dis{C_F^2}   \Bigg\{
           160 \zeta_3
\nonumber\\
&          + \frac{496}{9} \zeta_2
          \Bigg\}
       +  \log^3\l(\frac{q^2}{\mu_F^2}\r) \dis{C_F^2} \dis{C_A}   \Bigg\{
           \frac{352}{3} \zeta_2
          \Bigg\}
       +  \log^3\l(\frac{q^2}{\mu_F^2}\r) \dis{C_F^3}   \Bigg\{
           \frac{512}{3} \zeta_3
          \Bigg\}
\nonumber\\
&       +  \log^3\l(\frac{q^2}{\mu_F^2}\r) \dis{n_f} \dis{C_F^2}   \Bigg\{
          - \frac{64}{3} \zeta_2
          \Bigg\} 
\Bigg]
       + {\cal D}_0 \Bigg[ \dis{C_F} \dis{C_A^2}   \Bigg\{
          - \frac{594058}{729}
          - 384 \zeta_5
          + \frac{40144}{27} \zeta_3
\nonumber\\
&          + \frac{98224}{81} \zeta_2
          - \frac{352}{3} \zeta_2 \zeta_3
          - \frac{2992}{15} \zeta_2^2
          \Bigg\}
       +  \dis{C_F^2} \dis{C_A}   \Bigg\{
            \frac{6464}{27}
          + \frac{32288}{9} \zeta_3
          + \frac{6592}{27} \zeta_2
\nonumber\\
&          - 1472 \zeta_2 \zeta_3
          + \frac{1408}{3} \zeta_2^2
          \Bigg\}
       +  \dis{C_F^3}   \Bigg\{
           12288 \zeta_5
          - 1024 \zeta_3
          - 6144 \zeta_2 \zeta_3
          \Bigg\}
\nonumber\\
&       +  \dis{n_f} \dis{C_F} \dis{C_A}   \Bigg\{
           \frac{125252}{729}
          - \frac{2480}{9} \zeta_3
          - \frac{29392}{81} \zeta_2
          + \frac{736}{15} \zeta_2^2
          \Bigg\}
       +  \dis{n_f} \dis{C_F^2}   \Bigg\{
           \frac{842}{9}
\nonumber\\
&          - \frac{5728}{9} \zeta_3
          - \frac{1504}{27} \zeta_2
          - \frac{1472}{15} \zeta_2^2
          \Bigg\}
       +  \dis{n_f^2} \dis{C_F}   \Bigg\{
          - \frac{3712}{729}
          + \frac{320}{27} \zeta_3
          + \frac{640}{27} \zeta_2
          \Bigg\}
\nonumber\\
&       +  \log\l(\frac{q^2}{\mu_F^2}\r) \dis{C_F} \dis{C_A^2}   \Bigg\{
           \frac{62012}{81}
          - 352 \zeta_3
          - \frac{6016}{9} \zeta_2
          + \frac{352}{5} \zeta_2^2
          \Bigg\}
\nonumber\\
&       +  \log\l(\frac{q^2}{\mu_F^2}\r) \dis{C_F^2} \dis{C_A}   \Bigg\{
          - \frac{272}{3}
          - 2880 \zeta_3
          - \frac{10432}{9} \zeta_2
          + \frac{1824}{5} \zeta_2^2
          \Bigg\}
\nonumber\\
&       +  \log\l(\frac{q^2}{\mu_F^2}\r) \dis{C_F^3}   \Bigg\{
           128
          - 480 \zeta_3
          + 512 \zeta_2
          - \frac{7104}{5} \zeta_2^2
          \Bigg\}
\nonumber\\
&       +  \log\l(\frac{q^2}{\mu_F^2}\r) \dis{n_f} \dis{C_F} \dis{C_A}   \Bigg\{
          - \frac{16408}{81}
          + 192 \zeta_2
          \Bigg\}
       +  \log\l(\frac{q^2}{\mu_F^2}\r) \dis{n_f} \dis{C_F^2}   \Bigg\{
          - \frac{92}{3}
          + 640 \zeta_3
\nonumber\\
&          + \frac{1600}{9} \zeta_2
          \Bigg\}
       +  \log\l(\frac{q^2}{\mu_F^2}\r) \dis{n_f^2} \dis{C_F}   \Bigg\{
           \frac{800}{81}
          - \frac{128}{9} \zeta_2
          \Bigg\}
       +  \log^2\l(\frac{q^2}{\mu_F^2}\r) \dis{C_F} \dis{C_A^2}   \Bigg\{
          - \frac{7120}{27}
\nonumber\\
&          + \frac{176}{3} \zeta_2
          \Bigg\}
       +  \log^2\l(\frac{q^2}{\mu_F^2}\r) \dis{C_F^2} \dis{C_A}   \Bigg\{
          - \frac{112}{3}
          - 192 \zeta_3
          + \frac{1760}{3} \zeta_2
          \Bigg\}
\nonumber\\
&       +  \log^2\l(\frac{q^2}{\mu_F^2}\r) \dis{C_F^3}   \Bigg\{
           2432 \zeta_3
          - 192 \zeta_2
          \Bigg\}
       +  \log^2\l(\frac{q^2}{\mu_F^2}\r) \dis{n_f} \dis{C_F} \dis{C_A}   \Bigg\{
           \frac{2312}{27}
          - \frac{32}{3} \zeta_2
          \Bigg\}
\nonumber\\
&       +  \log^2\l(\frac{q^2}{\mu_F^2}\r) \dis{n_f} \dis{C_F^2}   \Bigg\{
          - \frac{8}{3}
          - \frac{320}{3} \zeta_2
          \Bigg\}
       +  \log^2\l(\frac{q^2}{\mu_F^2}\r) \dis{n_f^2} \dis{C_F}   \Bigg\{
          - \frac{160}{27}
          \Bigg\}
\nonumber\\
&       +  \log^3\l(\frac{q^2}{\mu_F^2}\r) \dis{C_F} \dis{C_A^2}   \Bigg\{
           \frac{968}{27}
          \Bigg\}
       +  \log^3\l(\frac{q^2}{\mu_F^2}\r) \dis{C_F^3}   \Bigg\{
          - 256 \zeta_2
          \Bigg\}
\nonumber\\
&       +  \log^3\l(\frac{q^2}{\mu_F^2}\r) \dis{n_f} \dis{C_F} \dis{C_A}   \Bigg\{
          - \frac{352}{27}
          \Bigg\}
       +  \log^3\l(\frac{q^2}{\mu_F^2}\r) \dis{n_f^2} \dis{C_F}   \Bigg\{
           \frac{32}{27}
          \Bigg\}
\Bigg]
\nonumber\\
&       + {\cal D}_1 \Bigg[ \dis{C_F} \dis{C_A^2}   \Bigg\{
           \frac{124024}{81}
          - 704 \zeta_3
          - \frac{12032}{9} \zeta_2
          + \frac{704}{5} \zeta_2^2
          \Bigg\}
       +  \dis{C_F^2} \dis{C_A}   \Bigg\{
          - \frac{544}{3}
          - 5760 \zeta_3
\nonumber\\
&          - \frac{20864}{9} \zeta_2
          + \frac{3648}{5} \zeta_2^2
          \Bigg\}
       +  \dis{C_F^3}   \Bigg\{
           256
          - 960 \zeta_3
          + 1024 \zeta_2
          - \frac{14208}{5} \zeta_2^2
          \Bigg\}
\nonumber\\
&       +  \dis{n_f} \dis{C_F} \dis{C_A}   \Bigg\{
          - \frac{32816}{81}
          + 384 \zeta_2
          \Bigg\}
       +  \dis{n_f} \dis{C_F^2}   \Bigg\{
          - \frac{184}{3}
          + 1280 \zeta_3
          + \frac{3200}{9} \zeta_2
          \Bigg\}
\nonumber\\
&       +  \dis{n_f^2} \dis{C_F}   \Bigg\{
           \frac{1600}{81}
          - \frac{256}{9} \zeta_2
          \Bigg\}
       +  \log\l(\frac{q^2}{\mu_F^2}\r) \dis{C_F} \dis{C_A^2}   \Bigg\{
          - \frac{28480}{27}
          + \frac{704}{3} \zeta_2
          \Bigg\}
\nonumber\\
&       +  \log\l(\frac{q^2}{\mu_F^2}\r) \dis{C_F^2} \dis{C_A}   \Bigg\{
          - \frac{24704}{27}
          + 512 \zeta_3
          + \frac{9856}{3} \zeta_2
          \Bigg\}
       +  \log\l(\frac{q^2}{\mu_F^2}\r) \dis{C_F^3}   \Bigg\{
           11008 \zeta_3
\nonumber\\
&          - 384 \zeta_2
          \Bigg\}
       +  \log\l(\frac{q^2}{\mu_F^2}\r) \dis{n_f} \dis{C_F} \dis{C_A}   \Bigg\{
          + \frac{9248}{27}
          - \frac{128}{3} \zeta_2
          \Bigg\}
       +  \log\l(\frac{q^2}{\mu_F^2}\r) \dis{n_f} \dis{C_F^2}   \Bigg\{
           \frac{3296}{27}
\nonumber\\
&          - \frac{1792}{3} \zeta_2
          \Bigg\}
       +  \log\l(\frac{q^2}{\mu_F^2}\r) \dis{n_f^2} \dis{C_F}   \Bigg\{
          - \frac{640}{27}
          \Bigg\}
       +  \log^2\l(\frac{q^2}{\mu_F^2}\r) \dis{C_F} \dis{C_A^2}   \Bigg\{
           \frac{1936}{9}
          \Bigg\}
\nonumber\\
&       +  \log^2\l(\frac{q^2}{\mu_F^2}\r) \dis{C_F^2} \dis{C_A}   \Bigg\{
           \frac{8576}{9}
          - 256 \zeta_2
          \Bigg\}
       +  \log^2\l(\frac{q^2}{\mu_F^2}\r) \dis{C_F^3}   \Bigg\{
          - 256
          - 2048 \zeta_2
          \Bigg\}
\nonumber\\
&       +  \log^2\l(\frac{q^2}{\mu_F^2}\r) \dis{n_f} \dis{C_F} \dis{C_A}   \Bigg\{
          - \frac{704}{9}
          \Bigg\}
       +  \log^2\l(\frac{q^2}{\mu_F^2}\r) \dis{n_f} \dis{C_F^2}   \Bigg\{
          - \frac{1280}{9}
          \Bigg\}
\nonumber\\
&       +  \log^2\l(\frac{q^2}{\mu_F^2}\r) \dis{n_f^2} \dis{C_F}   \Bigg\{
           \frac{64}{9}
          \Bigg\}
       +  \log^3\l(\frac{q^2}{\mu_F^2}\r) \dis{C_F^2} \dis{C_A}   \Bigg\{
          - \frac{704}{3}
          \Bigg\}
\nonumber\\
&       +  \log^3\l(\frac{q^2}{\mu_F^2}\r) \dis{n_f} \dis{C_F^2}   \Bigg\{
            \frac{128}{3}
          \Bigg\}
\Bigg]
       + {\cal D}_2 \Bigg[ \dis{C_F} \dis{C_A^2}   \Bigg\{
          - \frac{28480}{27}
          + \frac{704}{3} \zeta_2
          \Bigg\}
\nonumber\\
&       +  \dis{C_F^2} \dis{C_A}   \Bigg\{
          - \frac{10816}{9}
          + 1344 \zeta_3
          + \frac{11264}{3} \zeta_2
          \Bigg\}
       +  \dis{C_F^3}   \Bigg\{
           10240 \zeta_3
          \Bigg\}
\nonumber\\
&       +  \dis{n_f} \dis{C_F} \dis{C_A}   \Bigg\{
           \frac{9248}{27}
          - \frac{128}{3} \zeta_2
          \Bigg\}
       +  \dis{n_f} \dis{C_F^2}   \Bigg\{
           \frac{1696}{9}
          - \frac{2048}{3} \zeta_2
          \Bigg\}
       +  \dis{n_f^2} \dis{C_F}   \Bigg\{
          - \frac{640}{27}
          \Bigg\}
\nonumber\\
&       +  \log\l(\frac{q^2}{\mu_F^2}\r) \dis{C_F} \dis{C_A^2}   \Bigg\{
           \frac{3872}{9}
          \Bigg\}
       +  \log\l(\frac{q^2}{\mu_F^2}\r) \dis{C_F^2} \dis{C_A}   \Bigg\{
           \frac{8576}{3}
          - 768 \zeta_2
          \Bigg\}
\nonumber\\
&       +  \log\l(\frac{q^2}{\mu_F^2}\r) \dis{C_F^3}   \Bigg\{
          - 768
          - 4608 \zeta_2
          \Bigg\}
       +  \log\l(\frac{q^2}{\mu_F^2}\r) \dis{n_f} \dis{C_F} \dis{C_A}   \Bigg\{
          - \frac{1408}{9}
          \Bigg\}
\nonumber\\
&       +  \log\l(\frac{q^2}{\mu_F^2}\r) \dis{n_f} \dis{C_F^2}   \Bigg\{
          - \frac{1280}{3}
          \Bigg\}
       +  \log\l(\frac{q^2}{\mu_F^2}\r) \dis{n_f^2} \dis{C_F}   \Bigg\{
           \frac{128}{9}
          \Bigg\}
\nonumber\\
&       +  \log^2\l(\frac{q^2}{\mu_F^2}\r) \dis{C_F^2} \dis{C_A}   \Bigg\{
          - 1056
          \Bigg\}
       +  \log^2\l(\frac{q^2}{\mu_F^2}\r) \dis{n_f} \dis{C_F^2}   \Bigg\{
           192
          \Bigg\}
       +  \log^3\l(\frac{q^2}{\mu_F^2}\r) \dis{C_F^3}   \Bigg\{
           256
          \Bigg\} \Bigg]
\nonumber\\
&       + {\cal D}_3 \Bigg[ \dis{C_F} \dis{C_A^2}   \Bigg\{
           \frac{7744}{27}
          \Bigg\}
       +  \dis{C_F^2} \dis{C_A}   \Bigg\{
           \frac{17152}{9}
          - 512 \zeta_2
          \Bigg\}
       +  \dis{C_F^3}   \Bigg\{
          - 512
          - 3072 \zeta_2
          \Bigg\}
\nonumber\\
&       +  \dis{n_f} \dis{C_F} \dis{C_A}   \Bigg\{
          - \frac{2816}{27}
          \Bigg\}
       +  \dis{n_f} \dis{C_F^2}   \Bigg\{
          - \frac{2560}{9}
          \Bigg\}
       +  \dis{n_f^2} \dis{C_F}   \Bigg\{
           \frac{256}{27}
          \Bigg\}
\nonumber\\
&       +  \log\l(\frac{q^2}{\mu_F^2}\r) \dis{C_F^2} \dis{C_A}   \Bigg\{
          - \frac{14080}{9}
          \Bigg\}
       +  \log\l(\frac{q^2}{\mu_F^2}\r) \dis{n_f} \dis{C_F^2}   \Bigg\{
           \frac{2560}{9}
          \Bigg\}
\nonumber\\
&       +  \log^2\l(\frac{q^2}{\mu_F^2}\r) \dis{C_F^3}   \Bigg\{
           1024
          \Bigg\} \Bigg]
       + {\cal D}_4 \Bigg[ \dis{C_F^2} \dis{C_A}   \Bigg\{
          - \frac{7040}{9}
          \Bigg\}
       +  \dis{n_f} \dis{C_F^2}   \Bigg\{
           \frac{1280}{9}
          \Bigg\}
\nonumber\\
&       +  \log\l(\frac{q^2}{\mu_F^2}\r) \dis{C_F^3}   \Bigg\{
           1280
          \Bigg\} \Bigg]
       + {\cal D}_5 \Bigg[ \dis{C_F^3}   \Bigg\{
           512
          \Bigg\} \Bigg] \,.
\end{align}
The results at NLO $\l(\Delta^{H,{\rm SV}}_{b{\bar b},1}\r)$ and NNLO
$\l(\Delta^{H,{\rm SV}}_{b{\bar b},2}\r)$ match with the existing
ones~\cite{Harlander:2003ai}. At N$^3$LO level, only $\Delta^{H,{\rm
    SV}}_{b{\bar b},3}|_{{\cal D}_{j}}$ were
known~\cite{Ravindran:2005vv, Ravindran:2006cg}, remaining terms were
not available due to absence of the required quantities
$g^{H,2}_{b{\bar b},2}$, $g^{H,1}_{b{\bar b},3}$ from form factors and
$\overline{\cal G}^{H,2}_{b{\bar b},2}$, $\overline{\cal
  G}^{H,1}_{b{\bar b},3}$ from soft-collinear distributions. The
recent results of $g^{H,2}_{b{\bar b},2}$, $g^{H,1}_{b{\bar b},3}$
from~\cite{Gehrmann:2014vha}, $\overline{\cal G}^{H,2}_{b{\bar b},2}$
from~\cite{deFlorian:2012za} and $\overline{\cal G}^{H,1}_{b{\bar
    b},3}$ from~\cite{Ahmed:2014cla} are being employed to compute the
missing $\delta(1-z)$ part i.e. $\Delta^{H,{\rm SV}}_{b{\bar
    b},3}|_{\delta}$ which completes the full evaluation of the SV
cross section at N$^3$LO $\l(\Delta^{H,{\rm 
    SV}}_{b{\bar b},3}\r)$ and is presented for the first time
in~\cite{Ahmed:2014cha} by us. For the sake of 
completeness, we mention the leading order contribution which is
\begin{align}
\label{eq:bBH-Leading-Order}
\Delta^{H}_{b{\bar b},0} = \delta(1-z)
\end{align}
and the overall factor in Eq.~(\ref{eq:bBH-1}) comes out to be
\begin{align}
\label{eq:bBH-Leading-Order-1}
\sigma^{H,(0)}_{b{\bar b}} \l( \mu_F^2\r) = \frac{\pi \lambda^2\l(\mu_F^2\r)}{12 m_H^2}\,.
\end{align}
The above results are presented for the choice $\mu_R=\mu_F$. The
dependence of the SV cross section on renormalisation scale $\mu_{R}$
can be easily restored by employing the RG evolution of $a_s$ from
$\mu_F$ to $\mu_R$~\cite{Ahmed:2015sna}:
\begin{align}
\label{eq:bBH-asf2asr}
a_s\l(\mu_R^2 \r) &= a_s \l(\mu_F^2\r) \frac{1}{\omega} + a_s^2
  \l(\mu_F^2\r) \Bigg\{ \frac{1}{\omega^2} \l(-\eta_1 \log \omega\r)
  \Bigg\}
+ a_s^3 \l(\mu_F^2\r) \Bigg\{ \frac{1}{\omega^2}\l(\eta_1^2-\eta_2\r)
\nonumber\\
&  + \frac{1}{\omega^3} \l( -\eta_1^2+\eta_2 - \eta_1^2 \log \omega +
  \eta_1^2 \log^2 \omega \r) \Bigg\}
\end{align}
where
\begin{align}
\label{eq:bBH-asfasr-1}
&\omega \equiv 1 - \beta_0 a_s\l( \mu_F^2 \r) \log \l(
  \frac{\mu_F^2}{\mu_R^2} \r)\,,
\nonumber\\
&\eta_i \equiv \frac{\beta_i}{\beta_0}\,.
\end{align}
The above result of the evolution of the $a_s$ is a resummed one and
the fixed order result can be easily obtained by performing the series
expansion of this equation~(\ref{eq:bBH-asf2asr}).

\section{Numerical Impact of SV Cross Sections}
\label{sec:bBH-Numerics}

The numerical impact of our results can be studied using the exact LO, NLO, NNLO
$\Delta^{H}_{b{\bar b},i},~ i=0,1,2$ and the threshold N$^3$LO result
$\Delta^{H,{\rm SV}}_{b{\bar b},3}$.
We have used $\sqrt{s} = 14$ TeV for the LHC, the $Z$ boson mass $M_Z=91.1876$ GeV 
and Higgs boson mass $m_H$ = 125.5 GeV throughout. 
The strong coupling constant $\alpha_s (\mu_R^2)$ ($a_s=\alpha_s/4\pi$) is evolved 
using the 4-loop RG equations with 
$\alpha_s^{\text{N$^3$LO}} (m_Z ) = 0.117$ and for parton density sets we use 
MSTW 2008NNLO \cite{Martin:2009iq}.
The Yukawa coupling is evolved using 4 loop RG with
$\lambda(m_b)=\sqrt{2} m_b(m_b)/\nu$ and $m_b(m_b)=4.3$ GeV. 

\begin{figure}[htb]
 \centering
 \begin{minipage}[c]{0.48\textwidth}
\includegraphics[width=1.0\textwidth]{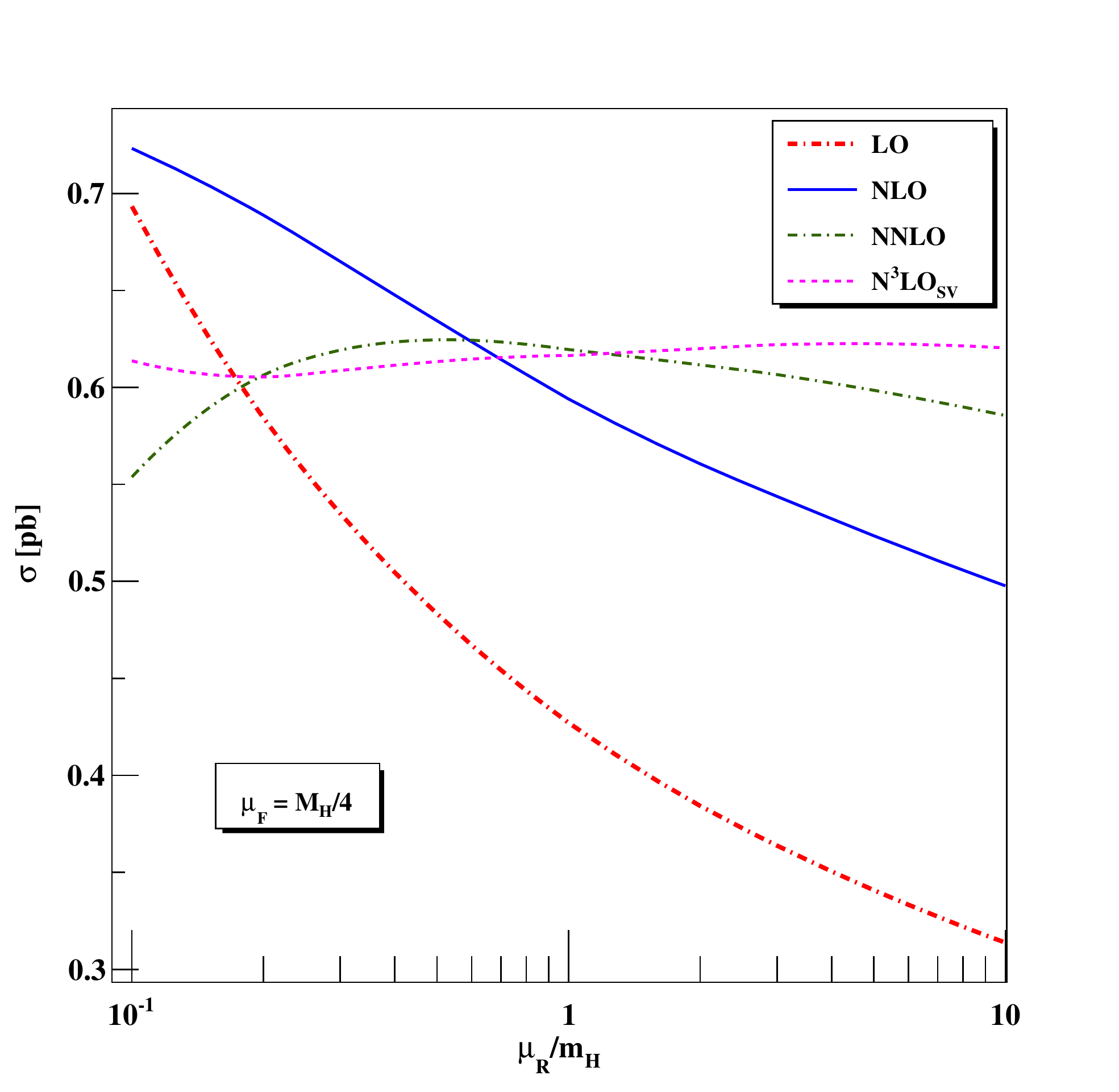}
\end{minipage}
\begin{minipage}[c]{0.48\textwidth}
\includegraphics[width=1.0\textwidth]{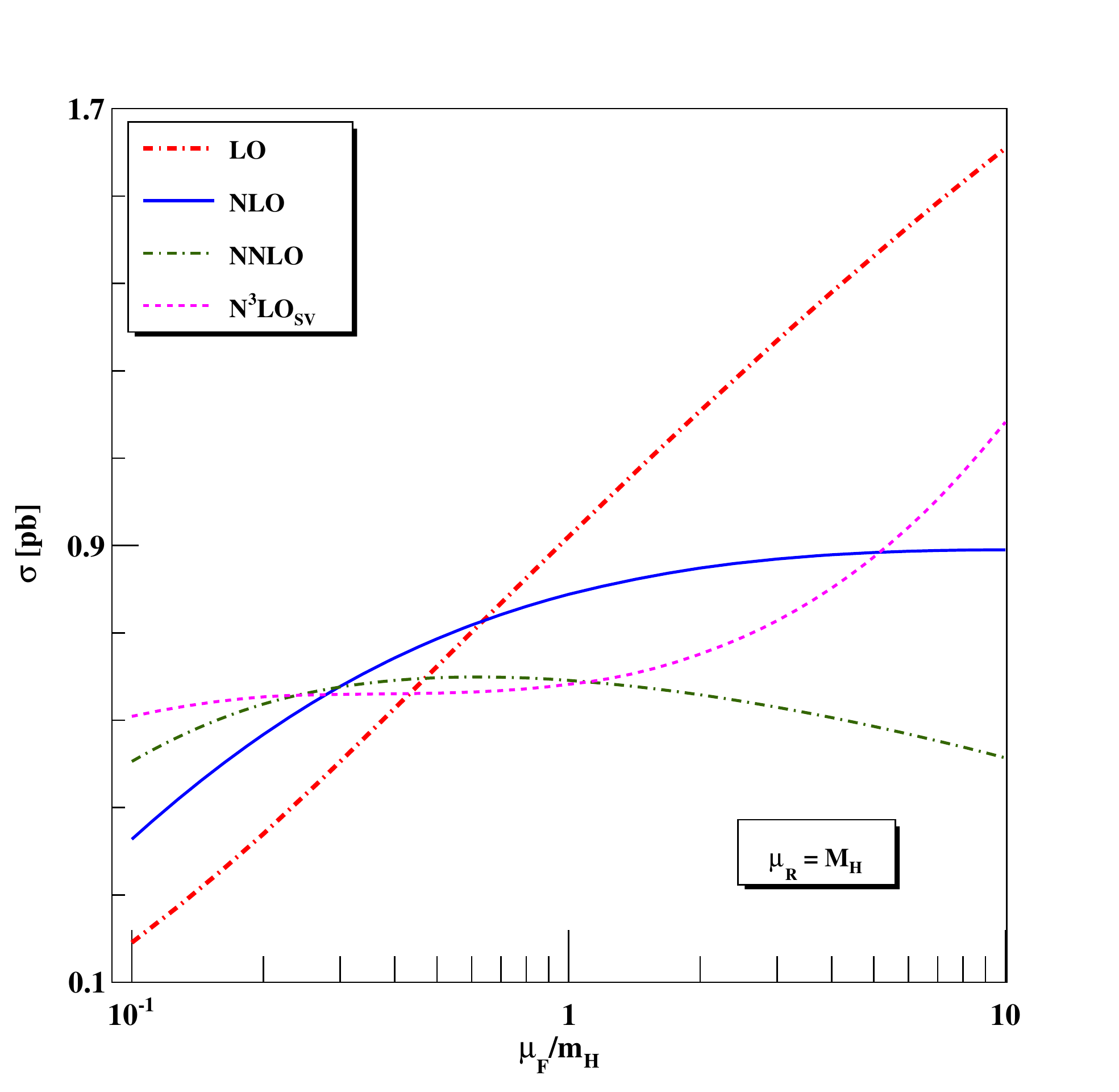}
\end{minipage}
\caption{\label{fig:murnmuf}
Total cross section for Higgs production in $b\bar{b}$ annihilation at
various orders in $a_s$ as a function of $\mu_R/m_H$ (left panel) and
of $\mu_F/m_H$ (right panel) 
at the LHC with $\sqrt{s}=14$ TeV.
}
\end{figure}

The renormalization scale dependence is studied by varying $\mu_{R}$
between $0.1 ~ m_H$ and $10 ~ m_H$  
keeping $\mu_{F}=m_{H}/4$ fixed. For the factorization scale, we have fixed $\mu_R=m_H$ and
varied $\mu_F$ between $0.1 ~ m_H$ and $10 ~ m_H$.  We find that the perturbation theory
behaves better if we include more and more higher order terms (see
Fig.\ref{fig:murnmuf}).  

 \section{Summary}
\label{sec:bBH-Summary}
 
To summarize, we have systematically developed a framework to compute
threshold contributions in
QCD to the production of Higgs boson in bottom anti-bottom annihilation subprocesses at the 
hadron colliders.  This formalism is applicable for any colorless particle. Factorization of UV, soft and collinear singularities and  
exponentiation of their sum allow us to obtain threshold corrections   
order by order in perturbation theory.
Using the recently obtained N${}^3$LO soft distribution function for Drell-Yan production
and the three loop Higgs form factor with bottom anti-bottom quarks, we have obtained threshold
N${}^3$LO corrections to Higgs production through bottom anti-bottom annihilation.  
We have also studied the stability of our result under renormalization and factorization scales.



\chapter{\label{chap:Rap}Rapidity Distributions of Drell-Yan and Higgs Boson
at Threshold in N$^{3}$LO QCD}

\textit{\textbf{The materials presented in this chapter are the result of an original research done in collaboration with Manoj K. Mandal, Narayan Rana and V. Ravindran, and these are based on the published article~\cite{Ahmed:2014uya}}}.
\\

\begingroup
\hypersetup{linkcolor=blue}
\minitoc
\endgroup



\def\D{{\cal D}}
\def\g{\overline {\cal G}}
\def\gm{\gamma}
\def\ep{\epsilon}
\def\zo{\overline{z}_1}
\def\zt{\overline{z}_2}
\def\C{\overline{C}}

\section{Prologue}
\label{sec:Rap-Intro}

The Drell-Yan production \cite{Drell:1970wh} of a pair of leptons at the LHC is one of the cleanest processes that can
be studied not only to test the SM to an 
unprecedented accuracy but also to probe physics beyond
the SM (BSM) scenarios in a very clear environment. Rapidity
distributions of $Z$ boson \cite{Affolder:2000rx} and charge 
asymmetries of leptons in $W$ boson decays \cite{Abe:1998rv} constrain various
parton densities and, in addition, possible excess events can provide
hints to BSM physics, namely R-parity violating supersymmetric models,
models with $Z'$ or with contact interactions and large
extra-dimension models.
One of the production mechanisms responsible for
discovering the Higgs boson of the SM at the LHC \cite{Aad:2012tfa, Chatrchyan:2012ufa} is
the gluon-gluon fusion through top quark
loop. Being a dominant one, it will continue to play a major role in studying the properties of the Higgs boson and its coupling to other SM particles. 
Distributions of transverse momentum and rapidity of the Higgs boson are going to be
very useful tools to achieve this task. Like the inclusive
rates \cite{Kubar-Andre:1978uy, Altarelli:1978id, Humpert:1980uv, Matsuura:1987wt, Matsuura:1988sm, Hamberg:1990np, Dawson:1990zj, Djouadi:1991tk, Spira:1995rr, Harlander:2001is, Catani:2001ic, Catani:2003zt, Harlander:2002wh, Anastasiou:2002yz, Ravindran:2003um}, the rapidity distribution 
of dileptons in DY production 
and of the Higgs boson in gluon-gluon fusion are also known to NNLO level in perturbative QCD due to seminal works
by Anastasiou \textit{et al.}~\cite{Anastasiou:2003yy}.
The quark and gluon form factors \cite{Moch:2005id, Moch:2005tm, Baikov:2009bg, Gehrmann:2010ue},
the mass factorization kernels \cite{Moch:2004pa}, and the renormalization 
constant \cite{Inami:1982xt, Chetyrkin:1997iv, Chetyrkin:1997un} for the effective operator describing the coupling of the 
Higgs boson with the SM fields in the infinite top quark mass limit up to three loop level in dimensional regularization
with space-time dimensions $n = 4 + \epsilon$ were found to be useful to obtain
the N$^3$LO threshold effects \cite{Moch:2005ky, Laenen:2005uz, Idilbi:2005ni, Ravindran:2005vv, Ravindran:2006cg} to the
inclusive Higgs boson and DY productions at the LHC, excluding $\delta(1-z)$ terms,
where the scaling parameter is $z=m_{l^+l^-}^2/\hat s$ for the DY process and 
$z=m_H^2/\hat s$ for the Higgs boson. 
Here, $m_{l^+l^-}$, $m_H$ and $\hat s$ are the
invariant mass of the dileptons, the mass of the Higgs boson,
and square of the center of mass energy of the partonic reaction responsible for the production mechanism, 
respectively.  
Recently, Anastasiou \textit{et al.}~\cite{Anastasiou:2014vaa} made an important contribution in computing 
the total rate for the Higgs boson production at N$^3$LO resulting from the threshold
region including the $\delta(1-z)$ term.  Their result, along with
three loop quark form factors and mass factorization kernels, was used to 
compute the DY cross section at N$^3$LO at threshold in \cite{Ahmed:2014cla}.

In this thesis, we will apply the formalism developed in \cite{Ravindran:2006bu} to obtain 
rapidity distributions of the dilepton pair and of the Higgs boson at N$^3$LO in the threshold
region using the available information that led to the computation 
of the N$^3$LO threshold corrections to the inclusive Higgs boson~\cite{Anastasiou:2014vaa} and DY productions~\cite{Ahmed:2014cla}. 

We begin by writing down the relevant interacting Lagrangian in Sec.~\ref{sec:Rap-Lag}. 
In the Sec.~\ref{sec:Rap-ThreResu}, we present the formalism of computing threshold QCD corrections to the differential rapidity distribution and in Sec.~\ref{sec:Rap-Res}, we present 
our results for the threshold N$^3$LO QCD corrections to the rapidity distributions of the dilepton pairs in DY and Higgs boson. The numerical impact in case of Higgs boson is discussed in brief in Sec.~\ref{sec:Rap-Numerics}. 
The numerical impact of threshold enhanced  N$^3$LO contributions
is demonstrated for the LHC energy $\sqrt{s} = 14$ TeV by studying 
the stability of the perturbation theory under factorization and renormalization scales.
Finally we give a brief summary of our findings in Sec.~\ref{sec:Rap-Summary}.

\section{The Lagrangian}
\label{sec:Rap-Lag}

In the SM, the scalar Higgs boson couples to gluons
only indirectly through a virtual heavy quark loop. This loop can be
integrated out in the limit of infinite quark mass. The resulting
effective Lagrangian encapsulates the interaction between a
scalar $\phi$ and QCD particles and reads: 
\begin{align}
\label{eq:Rap-Lag-H}
&{\cal L}^{H}_{\rm eff} = G_H \phi(x) O^{H}(x) 
\intertext{with} 
&O^H(x) \equiv - \frac{1}{4} G^a_{\mu\nu}(x) G^{a, \mu\nu}(x)\,,
\nonumber\\
&G_H \equiv - \frac{2^{5/4}}{3} a_s(\mu_R^2) G_F^{\frac{1}{2}} C_H
  \left( a_s(\mu_R^2), \frac{\mu_R^2}{m_t^2} \right)\,. 
\end{align}
$C_H(\mu_R^2)$ is the Wilson coefficient, given as a perturbative
expansion in the $\overline{MS}$ renormalised strong coupling constant
$a_s \equiv a_s(\mu_R^2)$, evaluated at the renormalisation scale
$\mu_R$. This is given by~\cite{Chetyrkin:1997un, Schroder:2005hy,
  Chetyrkin:2005ia} 
\begin{align}
\label{eq:Rap-Wilson}
C_H \left( a_s(\mu_R^2), \frac{\mu_R^2}{m_t^2} \right) &= 1 + a_s
                                                         \Bigg\{ 11
                                                         \Bigg\}
+ a_s^2 \Bigg\{ \frac{2777}{18} + 19 L_t + n_f \l(-\frac{67}{6} +
                                                         \frac{16}{3}
                                                         L_t\r) 
                                                         \Bigg\} 
 \nonumber\\
&+
 a_s^3 \Bigg\{ -\frac{2892659}{648} +
   \frac{897943}{144} \zeta_3 + \frac{3466}{9} L_t + 209 L_t^2 
\nonumber\\
&+ n_f \l( \frac{40291}{324} - \frac{110779}{216} \zeta_3 +
  \frac{1760}{27} L_t + 46 L_t^2 \r)
\nonumber\\
&+  n_f^2 \l(-\frac{6865}{486} + \frac{77}{27} L_t - \frac{32}{9} L_t^2 \r) 
  \Bigg\}
\end{align}
up to ${\cal O}(a_s^3)$ with $L_t= \log \l( \mu_R^2/m_t^2 \r)$ and
$n_f$ is the number of active light quark flavors. For the DY process, we work in the framework of exact
SM with $n_f=5$ number of active light quark flavors.

\section{Theoretical Framework for Threshold Corrections to Rapidity}
\label{sec:Rap-ThreResu}

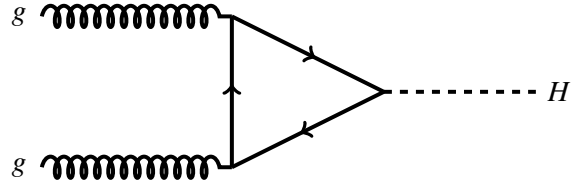
\begin{figure}[htb]
\begin{center}
\begin{tikzpicture}[line width=1.5 pt, scale=1]
\draw[gluon] (-2.5,0) -- (0,0);
\draw[gluon] (-2.5,-2) -- (0,-2);
\draw[fermion] (0,0) -- (2,-1);
\draw[fermion] (2,-1) -- (0,-2);
\draw[fermion] (0,-2) -- (0,0);
\draw[scalarnoarrow] (2,-1) -- (4,-1);
\node at (-2.8,0) {$g$};
\node at (-2.8,-2) {$g$};
\node at (4.3,-1) {$H$};
\end{tikzpicture}
\caption{Higgs boson production in gluon fusion}
\label{fig:Rap-ggH}
\end{center}
\end{figure}

The differential rapidity distribution for the production of a
colorless particle, 
namely, a Higgs boson through gluon fusion/bottom quark annihilation
or a pair of leptons in the DY at the hadron colliders can
be computed using 
\begin{align}
\label{eq:Rap-RapDefn}
\frac{d}{dY} \sigma_Y^{I} \left( \tau, q^2, Y \right) =
  \sigma^{I,(0)}_Y \left( \tau, q^2, \mu_R^2 \right) W^{I} \left(
  \tau, q^2, Y, \mu_R^2 \right)\,.
\end{align}
In the above expression, $Y$ stands for the rapidity which is defined
as
\begin{align}
\label{eq:Rap-RapDefn}
Y \equiv \frac{1}{2} \log \left( \frac{P_2.q}{P_1.q} \right)
\end{align}
where, $P_i$ and $q$ are the momentum of the incoming hadrons and the
colorless particle, respectively. The variable $\tau$ equals $q^2/s$
with
\begin{equation}
\label{eq:Rap-q2}
    q^{2} =  
\begin{cases}
    ~m_{H}^{2}& ~\text{for}~ I=H\, ,\\
    ~m_{l^+l^-}^{2}& ~\text{for}~ I={\rm DY}\, .
\end{cases}
\end{equation}
$m_{H}$ is the mass of the Higgs boson and $m_{l^+l^-}$is the invariant
mass of the final state dilepton pair ($l^{+}l^{-}$), which can be
$e^{+}e^{-},\mu^{+}\mu^{-}, \tau^{+}\tau^{-}$, in the DY production. 
$\sqrt{s}$ and $\sqrt{\hat{s}}$ stand for the hadronic and partonic
center of mass energy, respectively. Throughout this chapter, we
denote $I=H$ for the 
productions of the Higgs boson 
through gluon ($gg$) fusion (Fig.~\ref{fig:Rap-ggH}) and bottom quark ($b{\bar b}$)
annihilation (Fig.~\ref{fig:Rap-bBH}), whereas we write $I=$DY for the
production of a pair of 
leptons in the DY (Fig.~\ref{fig:Rap-qQll}). 
\begin{figure}[htb]
\begin{center}
\begin{tikzpicture}[line width=1.5 pt, scale=1]
\draw[fermion] (-2.5,0) -- (0,-1.3);
\draw[fermion] (0,-1.3) -- (-2.5,-2.6);
\draw[scalarnoarrow] (0,-1.3) -- (2,-1.3);
\node at (-2.8,0) {$b$};
\node at (-2.8,-2.6) {${\bar b}$};
\node at (2.3,-1.3) {$H$};
\end{tikzpicture}
\caption{Higgs boson production through bottom quark annihilation}
\label{fig:Rap-bBH}
\end{center}
\end{figure}
\begin{figure}[htb]
\begin{center}
\begin{tikzpicture}[line width=1.5 pt, scale=1]
\draw[fermion] (-2.5,0) -- (0,-1.3);
\draw[fermion] (0,-1.3) -- (-2.5,-2.6);
\draw[vector] (0,-1.3) -- (2,-1.3);
\draw[fermion] (2,-1.3) -- (4.5,0);
\draw[fermionbar] (2,-1.3) -- (4.5,-2.6);
\node at (-2.8,0) {$q$};
\node at (-2.8,-2.6) {${\bar q}$};
\node at (1.1,-0.7) {$\gamma^{*}/Z$};
\node at (4.8,0) {$l^{+}$};
\node at (4.8,-2.6) {$l^{-}$};
\end{tikzpicture}
\caption{Drell-Yan pair production}
\label{fig:Rap-qQll}
\end{center}
\end{figure}
In Eq.~(\ref{eq:Rap-RapDefn}), $\sigma^I_Y$ is defined through
\begin{equation}
\label{eq:Rap-sigmaIY}
    \sigma^I_Y \left( \tau, q^2, Y \right) =  
\begin{cases}
    ~&\sigma^I \left( \tau,q^2, Y \right) ~~\quad\text{for}~ I=H\, ,\\
    ~&\frac{d}{dq^2}\sigma^{I} \left( \tau,q^2, Y \right) ~\text{for}~
    I={\rm DY}\, .
\end{cases}
\end{equation}
where, $\sigma^I\left( \tau,q^2\right)$ is the inclusive production
cross section. $\sigma^{I, (0)}_Y$ is an overall prefactor extracted
from the leading order contribution. The other quantity $W^I$ is given
by
\begin{align}
\label{eq:Rap-WI}
W^I \left( \tau, q^2, Y, \mu_R^2 \right) &= \frac{\left( Z^I(\mu_R^2)
                                           \right)^2
                                           }{\sigma^{I,(0)}_{Y}}
                                           \sum\limits_{i,j=q,{\bar
                                           q},g} \int\limits_0^1 dx_1
                                           \int\limits_0^1 dx_2
                                           {\hat {\cal
                                           H}}_{ij}^{I}\l(x_{1},x_2 \r)
                                           \int\limits_0^1 dz
                                           \delta(\tau-z x_1 x_2) 
\nonumber\\
&~~~~\times\int
                                           dPS_{1+X} |{\hat {\cal M}}_{ij
                                           }^{I}|^2 \delta
                                           \left( Y-\frac{1}{2} \log
  \left( \frac{P_2.q}{P_1.q} \right) \right)\,,
\nonumber\\
&\equiv \sum\limits_{i,j=q,{\bar q},g} \int\limits_0^1 dx_1
                        \int\limits_0^1 dx_2
                        {\hat {\cal H}}_{ij}^{I}\l(x_{1},x_2 \r)
  \frac{1}{x_1 x_2}{\hat \Delta}^I_{Y,ij} \left( \tau, Y, {\hat a}_s,
  \mu^2, q^2, \mu_R^2, \epsilon \right)
\end{align}
where, we have introduced the dimensionless differential partonic
cross section ${\hat \Delta}^I_{Y,ij}$. $Z^I$ is the
overall operator UV renormalisation constant, $x_k ~(k=1,2)$ 
are the momentum fractions of the initial state partons i.e. $p_k=x_k
P_k$ and ${\hat {\cal H}}^I_{ij}$ stands for
\begin{align}
{\hat {\cal H}}_{ij}^{I}\l(x_{1},x_2\r) \equiv
  \begin{cases}
    ~& {\hat f}_i \left( x_1 \right) {\hat f}_j \left( x_2 \right) ~{\rm for}~
    I={\rm DY}\,,
\nonumber\\
    ~& {\hat f}_i \left( x_1 \right) {\hat f}_j \left( x_2 \right) ~{\rm for}~
    I=H ~{\rm through}~ b{\bar b} ~{\rm annihilation}\,,
\nonumber\\
~& x_1 {\hat f}_i \left( x_1 \right) x_2 {\hat f}_j \left( x_2 \right) ~{\rm for}~
I=H ~{\rm in}~ gg ~{\rm fusion}\,.
  \end{cases}
\end{align}
${\hat f}_i(x_k)$ is the unrenormalised PDF of the initial state partons $i$ 
with momentum fractions $x_k$.
$X$ is the remnants other than the colorless particle $I$, $dPS_{1+X}$
is the phase space element for the $I+X$ system and ${\hat {\cal
    M}}_{ij}^{I}$ represents the partonic level scattering matrix
element for the process $ij \rightarrow I$. The renormalised PDF, $f_i
\left( x_1, \mu_F^2 \right)$, renormalised at the factorisation scale
$\mu_F$, is related to the unrenormalised ones through
Altarelli-Parisi (AP) kernel:
\begin{align}
\label{eq:Rap-PDF-Renorm}
f_i \left( x_k, \mu_F^2 \right) = \sum\limits_{j=q,{\bar q},g}
  \int\limits_{x_k}^1 \frac{dz}{z} \Gamma_{ij} \left( {\hat a}_s,
  \mu^2, \mu_F^2, z, \epsilon \right) {\hat f}_j \left( \frac{x_k}{z} \right)
\end{align}
where the scale $\mu$ is introduced to keep the unrenormalised strong coupling
constant ${\hat a}_s$ dimensionless in space-time dimensions
$d=4+\epsilon$. ${\hat{a}_s} \equiv {\hat{g}}_{s}^{2}/16\pi^{2}$
is the unrenormalized strong coupling constant which is related to the
renormalized one $a_{s}(\mu_{R}^{2})\equiv a_{s}$ through the renormalization
constant $Z_{a_{s}}(\mu_{R}^{2}) \equiv Z_{a_{s}}$,
Eq.~(\ref{eq:bBH-ashatANDas}). The form of e
$Z_{a_s}$ is presented in Eq.~(\ref{eq:bBH-Zas}).  
Expanding the AP kernel in powers of ${\hat a}_s$, we
get
\begin{align}
\label{eq:Rap-APKernel-Expand}
\Gamma^I_{ij}({\hat a}_s, \mu^2, \mu_F^2, z, \epsilon) = \delta_{ij}\delta(1-z) +
  \sum_{k=1}^{\infty} 
    {\hat a}_{s}^{k}  S_{\ep}^{k} \l(\frac{\mu_{F}^{2}}{\mu^{2}}\r)^{k
    \frac{\ep}{2}}  {\hat \Gamma}^{I,(k)}_{ij}(z,\ep)\,.
\end{align}
${\hat \Gamma}^{I,(k)}_{ij}(z,\ep)$ in terms of the Altarelli-Parisi
splitting functions $P^{I,(k)}_{ij}\l(z,\mu_F^2\r)$ are presented in
the Appendix~(\ref{eq:App-Gamma-GenSoln}). Employing the
Eq.~(\ref{eq:Rap-PDF-Renorm}), we can write the renormalised ${\cal
  H}^I_{ij} \left( x_1, x_2, \mu_F^2 \right)$ as
\begin{align}
\label{eq:Rap-RenormH}
{\cal H}^I_{ij} \l( x_1, x_2, \mu_F^2 \r) &= \sum\limits_{k,l}
                                           \int\limits_{x_1}^1
                                            \frac{dy_1}{y_1}
                                                   \int\limits_{x_2}^1
                                                   \frac{dy_2}{y_2}
                                                   \Gamma^I_{ik}({\hat
                                                   a}_s, \mu^2,
                                                   \mu_F^2, y_1,
                                                   \epsilon)
                                                   {\hat{\cal
                                                   H}}^I_{kl} \l(
                                            \frac{x_1}{y_1},
                                            \frac{x_2}{y_2}\r)
                                            \Gamma^I_{jl}({\hat a}_s,
                                            \mu^2, \mu_F^2, y_2,
                                            \epsilon)\,.  
\end{align}
In addition to renormalising the PDF, the AP kernels absorb the
initial state collinear singularities present in the ${\hat
  \Delta}^I_{Y,ij}$ through
\begin{align}
\label{eq:Rap-RenormD}
\Delta^I_{Y,ij} \l( \tau, Y, q^2, \mu_R^2, \mu_F^2 \r)
&= \int \frac{dy_1}{y_1} \int \frac{dy_2}{y_2} \l( \Gamma^{I}
   \l( {\hat a}_s, \mu^2, \mu_F^2, y_1, \epsilon \r) \r)^{-1}_{ik}
{\hat \Delta}^I_{Y,kl} \l( \tau, Y, {\hat a}_s, \mu^2, q^2,
  \mu_R^2, \epsilon \r) 
\nonumber\\
& \times
\l( \Gamma^{I} \l( {\hat a}_s, \mu^2, \mu_F^2, y_2, \epsilon \r)
  \r)^{-1}_{jl} \,.
\end{align}
The $\Delta^I_{Y,ij}$ is free of UV, soft and collinear
singularities. With these we can express $W^I$ in terms of the
renormalised quantities. Before writing down the renormalised version
of the Eq.~(\ref{eq:Rap-WI}), we introduce two symmetric variables
$x_1^0$ and $x_2^0$ instead of $Y$ and $\tau$ through
\begin{align}
\label{eq:Rap-Def-x10-x20}
Y \equiv \frac{1}{2} \log \left( \frac{x_1^0}{x_2^0} \right)\,, \qquad
  \tau \equiv x_1^0 x_2^0\,.
\end{align}
In terms of these new variables, the contributions arising from
partonic subprocesses can be shown to depend on the ratios $z_j =
x_j^0/x_j$ which take the role of scaling variables at the partonic
level. In terms of these newly introduced variables, we get the
renormalised $W^I$ as
\begin{align}
\label{eq:Rap-Ren-WI}
W^I \left( x_1^0, x_2^0, q^2, \mu_R^2 \right) 
&= \sum\limits_{i,j=q,{\bar q}, g} \int\limits_{x_1^0}^1 \frac{dz_1}{z_1}
  \int\limits_{x_2^0}^1 \frac{dz_2}{z_2} {\cal H}^I_{ij} \l(
  \frac{x_1^0}{z_1} \frac{x_2^0}{z_2}, \mu_F^2 \r)
 \Delta^I_{Y,ij} \l( z_1, z_2, q^2, \mu_R^2, \mu_F^2 \r)\,.
\end{align}
The \textit{goal} of this chapter is to study the impact of the
contributions arising from the soft gluons to the differential
rapidity distributions of a colorless particle production at Hadron
colliders. The infrared safe contributions from the soft gluons is
obtained by adding the soft part of the distribution with the
UV renormalized virtual part and performing mass
factorisation using appropriate counter terms. This combination is
often called the soft-plus-virtual (SV) rapidity distribution whereas the
remaining portion is known as hard part. Hence, we write the rapidity
distribution by decomposing into two parts as  
\begin{align}
\label{eq:Rap-PartsOfDelta}
 &{\Delta}^{I}_{Y,ij} (z_1, z_2, q^{2}, \mu_{R}^{2}, \mu_F^2) = {\Delta}^{I,
   \text{SV}}_{Y,ij} (z_1, z_2, q^{2}, \mu_{R}^{2}, \mu_F^2) + {\Delta}^{I,
   \text{hard}}_{Y,ij} (z_1, z_2, q^{2}, \mu_{R}^{2}, \mu_F^2) \,.
\end{align}
The SV contributions
${\Delta}^{I, \text{SV}}_{Y, ij} (z_1, z_2, q^{2}, \mu_{R}^{2}, \mu_F^2)$
contains only the distributions of kind $\delta(1-z_1)$, $\delta(1-z_2)$ and
${\cal{D}}_{i}$, $\overline{\cal D}_{i}$ where the latter ones are
defined through  
\begin{align}
\label{eq:Rap-calD}
{\cal{D}}_{i} \equiv \left[ \frac{\ln^{i}(1-z_1)}{(1-z_1)}
  \right]_{+}\,, \qquad \overline{\cal{D}}_{i} \equiv \left[
  \frac{\ln^{i}(1-z_2)}{(1-z_2)} \right]_{+} \quad {\rm with} \quad
  i=0,1,2,\ldots\,.  
\end{align}
This is also known as the threshold contributions. On the other hand,
the hard part ${\Delta}^{I, \text{hard}}_{Y, ij}$ contains all the
regular terms in $z_1$ and $z_2$. The SV rapidity distribution in $z$-space is
computed in $d$-dimensions, as formulated in
\cite{Ravindran:2006bu}, using  
\begin{align}
\label{eq:Rap-Delta-Psi}
\Delta^{I, \sv}_{Y,ij} (z_1, z_2, q^2, \mu_{R}^{2}, \mu_F^2) = 
{\cal C} \exp \Big( \Psi^I_{Y,ij} \left(z_1, z_2, q^2, \mu_R^2, \mu_F^2, \epsilon
  \right)  \Big) \Big|_{\epsilon = 0} 
\end{align}
where, $\Psi^I_{Y, ij} \left(z_1, z_2, q^2, \mu_R^2, \mu_F^2, \epsilon
\right)$ is a 
finite distribution and ${\cal C}$ is the convolution defined
as  
\begin{equation}
\label{eq:bBH-conv}
 {\cal C} e^{f(z_1, z_2)} = \delta(1-z_1) \delta(1-z_2) + \frac{1}{1!} f(z_1,z_2) + \frac{1}{2!}
 f(z_1,z_2) \otimes f(z_1,z_2) + \cdots  \,.
\end{equation}
Here, $\otimes$ represents the double Mellin convolution with respect
to the pair of variables $z_1$, $z_2$ and $f(z_1,z_2)$ is a
distribution of the kind $\delta(1-z_j)$, ${\cal D}_i$ and
$\overline{\cal D}_{i}$. The $\Psi^I_{Y, ij} \left(z_1, z_2, q^2, \mu_R^2,
  \mu_F^2, \epsilon \right)$ is 
constructed from the form factors ${\cal F}^I_{ij} (\hat{a}_s, Q^2, \mu^2,
\epsilon)$ with $Q^{2}=-q^{2}$, the overall operator UV
renormalization constant $Z^I_{ij}(\hat{a}_s, 
\mu_R^2, \mu^2, \epsilon)$, 
the soft-collinear distribution $\Phi^I_{Y,ij}(\hat{a}_s,$ $q^2,$ $\mu^2,
z_1, z_2, \epsilon)$ 
arising from the real radiations in the partonic subprocesses and the
mass factorization kernels $\Gamma^{I}_{ij} (\hat{a}_s, \mu^2, \mu_F^2, z_{j},
\epsilon)$. In 
terms of the above-mentioned quantities it takes the following form,
as presented in \cite{Ravindran:2006bu, Ahmed:2014uya, Ahmed:2014era} 
\begin{align}
\label{eq:Rap-psi}
\Psi^{I}_{Y,ij} \left(z_1, z_2, q^2, \mu_R^2, \mu_F^2, \epsilon \right) 
= &\left( \ln \Big[ Z^I_{ij} (\hat{a}_s, \mu_R^2, \mu^2, \epsilon) \Big]^2 
+ \ln \Big|  {\cal F}^I_{ij} (\hat{a}_s, Q^2, \mu^2, \epsilon)
    \Big|^2 \right) \delta(1-z_1) \delta(1-z_2)     
\nonumber\\
& + 2 \Phi^I_{Y,ij} (\hat{a}_s, q^2, \mu^2, z_1, z_2, \epsilon) - {\cal C} \ln
  \Gamma^{I}_{ij} (\hat{a}_s, \mu^2, \mu_F^2, z_1,
  \epsilon)\delta(1-z_2) 
\nonumber\\
&
- {\cal C} \ln
  \Gamma^{I}_{ij} (\hat{a}_s, \mu^2, \mu_F^2, z_2,
  \epsilon)\delta(1-z_1)  \, . 
\end{align}
In this chapter, we will confine our discussion on the threshold
corrections to the Higgs boson production through gluon fusion and DY
pair productions. More precisely, our main \textit{goal} is to compute the SV corrections to the
rapidity distributions of these two processes at N$^{3}$LO QCD. In the
subsequent 
sections, we will demonstrate the methodology to get the ingredients,
Eq.~(\ref{eq:Rap-psi})  
to compute the SV rapidity distributions at N$^3$LO QCD.

\subsection{The Form Factor}
\label{ss:Rap-FF}

The quark and gluon form factors represent the QCD loop corrections to
the transition matrix element from an on-shell quark-antiquark pair or
two gluons to a color-neutral particle. For the processes under
consideration, we require gluon form factors in case of
scalar Higgs boson production in $gg$ fusion and quark form factors
for DY pair productions from $q{\bar q}$ annihilation (happens through
intermediate off-shell photon, $\gamma^{*}$ or $Z$-boson). The
unrenormalised quark form factors at ${\cal O}({\hat a}_{s}^{n})$ 
are defined through
\begin{align}
  \label{eq:Rap-DefFb}
  &{\hat{\cal F}}^{I,(n)}_{i\,{\bar i}} \equiv 
\frac{\langle{\hat{\cal
 M}}^{I,(0)}_{i\,{\bar i}}|{\hat{\cal M}}^{I,(n)}_{i\,{\bar
  i}}\rangle}{\langle{\hat{\cal  M}}^{I,(0)}_{i\,{\bar i}}|{\hat{\cal
  M}}^{I,(0)}_{i\,{\bar i}}\rangle}, \qquad n=0,1,2,3, \cdots 
\intertext{with}
&i~{\bar i} =
    \begin{cases}
      gg \qquad {\rm for ~H,}
\nonumber\\
      q{\bar q} \qquad {\rm for ~DY}\,.
    \end{cases}
\end{align}
In the above expressions
$|{\hat{\cal M}}^{I,(n)}_{i~{\bar i}}\rangle$ is the
${\cal O}({\hat a}_{s}^{n})$ contribution to the unrenormalised matrix
element for the production of the particle $I$ from on-shell $i~{\bar
  i}$ annihilation. In terms of these quantities, the full matrix
element and the full 
form factors can be written as a series expansion in ${\hat a}_{s}$ as
\begin{align}
  \label{eq:Rap-DefFlambda}
  |{\cal M}^{I}_{i~{\bar i}}\rangle \equiv \sum_{n=0}^{\infty} {\hat
  a}^{n}_{s} S^{n}_{\epsilon}
  \left( \frac{Q^{2}}{\mu^{2}} \right)^{n\frac{\epsilon}{2}}
  |{\hat{\cal M}}^{I,(n)}_{i~{\bar i}} \rangle \, , 
  \qquad \qquad 
  {\cal F}^{I}_{i~{\bar i}} \equiv
  \sum_{n=0}^{\infty} \left[ {\hat a}_{s}^{n} S_{\epsilon}^{n}
  \left( \frac{Q^{2}}{\mu^{2}} \right)^{n\frac{\epsilon}{2}}
    {\hat{\cal F}}^{I,(n)}_{i~{\bar i}}\right]\, ,
\end{align}
where $Q^{2}=-2\, p_{1}.p_{2}=-q^{2}$ and $p_i$ ($p_{i}^{2}=0$) are
the momenta of the external on-shell quarks or gluons. Gluon form
factors ${\cal F}^{H}_{gg}$ up to three loops in QCD were computed
in~\cite{Harlander:2000mg, Moch:2005tm, Gehrmann:2005pd,
  Baikov:2009bg, Lee:2010cga, Gehrmann:2010ue}. The quark form factors
${\cal F}^{\rm DY}_{q{\bar q}}$
up to three loops in QCD are available from~\cite{Kramer:1986sg,
  Matsuura:1987wt, Matsuura:1988sm, Moch:2005id, Gehrmann:2005pd,
  Moch:2005tm, Baikov:2009bg, Lee:2010cga, Gehrmann:2010ue}.

The form factor ${\cal F}^{I}_{i~{\bar i}}(\hat{a}_{s}, Q^{2}, \mu^{2}, \epsilon)$
satisfies the $KG$-differential equation \cite{Sudakov:1954sw,
  Mueller:1979ih, Collins:1980ih, Sen:1981sd, Magnea:1990zb} which is
a direct consequence of the facts that QCD amplitudes exhibit
factorisation property, gauge and renormalisation group (RG)
invariances:
\begin{equation}
  \label{eq:Rap-KG}
  Q^2 \frac{d}{dQ^2} \ln {\cal F}^{I}_{i~{\bar i}} (\hat{a}_s, Q^2,
  \mu^2, \epsilon) 
  = \frac{1}{2} \left[ K^{I}_{i~{\bar i}} \left(\hat{a}_s,
      \frac{\mu_R^2}{\mu^2}, \epsilon 
    \right)  + G^{I}_{i~{\bar i}} \left(\hat{a}_s, \frac{Q^2}{\mu_R^2},
      \frac{\mu_R^2}{\mu^2}, \epsilon \right) \right]\,. 
\end{equation}
In the above expression, all the poles in dimensional regularisation
parameter $\ep$ are captured in the $Q^{2}$ independent function
$K^{I}_{i~{\bar i}}$ and the quantities which are finite as
$\epsilon \rightarrow 0$ are encapsulated in $G^{I}_{i~{\bar i}}$. The
solution of the above $KG$ equation can be obtained
as~\cite{Ravindran:2005vv} (see also 
\cite{Ahmed:2014cla,Ahmed:2014cha})
\begin{align}
  \label{eq:Rap-lnFSoln}
  \ln {\cal F}^{I}_{i~{\bar i}}(\hat{a}_s, Q^2, \mu^2, \epsilon) =
  \sum_{k=1}^{\infty} {\hat a}_{s}^{k}S_{\epsilon}^{k}
  \left(\frac{Q^{2}}{\mu^{2}}\right)^{k 
  \frac{\epsilon}{2}}  \hat {\cal L}_{i~{\bar i}, k}^{I}(\epsilon)
\end{align}
with
\begin{align}
  \label{eq:Rap-lnFitoCalLF}
  \hat {\cal L}_{i~{\bar i},1}^{I}(\ep) =& { \frac{1}{\ep^2} }
                                           \Bigg\{-2 A^{I}_{{i~{\bar
                                           i}},1}\Bigg\} 
                                  + { \frac{1}{\ep}
                                  }
                                  \Bigg\{G^{I}_{{i~{\bar i}},1}
                                  (\ep)\Bigg\}\, ,
                                  \nonumber\\
  \hat {\cal L}_{{i~{\bar i}},2}^{I}(\ep) =& { \frac{1}{\ep^3} }
                                             \Bigg\{\beta_0
                                             A^{I}_{{i~{\bar
                                             i}},1}\Bigg\} 
                                  + {
                                  \frac{1}{\ep^2} }
                                  \Bigg\{-  {
                                  \frac{1}{2} }  A^{I}_{{i~{\bar i}},2}
                                  - \beta_0   G^{I}_{{i~{\bar i}},1}(\ep)\Bigg\}
                                  + { \frac{1}{\ep}
                                  } \Bigg\{ {
                                  \frac{1}{2} }  G^{I}_{{i~{\bar i}},2}(\ep)\Bigg\}\, ,
                                  \nonumber\\
  \hat {\cal L}_{{i~{\bar i}},3}^{I}(\ep) =& { \frac{1}{\ep^4}
                                  } \Bigg\{- {
                                  \frac{8}{9} }  \beta_0^2 A^{I}_{{i~{\bar i}},1}\Bigg\}
                                  + {
                                  \frac{1}{\ep^3} }
                                  \Bigg\{ { \frac{2}{9} } \beta_1 A^{I}_{{i~{\bar i}},1}
                                  + { \frac{8}{9} }
                                  \beta_0 A^{I}_{{i~{\bar i}},2}  + { \frac{4}{3} }
                                  \beta_0^2 G^{I}_{{i~{\bar i}},1}(\ep)\Bigg\}
                                  \nonumber\\
                                &
                                  + { \frac{1}{\ep^2} } \Bigg\{- {
                                  \frac{2}{9} } A^{I}_{{i{\bar i}},3} 
                                  - { \frac{1}{3} } \beta_1 G^{I}_{{i~{\bar i}},1}(\ep)
                                  - { \frac{4}{3} } \beta_0
                                  G^{I}_{{i~{\bar i}},2}(\ep)\Bigg\} 
                                  + { \frac{1}{\ep}
                                  } \Bigg\{  { \frac{1}{3} }
                                  G^{I}_{i~{\bar i},3}(\ep)\Bigg\}\, .
\end{align}
In Appendix~\ref{chpt:App-KGSoln}, the derivation of the above
solution is discussed in great details.
$A^{I}_{i~{\bar i}}$'s are called the cusp anomalous
dimensions. The constants $G^{I}_{{i~{\bar i}},i}$'s are the
coefficients of $a_{s}^{i}$ in the following 
expansions: 
\begin{align}
  \label{eq:Rap-GandAExp}
  G^{I}_{i~{\bar i}}\left(\hat{a}_s, \frac{Q^2}{\mu_R^2}, \frac{\mu_R^2}{\mu^2},
  \epsilon \right) &= G^{I}_{i~{\bar i}}\left(a_{s}(Q^{2}), 1, \epsilon \right)
                     + \int_{\frac{Q^{2}}{\mu_{R}^{2}}}^{1}
                     \frac{dx}{x} A^{I}_{i~{\bar i}}(a_{s}(x\mu_{R}^{2}))
                     \nonumber\\
                   &= \sum_{k=1}^{\infty}a_{s}^{k}(Q^{2})
                     G^{I}_{{i~{\bar i}},k}(\epsilon) +
                     \int_{\frac{Q^{2}}{\mu_{R}^{2}}}^{1} 
                     \frac{dx}{x} A^{I}_{i~{\bar i}}(a_{s}(x\mu_{R}^{2}))\,.
\end{align}
However, the solutions of the logarithm of the form factor involves
the unknown functions $G^{I}_{{i~{\bar i}},k}$ which are observed to fulfill
\cite{Ravindran:2004mb, Moch:2005tm} the following decomposition in
terms of collinear ($B^{I}_{i~{\bar i}}$), soft ($f^{I}_{i~{\bar i}}$) and UV
($\gamma^{I}_{i~{\bar i}}$) anomalous dimensions:
\begin{align}
  \label{eq:Rap-GIi}
  G^{I}_{{i~{\bar i}},k} (\ep) = 2 \left(B^{I}_{{i~{\bar i}},k} -
  \gamma^{I}_{{i~{\bar i}},k}\right)  + f^{I}_{{i~{\bar i}},k} +
  C^{I}_{{i~{\bar i}},k}  + \sum_{l=1}^{\infty} \epsilon^l g^{I,l}_{{i~{\bar i}},k} \, ,
\end{align}
where, the constants $C^{I}_{{i~{\bar i}},k}$ are given by
\cite{Ravindran:2006cg}
\begin{align}
  \label{eq:Rap-Cg}
  C^{I}_{{i~{\bar i}},1} &= 0\, ,
                \nonumber\\
  C^{I}_{{i~{\bar i}},2} &= - 2 \beta_{0} g^{I,1}_{{i~{\bar i}},1}\, ,
                \nonumber\\
  C^{I}_{{i~{\bar i}},3} &= - 2 \beta_{1} g^{I,1}_{{i~{\bar i}},1} - 2
                \beta_{0} \left(g^{I,1}_{{i~{\bar i}},2}  + 2
                           \beta_{0} g^{I,2}_{{i~{\bar i}},1}\right)\,
                           . 
\end{align}
In the above expressions, $X^{I}_{{i~{\bar i}},k}$ with $X=A,B,f$ and
$\gamma^{I}_{{i~{\bar i}}, k}$ are defined through the series expansion in powers
of $a_{s}$:
\begin{align}
  \label{eq:Rap-ABfgmExp}
  X^{I}_{i~{\bar i}} &\equiv \sum_{k=1}^{\infty} a_{s}^{k}
              X^{I}_{{i~{\bar i}},k}\,,
              \qquad \text{and} \qquad
              \gamma^{I}_{i~{\bar i}} \equiv \sum_{k=1}^{\infty}
                       a_{s}^{k} \gamma^{I}_{{i~{\bar i}},k}\,\,. 
\end{align}
$f_{i~{\bar i}}^{I}$ are introduced for the first time in the
article~\cite{Ravindran:2004mb} where it is shown to fulfill the
maximally non-Abelian property up to two loop level whose validity is
reconfirmed in~\cite{Moch:2005tm} at three loop:
\begin{align}
\label{eq:Rap-MaxNAf}
f^{I}_{q{\bar q}} = \frac{C_F}{C_A} f^{I}_{gg}\,.
\end{align}
This identity implies the soft anomalous dimensions for the production
of a colorless particle in quark annihilation are related to the same
appearing in the gluon fusion through a
simple ratio of quadratic Casimirs of SU(N) gauge group. The same property is also obeyed by the
cusp anomalous dimensions up to three loop level:
\begin{align}
\label{eq:Rap-MaxNAA}
A^{I}_{q{\bar q}} = \frac{C_F}{C_A} A^{I}_{gg}\,.
\end{align}
It is not clear whether this nice property holds true beyond this
order of perturbation theory. Moreover, due
to universality of the quantities denoted by $X$, these are
independent of the operators insertion. These are only dependent on the
initial state partons of any process:
\begin{align}
  \label{eq:Rap-IndOfX}
  X^{I}_{i~{\bar i}} = X_{i~{\bar i}}\,.
\end{align}
Moreover, these are independent of
the quark flavors.
Here, absence of $I$ represents the independence of the quantities on
the nature of colorless particles.
$f^{I}_{i~{\bar i}}$ can be found in \cite{Ravindran:2004mb, Moch:2005tm},
$A^{H}_{i~{\bar i}}$ in~\cite{Moch:2004pa, Moch:2005tm, Vogt:2004mw, Vogt:2000ci}
and $B^{H}_{i~{\bar i}}$ in \cite{Vogt:2004mw, Moch:2005tm} up to three loop
level. For readers' convenience we list them all up to three loop
level in the
Appendix~\ref{chpt:App-AnoDim}.
Utilising the results of these known quantities and comparing
the above expansions of $G^{I}_{{i~{\bar i}},k}(\ep)$ with
the results of the logarithm of the form factors, we extract the
relevant $g_{{i~{\bar i}},k}^{I,l}$ and $\gamma^{I}_{{i~{\bar i}},k}$'s up to three
loop level using Eq.~(\ref{eq:Rap-lnFSoln}),
(\ref{eq:Rap-lnFitoCalLF}) and (\ref{eq:Rap-GIi}). 
The relevant one loop terms for $I=H$ and $i~{\bar i}=gg$ are found to be 
\begin{align}
\label{eq:Rap-gk1}
g_{gg,1}^{H,1} &=  {{C_A}} \zeta_{2}\, ,
\quad\quad
g_{gg,1}^{H,2} =  {{C_A}} \Bigg\{1-\frac{7}{3} {\zeta_{3}}\Bigg\}\, ,
\quad\quad
g_{gg,1}^{H,3} =  {{C_A}} \Bigg\{\frac{47}{80}
{\zeta_2}^2-\frac{3}{2}\Bigg\}\, ,
\end{align}
the relevant two loop terms are
\begin{align}
\label{eq:Rap-gk2}
g_{gg,2}^{H,1} &= {{C_A}^2} \Bigg\{\frac{67}{3} {\zeta_2}-\frac{44}{3}
{\zeta_3}+\frac{4511}{81}\Bigg\} +  {{{C_A} {n_f}}}
                 \Bigg\{-\frac{10 }{3}{\zeta_2}-\frac{40}{3}
                 {\zeta_3}-\frac{1724}{81}\Bigg\} 
\nonumber\\
&+  {{{C_F} {n_f}}} \Bigg\{16 {\zeta_3}-\frac{67}{3}\Bigg\}\, ,
\nonumber\\
g_{gg,2}^{H,2} &= {{{C_A}^2}} \Bigg\{\frac{671}{120} {\zeta_2}^2+\frac{5}{3} {\zeta_2} 
{\zeta_3}-\frac{142}{9} {\zeta_2}+\frac{1139}{27} {\zeta_3}-39
                 {\zeta_5}-\frac{141677}{972}\Bigg\} +  {{{C_A}
                 {n_f}}} \Bigg\{\frac{259}{60}{\zeta_2}^2 
\nonumber\\
&+\frac{16}{9} {\zeta_2}+\frac{604}{27}
  {\zeta_3}+\frac{24103}{486}\Bigg\} +  {{{C_F} {n_f}}}
  \Bigg\{-\frac{16}{3} {\zeta_2}^2-\frac{7}{3}  
{\zeta_2}-\frac{92}{3} {\zeta_3}+\frac{2027}{36}\Bigg\}\, , 
\end{align}
and the required three loop term is
\begin{align}
\label{eq:Rap-gk3}
g_{gg,3}^{H,1} &= {{{C_A}^2 {n_f}}}\Bigg\{ -\frac{128}{45} {\zeta_2}^2-\frac{88}{9} 
{\zeta_2} {\zeta_3}-\frac{14225}{243} {\zeta_2}-\frac{11372}{81} 
{\zeta_3}+\frac{272}{3} 
{\zeta_5}-\frac{5035009}{2187}\Bigg\} 
\nonumber\\
&+  {{{C_F} {n_f}^2}} \Bigg\{-\frac{368}{45} {\zeta_2}^2-\frac{88}{9} 
{\zeta_2}-\frac{1376}{9} {\zeta_3}+\frac{6508}{27}\Bigg\} +
  {{{C_A} {C_F} {n_f}}} \Bigg\{\frac{1568}{45}
  {\zeta_2}^2+40{\zeta_2} {\zeta_3} 
\nonumber\\
&+\frac{503}{18} {\zeta_2}+\frac{20384}{27} 
{\zeta_3}+\frac{608}{3} 
{\zeta_5}-\frac{473705}{324}\Bigg\} +  {{{C_A} {n_f}^2}}
  \Bigg\{\frac{232}{45} {\zeta_2}^2+\frac{100}{27}  
{\zeta_2}+\frac{6992}{81} 
{\zeta_3}
\nonumber\\
&+\frac{912301}{4374}\Bigg\} +  {{{C_A^3}}}
  \Bigg\{-\frac{12352}{315} {\zeta_2}^3-\frac{5744}{45}  
{\zeta_2}^2-\frac{1496}{9} {\zeta_2} {\zeta_3}+\frac{221521}{486}
{\zeta_2}-\frac{104}{3} {\zeta_3}^2
\nonumber\\
&-\frac{57830}{27}
{\zeta_3}+\frac{3080}{3}
{\zeta_5} + \frac{39497339}{8748}\Bigg\} +  {{{C_F}^2 {n_f}}}
  \Bigg\{296 {\zeta_3}-480  
{\zeta_5}+\frac{304}{3}\Bigg\}\,.
\end{align}
Similarly for $I={\rm DY}$ and $i~{\bar i}=q{\bar q}$, we have for one loop
\begin{align}
\label{eq:Rap-gk1-DY}
g_{q{\bar q},1}^{{\rm DY},1} &=  {{{C_F}}} \left\{{\zeta_2}-8\right\}\, ,
\quad\quad
g_{q{\bar q},1}^{{\rm DY},2} = {{{C_F}}} \Bigg\{-\frac{3}{4} {\zeta_2}-\frac{7}{3}
{\zeta_3}+8\Bigg\}\, ,
\nonumber\\
g_{q{\bar q},1}^{{\rm DY},3} &= {{{C_F}}} \Bigg\{\frac{47}{80}
                               {\zeta_2}^2+{\zeta_2}+\frac{7}{4} 
{\zeta_3}-8\Bigg\}\, ,
\end{align}
for two loop we require
\begin{align}
\label{eq:Rap-gk2-DY}
g_{q{\bar q},2}^{{\rm DY},1} &= {{{C_F}^2}} \Bigg\{-\frac{88}{5}
                               {\zeta_2}^2+58 {\zeta_2}-60 
{\zeta_3}-\frac{1}{4}\Bigg\} +  {{{C_A} {C_F}}}
                               \Bigg\{\frac{88}{5}
                               {\zeta_2}^2-\frac{575}{18} 
{\zeta_2}+\frac{260}{3}
{\zeta_3}-\frac{70165}{324}\Bigg\}
\nonumber\\
&+ {{{C_F} {n_f}}} \Bigg\{\frac{37}{9} {\zeta_2}-\frac{8}{3}
{\zeta_3}+\frac{5813}{162}\Bigg\}\, ,
\nonumber\\
g_{q{\bar q},2}^{{\rm DY},2} &= {{{C_F}^2}} \Bigg\{\frac{108}{5} {\zeta_2}^2-28 {\zeta_2}
{\zeta_3}-\frac{437}{4} {\zeta_2}+184 {\zeta_3}+12
{\zeta_5}-\frac{109}{16}\Bigg\} +  {{{C_A} {C_F}}} \Bigg\{-\frac{653}{24} {\zeta_2}^2
\nonumber\\
&+\frac{89}{3}
{\zeta_2} {\zeta_3}+\frac{7297}{108} {\zeta_2}-\frac{12479}{54}
{\zeta_3}-51 {\zeta_5}+\frac{1547797}{3888}\Bigg\} 
\nonumber\\
&+  {{{C_F}
  {n_f}}} \Bigg\{\frac{7}{12} {\zeta_2}^2-\frac{425}{54} 
{\zeta_2}+\frac{301}{27}
{\zeta_3}
-\frac{129389}{1944}\Bigg\}
\end{align}
and the only required three loop term is
\begin{align}
\label{eq:Rap-gk3-DY}
g_{q{\bar q},3}^{{\rm DY},1} &= {{{C_F}^3}} \Bigg\{\frac{21584}{105} {\zeta_2}^3-534
{\zeta_2}^2+840 {\zeta_2} {\zeta_3}-206 {\zeta_2}+48
{\zeta_3}^2-2130 {\zeta_3}+1992
{\zeta_5}-\frac{1527}{4}\Bigg\}
\nonumber\\
&+  {{{C_A} {C_F}^2}} \Bigg\{-\frac{15448}{105}
{\zeta_2}^3+\frac{2432}{45} {\zeta_2}^2-\frac{3448}{3} {\zeta_2}
{\zeta_3}+\frac{55499}{18} {\zeta_2}+296
{\zeta_3}^2-\frac{23402}{9} {\zeta_3}
\nonumber\\
&-\frac{3020}{3}
{\zeta_5}+\frac{230}{3}\Bigg\} +  {{{C_F}^2 {n_f}}}
  \Bigg\{-\frac{704}{45} {\zeta_2}^2-\frac{152}{3} 
{\zeta_2} {\zeta_3}-\frac{7541}{18} {\zeta_2}+\frac{19700}{27}
{\zeta_3}-\frac{368}{3}
{\zeta_5}
\nonumber\\
&+\frac{73271}{162}\Bigg\} +  {{{C_A}^2 {C_F}}} \Bigg\{-\frac{6152}{63}
{\zeta_2}^3+\frac{37271}{90} {\zeta_2}^2+\frac{1786}{9} {\zeta_2}
{\zeta_3}-\frac{1083305}{486} {\zeta_2}-\frac{1136}{3}
{\zeta_3}^2
\nonumber\\
&+\frac{85883}{18} {\zeta_3}+\frac{688}{3}
{\zeta_5}-\frac{48902713}{8748}\Bigg\} +  {{{C_F} {n_f}^2}}
  \Bigg\{-\frac{40}{9} {\zeta_2}^2-\frac{3466}{81} 
{\zeta_2}+\frac{536}{81}
{\zeta_3}
\nonumber\\
&-\frac{258445}{2187}\Bigg\}
+ {{{C_A} {C_F} {n_f}}} \Bigg\{-\frac{1298}{45}
{\zeta_2}^2+\frac{392}{9} {\zeta_2} {\zeta_3}+\frac{155008}{243}
{\zeta_2}-\frac{68660}{81} {\zeta_3}-72
{\zeta_5}
\nonumber\\
&+\frac{3702974}{2187}\Bigg\}
+ {{{C_F} {n_{f,v}} \left(\frac{N^2-4}{N}\right) }}
  \Bigg\{-\frac{6}{5} {\zeta_2}^2+30 
{\zeta_2}+14 {\zeta_3}-80 {\zeta_5}+12\Bigg\}\,.
\end{align}
$n_{f,v}$ is proportional to the charge weighted sum of the quark
flavors~\cite{Gehrmann:2010ue}. The other constants $\gamma^{I}_{i~{\bar i},k}$,
appearing in the Eq.~(\ref{eq:Rap-GIi}), up to three loop
($k=3$) are obtained as  
\begin{align}
\label{eq:Rap-gamma}
&\gamma_{gg,1}^{H} = \beta_{0}\, ,
\quad\quad
\gamma_{gg,2}^{H} = 2 \beta_{1}\, , 
\quad\quad
\gamma_{gg,3}^{H} = 3 \beta_{2}\, 
\nonumber\\
{\rm and} \quad\quad
&\gamma_{q{\bar q}}^{\rm DY} = 0\,.
\end{align}
$\beta_i$ are the coefficient of QCD-$\beta$ function, presented in
Eq.~(\ref{eq:bBH-beta}). These will be utilised in the next subsection
to determine the overall operator renormalisation constants.

\subsection{Operator Renormalisation Constant}
\label{ss:Rap-OOR}

The strong coupling constant renormalisation through $Z_{a_{s}}$ may
not be sufficient to make the form factor ${\cal F}^{I}_{i~{\bar i}}$ completely UV
finite, one needs to perform additional renormalisation to remove the
residual UV divergences which is reflected through the presence of
non-zero $\gamma^{I}_{i~{\bar i}}$. Due to non-zero $\gamma^H_{gg}$ in
Eq.~(\ref{eq:Rap-gamma}), overall UV renormalisation is required for
the Higgs boson production in gluon fusion. However, for DY this is
not required. This additional
renormalisation is called the overall operator renormalisation which
is performed through the constant $Z^{I}_{i~{\bar i}}$. This is determined by
solving the underlying RG equation:
\begin{align}
  \label{eq:Rap-ZRGE}
  \mu_{R}^{2} \frac{d}{d\mu_{R}^{2}} \ln Z^{I}_{i~{\bar i}} \left( {\hat a}_{s},
  \mu_{R}^{2}, \mu^{2}, \epsilon \right) = \sum_{k=1}^{\infty}
  a_{s}^{k}(\mu_R^2) \gamma^{I}_{i~{\bar i},k}\,. 
\end{align}
Using the results of $\gamma^{I}_{i~{\bar i},k}$ from Eq.~(\ref{eq:Rap-gamma}) and
solving the above RG equation following the methodology described in
the Appendix~\ref{chpt:App-SolRGEZas}, we obtain the following overall renormalisation
constant up to three loop level:
\begin{align}
\label{eq:Rap-OOR-Soln}
Z^I_{i~{\bar i}} &= 1+ \sum\limits_{k=1}^{\infty} {\hat a}_s^k
                   S_{\epsilon}^k \left( \frac{\mu_R^2}{\mu^2} 
            \right)^{k\frac{\epsilon}{2}} {\hat Z}_{i~{\bar i}}^{I,(k)}
\end{align}
where,
\begin{align}
\label{eq:bBH-OOR-Soln-1}
{\hat Z}_{gg}^{H,(1)} &= \frac{1}{\epsilon} \Bigg\{ 2 \beta_0
                        \Bigg\}\,,
\nonumber\\
{\hat Z}_{gg}^{H,(2)} &= \frac{1}{\epsilon} \Bigg\{ 2 \beta_1
                        \Bigg\}\,,
\nonumber\\
{\hat Z}_{gg}^{H,(3)} &= \frac{1}{\epsilon^2} \Bigg\{ - 2 \beta_0
                        \beta_1 \Bigg\} + \frac{1}{\epsilon} \Bigg\{ 2
                        \beta_2\Bigg\}\,
\nonumber\\
{\rm and} \qquad {\hat Z}_{q{\bar q}}^{\rm DY} &= 1\,.
\end{align}

\subsection{Mass Factorisation Kernel}
\label{ss:Rap-MFK}

The UV finite form factor contains additional divergences arising from
the soft and collinear regions of the loop momenta. In this section,
we address the issue of collinear divergences and describe a
prescription to remove them. The collinear singularities that arise in
the massless limit of partons are removed by absorbing the divergences
in the bare PDF through
renormalisation of the PDF. This prescription is called the mass
factorisation (MF) and is performed at the factorisation scale
$\mu_F$. In the process of performing this, one needs to introduce
mass factorisation kernels 
$\Gamma^I_{ij}(\hat{a}_s, \mu^2, \mu_F^2, z_j, \epsilon)$ which
essentially absorb the collinear singularities. More specifically, MF
removes the collinear singularities arising from the collinear
configuration associated with the initial state partons. The final
state collinear singularities are guaranteed to go away once the phase
space integrals are performed after summing over the contributions from
virtual and real emission diagrams, thanks to
Kinoshita-Lee-Nauenberg theorem.
The kernels satisfy the following RG equation :
\begin{align}
  \label{eq:Rap-kernelRGE}
  \mu_F^2 \frac{d}{d\mu_F^2} \Gamma^I_{ij}(z_j,\mu_F^2,\epsilon) =
  \frac{1}{2} \sum\limits_{k} P^I_{ik} \left(z_j,\mu_F^2\right) \otimes
  \Gamma^I_{kj} \left(z_j,\mu_F^2,\epsilon \right)  
\end{align}
where, $P^I\left(z_j,\mu_{F}^{2}\right)$ are Altarelli-Parisi splitting
functions (matrix valued). Expanding $P^{I}\left(z_j,\mu_{F}^{2}\right)$ and
$\Gamma^{I}(z_j,\mu_F^2,\epsilon)$ in powers of the strong coupling constant
we get
\begin{align}
  \label{eq:Rap-APexpand}
  &P^{I}_{ij}(z_j,\mu_{F}^{2}) = \sum_{k=1}^{\infty}
    a_{s}^{k}(\mu_{F}^{2})P^{I,(k-1)}_{ij}(z)\,  
    \intertext{and}
  &\Gamma^I_{ij}(z,\mu_F^2,\epsilon) = \delta_{ij} \delta(1-z) + \sum_{k=1}^{\infty}
    {\hat a}_{s}^{k}  S_{\ep}^{k} \l(\frac{\mu_{F}^{2}}{\mu^{2}}\r)^{k
    \frac{\ep}{2}}  {\hat \Gamma}^{I,(k)}_{ij}(z,\ep)\, .
\end{align}
The RG equation of $\Gamma^{I}(z,\mu_F^2,\epsilon)$,
Eq.~(\ref{eq:Rap-kernelRGE}), can be solved in dimensional regularisation
in powers of ${\hat a}_{s}$.  In the $\overline{MS}$ scheme, the
kernel contains only the poles in $\ep$. The solutions~\cite{Ravindran:2005vv} up to the
required order $\Gamma^{I,(3)}(z,\epsilon)$ in terms of $P^{I,(k)}(z)$
are presented in the Appendix~(\ref{eq:App-Gamma-GenSoln}). 
The relevant ones up
to three loop, $P^{I,(0)}(z), P^{I,(1)}(z) ~\text{and}~ P^{I,(2)}(z)$ are
computed in the articles~\cite{Moch:2004pa, Vogt:2004mw}. For the SV
cross section only the diagonal parts of the splitting functions
$P^{I,(k)}_{ij}(z)$ and kernels $\Gamma^{I,(k)}_{ij}(z,\ep)$
contribute since the diagonal elements of $P^{I,(k)}_{ij}(z)$ contain
$\delta(1-z)$ and ${\cal D}_{0}$ whereas the off-diagonal elements are
regular in the limit $z \rightarrow 1$.
The most remarkable fact is that these quantities are universal, independent of
the operators insertion. Hence, for the processes under consideration,
we make use of the existing process independent results of kernels and splitting
functions:
\begin{align}
\label{eq:Rap-Gamma-P-ProcessInd}
\Gamma^H_{ij} = \Gamma^{\rm DY}_{ij} = \Gamma^I_{ij} = \Gamma_{ij}
  \qquad \text{and} \qquad P^H_{ij} = P^{\rm DY}_{ij} = P^I_{ij} = P_{ij}\,.
\end{align}
The absence of $I$ represents the independence of these quantities on
$I$. In the next subsection, we discuss the only remaining ingredient,
namely, the soft-collinear distribution.

\subsection{Soft-Collinear Distribution for Rapidity}
\label{ss:Rap-SCD}

The resulting expression from form factor along with the operator
renormalisation constant and mass factorisation kernel is not
completely finite, it contains some residual divergences which get
cancelled against the contribution arising from soft gluon
emissions. Hence, the finiteness of $\Delta_{Y,i~{\bar i}}^{I, \sv}$ in
Eq.~(\ref{eq:Rap-Delta-Psi}) in the limit 
$\ep \rightarrow 0$ demands that the soft-collinear distribution,
$\Phi^I_{Y,i~{\bar i}} (\hat{a}_s, q^2, \mu^2, z_1, z_2, \epsilon)$, has pole structure
in $\ep$ similar to that of residual divergences. In
article~\cite{Ravindran:2006bu}, it was
shown that $\Phi^{I}_{Y,i~{\bar i}}$ must obey $KG$ type integro-differential
equation, which we call ${\overline{KG}_Y}$ equation, to remove that
residual divergences:
\begin{align}
  \label{eq:Rap-KGbarEqn}
  q^2 \frac{d}{dq^2} \Phi^I_{Y,i~{\bar i}}\left(\hat{a}_s, q^2, \mu^2,
  z_1, z_2,
    \ep\right)   = \frac{1}{2} \left[ \overline K^I_{Y,i~{\bar i}}
  \left(\hat{a}_s, \frac{\mu_R^2}{\mu^2}, z_1, z_2, 
      \ep \right)  + \overline G^I_{Y,i~{\bar i}} \left(\hat{a}_s,
      \frac{q^2}{\mu_R^2},  \frac{\mu_R^2}{\mu^2}, z_1, z_2, \ep \right) \right] \, .
\end{align}
${\overline K}^I_{Y,i~{\bar i}}$ and ${\overline G}^I_{Y,i~{\bar i}}$
play similar roles as those of $K^I_{i~{\bar i}}$ 
and $G^I_{i~{\bar i}}$ in Eq.~(\ref{eq:Rap-KG}), respectively. Also,
$\Phi^I_{Y,i~{\bar i}} (\hat{a}_s, q^2, \mu^2, z, \ep)$ being independent of
$\mu_{R}^{2}$ satisfy the RG equation
\begin{align}
  \label{eq:Rap-RGEphi}
  \mu_{R}^{2}\frac{d}{d\mu_{R}^{2}}\Phi^I_{Y,i~{\bar i}} (\hat{a}_s,
  q^2, \mu^2, z_1, z_2, \epsilon) = 0\, .
\end{align}
This RG invariance and the demand of cancellation of all the residual
divergences arising from ${\cal F}^I_{i~{\bar i}}, Z^I_{i~{\bar i}}$
and $\Gamma^I_{i~{\bar i}}$ 
against $\Phi^{I}_{Y,i~{\bar i}}$ implies the solution of the ${\overline {KG}}_{Y}$
equation as~\cite{Ravindran:2006bu}
\begin{align}
  \label{eq:Rap-PhiSoln}
&\Phi^I_{Y,i~{\bar i}}(\hat{a}_s, q^2, \mu^2, z_1, z_2, \epsilon) =
  \sum_{k=1}^{\infty} {\hat a}_{s}^{k}S_{\epsilon}^{k} \left(\frac{q^{2}}{\mu^{2}}\right)^{k
  \frac{\epsilon}{2}}  \hat {\Phi}^{I}_{Y,i~{\bar i},k}(z_1, z_2, \epsilon)
\intertext{with}
&\hat {\Phi}^{I}_{Y,i~{\bar i},k}(z_1, z_2, \epsilon) = \Bigg\{ (k \epsilon)^2
                                          \frac{1}{4 (1-z_1) (1-z_2)}
                                \left[
  (1-z_1) (1-z_2) \right]^{k \frac{\epsilon}{2}}\Bigg\}
                                \hat{\Phi}^I_{Y,i~{\bar i},k}(\epsilon)\,,
\nonumber\\
&\hat {\Phi}^{I}_{Y,i~{\bar i},k}(\epsilon) = {\hat{\cal
  L}}^I_{i{\bar i},k} \left( A^I_{l}
  \rightarrow -A^I_{l}, G^I_{l} \rightarrow \overline{\cal
                                  G}^I_{Y,i~{\bar i},l}(\epsilon) \right)\,.
\end{align} 
where, ${\hat {\cal L}}_{i~{\bar i},k}^{I}(\ep)$ are defined in
Eq.~(\ref{eq:Rap-lnFitoCalLF}). In
Appendix~\ref{chpt:App-Rap-Soft-Col-Dist}, the derivation of this
solution is depicted in great details.
The $z_j$-independent constants ${\overline{\cal G}}^{I}_{Y,i~{\bar i},l}(\ep)$ can
be obtained by comparing the poles as well as non-pole terms in $\ep$
of ${\hat \Phi}^{I}_{Y,i~{\bar i},k}(\ep)$ with those arising from form factor,
overall renormalisation constant and splitting functions. We find
\begin{align}
  \label{eq:Rap-calGexpans}
&  \overline {\cal G}^{I}_{Y,i~{\bar i},k}(\ep)= -f^I_{i~{\bar i},k} +
                                                \overline{C}^I_{Y,i~{\bar
                                                i}, k} 
                                       + \sum\limits_{l=1}^{\infty}
                                       \epsilon^l \overline{\cal
                                       G}^{I,l}_{Y,i~{\bar i},k}
\intertext{where}
&\overline{C}^I_{Y,i~{\bar i},1} = 0\,,
\nonumber\\
&\overline{C}^I_{Y,i~{\bar i},2} = - 2 \beta_0 \overline{\cal G}^{I,1}_{Y,i~{\bar i},1}\,,
\nonumber\\
&\overline{C}^I_{Y,i~{\bar i},3} = - 2 \beta_1 \overline{\cal G}^{I,1}_{Y,i~{\bar i},1} - 2
  \beta_0 \l( \overline{\cal G}^{I,1}_{Y,i~{\bar i},2} + 2 \beta_0 \overline{\cal
  G}^{I,2}_{Y,i~{\bar i},1} \r)\,.
\end{align}

In-depth understanding about the pole structures including the
single pole~\cite{Ravindran:2004mb} of the form factors, overall operator
renormalisation constants and mass factorisation kernels helps us to
predict all the poles of the soft-collinear
distribution.
However, to determine the finite part of the SV corrections to the
rapidity distribution, we 
need the coefficients of $\epsilon^{k}~(k \ge 1)$, $\overline{\cal
  G}^{I,k}_{Y,i~{\bar i},l}$. Now, we address the question of
determining those constants. This is achieved with the help of an
identity which has been found:
\begin{align}
\label{eq:Rap-Reln-CS-Rap-1}
\int\limits_0^1 dx_1^0 \int\limits_0^1 dx_2^0 \left( x_1^0 x_2^0
  \right)^{{\cal N}-1} \frac{d\sigma^I_{ij}}{dY} = \int\limits_0^1
  d\tau \,\tau^{{\cal N}-1} \sigma^I_{ij}\,.
\end{align}
In the large ${\cal N}$ limit i.e. ${\cal N}
\rightarrow \infty$ the above identity relates~\cite{Ravindran:2006bu} the
$\hat{\Phi}^I_{Y,i~{\bar i},k}(\epsilon)$ with the corresponding
$\hat{\Phi}^I_{i~{\bar i},k}(\epsilon)$ appearing in the computation of
SV cross section, Eq.~(\ref{eq:bBH-PhiSoln}):
\begin{align}
\label{eq:Rap-Reln-CS-Rap-2}
\hat{\Phi}^I_{Y,i~{\bar i},k}(\epsilon) =
  \frac{\Gamma(1+k\epsilon)}{\Gamma^2(1+k \frac{\epsilon}{2})}
  \hat{\Phi}^I_{i~{\bar i},k}(\epsilon)\,.  
\end{align}
Hence, the computation of soft-collinear distribution for the
inclusive production cross section is sufficient to determine the
corresponding one for the rapidity distribution. All the properties
satisfied by $\hat{\Phi}^I_{i~{\bar i},k}(\epsilon)$ are obeyed by
$\hat{\Phi}^I_{Y,i~{\bar i},k}(\epsilon)$ too, see Sec.~\ref{ss:bBH-SCD}
for all the details. Utilising the
relation~(\ref{eq:Rap-Reln-CS-Rap-2}), the relevant constants $\overline {\cal G}_{Y,i~{\bar i},k}^{{I},l}$
to determine 
$\hat{\Phi}^I_{Y,i~{\bar i},k}(\epsilon)$ up to N$^3$LO level are found to be
\begin{align}
\label{eq:Rap-calGres}
\overline {\cal G}_{Y,q{\bar q},1}^{{\rm DY},1} &= C_{F} \Bigg\{ -
                                                  \zeta_{2} \Bigg\} \,, 
\nonumber\\
\overline {\cal G}_{Y,q{\bar q},1}^{{\rm DY},2} &= C_{F} \Bigg\{
                                                  \frac{1}{3}
                                                  \zeta_{3} \Bigg\}
                                                  \,, 
\nonumber\\
\overline {\cal G}_{Y,q{\bar q},1}^{{\rm DY},3} &= C_{F} \Bigg\{
                                                  \frac{1}{80}
                                                  \zeta_{2}^{2}
                                                  \Bigg\} \,, 
\nonumber\\
\overline {\cal G}_{Y,q{\bar q},2}^{{\rm DY},1} &= {C_A} {C_F} \Bigg\{
                                                  -4
                                                  {\zeta_2}^2-\frac{67}{3}
                                                  {\zeta_2} -\frac{44
                                                  }{3} {\zeta_3} +
                                                  \frac{2428}{81}
                                                  \Bigg\}  
                                + C_{F} n_{f} \Bigg\{ \frac{8}{3}
                                                  \zeta_{3} +
                                                  \frac{10}{3}
                                                  \zeta_{2} -
                                                  \frac{328}{81}
                                                  \Bigg\} \,, 
\nonumber\\
\overline {\cal G}_{Y,q{\bar q},2}^{{\rm DY},2} &= {C_A} {C_F} \Bigg\{
                                                  -\frac{319 }{120}
                                                  {\zeta_2}^2 -
                                                  \frac{71 \ 
                                }{3} {\zeta_2} {\zeta_3} + \frac{202
                                                  }{9} {\zeta_2} +
                                                  \frac{469 \ 
                                }{27} {\zeta_3} + 43 {\zeta_5}-\frac{7288}{243} \Bigg\} 
\nonumber\\ 
                             &  + {C_F} {n_f} \Bigg\{ \frac{29 }{60}
                               {\zeta_2}^2 - \frac{28 \ 
                                }{9} {\zeta_2} - \frac{70 }{27}
                               {\zeta_3} + \frac{976}{243} \Bigg\} \,, 
\nonumber\\%
\overline {\cal G}_{Y,q{\bar q},3}^{{\rm DY},1} &= {C_A}^2 {C_F}
                                                  \Bigg\{ \frac{17392
                                                  }{315} {\zeta_2}^3 +
                                                  \frac{1538 \ 
                                }{45} {\zeta_2}^2 + \frac{4136 }{9}
                                                  {\zeta_2} {\zeta_3}
                                                  - \frac{379417 \ 
                                }{486} {\zeta_2} + \frac{536 }{3} {\zeta_3}^2 
                                - 936 {\zeta_3} 
\nonumber\\ 
                             & - \frac{1430 \
                               }{3} {\zeta_5} + \frac{7135981}{8748} \Bigg\} 
                             + {C_A} {C_F} {n_f} \Bigg\{ -\frac{1372 \
                               }{45} {\zeta_2}^2 -\frac{392}{9}
                               {\zeta_2} {\zeta_3} + \frac{51053 \ 
                               }{243} {\zeta_2} 
\nonumber\\
                             & + \frac{12356}{81} {\zeta_3} + \frac{148 \ 
                               }{3} {\zeta_5} - \frac{716509}{4374} \Bigg\}
                             + {C_F} {n_f}^2 \Bigg\{ \frac{152}{45}
                               {\zeta_2}^2 - \frac{316 \ 
                               }{27} {\zeta_2} - \frac{320 }{81}
                               {\zeta_3} + \frac{11584}{2187} \Bigg\} 
\nonumber\\
                            & + {C_F}^2 {n_f} \Bigg\{ \frac{152}{15}
                              {\zeta_2}^2 - 40 {\zeta_2} \ 
                                {\zeta_3}+\frac{275 }{6} {\zeta_2} + \frac{1672 \
                                }{27} {\zeta_3} + \frac{112 }{3}
                              {\zeta_5} - \frac{42727}{324} \Bigg\} \,
                              . 
\end{align}
The corresponding constants for the Higgs boson production in gluon
fusion can be obtained by employing the identity
\begin{align}
\label{eq:Rap-calG-MaxNA}
{\g}^{H,k}_{Y, gg, i} = \frac{C_A}{C_F} {\g}^{{\rm DY},k}_{Y, q{\bar q},i}\,.
\end{align}
The results up to ${\cal O}(a_s^2)$ were present in the
literature~\cite{Ravindran:2006bu} and the term at ${\cal O}(a_s^3)$
is computed for the first time by us in the article~\cite{Ahmed:2014uya}.
Using these, the $\Phi^I_{Y,i~{\bar i}}$ can be obtained which are
presented up to three loops in the Appendix~\ref{app:ss-RapSCD-Res}.
This completes all the ingredients required to compute the SV
correction to the rapidity distributions up to N$^3$LO that are
provided in the next section.

\section{Results of the SV Rapidity Distributions}
\label{sec:Rap-Res}

In this section, we present our findings of the SV rapidity
distributions at N$^3$LO along with the results of the previous
orders. Expanding the SV rapidity distribution ,
Eq.~(\ref{eq:Rap-Delta-Psi}), in powers of $a_s(\mu_F^2)$, we obtain
\begin{align}
\label{eq:Rap-SV-Expand}
&\Delta^{I,{\rm SV}}_{Y,i~{\bar i}} \left( z_1, z_2, q^2, \mu_F^2
  \right) = \sum\limits_{k=1}^{\infty} a_s^k(\mu_F^2) \Delta^{I,{\rm
  SV}}_{Y,i~{\bar i},k} \left( z_1, z_2, q^2, \mu_F^2 \right)
\end{align}
where,
\begin{align}
\label{eq:Rap-SV-Expand-1}
\Delta^{I,{\rm SV}}_{Y,i~{\bar i},k} &= 
\Delta^{I,{\rm SV}}_{Y,i~{\bar i},k}|_{\delta\delta} \delta(1-z_1)
                                       \delta(1-z_2) 
+ \sum\limits_{j=0}^{\infty} \Delta^{I,{\rm SV}}_{Y,i~{\bar
                                       i},k}|_{\delta{\cal D}_j}
                                       \delta(1-z_2) {\cal D}_j 
\nonumber\\
&+ \sum\limits_{j=0}^{\infty} \Delta^{I,{\rm SV}}_{Y,i~{\bar
                                       i},k}|_{\delta \overline{\cal D}_j}
                                       \delta(1-z_2) \overline{\cal
                                       D}_j
+ \sum\limits_{j \circledS l} \Delta^{I,{\rm SV}}_{Y,i~{\bar
                                       i},k}|_{{\cal D}_j \overline{\cal D}_l}
                                       {\cal D}_j \overline{\cal D}_l\,.
\end{align}
The symbol $j \circledS l$ implies $j, l \geq 0$ and $ j + l \leq (2 k - 2)$.
Terms proportional to ${\cal D}_j$ and/or $\overline {\cal D}_j$ in
Eq.~(\ref{eq:Rap-SV-Expand-1})  
were obtained in \cite{Ravindran:2006bu} and the first term is possible to calculate 
as the results for the threshold N$^3$LO QCD corrections to the
production cross section are now available for
DY \cite{Ahmed:2014cla} and the Higgs boson \cite{Anastasiou:2014vaa}
productions. By setting $\mu_R^2=\mu_F^2$ we present the results. For
$I=H$ and $i~{\bar i}=gg$, we obtain~\cite{Ahmed:2014uya}  

For sake of completeness, we mention the leading order contribution
which is
\begin{align}
\label{eq:Rap-DeltaSV0}
\Delta^{I}_{Y,i~{\bar i},0} = \delta(1-z_1) \delta(1-z_2)\,.
\end{align}
The above results are presented for the choice $\mu_R=\mu_F$. The
dependence of the SV rapidity distributions on renormalisation scale $\mu_{R}$
can be easily restored by employing the RG evolution of $a_s$ from
$\mu_F$ to $\mu_R$~\cite{Ahmed:2015sna} using Eq.~(\ref{eq:bBH-asf2asr}).
In the next Sec.~\ref{sec:Rap-Numerics}, we discuss the numerical
impact of the N$^3$LO SV correction to the Higgs rapidity distribution
at the LHC.

\section{Numerical Impact of SV Rapidity Distributions}
\label{sec:Rap-Numerics}

\begin{figure}[htb]
\centering
\includegraphics[width=1\textwidth]{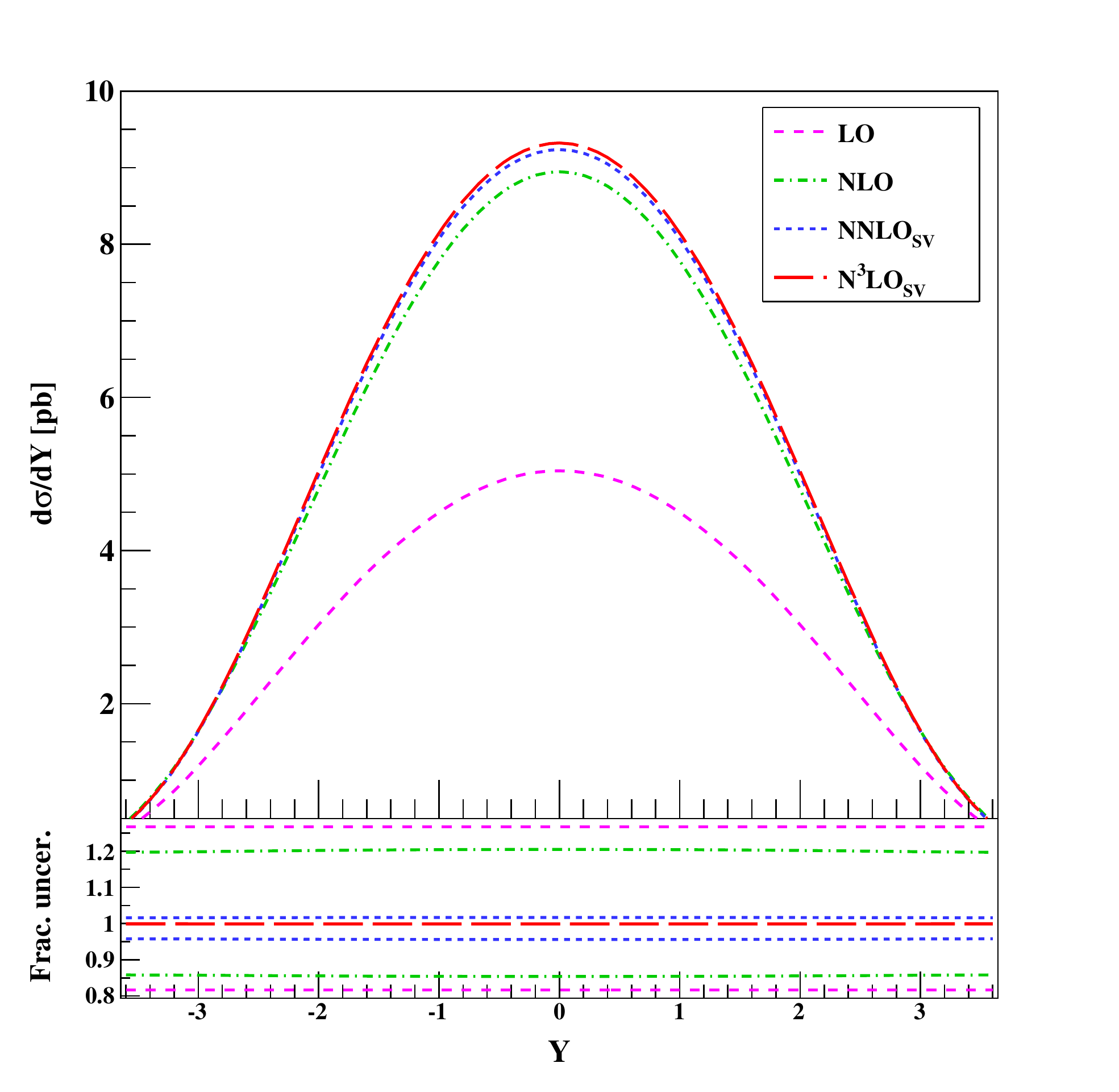}
\caption{
\label{fig:Rap-rapidity_distr}
Rapidity distribution of Higgs boson}
\vspace{-0.5cm}
\end{figure}

In this section, we confine ourselves to the numerical impact of the
SV rapidity distributions of the Higgs boson production through gluon
fusion. We present the relative contributions in percentage
of the pure N$^3$LO terms in Eq.~(\ref{eq:Rap-SV-Expand-1}) 
with respect to $\Delta^{H,{\rm SV}}_{Y,gg,3}$, for rapidity $Y$ = 0
in Table~\ref{table:Rap-perc-1} and \ref{table:Rap-perc-2}.
\begin{table}[h!]
\centering
\begin{tabular}{ c  c  c  c  c  c  c  c  c  c  c }
    \hline\hline
    &
    $~\delta \delta~$ &
    $~\delta \overline{\D}_0~$ & $~\delta \overline{\D}_1~$ & $~\delta \overline{\D}_2~$ & 
    $~\delta \overline{\D}_3~$ & $~\delta \overline{\D}_4~$ & $~\delta \overline{\D}_5~$ &
    $~\D_0 \overline{\D}_0~$ & $~\D_0 \overline{\D}_1~$ 
  \\      
    \hline
    \% &
    73.3 & 16.0 & 9.1 & 31.4 & 1.0 & -9.9 & -23.1 &
    -13.7 & -10.7
  \\
    \hline\hline
  \end{tabular}
\vspace*{5mm}
\caption{Relative contributions of pure N$^3$LO terms.}
 \label{table:Rap-perc-1}
\end{table}
\begin{table}[h!]
\centering
\begin{tabular}{c  c  c  c  c  c  c  c }
    \hline\hline
    & $~\D_0 \overline{\D}_2~$ &
    $~\D_0 \overline{\D}_3~$ & $~\D_0 \overline{\D}_4~$ & 
    $~\D_1 \overline{\D}_1~$ & $~\D_1 \overline{\D}_2~$ & $~\D_1 \overline{\D}_3~$ &
    $~\D_2 \overline{\D}_2~$ \\      
    \hline
    \% & -0.3 &
    3.1 & 7.3 & -0.2 & 3.8 & 8.6 & 4.2 \\
    \hline\hline
  \end{tabular}
\vspace*{5mm}
\caption{Relative contributions of pure N$^3$LO terms.}
 \label{table:Rap-perc-2}
\end{table}
The notation $\D_i \overline{\D}_j$ corresponds to the sum of the
contributions coming from  
$\D_i \overline{\D}_j$ and $\D_j \overline{\D}_i$.
We have used $\sqrt{s} = 14$ TeV for the LHC, $G_F = 4541.68$ pb, the $Z$ boson 
mass $m_Z$ = 91.1876 GeV, 
top quark mass $m_t$ = 173.4 GeV
and the Higgs boson mass $m_H$ = 125.5 GeV throughout. 
For the Higgs boson production, we use the effective theory where top
quark is integrated out in the large $m_t$ limit. 
The strong coupling constant $\alpha_s (\mu_R^2)$ is evolved 
using the 4-loop RG equations with 
$\alpha_s^{\text{N$^3$LO}} (m_Z ) = 0.117$ and for parton density sets we use 
MSTW 2008NNLO \cite{Martin:2009iq}, as N$^3$LO evolution kernels are not yet available. 
In \cite{Forte:2013mda}, Forte \textit{et al.} pointed out that the
Higgs boson cross sections will remain unaffected with this
shortcoming. However, for the DY process, it is not clear whether the
same will be true. 
We find that the contribution from the $\delta(1-z_1) \delta (1-z_2)$
part is the largest.
Impact of the threshold NNLO and N$^3$LO contributions to the Higgs boson rapidity distribution
at the LHC is presented in Fig.~\ref{fig:Rap-rapidity_distr}.
The dependence on the renormalization and factorization scales can by studied by
varying them in the range $m_H/2 <\mu_R,\mu_F<2 m_H$. We find
that the inclusion of the  
threshold correction at N$^3$LO further reduces their dependence.  
For the inclusive Higgs boson production, we find that
about 50\% of exact NNLO contribution comes from   
threshold NLO and NNLO terms.  It increases to 80\% if we use exact
NLO and threshold NNLO terms. 
Hence, it is expected that the rapidity distribution of the Higgs boson
will receive a significant contribution from the threshold region compared to
inclusive rate due to the soft emission over the entire range of $Y$.
Our numerical study with threshold enhanced NNLO rapidity distribution
confirms our expectation.  
Comparing our threshold NNLO results
against exact NNLO distribution using the FEHiP \cite{Anastasiou:2005qj} code , 
we find that about $90\%$ of
exact NNLO distribution comes from the threshold region as can be seen
from Table~\ref{table:Rap-nnlo-1} and \ref{table:Rap-nnlo-2}, in accordance with~\cite{Becher:2007ty},
where it was shown that for low $\tau~(m_H^2/s \approx 10^{-5})$
values the threshold terms are dominant, thanks to the inherent
property of the matrix element, which receives the largest radiative
corrections from the phase-space points corresponding to Born
kinematics.   
\begin{table}[h!]
\centering
\begin{tabular}{ l  c  c  c  c  c  }
    \hline\hline
    $Y$ & ~~0.0~~ & ~~0.4~~ & ~~0.8~~ & ~~1.2~~ & ~~1.6~~  \\
    \hline
    NNLO & 11.21 & 10.96 & 10.70 & 9.13 & 7.80  \\
    NNLO$_{\rm SV}$  & 9.81 & 9.61 & 8.99 & 8.00 & 6.71  \\
    NNLO$_{\rm SV}$(A)  &  10.67 & 10.46 & 9.84 & 8.82 & 7.48  \\
    N$^3$LO$_{\rm SV}$  & 11.62 & 11.36 & 11.07 & 9.44 & 8.04 \\
    N$^3$LO$_{\rm SV}$(A) & 11.88 & 11.62 & 11.33 & 9.70 & 8.30 \\
    $K$3 & 2.31 & 2.29 & 2.36 & 2.21 & 2.17  \\
    \hline\hline
 \end{tabular}
\vspace*{5mm}
 \caption{Contributions of exact NNLO, NNLO$_{\rm SV}$, N$^3$LO$_{\rm SV}$, and $K3$.}
 \label{table:Rap-nnlo-1}
\end{table}
\begin{table}[h!]
\centering
\begin{tabular}{ l  c  c  c  c  c }
    \hline\hline
    $Y$ & ~~2.0~~ & ~~2.4~~ & ~~2.8~~ & ~~3.2~~ & ~~3.6~~ \\
    \hline
    NNLO & 6.10 & 4.23 & 2.66 & 1.40 & 0.54 \\
    NNLO$_{\rm SV}$ & 5.21 & 3.66 & 2.25 & 1.14 & 0.42 \\
    NNLO$_{\rm SV}$(A)  & 5.90 & 4.24 & 2.69 & 1.42 & 0.56 \\
    N$^3$LO$_{\rm SV}$  & 6.27 & 4.33 & 2.70 & 1.40 & 0.53 \\
    N$^3$LO$_{\rm SV}$(A) & 6.51 & 4.54 & 2.88 & 1.53 & 0.60 \\
    $K$3 & 2.07 & 1.89 & 1.70 & 1.63 & 1.51 \\
    \hline\hline
 \end{tabular}
\vspace*{5mm}
 \caption{Contributions of exact NNLO, NNLO$_{\rm SV}$, N$^3$LO$_{\rm SV}$, and $K3$.}
 \label{table:Rap-nnlo-2}
\end{table}
Here we have used the exact results up to the NLO level. 
Because of an inherent ambiguity in the definition of 
the partonic cross section at threshold one can multiply a factor $z
g(z)$, where $z=\tau/x_1 x_2$ and $\lim_{z \rightarrow 1} g(z) = 1$,
with the partonic flux and divide the same in the partonic cross
section for an inclusive rate. In~\cite{Catani:2003zt,Kramer:1996iq}
this was exploited to take into account the subleading collinear logs
also, thereby making the threshold approximation a better
one. Recently, Anastasiou \textit{et al.} used this
in~\cite{Anastasiou:2014vaa} to modify the partonic flux keeping the
partonic cross section unaltered to improve the threshold effects.
Following \cite{Anastasiou:2014vaa,Herzog:2014wja}, we  
introduce
$G (z_1,z_2)$ such that $\lim_{z_1,z_2 \rightarrow 1} G = 1$ in Eq.~(\ref{eq:Rap-Ren-WI}):
\begin{align}
\label{eq:Rap-Ren-WI}
W^I \left( x_1^0, x_2^0, q^2, \mu_R^2 \right) 
&= \sum\limits_{i,j=q,{\bar q}, g} \int\limits_{x_1^0}^1 \frac{dz_1}{z_1}
  \int\limits_{x_2^0}^1 \frac{dz_2}{z_2} {\cal H}^I_{ij} \l(
  \frac{x_1^0}{z_1} \frac{x_2^0}{z_2}, \mu_F^2 \r) G(z_1, z_2)
\nonumber\\
& \times \lim\limits_{z_1,z_2 \to 1} \l[ \frac{\Delta^I_{Y,ij} \l(
  z_1, z_2, q^2, \mu_R^2, \mu_F^2 \r)}{G(z_1,z_2)} \r] \,.
\end{align} 
We also find that with the choice $G(z_1,z_2)=z_1^2 z_2^2$, the threshold NNLO results  are
remarkably close to the exact ones for the entire range of $Y$ [see
Table~\ref{table:Rap-nnlo-1} and \ref{table:Rap-nnlo-2}, denoted by $(A)$].  
This clearly demonstrates the dominance
of threshold contributions to rapidity distribution of the Higgs boson production at the NNLO level.  
Assuming that the trend will not change drastically beyond NNLO, we present numerical values for
N$^3$LO distributions for $G(z_1,z_2)=1, z_1^2 z_2^2$, respectively, as N$^3$LO$_{\rm SV}$ and N$^3$LO$_{\rm SV}(A)$
in Table~\ref{table:Rap-nnlo-1} and \ref{table:Rap-nnlo-1}.  
The threshold N$^3$LO terms give $6 \% (Y = 0)$ to $12 \% (Y = 3.6)$ additional 
correction over the NNLO contribution to the inclusive DY production.
Finally, in Table~\ref{table:Rap-nnlo-1} and \ref{table:Rap-nnlo-2}, we have presented $K3 =$
N$^3$LO$_{\rm SV}$/LO as a function of $Y$ in order to demonstrate  
the sensitivity of higher order effects to the rapidity $Y$. 

\section{Summary}
\label{sec:Rap-Summary}

To summarize, we present the full threshold enhanced N$^3$LO QCD corrections to 
rapidity distributions of the dilepton pair in the DY process and of
the Higgs boson in gluon fusion at the LHC. These are the most
accurate results for these observables available in the literature.
We show that the infrared structure of QCD amplitudes, 
in particular, their factorization properties,  
along with Sudakov resummation of soft gluons and renormalization group invariance provide an elegant
framework to compute these threshold corrections systematically for rapidity distributions 
order by order in QCD perturbation theory.   
The recent N$^3$LO results for inclusive DY and Higgs boson production cross sections
at the threshold provide crucial ingredients to obtain $\delta(1-z_1) \delta(1-z_2)$
contribution of their rapidity distributions for the first time.  
We find that this contribution numerically dominates over the rest of
the terms in $\Delta^{H,{\rm SV}}_{Y,gg,3}$ at 
the LHC. Inclusion of N$^3$LO contributions reduces the scale dependence further.  
We also demonstrate the dominance of the threshold contribution to
rapidity distributions by comparing it against the exact NNLO for two
different choices of $G(z_1,z_2)$. Finally, we find that threshold
N$^3$LO rapidity distribution with $G(z_1,z_2)=1,z_1^2 z_2^2$ shows a
moderate effect over NNLO distribution. 

\section{Outlook-Beyond N$^3$LO}
\label{sec:Rap-Outlook}

The results presented above is the most accurate one existing in the
literature. However, in coming future, we may need to go beyond this
threshold N$^3$LO in hope of making more precise theoretical predictions. The
immediate step would be to compute the complete N$^3$LO QCD
corrections to the differential rapidity distributions. No doubt, this
is an extremely challenging goal! Presently, though we are incapable of computing
this result, we can obtain the general form of the threshold N$^4$LO
QCD corrections to the rapidity distributions! However, due to
unavailability of the quantities, namely, form factors, anomalous
dimensions, soft-collinear distributions at 4-loop level, we are
unable to estimate the contributions arising from this. Nevertheless, the general
form of this contribution is available to the authors which can be utilised to make the predictions once
the missing ingredients become available in future. 
\chapter{\label{chap:Multiloop}A Diagrammatic Approach To Compute
  Multiloop Amplitude}

\begingroup
\hypersetup{linkcolor=blue}
\minitoc
\endgroup



\section{Prologue}
\label{sec:Multi-pro}

The scattering amplitudes play the most crucial role in any quantum
field theory. These are the gateway to unveil the elegant structures
associated with the quantum world. At the phenomenological level,
they are the main ingredients in predicting the observables at high energy
colliders for the processes within and beyond the SM. Hence, the efficient evaluation
of the scattering amplitudes is of prime importance at theoretical
as well as experimental level. However, in perturbative QFT, the theoretical predictions based
on the leading order calculation happens to be unreliable. One must go
beyond the leading order to make the predictions more accurate and
reliable. While considering the effects arising from the higher
orders, the contributions coming from the QCD radiations dominate
substantially, in particular, at high energy colliders like Tevatron
or LHC. In this thesis, we are concentrating only on the corrections arising
from the QCD sector. In the process of computing these higher order
QCD corrections, one has to carry out three different types of
contributions, namely, virtual, real and real-virtual processes. Upon clubbing
together all the three contributions appropriately, finite result for
any observable is obtained. As very much expected, the complexity involved in
the calculations grows very
rapidly as we go towards higher and higher orders in perturbation
theory, where more and more different pieces interfere with each other
that eventually contribute to the final physical observables. 

In this Chapter, we will confine our discussion only to the higher
order QCD virtual or loop corrections. There exists at least two
different formalisms to compute these.
\begin{enumerate}
\item\label{item:1} \textit{Diagrammatic approach:} one directly evaluates all
  the relevant Feynman diagrams appearing at the perturbative order under
  consideration.
\item\label{item:2} \textit{Unitary-based approach:} the unitary
  properties of the scattering amplitudes are employed extensively to
  avoid the direct evaluation of all the Feynman diagrams. 
\end{enumerate}
Despite the spectacular beauty of the unitary based approach, its
applicability to the computation of the amplitudes remains confined mostly to one
loop or only few multiloop problems. Its generalisation to any
multiloop computation is still unavailable in the literature. In these
more complicated scenarios, the first methodology of directly evaluating
the Feynman diagrams is more effective and is therefore employed more often.

\section{Feynman Diagrams and Simplifications}
\label{sec:Multi-dia}

For any generic scattering process in QFT, we can expand any
observable in powers of all the coupling
constants present in the underlying Lagrangian. Feynman diagrams are the diagrammatic representations of
this expansion. In this thesis, we confine our discussion into QCD. Let us
consider a scattering process involving $E$ external particles with
momenta $p_1, p_2, \cdots, p_E$. Without loss of generality, we
consider the cross-section which can be expanded in powers of strong
coupling constant:
\begin{align}
\label{eq:Multi-CSexpand}
\sigma_E \l(p_1, p_2, \cdots, p_E\r) = \sum\limits_{l=0}^{\infty}
  a_s^l\sigma_E^{(l)}\l(p_1, p_2, \cdots, p_E\r)\,. 
\end{align}
For the sake of simplicity, we suppress all the dependence on quantum
numbers of the external particles. The index $l$ denotes the order of
perturbative expansion. The cross section for $l=0$ is called the leading
order (LO), $l=1$ is next-to-leading order (NLO) and so on.
The cross section at at each perturbative
order, $\sigma^{(l)}_E$, is related to the scattering matrix elements through
\begin{align}
\label{eq:Multi-Sigma-MatrixEle}
\sigma_E^{(0)} &= K \int | \, | {\cal M}_E^{(0)} \rangle \, |^2 \, d
                 \Phi_E\,,
\nonumber\\
\sigma_E^{(1)} &= K \int 2 \, \text{Re} \, ( \, \langle {\cal
                 M}_E^{(0)} | {\cal M}_E^{(1)} \rangle \, ) \, d
                 \Phi_E  
+ K \int | \, | {\cal M}_{E + 1}^{(0)} \rangle \, |^2 \, d \Phi_{E +
                 1}\,,
\nonumber\\
\sigma_E^{(2)} &= K \int 2 \, \text{Re} \, ( \, \langle {\cal
                 M}_E^{(0)} | {\cal M}_E^{(2)} \rangle \, ) d \Phi_E 
+  K \int 2 \, \text{Re} \, ( \, \langle {\cal M}_{E + 1}^{(0)} | {\cal
                 M}_{E + 1}^{(1)} \rangle \, ) \, d \Phi_{E + 1} 
\nonumber\\
&+ K \int | \, | {\cal M}_{E + 2}^{(0)} \rangle \, |^2 \, d \Phi_{E +
   2}
\nonumber\\
\text{and}&~ \text{so on.}
\end{align}
In the above set of equations, $| {\cal M}_E^{(l)} \rangle$ is the scattering
amplitude at $l^{\rm th}$ order in perturbation theory involving $E$
number of external particles. The quantity $d\Phi_E$ is the phase
space element. "Re" denotes the real part of the amplitude and $K$ is
an overall constant containing various factors. The amplitudes with $E$
number of external particles and $l \ge 1$ represent the contributions
arising from the virtual Feynman diagrams, whereas amplitudes with
more than $E$ number of external particles come from the real emission
diagrams. In this chapter, we address the issue of evaluating the
virtual diagrams. The scattering matrix element can also be expanded perturbatively in
powers of $a_s$ as
\begin{align}
\label{eq:Multi-MatrixEle-Expand}
| {\cal M}_E \rangle = \sum\limits_{l=0}^{\infty}
  a_s^l | {\cal M}_E^{(l)} \rangle\,. 
\end{align}
Each term in the right hand side can be represented through a set of
Feynman diagrams of same perturbative order. In this chapter, we will
explain the prescription to evaluate the contribution to the matrix
element arising from the virtual diagrams.

The evaluation of the scattering matrix element at any particular order begins with the
generation of associated Feynman diagrams. We make use of a package,
named, QGRAF~\cite{Nogueira:1991ex} to accomplish this job. QGRAF does
not provide the graphical representation of the 
Feynman diagrams, rather it generates those symbolically. We use our
in-house codes written in FORM~\cite{Vermaseren:2000nd} to convert the raw
output into a format for further computation. Employing
the Feynman rules derived from the underlying Lagrangian,
which are the languages establishing the connection between the diagrams
and the corresponding formal mathematical expressions, we obtain the
amplitude. The raw amplitude contains series of
Dirac gamma matrices, QCD color factors, Dirac and Lorentz
indices. We simplify those using our in-house codes. We perform the
color simplification in SU(N) gauge theory and follow dimensional
regularisation where the space-time dimension is considered to be
$d=4+\epsilon$. The amplitude, beyond
leading order, consists of a set of tensorial Feynman integrals.  
Instead of handling the tensorial integrals, we
multiply the amplitude with appropriate projectors to convert those to
scalar integrals. Hence, the problem essentially boils down to solving
those scalar integrals. Often, at any typical order in perturbation
theory, this involves hundreds or thousands of different scalar loop integrals. Of
course, start evaluating all of these integrals is not a practical way
of dealing with the problem. Remarkably, it has been found that the
appeared integrals are not independent of each other, they can be
related through some set of identities! This drastically reduces the
independent integrals which ultimately need to be computed. In the next section, we will
elaborate this procedure.

\section{Reduction to Master Integrals}
\label{sec:Multi-Reduction}

The dimensionally regularised Feynman loop integrals do satisfy a
large number of relations, which allow one to express most of those
integrals in terms of a much smaller subset of independent integrals
(where ''independent'' is to be understood in the sense of the
identities introduced below), which are now commonly referred to as
the Master Integrals (MIs). For a detailed review on this,
see~\cite{Gehrmann:1999as, Argeri:2007up}. These identities are known
as \textbf{integration-by-parts} and \textbf{Lorentz invariance} identities.

\subsection{Integration-by-Parts Identities (IBP)}
\label{ss:Multi-IBP}

The integration-by-parts identities~\cite{Tkachov:1981wb,
  Chetyrkin:1981qh} are the most important class of identities which
establish the relations among the dimensionally regularised scalar
Feynman integrals. These can be seen as a generalisation of Gauss'
divergence theorem in $d$-dimensions. They are based on the fact that,
given a Feynman integral which is a function of space-time dimensions $d$, there
always exists a value of $d$ in the complex plane where the integral is
well defined and consequently convergent. The necessary condition for the
convergence of an integral is the integrand be zero at the
boundaries. This condition can be rephrased as, the integral of the
total derivative with respect to any loop momenta vanishes, that is 
\begin{align}
\label{eq:Multi-IBP-1}
\int \prod\limits_{j=1}^l \frac{d^dk_j}{(2\pi)^{d}}
  \frac{\partial}{\partial k_{i}^{\mu}} \left( v_{s}^{\mu}
  \frac{1}{\cD_1^{b_1} \cdots
  \cD_\beta^{b_\beta}} \right) = 0\,.
\end{align}
In the expression, $k_j$ are the loop momenta, $v_s^{\mu}$ can be loop
or external momenta, $v_s^{\mu}=\{ k_1^{\mu}, \cdots, k_l^{\mu};
p_1^{\mu}, \cdots, p_E^{\mu}\}$. $\cD_i$ are the propagators that
depend on the masses, loop and external momenta. To begin with, a
diagram contains a set of propagators as well as scalar products
involving the loop and external momenta. However, we can express all the
scalar products involving loop momenta in terms of propagators. This
is possible since any Lorentz scalar can be written either in terms of
scalar products or propagators. Both of the representations are
equivalent. For our convenience, we choose to work in the propagator
representation. Performing the differentiation on
the left hand side of the above Eq.~(\ref{eq:Multi-IBP-1}), one obtains
set of IBP identities.
%
Let us demonstrate the role of IBP identities through an one-loop
example.

\begin{itemize}
\item \textbf{Example:} We consider an one loop box diagram, depicted
  through Fig.~\ref{fig:Multi-IBP-Dia} where, all
  the external legs are taken to be massless, for simplicity, and the
  momentum $q=p_1+p_2+p_3$.
\begin{figure}
\begin{center}
\begin{tikzpicture}[line width=0.6 pt, scale=0.7]
\draw (-4,0) -- (4,0);
\draw (-4,-3) -- (4, -3);
\draw (-2,0) -- (-2,-3);
\draw (2,0) -- (2,-3);
\draw[->] (-3.5,0.3) -- (-3,0.3);
\node at (-3.3,0.8) {$q$};
\draw[->] (-0.2,0.3) -- (0.3,0.3);
\node at (0,0.8) {$k_{1}$};
\draw[->] (3,0.3) -- (3.5,0.3);
\node at (3.2,0.8) {$p_{1}$};
\draw[->] (3,-3.3) -- (3.5,-3.3);
\node at (3.2,-3.8) {$p_{2}$};
\draw[<-] (-0.2,-3.3) -- (0.3,-3.3);
\node at (0,-3.8) {$k_{1}-p_1-p_2$};
\draw[<-] (-3.5,-3.3) -- (-3,-3.3);
\node at (-3.3,-3.8) {$p_{3}$};
\draw[->] (2.3,-1.3) -- (2.3,-1.8);
\node at (3.5,-1.5) {$k_{1}-p_{1}$};
\draw[<-] (-2.3,-1.3) -- (-2.3,-1.8);
\node at (-4.7,-1.5) {$k_{1}-p_{1}-p_2-p_3$};
\end{tikzpicture}
\caption{One loop box}
\label{fig:Multi-IBP-Dia} 
\end{center}
\end{figure}
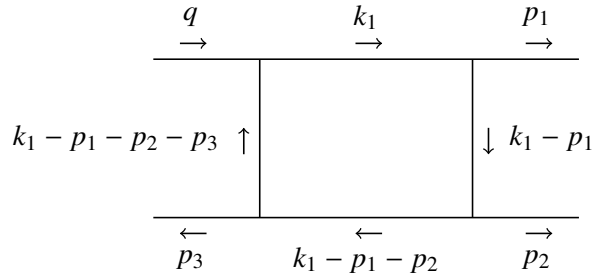
The corresponding dimensionally regularised Feynman integral can be
cast into the following form
\begin{align}
\label{eq:Multi-IBP-Ex-1}
&\int \frac{d^d k}{(2 \pi)^d} \frac{1}{\cD_1^{b_1} \, \cD_2^{b_2} \,
  \cD_3^{b_3} \, \cD_4^{b_4}} \equiv I\l[b_1, b_2, b_3, b_4\r] 
\intertext{with}
&\cD_1 \equiv k_1\,,
\nonumber\\
&\cD_2 \equiv (k_1 - p_1)\,, 
\nonumber\\
&\cD_3 \equiv (k_1 - p_1 - p_2)\,, 
\nonumber\\
&\cD_4 \equiv (k_1 - p_1 - p_2 - p_3)\,.
\end{align}
We can obtain 4-set of IBP identities for each choice of the set
$\{b_1, b_2, b_3, b_4\}$. For a choice of $v_s^{\mu}=p_1^{\mu}$ in
Eq.~(\ref{eq:Multi-IBP-1}), we obtain the corresponding IBP identities
as
\begin{align}
\label{eq:Multi-IBP-Ex-2}
0 = \int \frac{d^d k}{(2 \pi)^d} &\bigg[b_1 \l( - 1 +
                                   \frac{\cD_2}{\cD_1}\r) + b_2 \l(1 -
                                   \frac{\cD_1}{\cD_2}\r) - b_3
                                   \l(\frac{\cD_1}{\cD_3} -
                                   \frac{\cD_2}{\cD_3} -
                                   \frac{s}{\cD_3}\r)  
\nonumber\\
& - b_4 \l(\frac{\cD_1}{\cD_4} - \frac{\cD_2}{\cD_4} - \frac{s}{\cD_4}
  - \frac{u}{\cD_4}\r) \bigg] \frac{1}{\cD_1^{b_1} \,
  \cD_2^{b_2} \, \cD_3^{b_3} \, \cD_4^{b_4}}\,.  
\end{align}
It can be symbolically expressed as
\begin{align}
\label{eq:Multi-IBP-Ex-3}
0 &= b_1 ( - 1 + 1^+ 2^-) + b_2 (1 - 2^+ 1^-) - b_3 (3^+ 1^- - 3^+ 2^- - s
  \, 3^+) 
\nonumber\\
&- b_4 (4^+ 1^- - 4^+ 2^- - s \, 4^+ - u \, 4^+)
\end{align}
where, we have made use of the convention as $1^+ 2^- I[b_1, b_2, b_3, b_4] =
I[b_1 + 1, b_2 - 1, b_3, b_4]$ and the associated Mandelstam variables
are defined as $s \equiv (p_1+p_2)^2=2p_1.p_2$, $t \equiv
(p_2+p_3)^2=2p_2.p_3$, $u \equiv (p_1+p_3)^2=2p_1.p_3$. From the
Eq.~(\ref{eq:Multi-IBP-Ex-3}), it is clear 
that the IBP identities provide recursion relations among the
integrals of a topology and/or its sub-topologies. Similarly, we
can get the IBP identities corresponding to other external as well as
internal momenta. Upon employing all of these identities, it can be
shown that there exists only three MIs, which are
$I[1,0,1,0]$, $I[1,0,0,1]$ and $I[1,1,1,1]$. Hence, at the end we need
to evaluate only three independent integrals corresponding to the
problem under consideration. For higher loop and more number of
external legs, the IBP identities often become too clumsy to handle
manually. Hence, these identities are generated systematically with
the help of some computer algorithms in some packages, such as
AIR~\cite{Anastasiou:2004vj}, FIRE~\cite{Smirnov:2008iw}, 
REDUZE~\cite{Studerus:2009ye,vonManteuffel:2012np},
LiteRed~\cite{Lee:2012cn,Lee:2013mka}.

\end{itemize}

\subsection{Lorentz Invariant Identities (LI)}
\label{ss:Multi-LI}

The Lorentz invariance of the scalar Feynman integrals can be used in
order to obtain more set of identities among the integrals, which are
known as Lorentz invariant identities~\cite{Gehrmann:1999as}:
\begin{align}
\label{eq:Multi-LI-1}
p_{j}^{[\mu} p_{k}^{\nu]} \sum_i p_{i,[\mu} \frac{\partial}{\partial p_i^{\nu]}} I(p_i) = 0\,.
\end{align}
It has been recently found~\cite{Lee:2008tj} that the LI identities
are not independent from IBP ones, since these can be reproduced
generating and solving larger systems of IBPs. However, use of LI
identities along with the IBP helps to speed up the solution. Hence,
in almost all of the computer codes for performing automated reduction
to MIs, LI identities are therefore extensively used.

Employing the IBP and LI identities, we obtain a set of MIs which
ultimately need to be evaluated. Upon evaluation of the MIs, we can
obtain the final unrenormalised result of the virtual
corrections. Often these contain UV as well as soft and collinear
divergences. The UV divergences are removed through UV
renormalisation. The UV renormalised result of the virtual corrections exhibit
a universal infrared pole structures which serve a crucial check on
the correctness of the computation. In the next chapter, we employ
this methodology to compute the three loop quark and gluon form
factors in QCD
for the production of a pseudo-scalar.

\section{Summary}
\label{ss:Multi-Summary}

We have discussed the techniques largely used for the computations of the multiloop amplitudes which is mostly based on the IBP and LI identities. These are employed in the computer codes to automatise the reduction process. Among some packages, we utilise LiteRed~\cite{Lee:2013mka, Lee:2012cn} for our computations. In these articles~\cite{Ahmed:2014gla, Ahmed:2014pka, Ahmed:2015qia, Ahmed:2015qpa}, we have applied this methodology successfully to compute the 2- and 3-loop QCD corrections. In the next chapter, we will present the computation of 3-loop QCD form factors for the pseudo-scalar production where we have essentially made use of the methodology discussed in this chapter.

\chapter{\label{chap:pScalar}Pseudo-Scalar Form Factors at Three Loops
  in QCD}

\textit{\textbf{The materials presented in this chapter are the result of an original research done in collaboration with Thomas Gehrmann, M. C. Kumar, Prakash Mathews, Narayan Rana and V. Ravindran, and these are based on the published articles~\cite{Ahmed:2015qpa, Ahmed:2015pSSV}}}.
\\

\begingroup
\hypersetup{linkcolor=blue}
\minitoc
\endgroup



\section{Prologue}
\setcounter{equation}{0}
\label{sec:intro}

Form factors are the matrix elements of local composite operators
between physical states. In the calculation of scattering cross
sections, they provide the purely virtual corrections.  For example,
in the context of hard scattering processes such as
Drell-Yan~\cite{Altarelli:1979ub,Hamberg:1990np} and the Higgs boson
production in gluon
fusion~\cite{Dawson:1990zj,Djouadi:1991tka,Graudenz:1992pv,Spira:1995rr,
  Djouadi:1995gt, Spira:1997dg,Catani:2001ic, Harlander:2002wh,
  Anastasiou:2002yz, Ravindran:2003um, Ravindran:2004mb,
  Harlander:2005rq,Anastasiou:2015ema}, the form factors corresponding
to the vector current operator $\overline \psi \gamma_\mu \psi$ and
the gluonic operator $G^{a}_{\mu \nu} G^{a,\mu\nu}$ contribute,
respectively.  Here $\psi$ is the fermionic field operator and
$G^{a}_{\mu \nu}$ is the field tensor of the non-Abelian gauge field
$A_\mu^a$ corresponding to the gauge group SU(N).  In QCD the form factors can be computed order by order
in the strong coupling constant using perturbation theory.  Beyond
leading order, the UV renormalisation of the form
factors involves the renormalisation of the composite operator itself,
besides the standard procedure for coupling constant and external
fields.

The resulting UV finite form factors still contain divergences of
infrared origin, namely, soft and collinear divergences due to
the presence of massless gluons and quarks/ antiquarks in the theory.
The inclusive hard scattering cross sections require, in addition to
the form factor, the real-emission partonic subprocesses as well as
suitable mass factorisation kernels for incoming partons.  The soft
divergences in the form factor resulting from the gluons cancel
against those present in the real emission processes and the mass
factorisation kernels remove the remaining collinear divergences
rendering the hadronic inclusive cross section IR finite.  While these
IR divergences cancel among various parts in the perturbative
computations, they can give rise to logarithms involving physical
scales and kinematic scaling variables of the processes under study.
In kinematical regions where these logarithms become large, they may
affect the convergence and reliability of the perturbation series
expansion in powers of the coupling constant.  The solution for this
problem goes back to the pioneering work by
Sudakov~\cite{Sudakov:1954sw} on the asymptotic behaviour of the form
factor in Quantum Electrodynamics: all leading logarithms can be
summed up to all orders in perturbation theory.  Later on, this
resummation was extended to non-leading
logarithms~\cite{Collins:1980ih} and systematised for non-Abelian
gauge theories~\cite{Sen:1981sd}.  Ever since, form factors have been
central to understand the underlying structure of amplitudes in gauge
theories.

The infrared origin of universal logarithmic corrections to form
factors~\cite{Magnea:1990zb} and scattering amplitudes results in a
close interplay between resummation and infrared pole structure.
Working in dimensional regularisation in $d=4+\epsilon$ dimensions,
these poles appear as inverse powers in the Laurent expansion in
$\epsilon$. In a seminal paper, Catani~\cite{Catani:1998bh} proposed a
universal formula for the IR pole structure of massless two-loop QCD
amplitudes of arbitrary multiplicity (valid through to double pole
terms).  This formula was later on justified systematically from
infrared factorization~\cite{Sterman:2002qn}, thereby also revealing
the structure of the single poles in terms of the anomalous dimensions
for the soft radiation.  In \cite{Ravindran:2004mb}, it was shown that
the single pole term in quark and gluon form factors up to two loop
level can be shown to decompose into UV ($\gamma_{I}, I=q,g$) and
universal collinear ($B_I$), color singlet soft ($f_I$) anomalous
dimensions, later on observed to hold even at three loop level
in~\cite{Moch:2005tm}.  An all loop conjecture for the pole structure
of the on-shell multi-loop multi-leg amplitudes in SU(N) gauge theory
with $n_f$ light flavors in terms of cusp ($A_I$), collinear ($B_{I}$)
and soft anomalous dimensions ($\Gamma_{IJ},f_I$ - colour non-singlet
as well as singlet) was proposed by Becher and
Neubert~\cite{Becher:2009cu} and Gardi and Magnea~\cite{Gardi:2009qi},
generalising the earlier results~\cite{Catani:1998bh,Sterman:2002qn}.
The validity of this conjecture beyond three loops depends on the
presence/absence of non-trivial colour correlations and crossing
ratios involving kinematical invariants~\cite{Almelid:2015jia} and
there exists no all-order proof at present. The three-loop expressions
for cusp, collinear and colour singlet soft anomalous dimensions were
extracted~\cite{Moch:2005ba,Laenen:2005uz} from the three loop flavour
singlet~\cite{Vogt:2004mw} and non-singlet~\cite{Moch:2004pa}
splitting functions, thereby also predicting~\cite{Moch:2005tm} the
full pole structure of the three-loop form factors.

The three-loop quark and gluon form factors through to finite terms
were computed
in~\cite{Baikov:2009bg,Gehrmann:2010ue,Gehrmann:2011xn,Gehrmann:2014vha}
and subsequently extended to higher powers in the $\epsilon$
expansion~\cite{Gehrmann:2010tu}.  These results were enabled by
modern techniques for multi-loop calculations in quantum field theory,
in particular integral reduction methods.  These are based on
IBP~\cite{Tkachov:1981wb,Chetyrkin:1981qh} and
LI~\cite{Gehrmann:1999as} identities which reduce
the set of thousands of multi-loop integrals to the one with few
integrals, so called MIs in dimensional
regularisation. To solve these large systems of IBP and LI identities,
the Laporta algorithm~\cite{Laporta:2001dd}, which is based on
lexicographic ordering of the integrals, is the main tool of
choice. It has been implemented in several computer algebra
codes~\cite{Anastasiou:2004vj, Smirnov:2008iw, Studerus:2009ye,
  vonManteuffel:2012np, Lee:2012cn,Lee:2013mka}.  The MIs relevant to
the form factors are single-scale three-loop vertex functions, for
which analytical expressions were derived in
Refs.~\cite{Gehrmann:2005pd, Gehrmann:2006wg, Heinrich:2007at,
  Heinrich:2009be, Lee:2010cga, Gehrmann:2010ue}.
  
Recently, some of us have applied these state-of-the-art methods to
accomplish the task of computing spin-2 quark and gluon form factors
up to three loops~\cite{Ahmed:2015qia} level in SU(N) gauge theory
with $n_f$ light flavours.  These form factors are ingredients to the
precise description of production cross sections for graviton
production, that are predicted in extensions of the SM.
In addition, the spin-2 form factors relate to operators with higher
tensorial structure and thus provide the opportunity to test the
versatility and robustness of calculational techniques for the vertex
functions at three loop level.  The results~\cite{Ahmed:2015qia} also
confirm the universality of the UV and IR structure of the gauge
theory amplitudes in dimensional regularisation.

In the present work, we derive the three-loop corrections to the quark
and gluon form factors for pseudo-scalar operators. These operators
appear frequently in effective field theory descriptions of extensions
of the SM. Most notably, a pseudo-scalar state coupling to
massive fermions is an inherent prediction of any two-Higgs doublet
model~\cite{Fayet:1974pd,Fayet:1976et, Fayet:1977yc,
  Dimopoulos:1981zb, Sakai:1981gr,Inoue:1982pi, Inoue:1983pp,
  Inoue:1982ej}.  In the limit of infinite fermion mass, this gives
rise to the operator insertions considered here.  The recent discovery
of a Standard-Model-like Higgs boson at the LHC~\cite{Aad:2012tfa,
  Chatrchyan:2012xdj} has not only revived the interest in such Higgs
bosons but also prompted the study of the properties of the discovered
boson to identify either with lightest scalar or pseudo-scalar Higgs
bosons of extended models. Such a study requires precise predictions
for their production cross sections. In the context of a CP-even
scalar Higgs boson, results for the inclusive production cross section
in the gluon fusion are available up to N$^{3}$LO
QCD~\cite{Anastasiou:2002yz,Harlander:2002wh,Ravindran:2003um,
  Anastasiou:2015ema}, based on an effective scalar coupling that
results from integration of massive quark loops that mediate the
coupling of the Higgs boson to
gluons~\cite{Ellis:1975ap,Shifman:1979eb,Kniehl:1995tn}.  On the other
hand for the CP-odd pseudo-scalar, only NNLO QCD
results~\cite{Kauffman:1993nv,Djouadi:1993ji,Harlander:2002vv,Anastasiou:2002wq,
  Ravindran:2003um} in the effective theory~\cite{Chetyrkin:1998mw}
are known.  The exact quark mass dependence for scalar and
pseudo-scalar production is known to NLO
QCD~\cite{Spira:1993bb,Spira:1995rr}, and is usually included through
a re-weighting of the effective theory results. Soft gluon resummation
of the gluon fusion cross section has been performed to N$^3$LL for
the scalar
case~\cite{Catani:2003zt,Moch:2005ky,Ravindran:2005vv,Ravindran:2006cg,
  Idilbi:2005ni,Ahrens:2008nc,deFlorian:2009hc,Bonvini:2014joa,Catani:2014uta}
and to NNLL for the pseudo-scalar case~\cite{deFlorian:2007sr}. A
generic threshold resummation formula valid to N$^3$LL accuracy for
colour-neutral final states was derived in~\cite{Catani:2014uta},
requiring only the virtual three-loop amplitudes as process-dependent
input.  The numerical impact of soft gluon resummation in scalar and
pseudo-scalar Higgs boson production and its combination with mass
corrections is reviewed comprehensively in~\cite{Schmidt:2015cea}.
The three-loop corrections to the pseudo-scalar form factors computed
in this thesis are an important ingredient to the N$^3$LO and N$^3$LL
gluon fusion cross sections~\cite{Ahmed:2015pSSV} for pseudo-scalar
Higgs boson production, thereby enabling predictions at the same level
of precision that is attained in the scalar case.
   
The framework of the calculation is outlined in
Section~\ref{sec:frame}, where we describe the effective
theory~\cite{Chetyrkin:1998mw}.  Due to the pseudo-scalar coupling,
one is left with two effective operators with same quantum number and
mass dimensions, which mix under renormalisation.  Since these
operators contain the Levi-Civita tensor as well as $\gamma_5$, the
computation of the matrix elements requires additional care in
$4+\epsilon$ dimensions where neither Levi-Civita tensor nor
$\gamma_5$ can be defined unambiguously.  We use the prescription by
't Hooft and Veltman~\cite{'tHooft:1972fi,Larin:1993tq} to define
$\gamma_5$.  We describe the calculation in Section~\ref{sec:FF},
putting particular emphasis on the UV renormalisation.  Exploiting the
universal IR pole structure of the form factors, we determine the UV
renormalisation constants and mixing of the effective operators up to
three loop level.  We also show that the finite renormalisation
constant, known up to three loops~\cite{Larin:1993tq}, required to
preserve one loop nature of the chiral anomaly, is consistent with
anomalous dimensions of the overall renormalisation constants. As a
first application of our form factors, we compute the hard matching
functions for N$^3$LL resummation in soft-collinear effective theory
(SCET) in Section~\ref{sec:scet}.  Section~\ref{sec:conc} summarises
our results and contains an outlook on future applications to
precision phenomenology of pseudo-scalar Higgs production.


\section{Framework of the Calculation}
\label{sec:frame}
\subsection{The Effective Lagrangian}
\label{sec:ThreResu}

A pseudo-scalar Higgs boson couples to gluons only indirectly through
a virtual heavy quark loop. This loop can be integrated out in the
limit of infinite quark mass.  The resulting effective
Lagrangian~\cite{Chetyrkin:1998mw} encapsulates the interaction
between a pseudo-scalar $\Phi^A$ and QCD particles and reads:
\begin{align} {\cal L}^{A}_{\rm eff} = \Phi^{A}(x) \Big[ -\frac{1}{8}
  {C}_{G} O_{G}(x) - \frac{1}{2} {C}_{J} O_{J}(x)\Big]
\end{align}
where the operators are defined as
\begin{equation}
  O_{G}(x) = G^{\mu\nu}_a \tilde{G}_{a,\mu
    \nu} \equiv  \epsilon_{\mu \nu \rho \sigma} G^{\mu\nu}_a G^{\rho
    \sigma}_a\, ,\qquad
  O_{J}(x) = \partial_{\mu} \left( \bar{\psi}
    \gamma^{\mu}\gamma_5 \psi \right)  \,.
  \label{eq:operators}
\end{equation}
The Wilson coefficients $C_G$ and $C_J$ are obtained by integrating
out the heavy quark loop, and $C_G$ does not receive any QCD
corrections beyond one loop due to the Adler-Bardeen
theorem~\cite{Adler:1969gk}, while $C_J$ starts only at second order
in the strong coupling constant. Expanded in
$a_s \equiv {g}_{s}^{2}/(16\pi^{2}) = \alpha_s/(4\pi)$,
they read
\begin{align}
  \label{eq:const}
  & C_{G} = -a_{s} 2^{\frac{5}{4}} G_{F}^{\frac{1}{2}} {\rm \cot} \beta
    \nonumber\\
  & C_{J} = - \left[ a_{s} C_{F} \left( \frac{3}{2} - 3\ln
    \frac{\mu_{R}^{2}}{m_{t}^{2}} \right) + a_s^2 C_J^{(2)} + \cdots \right] C_{G}\, .
\end{align}
In the above expressions, $G^{\mu\nu}_{a}$ and $\psi$ represent
gluonic field strength tensor and light quark fields, respectively and
$G_{F}$ is the Fermi constant and ${\rm \cot}\beta$ is the mixing angle in a generic
Two-Higgs-Doublet model.
$a_{s} \equiv a_{s} \left( \mu_{R}^{2} \right)$ is the strong coupling
constant renormalised at the scale $\mu_{R}$ which is related to the
unrenormalised one, ${\hat a}_{s} \equiv {\hat g}_{s}^{2}/(16\pi^{2})$
through

\begin{align}
  \label{eq:asAasc}
  {\hat a}_{s} S_{\epsilon} = \left( \frac{\mu^{2}}{\mu_{R}^{2}}  \right)^{\epsilon/2}
  Z_{a_{s}} a_{s}
\end{align}
with
$S_{\epsilon} = {\rm exp} \left[ (\gamma_{E} - \ln 4\pi)\epsilon/2
\right]$
and $\mu$ is the scale introduced to keep the strong coupling constant
dimensionless in $d=4+\epsilon$ space-time dimensions.  The
renormalisation constant $Z_{a_{s}}$~\cite{Tarasov:1980au} is given by
\begin{align}
  \label{eq:Zas}
  Z_{a_{s}}&= 1+ a_s\left[\frac{2}{\epsilon} \beta_0\right]
             + a_s^2 \left[\frac{4}{\epsilon^2 } \beta_0^2
             + \frac{1}{\epsilon}  \beta_1 \right]
             + a_s^3 \left[\frac{8}{ \epsilon^3} \beta_0^3
             +\frac{14}{3 \epsilon^2}  \beta_0 \beta_1 +  \frac{2}{3
             \epsilon}   \beta_2 \right]
\end{align}
up to ${\cal O}(a_{s}^{3})$. $\beta_{i}$ are the coefficients of the
QCD $\beta$ functions which are given by~\cite{Tarasov:1980au}
and presented in Eq.~(\ref{eq:bBH-beta}).
%

\subsection{Treatment of $\gamma_5$ in Dimensional Regularization}
\label{sec:gamma5}

Higher order calculations of chiral quantities in dimensional regularization face the problem of
defining a generalization of
the inherently four-dimensional objects  $\gamma_5$ and $\varepsilon^{\mu\nu\rho\sigma}$
to values of $d\neq 4$. 
In this thesis, we have followed the most practical and
self-consistent definition of $\gamma_{5}$ for multiloop calculations
in dimensional regularization which was introduced by 't~Hooft and
Veltman through \cite{'tHooft:1972fi}
\begin{align}
  \gamma_5 = i \frac{1}{4!} \varepsilon_{\nu_1 \nu_2 \nu_3 \nu_4}
  \gamma^{\nu_1}  \gamma^{\nu_2} \gamma^{\nu_3} \gamma^{\nu_4} \,.
\end{align}
Here, $\varepsilon^{\mu\nu\rho\sigma}$ is the Levi-Civita tensor which is contracted as
\begin{align}
  \label{eqn:LeviContract}
  \varepsilon_{\mu_1\nu_1\lambda_1\sigma_1}\,\varepsilon^{\mu_2\nu_2\lambda_2\sigma_2}=
  \large{\left |
  \begin{array}{cccc}
    \delta_{\mu_1}^{\mu_2} &\delta_{\mu_1}^{\nu_2}&\delta_{\mu_1}^{\lambda_2} & \delta_{\mu_1}^{\sigma_2}\\
    \delta_{\nu_1}^{\mu_2}&\delta_{\nu_1}^{\nu_2}&\delta_{\nu_1}^{\lambda_2}&\delta_{\nu_1}^{\sigma_2}\\
    \delta_{\lambda_1}^{\mu_2}&\delta_{\lambda_1}^{\nu_2}&\delta_{\lambda_1}^{\lambda_2}&\delta_{\lambda_1}^{\sigma_2}\\
    \delta_{\sigma_1}^{\mu_2}&\delta_{\sigma_1}^{\nu_2}&\delta_{\sigma_1}^{\lambda_2}&\delta_{\sigma_1}^{\sigma_2}
  \end{array}
                                                                                       \right |}
\end{align}
and all the Lorentz indices are considered to be $d$-dimensional~\cite{Larin:1993tq}. 
In this scheme, a finite renormalisation of the 
axial vector current is required in order to fulfill chiral Ward identities and the Adler-Bardeen theorem. We discuss this 
in detail in Section~\ref{ss:UV} below. 

\section{Pseudo-scalar Quark and Gluon Form Factors}
\label{sec:FF}

The quark and gluon form factors describe the QCD loop corrections to
the transition matrix element from a color-neutral operator $O$ to an
on-shell quark-antiquark pair or to two gluons.  For the pseudo-scalar
interaction, we need to consider the two operators $O_{G}$ and
$O_{J}$, defined in Eq.~(\ref{eq:operators}), thus yielding in total
four form factors.
We define the unrenormalised gluon form factors at
${\cal O}({\hat a}_{s}^{n})$ as

\begin{align}
  \label{eq:DefFg}
  {\hat{\cal F}}^{G,(n)}_{g} \equiv \frac{\langle{\hat{\cal
  M}}^{G,(0)}_{g}|{\hat{\cal M}}^{G,(n)}_{g}\rangle}{\langle{\hat{\cal
  M}}^{G,(0)}_{g}|{\hat{\cal M}}^{G,(0)}_{g}\rangle}\, ,
  \qquad \qquad
  {\hat{\cal F}}^{J,(n)}_{g} \equiv \frac{\langle{\hat{\cal
  M}}^{G,(0)}_{g}|{\hat{\cal M}}^{J,(n+1)}_{g}\rangle}{\langle{\hat{\cal
  M}}^{G,(0)}_{g}|{\hat{\cal M}}^{J,(1)}_{g}\rangle}
\end{align}
and similarly the unrenormalised quark form factors through
\begin{align}
  \label{eq:DefFq}
  {\hat{\cal F}}^{G,(n)}_{q} \equiv \frac{\langle{\hat{\cal
  M}}^{J,(0)}_{q}|{\hat{\cal M}}^{G,(n+1)}_{q}\rangle}{\langle{\hat{\cal
  M}}^{J,(0)}_{q}|{\hat{\cal M}}^{G,(1)}_{q}\rangle}\, ,
  \qquad \qquad
  {\hat{\cal F}}^{J,(n)}_{q} \equiv \frac{\langle{\hat{\cal
  M}}^{J,(0)}_{q}|{\hat{\cal M}}^{J,(n)}_{q}\rangle}{\langle{\hat{\cal
  M}}^{J,(0)}_{q}|{\hat{\cal M}}^{J,(0)}_{q}\rangle}
\end{align}
where, $n=0, 1, 2, 3, \ldots$\,. In the above expressions
$|{\hat{\cal M}}^{\lambda,(n)}_{\beta}\rangle$ is the
${\cal O}({\hat a}_{s}^{n})$ contribution to the unrenormalised matrix
element for the transition from the bare operator $[O_{\lambda}]_B$
$(\lambda = G,J)$ to a quark-antiquark pair ($\beta=q$) or to two
gluons ($\beta=g$).  The expansion of these quantities in powers of
${\hat a}_{s}$ is performed through

\begin{tabular}{p{5cm}p{8.5cm}}
  \begin{align*}
    |{\cal M}^{\lambda}_{\beta}\rangle \equiv \sum_{n=0}^{\infty} {\hat
    a}^{n}_{s} S^{n}_{\epsilon}
     \left( \frac{Q^{2}}{\mu^{2}} \right)^{n\frac{\epsilon}{2}}
    |{\hat{\cal M}}^{\lambda,(n)}_{\beta} \rangle
  \end{align*}
  &
    \begin{equation}
      \label{eq:Mexp}
      \hspace{-0.5cm}\text{and}\qquad
      {\cal F}^{\lambda}_{\beta} \equiv
      \sum_{n=0}^{\infty} \left[ {\hat a}_{s}^{n}
        \left( \frac{Q^{2}}{\mu^{2}} \right)^{n\frac{\epsilon}{2}}
        S_{\epsilon}^{n}  {\hat{\cal F}}^{\lambda,(n)}_{\beta}\right]\,.
    \end{equation}
\end{tabular}
\\
where, $Q^{2}=-2\, p_{1}.p_{2}$ and $p_{i}'s$ $(p_{i}^{2}=0)$ are the
momenta of the external quarks and gluons. Note that
$|{\hat{\cal M}}^{G,(n)}_{q}\rangle$ and
$|{\hat{\cal M}}^{J,(n)}_{g}\rangle$ start from $n=1$ i.e. from one
loop level.

\subsection{Calculation of the Unrenormalised Form Factors}
\label{ss:CalcFF}

The calculation of the unrenormalised pseudo-scalar form factors up to
three loops follows closely the steps used in the derivation of the
three-loop scalar and vector form factors
\cite{Gehrmann:2010ue,Gehrmann:2014vha}.  The Feynman diagrams for all
transition matrix elements (Eq.~(\ref{eq:DefFg}),
Eq.~(\ref{eq:DefFq})) are generated using
QGRAF~\cite{Nogueira:1991ex}. The numbers of diagrams contributing to
three loop amplitudes are 1586 for
$|{\hat{\cal M}}^{G,(3)}_{g}\rangle$, 447 for
$|{\hat{\cal M}}^{J,(3)}_{g}\rangle$, 400 for
$|{\hat{\cal M}}^{G,(3)}_{q}\rangle$ and 244 for
$|{\hat{\cal M}}^{J,(3)}_{q}\rangle$ where all the external particles
are considered to be on-shell. The raw output of QGRAF is converted to
a format suitable for further manipulation. A set of in-house routines
written in the symbolic manipulating program FORM
\cite{Vermaseren:2000nd} is utilized to perform the simplification of
the matrix elements involving Lorentz and color indices. Contributions
arising from ghost loops are taken into account as well since we use
Feynman gauge for internal gluons. For the external on-shell gluons,
we ensure the summing over only transverse polarization states by
employing an axial polarization sum:
\begin{equation}
  \label{eq:PolSum}
  \sum_{s} {\varepsilon^{\mu}}^{\, *}(p_{i},s)
  \varepsilon^{\nu}(p_{i},s)  = - \eta^{\mu\nu} + \frac{p_{i}^{\mu}
    q_{i}^{\nu}  + q_{i}^{\mu} p_{i}^{\nu}}{p_{i}.q_{i}} \quad ,
\end{equation}
where $p_{i}$ is the $i^{\rm th}$-gluon momentum, $q_{i}$ is the
corresponding reference momentum which is an arbitrary light like
4-vector and $s$ stands for spin (polarization) of gluons. We choose
$q_{1}=p_{2}$ and $q_{2}=p_{1}$ for our calculation. Finally, traces
over the Dirac matrices are carried out in $d$ dimensions.

The expressions involve thousands of three-loop scalar
integrals. However, they are expressible in terms of a much smaller
set of scalar integrals, called master integrals (MIs), by use of
IBP~\cite{Tkachov:1981wb, Chetyrkin:1981qh} and LI~\cite{Gehrmann:1999as} identities.  These
identities follow from the Poincare invariance of the integrands, they
result in a large linear system of equations for the integrals
relevant to given external kinematics at a fixed loop-order. The LI
identities are not linearly independent from the IBP
identities~\cite{Lee:2008tj}, their inclusion does however help to 
accelerate the solution of the system of equations. By employing
lexicographic ordering of these integrals (Laporta
algorithm,~\cite{Laporta:2001dd}), a reduction to MIs is accomplished.
Several implementations of the Laporta algorithm exist in the
literature: AIR~\cite{Anastasiou:2004vj}, FIRE~\cite{Smirnov:2008iw},
Reduze2~\cite{vonManteuffel:2012np, Studerus:2009ye} and
LiteRed~\cite{Lee:2013mka, Lee:2012cn}.  In the context of the present
calculation, we used LiteRed~\cite{Lee:2013mka, Lee:2012cn} to perform
the reductions of all the integrals to MIs.

Each three-loop Feynman integral is expressed in terms of a list of
propagators involving loop momenta that can be attributed to one of
the following three sets (auxiliary
topologies,~\cite{Gehrmann:2010ue})
\begin{align}
  \label{eq:Basis}
  {\rm A}_1 &: \{ \cD_1, \cD_2, \cD_3, \cD_{12}, \cD_{13},
              \cD_{23}, \cD_{1;1}, \cD_{1;12}, \cD_{2;1}, \cD_{2;12}, \cD_{3;1},
              \cD_{3;12} \}
              \nonumber\\
  {\rm A}_2 &: \{ \cD_1, \cD_2, \cD_3, \cD_{12}, \cD_{13}, \cD_{23},
              \cD_{13;2}, \cD_{1;12}, \cD_{2;1}, \cD_{12;2},
              \cD_{3;1}, \cD_{3;12} \}
              \nonumber\\
  {\rm A}_3 &: \{ \cD_1, \cD_2, \cD_3, \cD_{12}, \cD_{13}, \cD_{123},
              \cD_{1;1}, \cD_{1;12}, \cD_{2;1}, \cD_{2;12}, \cD_{3;1},
              \cD_{3;12} \}\, .
\end{align}
In the above sets
\begin{align*}
  \cD_{i} = k_{i}^2, \cD_{ij} = (k_i-k_j)^2, \cD_{ijl} = (k_i-k_j-k_l)^2,
\end{align*}
\vspace{-0.8cm}
\begin{align*}
  \cD_{i;j} = (k_i-p_j)^2, \cD_{i;jl} = (k_i-p_j-p_l)^2, \cD_{ij;l} = (k_i-k_j-p_l)^2 
\end{align*}
To accomplish this, we have used the package
Reduze2~\cite{vonManteuffel:2012np, Studerus:2009ye}.  In each set in
Eq.~(\ref{eq:Basis}), $\cD'$s are linearly independent and form a
complete basis in a sense that any Lorentz-invariant scalar product
involving loop momenta and external momenta can be expressed uniquely
in terms of $\cD'$s from that set.

As a result, we can express the unrenormalised form factors in terms
of 22 topologically different master integrals (MIs) which can be
broadly classified into three different types: genuine three-loop
integrals with vertex functions ($A_{t,i}$), three-loop propagator
integrals ($B_{t,i}$) and integrals which are product of one- and
two-loop integrals ($C_{t,i}$). 
Defining a generic three loop
master integral through
\begin{align}
  A_{i, m_{1}^i m_{2}^i \cdots m_{12}^i} = \int \frac{d^d k_1}{(2 \pi)^d} 
  \int \frac{d^d k_2}{(2 \pi)^d} 
  \int \frac{d^d k_3}{(2 \pi)^d} 
  \frac{1}{\prod_{j} D_j^{m_j^i} } ,   \quad \quad \quad i=1,2,3
\end{align}
where $D_j$ is the $j^{\rm th}$ element of the basis set $A_i$. We
identify the resulting master integrals appeared in our computation to
those given in \cite{Gehrmann:2010ue} and they are listed in the
figures below. 
%
%
%
\begin{figure}[h!]
  \centering \hspace{-0.5cm}
  \begin{subfigure}[b]{0.43\textwidth}
    \includegraphics[width=\textwidth]{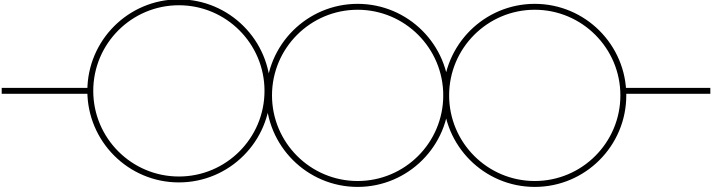}
    \caption*{$B_{6,1} \equiv A_{1,111000010101}$}
  \end{subfigure}
  \quad\quad
  \begin{subfigure}[b]{0.23\textwidth}
    \includegraphics[width=\textwidth]{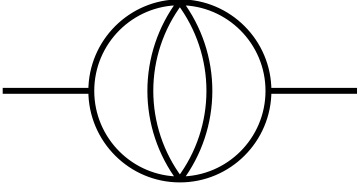}
    \caption*{$B_{6,2} \equiv A_{1,011110000101}$}
  \end{subfigure}
  \quad\quad
  \begin{subfigure}[b]{0.23\textwidth}
    \includegraphics[width=\textwidth]{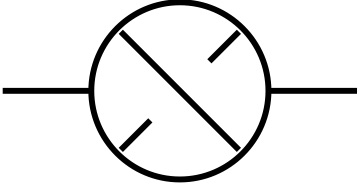}
    \caption*{$B_{8,1} \equiv A_{3,011111010101}$}
  \end{subfigure}
\end{figure}
%
%
\begin{figure}[h!]
  \centering \hspace{-1cm}
  \begin{subfigure}[b]{0.23\textwidth}
    \includegraphics[width=\textwidth]{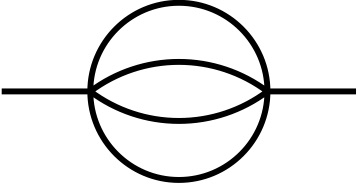}
    \caption*{$B_{4,1} \equiv A_{1,001101010000} $}
  \end{subfigure}
  \quad\quad\quad
  \begin{subfigure}[b]{0.35\textwidth}
    \includegraphics[width=\textwidth]{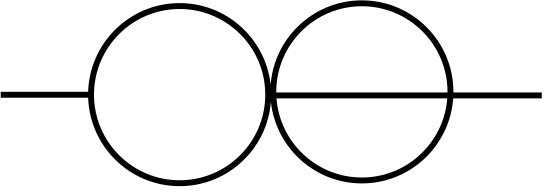}
    \caption*{$B_{5,1} \equiv A_{1,011010010100}$}
  \end{subfigure}
  \quad\quad
  \begin{subfigure}[b]{0.23\textwidth}
    \includegraphics[width=\textwidth]{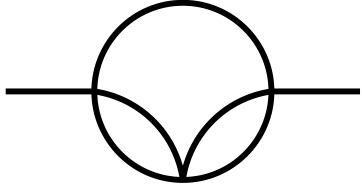}
    \caption*{$B_{5,2} \equiv A_{1,001011010100}$}
  \end{subfigure}
\end{figure}
%
%
\begin{figure}[h!]
  \centering \hspace{-1cm}
  \begin{subfigure}[b]{0.31\textwidth}
    \includegraphics[width=\textwidth]{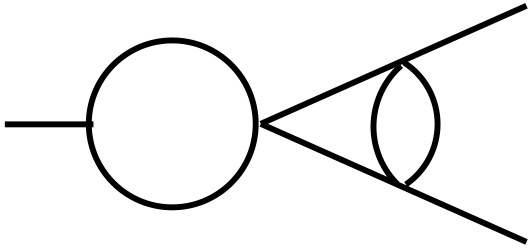}
    \caption*{$C_{6,1} \equiv A_{1,011100100101}$}
  \end{subfigure}
  \quad\quad
  \begin{subfigure}[b]{0.31\textwidth}
    \includegraphics[width=\textwidth]{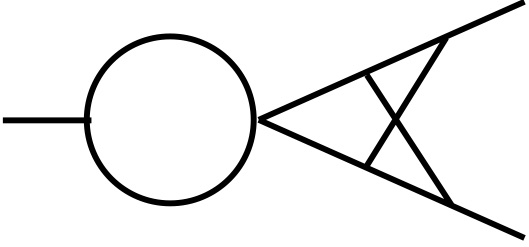}
    \caption*{$C_{8,1} \equiv A_{2,111100011101}$}
  \end{subfigure}
  \quad\quad\quad
  \begin{subfigure}[b]{0.22\textwidth}
    \includegraphics[width=\textwidth]{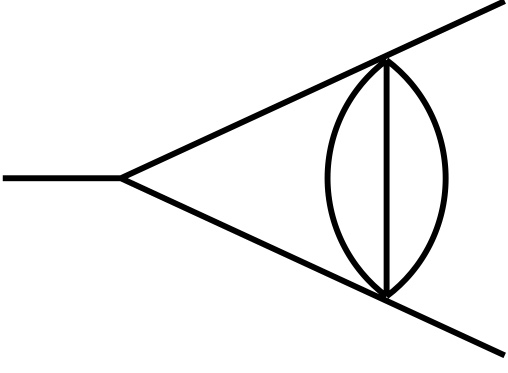}
    \caption*{$A_{5,1} \equiv A_{1,001101100001}$}
  \end{subfigure}
\end{figure}
%
%
\begin{figure}[h!]
  \centering
  \begin{subfigure}[b]{0.22\textwidth}
    \includegraphics[width=\textwidth]{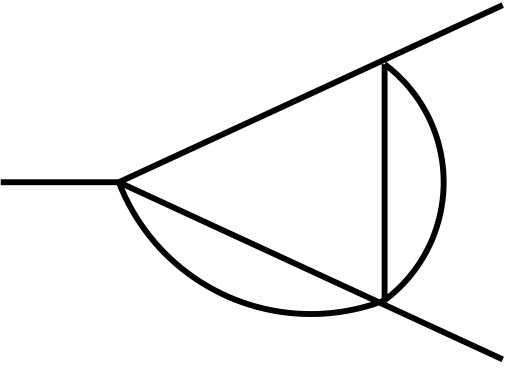}
    \caption*{$A_{5,2} \equiv A_{1,001011011000}$}
  \end{subfigure}
  \qquad\qquad
  \begin{subfigure}[b]{0.22\textwidth}
    \includegraphics[width=\textwidth]{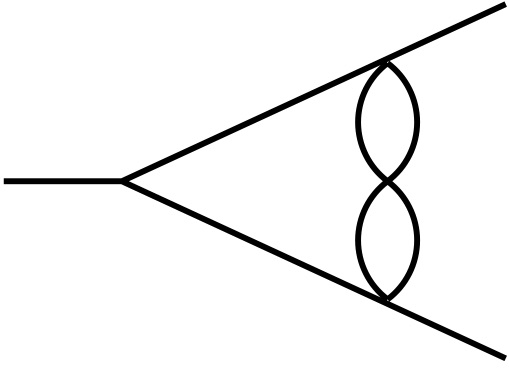}
    \caption*{$A_{6,1} \equiv A_{1,010101100110}$}
  \end{subfigure}
  \hspace{0.3cm} \quad~~~
  \begin{subfigure}[b]{0.23\textwidth}
    \includegraphics[width=\textwidth]{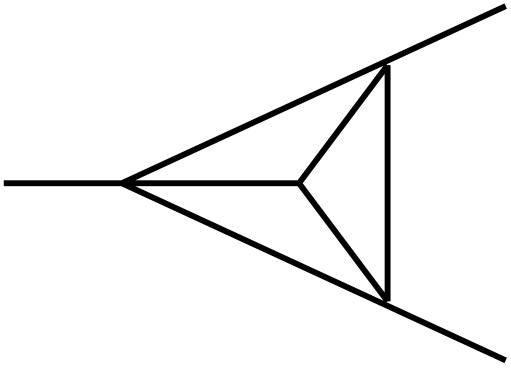}
    \caption*{$A_{6,2} \equiv A_{1,001111011000}$}
  \end{subfigure}
  ~~
\end{figure}
%
%
\begin{figure}[h!]
  \centering \hspace{-1.8cm}
  \begin{subfigure}[b]{0.22\textwidth}
    \includegraphics[width=\textwidth]{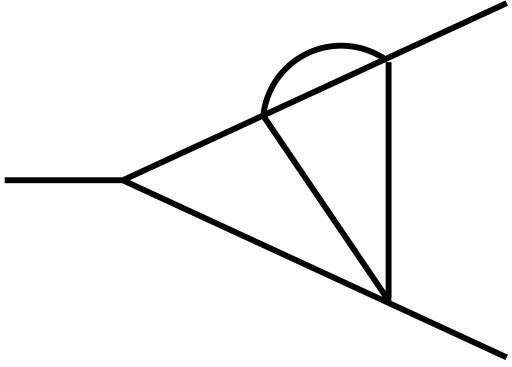}
    \caption*{$A_{6,3} \equiv A_{1,001110100101}$}
  \end{subfigure}
  \qquad\qquad
  \begin{subfigure}[b]{0.22\textwidth}
    \includegraphics[width=\textwidth]{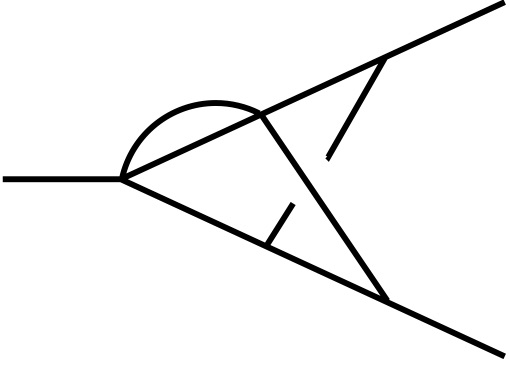}
    \caption*{$A_{7,1} \equiv A_{2,011110011100}$}
  \end{subfigure}
  \quad\quad\quad
  \begin{subfigure}[b]{0.22\textwidth}
    \includegraphics[width=\textwidth]{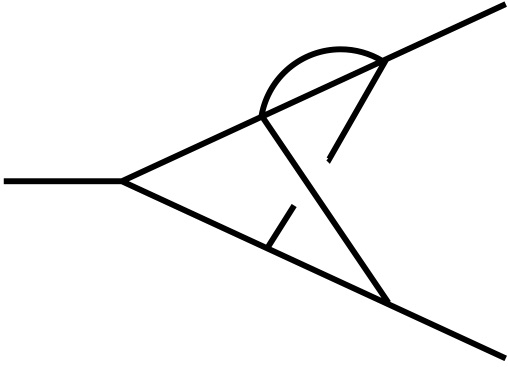}
    \caption*{$A_{7,2} \equiv A_{2,011011001101}$}
  \end{subfigure}
  \hspace{-1.3cm}
\end{figure}
%
%
\begin{figure}[h!]
  \centering \hspace{-1.8cm}
  \begin{subfigure}[b]{0.22\textwidth}
    \includegraphics[width=\textwidth]{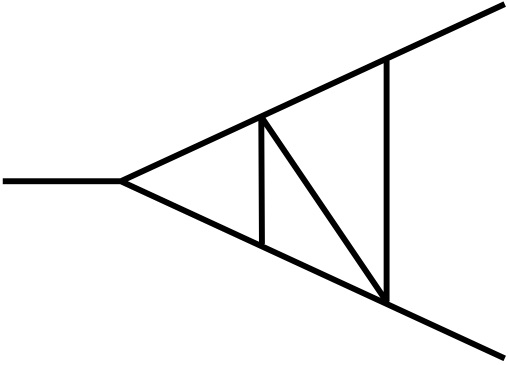}
    \caption*{$A_{7,3} \equiv A_{1,011011110100}$}
  \end{subfigure}
  \qquad\qquad
  \begin{subfigure}[b]{0.22\textwidth}
    \includegraphics[width=\textwidth]{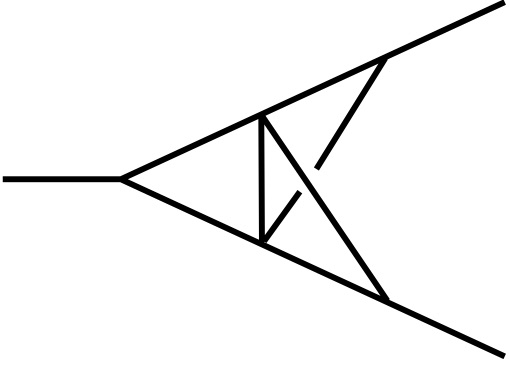}
    \caption*{$A_{7,4} \equiv A_{2,011110001101}$}
  \end{subfigure}
\end{figure}
%
%
\begin{figure}[h!]
  \centering \hspace{-1.8cm}
  \begin{subfigure}[b]{0.22\textwidth}
    \includegraphics[width=\textwidth]{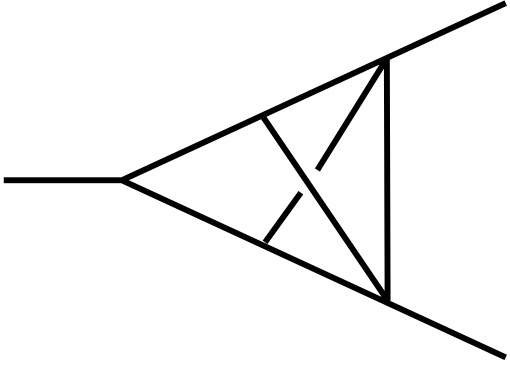}
    \caption*{$A_{7,5} \equiv A_{2,011011010101}$}
  \end{subfigure}
  \qquad\qquad
  \begin{subfigure}[b]{0.22\textwidth}
    \includegraphics[width=\textwidth]{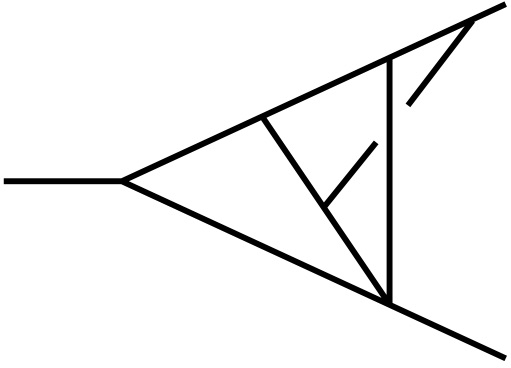}
    \caption*{$A_{8,1} \equiv A_{2,001111011101}$}
  \end{subfigure}
\end{figure}
%
%
\begin{figure}[h!]
  \centering \hspace{-1.8cm}
  \begin{subfigure}[b]{0.22\textwidth}
    \includegraphics[width=\textwidth]{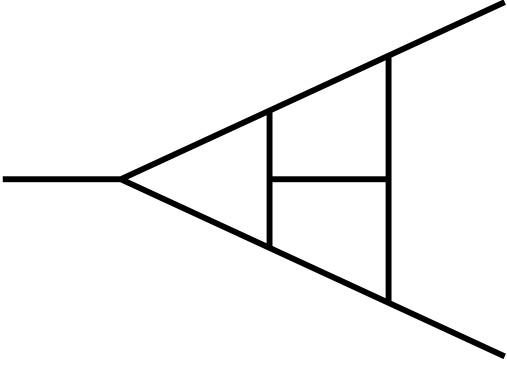}
    \caption*{$A_{9,1} \equiv A_{1,011111110110}$}
  \end{subfigure}
  \qquad\qquad
  \begin{subfigure}[b]{0.22\textwidth}
    \includegraphics[width=\textwidth]{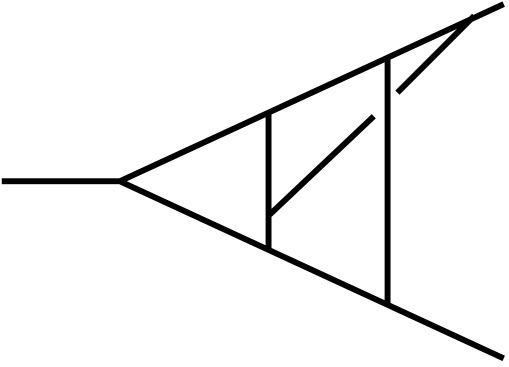}
    \caption*{$A_{9,2} \equiv A_{2,011111011101}$}
  \end{subfigure}
  \quad\quad\quad
  \begin{subfigure}[b]{0.22\textwidth}
    \includegraphics[width=\textwidth]{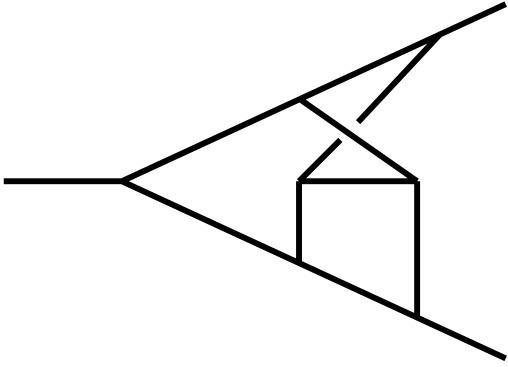}
    \caption*{$A_{9,4} \equiv A_{2,111011111100}$}
  \end{subfigure}
\end{figure}

These integrals were computed
analytically as Laurent series in $\epsilon$
in~\cite{Gehrmann:2005pd,Gehrmann:2006wg,Heinrich:2007at,Heinrich:2009be,Lee:2010cga}
and are collected in the appendix of~\cite{Gehrmann:2010ue}.
Inserting those, we obtain the final expressions for the
unrenormalised (bare) form factors that are listed in
Appendix~\ref{App:pScalar-Results}.

\subsection{UV Renormalisation}
\label{ss:UV}

To obtain ultraviolet-finite expressions for the form factors, a
renormalisation of the coupling constant and of the operators is
required. The UV renormalisation of the operators
$\left[ O_{G} \right]_{B}$ and $\left[ O_{J} \right]_{B}$ involves
some non-trivial prescriptions. These are in part related to the
formalism used for the $\gamma_{5}$ matrix, section~\ref{sec:gamma5}
above.

This formalism fails to preserve the anti-commutativity of
$\gamma_{5}$ with $\gamma^{\mu}$ in $d$ dimensions. In addition, the
standard properties of the axial current and Ward identities, which
are valid in a basic regularization scheme like the one of
Pauli-Villars, are violated as well. As a consequence, one fails to
restore the correct renormalised axial current, which is defined as
\cite{Larin:1993tq, Akyeampong:1973xi}
\begin{align}
  \label{eq:J5}
  J^{\mu}_{5} \equiv \bar{\psi}\gamma^{\mu}\gamma_{5}\psi = i
  \frac{1}{3!} \varepsilon^{\mu\nu_{1}\nu_{2}\nu_{3}} \bar{\psi}
  \gamma_{\nu_{1}} \gamma_{\nu_{2}}\gamma_{\nu_{3}} \psi
\end{align}
in dimensional regularization. To rectify this, one needs to introduce
a finite renormalisation constant $Z^{s}_{5}$
\cite{Adler:1969gk,Kodaira:1979pa} in addition to the standard overall
ultraviolet renormalisation constant $Z^{s}_{\overline{MS}}$ within
the $\overline{MS}$-scheme:
\begin{align}
  \label{eq:J5Ren}
  \left[ J^{\mu}_{5} \right]_{R} = Z^{s}_{5} Z^{s}_{\overline{MS}} \left[ J^{\mu}_{5} \right]_{B}\,.
\end{align}
By evaluating the appropriate Feynman diagrams explicitly,
$Z^{s}_{\overline{MS}}$ can be computed, however the finite
renormalisation constant is not fixed through this calculation. To
determine $Z^{s}_{5}$ one has to demand the conservation of the one
loop character \cite{Adler:1969er} of the operator relation of the
axial anomaly in dimensional regularization:
\begin{align}
  \label{eq:Anomaly}
  \left[ \partial_{\mu}J^{\mu}_{5} \right]_{R} &= a_{s} \frac{n_{f}}{2} \left[ G\tilde{G} \right]_{R}
                                                 \nonumber\\
  \text{i.e.}~~~ \left[ O_{J} \right]_{R} &= a_{s} \frac{n_{f}}{2} \left[ O_{G} \right]_{R}\,.
\end{align}
The bare operator $\left[ O_{J} \right]_{B}$ is renormalised
multiplicatively exactly in the same way as the axial current
$J^{\mu}_{5}$ through
\begin{align}
  \label{eq:OJRen}
  \left[ O_{J} \right]_{R} = Z^{s}_{5} Z^{s}_{\overline{MS}} \left[ O_{J}\right]_{B}\,,
\end{align}
whereas the other one $\left[ O_{G} \right]_{B}$ mixes under the
renormalisation through
\begin{align}
  \left[ O_{G} \right]_{R} = Z_{GG} \left[ O_{G}\right]_B +
  Z_{GJ} \left[ O_{J} \right]_B
\end{align}
with the corresponding renormalisation constants $Z_{GG}$ and
$Z_{GJ}$. The above two equations can be combined to express them
through the matrix equation
\begin{align}
  \label{eq:OpMat}
  \left[ O_{i} \right]_{R} &= Z_{ij} \left[  O_{j}\right]_{B} 
\end{align}
with
\begin{align}
  \label{eq:ZMat}
  i,j &= \{G, J\}\,, 
        \nonumber\\
  O \equiv
  \begin{bmatrix}
    O_{G}\\
    O_{J}
  \end{bmatrix}
  \qquad\quad &\text{and}  \qquad\quad
                Z \equiv
                \begin{bmatrix}
                  Z_{GG} & Z_{GJ}\\
                  Z_{JG} & Z_{JJ}
                \end{bmatrix}\,.
\end{align}
In the above expressions
\begin{align}
  \label{eq:ZJGZJJ}
  Z_{JG} &= 0 \qquad \text{to all orders in perturbation theory}\,,
           \nonumber\\
  Z_{JJ} &\equiv Z^{s}_{5} Z^{s}_{\overline{MS}}\,.
\end{align}
We determine the above-mentioned renormalisation constants
$Z^{s}_{\overline{MS}},Z_{GG},Z_{GJ}$ up to
${\cal{O}}\left( a_{s}^{3} \right)$ from our calculation of the bare
on-shell pseudo-scalar form factors described in the previous
subsection. This procedure provides a completely independent approach
to their original computation, which was done in the operator product
expansion~\cite{Zoller:2013ixa}.

Our approach to compute those $Z_{ij}$ is based on the infrared
evolution equation for the form factor, and will be detailed in
Section~\ref{ss:IR} below.  Moreover, we can fix $Z^{s}_{5}$ up to
${\cal O}(a_{s}^{2})$ by demanding the operator relation of the axial
anomaly (Eq.~(\ref{eq:Anomaly})).  Using these overall operator
renormalisation constants along with strong coupling constant
renormalisation through $Z_{a_{s}}$, Eq.~(\ref{eq:Zas}), we obtain the
UV finite on-shell quark and gluon form factors.

To define the UV renormalised form factors, we introduce a quantity
${\cal{S}}^{\lambda}_{\beta}$, constructed out of bare matrix
elements, through
\begin{align}
  \label{eq:CalSG}
  {\cal{S}}^{G}_{g} &\equiv Z_{GG} \langle {\hat{\cal
                      M}}^{G,(0)}_{g}|{{\cal M}}^{G}_{g}\rangle + Z_{GJ} \langle {\hat{\cal
                      M}}^{G,(0)}_{g}|{{\cal M}}^{J}_{g}\rangle 
                      \nonumber\\
  \intertext{and}
  {\cal{S}}^{G}_{q} &\equiv Z_{GG} \langle {\hat{\cal
                      M}}^{J,(0)}_{q}|{{\cal M}}^{G}_{q}\rangle + Z_{GJ} \langle {\hat{\cal
                      M}}^{J,(0)}_{q}|{{\cal M}}^{J}_{q}\rangle \,.
\end{align}
Expanding the quantities appearing on the right hand side of the above
equation in powers of $a_{s}$ :
\begin{align}
  \label{eq:MZExpRenas}
  |{\cal M}^{\lambda}_{\beta}\rangle &= \sum_{n=0}^{\infty} {a}^{n}_{s}
                                       |{\cal M}^{\lambda,(n)}_{\beta}\rangle\,,
                                       \nonumber\\
  Z_{I} &= \sum_{n=0}^{\infty} a^{n}_{s} Z^{(n)}_{I} \qquad \text{with}
          \qquad I=GG, GJ\,\,\,,
\end{align}
we can write \\
\begin{equation}
  {\cal S}^{G}_{g} = \sum_{n=0}^{\infty} a^{n}_{s} {\cal S}^{G,(n)}_{g}\qquad
  \text{and}\qquad {\cal S}^{G}_{q} = \sum_{n=1}^{\infty} a^{n}_{s} {\cal S}^{G,(n)}_{q}\,.
  \label{eq:CalSG}
\end{equation}
\\
Then the UV renormalised form factors corresponding to $O_{G}$ are
defined as
\begin{align}
  \label{eq:RenFFG}
  \left[ {\cal F}^{G}_{g} \right]_{R} \equiv \frac{{\cal S}^{G}_{g}}{{\cal
  S}^{G,(0)}_{g}} 
  &=
    Z_{GG} {\cal
    F}^{G}_{g} + Z_{GJ} {\cal F}^{J}_{g}
    \frac{\langle {\cal
    M}^{G,(0)}_{g}|{\cal M}^{J,(1)}_{g}\rangle}{\langle {\cal
    M}^{G,(0)}_{g}|{\cal M}^{G,(0)}_{g}\rangle} 
    \nonumber\\
  &\equiv 1 + \sum^{\infty}_{n=1} a^{n}_{s} \left[ {\cal
    F}^{G,(n)}_{g} \right]_{R} \,,
    \nonumber\\ \nonumber\\
  \left[ {\cal F}^{G}_{q}
  \right]_{R}  \equiv \frac{{\cal S}^{G}_{q}}{a_{s} {\cal
  S}^{G,(1)}_{q}} 
  &= \frac{Z_{GG} {\cal F}^{G}_{q} \langle {\cal
    M}^{J,(0)}_{q}|{\cal M}^{G,(1)}_{q}\rangle + Z_{GJ} {\cal
    F}^{J}_{q} \langle {\cal
    M}^{J,(0)}_{q}|{\cal M}^{J,(0)}_{q}\rangle}{a_{s} \left[ \langle {\cal
    M}^{J,(0)}_{q}|{\cal M}^{G,(1)}_{q}\rangle + Z^{(1)}_{GJ} \langle {\cal
    M}^{J,(0)}_{q}|{\cal M}^{J,(0)}_{q}\rangle \right]} 
    \nonumber\\
  &
    \equiv 1 + \sum^{\infty}_{n=1} a^{n}_{s} 
    \left[ {\cal
    F}^{G,(n)}_{q} \right]_{R} \,
\end{align}
where
\begin{align}
  \label{eq:SGg0SGq1}
  {\cal
  S}^{G,(0)}_{g} &= \langle {\cal
                   M}^{G,(0)}_{g}|{\cal M}^{G,(0)}_{g}\rangle \,,
                   \nonumber\\
  {\cal
  S}^{G,(1)}_{q} &= 
                   \langle {\cal M}^{J,(0)}_{q}|{\cal M}^{G,(1)}_{q}\rangle 
                   + Z^{(1)}_{GJ}\langle {\cal M}^{J,(0)}_{q}|{\cal M}^{J,(0)}_{q}\rangle\,.
\end{align}
Similarly, for defining the UV finite form factors for the other
operator $O_{J}$ we introduce

\begin{align}
  \label{eq:CalSJ}
  {\cal S}^{J}_{g} &\equiv Z^{s}_{5} Z^{s}_{\overline{MS}} \langle
                     {\hat{\cal M}}^{G,(0)}_{g}| {\cal M}^{J}_{g} \rangle\,
                     \nonumber\\
  \intertext{and}
  {\cal S}^{J}_{q} &\equiv Z^{s}_{5} Z^{s}_{\overline{MS}} \langle
                     {\hat{\cal M}}^{J,(0)}_{q}| {\cal M}^{J}_{q} \rangle\,.
\end{align}
Expanding $Z^{s}_{\overline{MS}}$ and
$|{\cal{M}}^{\lambda}_{\beta}\rangle$ in powers of $a_{s}$, following
Eq.~(\ref{eq:MZExpRenas}), we get
\begin{align}
  \label{eq:CalSJExpand}  
  {\cal S}^{J}_{g} = \sum_{n=1}^{\infty} a^{n}_{s} {\cal S}^{J,(n)}_{g}
  \qquad\quad \text{and} \qquad\quad
  {\cal S}^{J}_{q} = \sum_{n=0}^{\infty} a^{n}_{s} {\cal S}^{J,(n)}_{q}\,.
\end{align}
\\
With these we define the UV renormalised form factors corresponding to
$O_{J}$ through
\begin{align}
  \label{eq:RenFFJ}
  \left[ {\cal F}^{J}_{g} \right]_{R} &\equiv \frac{{\cal S}^{J}_{g}}{a_{s}{\cal
                                        S}^{J,(1)}_{g}} = Z^{s}_{5}
                                        Z^{s}_{\overline{MS}} {\cal
                                        F}^{J}_{g} \equiv 1 +
                                        \sum^{\infty}_{n=1} a^{n}_{s} \left[ {\cal
                                        F}^{J,(n)}_{g} \right]_{R} \,,
                                        \nonumber\\
  \left[ {\cal F}^{J}_{q}
  \right]_{R}  &\equiv \frac{{\cal S}^{J}_{q}}{{\cal
                 S}^{J,(0)}_{q}} = Z^{s}_{5}
                 Z^{s}_{\overline{MS}} {\cal
                 F}^{J}_{q} = 1 + \sum^{\infty}_{n=1} a^{n}_{s} 
                 \left[ {\cal
                 F}^{J,(n)}_{q} \right]_{R} \,
\end{align}
where
\begin{align}
  \label{eq:SGg0SGq1}
  {\cal
  S}^{J,(1)}_{g} &= \langle {\cal
                   M}^{G,(0)}_{g}|{\cal M}^{J,(1)}_{g}\rangle \,,
                   \nonumber\\
  {\cal
  S}^{J,(0)}_{q} &= 
                   \langle {\cal M}^{J,(0)}_{q}|{\cal M}^{J,(0)}_{q}\rangle\,.
\end{align}
The finite renormalisation constant $Z^{s}_{5}$ is multiplied in
Eq.~(\ref{eq:CalSJ}) to restore the axial anomaly equation in
dimensional regularisation. We determine all required renormalisation
constants from consistency conditions on the universal structure of
the infrared poles of the renormalised form factors in the next
section, and use these constants to derive the UV-finite form factors
in Section~\ref{ss:Ren}.

\subsection{Infrared Singularities and Universal Pole Structure}
\label{ss:IR}

The renormalised form factors are ultraviolet-finite, but still
contain divergences of infrared origin.  In the calculation of
physical quantities (which fulfill certain infrared-safety
criteria~\cite{Sterman:1977wj}), these infrared singularities are
cancelled by contributions from real radiation processes that yield
the same observable final state, and by mass factorization
contributions associated with initial-state partons.  The pole
structures of these infrared divergences arising in QCD form factors
exhibit some universal behaviour. The very first successful proposal
along this direction was presented by Catani~\cite{Catani:1998bh} (see
also \cite{Sterman:2002qn}) for one and two-loop QCD amplitudes using
the universal subtraction operators. The factorization of the single
pole in quark and gluon form factors in terms of soft and collinear
anomalous dimensions was first revealed in \cite{Ravindran:2004mb} up
to two loop level whose validity at three loop was later established
in the article \cite{Moch:2005tm}. The proposal by Catani was
generalized beyond two loops by Becher and
Neubert~\cite{Becher:2009cu} and by Gardi and
Magnea~\cite{Gardi:2009qi}. Below, we outline this behaviour in the
context of pseudo-scalar form factors up to three loop level,
following closely the notation used in~\cite{Ravindran:2005vv}.

The unrenormalised form factors
${\cal F}^{\lambda}_{\beta}(\hat{a}_{s}, Q^{2}, \mu^{2}, \epsilon)$
satisfy the so-called $KG$-differential equation \cite{Sudakov:1954sw,
  Mueller:1979ih, Collins:1980ih, Sen:1981sd} which is dictated by the
factorization property, gauge and renormalisation group (RG)
invariances:
\begin{equation}
  \label{eq:KG}
  Q^2 \frac{d}{dQ^2} \ln {\cal F}^{\lambda}_{\beta} (\hat{a}_s, Q^2, \mu^2, \epsilon)
  = \frac{1}{2} \left[ K^{\lambda}_{\beta} (\hat{a}_s, \frac{\mu_R^2}{\mu^2}, \epsilon
    )  + G^{\lambda}_{\beta} (\hat{a}_s, \frac{Q^2}{\mu_R^2}, \frac{\mu_R^2}{\mu^2}, \epsilon ) \right]
\end{equation}
where all poles in the dimensional regulator $\ep$ are contained in
the $Q^{2}$ independent function $K^{\lambda}_{\beta}$ and the finite
terms in $\epsilon \rightarrow 0$ are encapsulated in
$G^{\lambda}_{\beta}$. RG invariance of the form factor implies
\begin{equation}
  \label{eq:KIA}
  \mu_R^2 \frac{d}{d\mu_R^2} K^{\lambda}_{\beta}(\hat{a}_s, \frac{\mu_R^2}{\mu^2},
  \epsilon )  = - \mu_R^2 \frac{d}{d\mu_R^2} G^{\lambda}_{\beta}(\hat{a}_s,
  \frac{Q^2}{\mu_R^2},  \frac{\mu_R^2}{\mu^2}, \epsilon ) 
  = - A^{\lambda}_{\beta} (a_s (\mu_R^2)) = - \sum_{i=1}^{\infty}  a_s^i (\mu_R^2) A^{\lambda}_{\beta,i} 
\end{equation}
where, $A^{\lambda}_{\beta,i}$ on the right hand side are the $i$-loop
cusp anomalous dimensions. It is straightforward to solve for
$K^{\lambda}_{\beta}$ in Eq.~(\ref{eq:KIA}) in powers of bare strong
coupling constant $\ashat$ by performing the following expansion
\begin{align}
  K^{\lambda}_{\beta}\l({\hat a}_{s}, \frac{\mu_{R}^{2}}{\mu^{2}}, \ep\r)  =
  \sum_{i=1}^{\infty}  {\hat a}_{s}^{i}
  \l(\frac{\mu_{R}^{2}}{\mu^{2}}\r)^{i\frac{\ep}{2}}  S_{\ep}^{i} K^{\lambda}_{\beta,i}(\ep)\, .
\end{align}
The solutions $K^{\lambda}_{\beta,i}(\ep)$ consist of simple poles in
$\ep$ with the coefficients consisting of $A_{\beta, i}^{\lambda}$ and
$\beta_{i}$. These can be found in \cite{Ravindran:2005vv,
  Ravindran:2006cg}. On the other hand, the RGE of
$G^{\lambda}_{\beta,i}(\hat{a}_s, \frac{Q^2}{\mu_R^2},
\frac{\mu_R^2}{\mu^2}, \epsilon )$
can be solved. The solution contains two parts, one is dependent on
$\mu_{R}^{2}$ whereas the other part depends only the boundary point
$\mu^{2}_{R}=Q^{2}$. The $\mu_{R}^{2}$ dependent part can eventually
be expressed in terms of $A^{\lambda}_{\beta}$:
\begin{align}
  \label{eq:GSoln}
  G^{\lambda}_{\beta}(\hat{a}_s, \frac{Q^2}{\mu_R^2},
  \frac{\mu_R^2}{\mu^2}, \epsilon ) = G^{\lambda}_{\beta}({a}_s(Q^{2}),
  1, \epsilon ) + \int_{\frac{Q^{2}}{\mu_{R}^{2}}}^{1} \frac{dx}{x}  
  A^{\lambda}_{\beta}(a_{s}\l(x\mu_{R}^{2})\r)\,. 
\end{align}
The boundary term can be expanded in powers of $a_{s}$ as
\begin{align}
  G^{\lambda}_{\beta}(a_s(Q^2), 1, \epsilon) = \sum_{i=1}^{\infty} a_s^i(Q^2) G^{\lambda}_{\beta,i}(\epsilon)\, .
\end{align}
The solutions of $K^{\lambda}_{\beta}$ and $G^{\lambda}_{\beta}$
enable us to solve the $KG$ equation (Eq.~(\ref{eq:KG})) and thereby
facilitate to obtain the
$\ln {\cal F}^{\lambda}_{\beta}(\hat{a}_s, Q^2, \mu^2, \ep)$ in terms
of $A^{\lambda}_{\beta, i}, G^{\lambda}_{\beta, i}$ and $\beta_{i}$
which is given by~\cite{Ravindran:2005vv}
\begin{align}
  \label{eq:lnFSoln}
  \ln {\cal F}^{\lambda}_{\beta}(\hat{a}_s, Q^2, \mu^2, \ep) =
  \sum_{i=1}^{\infty} {\hat a}_{s}^{i} \l(\frac{Q^{2}}{\mu^{2}}\r)^{i
  \frac{\ep}{2}} S_{\ep}^{i} \hat {\cal L}_{\beta,i}^{\lambda}(\ep)
\end{align}
with
\begin{align}
  \label{eq:lnFitoCalLF}
  \hat {\cal L}_{\beta,1}^{\lambda}(\ep) =& { \frac{1}{\ep^2} } \Bigg\{-2 A^{\lambda}_{\beta,1}\Bigg\}
                                            + { \frac{1}{\ep}
                                            }
                                            \Bigg\{G^{\lambda}_{\beta,1}
                                            (\ep)\Bigg\}\, ,
                                            \nonumber\\
  \hat {\cal L}_{\beta,2}^{\lambda}(\ep) =& { \frac{1}{\ep^3} } \Bigg\{\beta_0 A^{\lambda}_{\beta,1}\Bigg\}
                                            + {
                                            \frac{1}{\ep^2} }
                                            \Bigg\{-  {
                                            \frac{1}{2} }  A^{\lambda}_{\beta,2}
                                            - \beta_0   G^{\lambda}_{\beta,1}(\ep)\Bigg\}
                                            + { \frac{1}{\ep}
                                            } \Bigg\{ {
                                            \frac{1}{2} }  G^{\lambda}_{\beta,2}(\ep)\Bigg\}\, ,
                                            \nonumber\\
  \hat {\cal L}_{\beta,3}^{\lambda}(\ep) =& { \frac{1}{\ep^4}
                                            } \Bigg\{- {
                                            \frac{8}{9} }  \beta_0^2 A^{\lambda}_{\beta,1}\Bigg\}
                                            + {
                                            \frac{1}{\ep^3} }
                                            \Bigg\{ { \frac{2}{9} } \beta_1 A^{\lambda}_{\beta,1}
                                            + { \frac{8}{9} }
                                            \beta_0 A^{\lambda}_{\beta,2}  + { \frac{4}{3} }
                                            \beta_0^2 G^{\lambda}_{\beta,1}(\ep)\Bigg\}
                                            \nonumber\\
                                          &
                                            + { \frac{1}{\ep^2} } \Bigg\{- { \frac{2}{9} } A^{\lambda}_{\beta,3}
                                            - { \frac{1}{3} } \beta_1 G^{\lambda}_{\beta,1}(\ep)
                                            - { \frac{4}{3} } \beta_0 G^{\lambda}_{\beta,2}(\ep)\Bigg\}
                                            + { \frac{1}{\ep}
                                            } \Bigg\{  { \frac{1}{3} } G^{\lambda}_{\beta,3}(\ep)\Bigg\}\, .
\end{align}
All these form factors are observed to satisfy \cite{Ravindran:2004mb,
  Moch:2005tm} the following decomposition in terms of collinear
($B^{\lambda}_{\beta}$), soft ($f^{\lambda}_{\beta}$) and UV
($\gamma^{\lambda}_{\beta}$) anomalous dimensions:
\begin{align}
  \label{eq:GIi}
  G^{\lambda}_{\beta,i} (\ep) = 2 \left(B^{\lambda}_{\beta,i} -
  \gamma^{\lambda}_{\beta,i}\right)  + f^{\lambda}_{\beta,i} +
  C^{\lambda}_{\beta,i}  + \sum_{k=1}^{\infty} \epsilon^k g^{\lambda,k}_{\beta,i} \, ,
\end{align}
where the constants $C^{\lambda}_{\beta,i}$ are given by
\cite{Ravindran:2006cg}
\begin{align}
  \label{eq:Cg}
  C^{\lambda}_{\beta,1} &= 0\, ,
                          \nonumber\\
  C^{\lambda}_{\beta,2} &= - 2 \beta_{0} g^{\lambda,1}_{\beta,1}\, ,
                          \nonumber\\
  C^{\lambda}_{\beta,3} &= - 2 \beta_{1} g^{\lambda,1}_{\beta,1} - 2
                          \beta_{0} \left(g^{\lambda,1}_{\beta,2}  + 2 \beta_{0} g^{\lambda,2}_{\beta,1}\right)\, .
\end{align}
In the above expressions, $X^{\lambda}_{\beta,i}$ with $X=A,B,f$ and
$\gamma^{\lambda}_{\beta, i}$ are defined through
\begin{align}
  \label{eq:ABfgmExp}
  X^{\lambda}_{\beta} &\equiv \sum_{i=1}^{\infty} a_{s}^{i}
                        X^{\lambda}_{\beta,i}\,,
                        \qquad \text{and} \qquad
                        \gamma^{\lambda}_{\beta} \equiv \sum_{i=1}^{\infty} a_{s}^{i} \gamma^{\lambda}_{\beta,i}\,\,.
\end{align}
Within this framework, we will now determine this universal structure
of IR singularities of the pseudo-scalar form factors. This
prescription will be used subsequently to determine the overall
operator renormalisation constants.

We begin with the discussion of form factors corresponding to
$O_{J}$. The results of the form factors ${\cal F}^{J}_{\beta}$ for
$\beta=q,g$, which have been computed up to three loop level in this
article are being used to extract the unknown factors,
$\gamma^{J}_{\beta,i}$ and $g^{J,k}_{\beta,i}$, by employing the $KG$
equation. Since the ${\cal F}^{J}_{\beta}$ satisfy $KG$ equation, we
can obtain the solutions Eq.~(\ref{eq:lnFSoln}) along with
Eq.~(\ref{eq:lnFitoCalLF}) and Eq.~(\ref{eq:GIi}) to examine our
results against the well known decomposition of the form factors in
terms of the quantities $X^{J}_{\beta}$. These are universal, and
appear also in the vector and scalar quark and gluon form
factors~\cite{Moch:2005tm, Ravindran:2004mb}. They are
known~\cite{Vogt:2004mw, Catani:1990rp, Vogt:2000ci, Ravindran:2004mb,
  Ahmed:2014cha} up to three loop level in the literature.  Using
these in the above decomposition, we obtain
$\gamma^{J}_{\beta,i}$. The other process dependent constants, namely,
$g^{J,k}_{\beta,i}$ can be obtained by comparing the coefficients of
$\epsilon^{k}$ in Eq.~(\ref{eq:lnFitoCalLF}) at every order in
${\hat a}_{s}$. We can get the quantities $\gamma^{J}_{g,i}$ and
$g^{J,k}_{g,i}$ up to two loop level, since this process starts at one
loop. From gluon form factors we get
\begin{align}
  \label{eq:gmJqQ}
  \gamma^{J}_{g,1} &= 0\,,
                     \nonumber\\
  \gamma^{J}_{g,2} &= {\dis{C_{A} C_{F}}} \Bigg\{- \frac{44}{3} \Bigg\}+
                     {\dis{C_{F} n_{f}}} \Bigg\{- \frac{10}{3}
                     \Bigg\}\,. \hspace{6cm}
\end{align}
Similarly, from the quark form factors we obtain
\begin{align}
  \label{eq:gmJqQ}
  \gamma^{J}_{q,1} &= 0\,,
                     \nonumber\\
  \gamma^{J}_{q,2} &= {\dis{C_{A} C_{F}}} \Bigg\{- \frac{44}{3} \Bigg\}+
                     {\dis{C_{F} n_{f}}} \Bigg\{- \frac{10}{3} \Bigg\}\,, 
                     \nonumber\\
  \gamma^{J}_{q,3} &= {\dis{C^{2}_{A} C_{F}}} \Bigg\{ -
                     \frac{3578}{27}\Bigg\} +
                     {\dis{C^{2}_{F}n_{f}}} \Bigg\{\frac{22}{3}\Bigg\}
                     - {\dis{C_{F} n^{2}_{f}}}
                     \Bigg\{\frac{26}{27}\Bigg\} + {\dis{C_{A}
                     C^{2}_{F}}} \Bigg\{\frac{308}{3}\Bigg\} 
                     \nonumber\\
                   &+
                     {\dis{C_{A} C_{F} n_{f}}} \Bigg\{-\frac{149}{27}\Bigg\}\,.
\end{align}
Note that $\gamma^{J}_{q,i} = \gamma^{J}_{g,i}$ which is expected
since these are the UV anomalous dimensions associated with the same
operator $[O_{J}]_{B}$. The $\gamma^{J}_{\beta,i}$ are further used to
obtain the overall operator renormalisation constant
$Z^{s}_{\overline{MS}}$ through the RGE:
\begin{align}
  \label{eq:RGEZMS}
  \mu_{R}^{2}\frac{d}{d\mu_{R}^{2}}
  \ln Z^{\lambda}(a_{s},\mu_{R}^{2},\epsilon) = \sum_{i=1}^{\infty}
  a_{s}^{i} \gamma^{\lambda}_{i}.
\end{align}
\\
The general solution of the RGE is obtained as
\begin{align}
  \label{eq:GenSolZI}
  Z^{\lambda} &= 1 + a_s \Bigg[ \frac{1}{\epsilon} 2  {\gamma^{\lambda}_{1}} \Bigg]  +
                a_s^2 \Bigg[  \frac{1}{\epsilon^2} \Bigg\{ 2 \beta_0
                {\gamma^{\lambda}_{1}}  + 2 ({\gamma^{\lambda}_{1}})^2 \Bigg\} + 
                \frac{1}{\epsilon} {\gamma^{\lambda}_{2}}  \Bigg] 
                + a_s^3 \Bigg[
                \frac{1}{\epsilon^{3}} \Bigg\{ 8 \beta_0^2 {\gamma^{\lambda}_{1}}  +
                4 \beta_0 ({\gamma^{\lambda}_{1}})^2 
                \nonumber\\
              &+
                \frac{4 ({\gamma^{\lambda}_{1}})^3}{3} \Bigg\} +  \frac{1}{\epsilon^{2}} \Bigg\{ \frac{4
                \beta_1 {\gamma^{\lambda}_{1}}}{3} + \frac{4 \beta_0
                {\gamma^{\lambda}_{2}}}{3}  + 2 {\gamma^{\lambda}_{1}}
                {\gamma^{\lambda}_{2}} \Bigg\} + 
                \frac{1}{\epsilon} \Bigg\{ \frac{2 {\gamma^{\lambda}_{3}}}{3} \Bigg\}
                \Bigg]\,.
\end{align}
By substituting the results of $\gamma^{J}_{\beta,i}$ in the above
solution we get $Z^{s}_{\overline{MS}}$ up to ${\cal O}(a_{s}^{3})$:

\begin{align}
  \label{eq:ZMS}
  Z^{s}_{\overline{MS}} &= 1 + a^{2}_{s} \Bigg[C_{A} C_{F} \Bigg\{-
                          \frac{44}{3 \epsilon} \Bigg\} + C_{F} n_{f} \Bigg\{ -
                          \frac{10}{3 \epsilon} \Bigg\}
                          \Bigg]
                          + a^{3}_{s} \Bigg[ C_{A}^2 C_{F} \Bigg\{ -
                          \frac{1936}{27 \epsilon^2} - \frac{7156}{81 \epsilon}
                          \Bigg\}
                          \nonumber\\
                        &+  C_{F}^2 n_{f} \Bigg\{ \frac{44}{9 \epsilon} \Bigg\} + 
                          C_{F} n_{f}^2 \Bigg\{ \frac{80}{27 \epsilon^2} - \frac{52}{81 \epsilon} \Bigg\}  + 
                          C_{A}  C_{F}^2 \Bigg\{ \frac{616}{9 \epsilon}\Bigg\} +
                          C_{A} C_{F} n_{f} \Bigg\{ - \frac{88}{27 \epsilon^2} -
                          \frac{298}{81 \epsilon} \Bigg\}
                          \Bigg],
\end{align}
which agrees completely with the known result in \cite{Larin:1993tq}.
In order to restore the axial anomaly equation in dimensional
regularization (see Section~\ref{ss:UV} above), we must multiply the
$Z^{s}_{\overline{MS}} \left[ O_{J} \right]_{B}$ by a finite
renormalisation constant $Z^{s}_{5}$, which reads \cite{Larin:1993tq}
\begin{align}
  \label{eq:Z5s}
  Z^{s}_{5} = 1 + a_{s} \{-4 C_{F}\} + a^{2}_{s} \left\{ 22 C^{2}_{F} -
  \frac{107}{9} C_{A} C_{F} + \frac{31}{18} C_{F} n_{f} \right\}\,.
\end{align}
Following the computation of the operator mixing constants below, we
will be able to verify explicitly that this expression yields the
correct expression for the axial anomaly.

Now, we move towards the discussion of $O_{G}$ form factors. Similar
to previous case, we consider the form factors
$Z_{GG}^{-1} [ {\cal F}^{G}_{\beta} ]_{R}$, defined through
Eq.~(\ref{eq:RenFFG}), to extract the unknown constants,
$\gamma^{G}_{\beta,i}$ and $g^{G,k}_{\beta,i}$, by utilizing the $KG$
differential equation. Since, $[{\cal F}^G_{\beta}]_{R}$ is UV finite,
the product of $Z_{GG}^{-1}$ with $[{\cal F}^G_{\beta}]_{R}$ can
effectively be treated as unrenormalised form factor and hence we can
demand that $Z_{GG}^{-1} [{\cal F}^{G}_{\beta}]_{R}$ satisfy $KG$
equation.  Further we make use of the solutions Eq.~(\ref{eq:lnFSoln})
in conjunction with Eq.~(\ref{eq:lnFitoCalLF}) and Eq.~(\ref{eq:GIi})
to compare our results against the universal decomposition of the form
factors in terms of the constants $X^{G}_{\beta}$. Upon substituting
the existing results of the quantities
$A^{G}_{\beta,i}, B^{G}_{\beta,i}$ and $f^{G}_{\beta,i}$ up to three
loops, which are obtained in case of quark and gluon form factors, we
determine the anomalous dimensions $\gamma^{G}_{\beta,i}$ and the
constants $g^{G,k}_{\beta,i}$. However, it is only possible to get the
factors $\gamma^{G}_{q,i}$ and $g^{G,k}_{q,i}$ up to two loops because
of the absence of a tree level amplitude in the quark initiated
process for the operator $O_{G}$. Since $[{\cal F}^G_{\beta}]_{R}$ are
UV finite, the anomalous dimensions $\gamma^{G}_{\beta,i}$ must be
equal to the anomalous dimension corresponding to the renormalisation
constant $Z_{GG}$. This fact is being used to determine the overall
renormalisation constants $Z_{GG}$ and $Z_{GJ}$ up to three loop level
where these quantities are parameterized in terms of the newly
introduced anomalous dimensions $\gamma_{ij}$ through the matrix
equation
\begin{align}
  \label{eq:ZijDefn}
  \mu_{R}^{2}\frac{d}{d\mu_{R}^{2}}Z_{ij} \equiv \gamma_{ik} Z_{kj}\,
  \qquad \text{with} \qquad i,j,k={G,J}
\end{align}
This can be equivalently written as
\begin{align}
  \label{eq:ZijDefn-1}
\gamma_{ij} = \left( \mu_{R}^{2}\frac{d}{d\mu_{R}^{2}}Z_{ik} \right)  \left( Z^{-1} \right)_{kj}\,.
\end{align}
The general solution (See Example 2 in Appendix~\ref{chpt:App-SolRGEZas}) of the RGE up to $a_{s}^{3}$ is obtained as
\begin{align}
  \label{eq:ZCoupSoln}
  Z_{ij} &= \delta_{ij} 
           + {a}_{s} \Bigg[ \frac{2}{\epsilon}
           \gamma_{ij,1} \Bigg] 
           + {a}_{s}^{2} \Bigg[
           \frac{1}{\epsilon^{2}} \Bigg\{  2
           \beta_{0} \gamma_{ij,1} + 2 \gamma_{ik,1} \gamma_{kj,1}  \Bigg\} + \frac{1}{\epsilon} \Bigg\{ \gamma_{ij,2}\Bigg\}
           \Bigg] 
           + {a}_{s}^{3} \Bigg[ \frac{1}{\epsilon^{3}} \Bigg\{
           \frac{8}{3} \beta_{0}^{2} \gamma_{ij,1} 
           \nonumber\\
         &+ 4 \beta_{0} \gamma_{ik,1}
           \gamma_{kj,1} + \frac{4}{3} \gamma_{ik,1} \gamma_{kl,1}
           \gamma_{lj,1} \Bigg\} + \frac{1}{\epsilon^{2}} \Bigg\{ \frac{4}{3} \beta_{1} \gamma_{ij,1} +
           \frac{4}{3} \beta_{0} \gamma_{ij,2} 
           + \frac{2}{3}
           \gamma_{ik,1} \gamma_{kj,2} 
           \nonumber\\
         &+ \frac{4}{3}
           \gamma_{ik,2} \gamma_{kj,1} \Bigg\} + \frac{1}{\epsilon}
           \Bigg\{ \frac{2}{3} \gamma_{ij,3} \Bigg\} \Bigg]
\end{align}
where, $\gamma_{ij}$ is expanded in powers of $a_{s}$ as
\begin{align}
  \label{eq:gammaijExp}
  \gamma_{ij} = \sum_{n=1}^{\infty} a_{s}^{n} \gamma_{ij,n}\,.
\end{align}
Demanding the vanishing of $\gamma^{G}_{\beta,i}$, we get
\begin{align}
  \label{eq:gammaGG}
  \gamma_{GG} &= a_{s} \Bigg[ \frac{11}{3} C_{A} - \frac{2}{3} n_{f}\Bigg] +
                a_{s}^{2} \Bigg[ \frac{34}{3} C_{A}^{2} - \frac{10}{3} C_{A} n_{f} -
                2 C_{F} n_{f} \Bigg]
                + a_{s}^{3} \Bigg[ \frac{2857}{54} C_{A}^3 - \frac{1415}{54} C_{A}^2
                n_{f}  
                \nonumber\\
              &- \frac{205}{18} C_{A} C_{F} n_{f} + C_{F}^2 n_{f} +
                \frac{79}{54} C_{A} n_{f}^2 + \frac{11}{9} C_{F} n_{f}^2\Bigg]\,,
                \nonumber\\
  \gamma_{GJ} &= a_{s} \Bigg[ - 12 C_{F} \Bigg] + a_{s}^{2} \Bigg[
                - \frac{284}{3} C_{A} C_{F}  + 36 C_{F}^2 +
                \frac{8}{3} C_{F} n_{f} \Bigg]
                + a_{s}^{3} \Bigg[ - \frac{1607}{3} C_{A}^2 C_{F} 
                \nonumber\\
              &+ 461 C_{A}  C_{F}^2 - 126 C_{F}^3  - \frac{164}{3} C_{A} C_{F} n_{f} + 
                214 C_{F}^2 n_{f} + \frac{52}{3} C_{F} n_{f}^2
                + 288 C_{A} C_{F}
                n_{f} \zeta_3  
                \nonumber\\
              &- 288 C_{F}^2 n_{f} \zeta_3
                \Bigg]\,.
\end{align}
In addition to the demand of vanishing $\gamma^{G}_{\beta,i}$, it is
required to use the results of $\gamma_{JJ}$ and $\gamma_{JG}$, which
are implied by the definition, Eq.~(\ref{eq:ZijDefn}), up to
${\cal O}(a_{s}^{2})$ to determine the above-mentioned $\gamma_{GG}$
and $\gamma_{GJ}$ up to the given order. This is a consequence of the
fact that the operators mix under UV renormalisation. Following
Eq.~(\ref{eq:ZijDefn}) along with Eq.~(\ref{eq:ZJGZJJ}),
Eq.~(\ref{eq:ZMS}) and Eq.~(\ref{eq:Z5s}), we obtain
\begin{align}
  \label{eq:gammaJJJG}
  \gamma_{JJ} &= a_{s} \Bigg[ - \epsilon 2 C_{F}  \Bigg] + a_{s}^{2}
                \Bigg[ \epsilon \Bigg\{ - \frac{107}{9} C_A C_F + 14
                C_F^2 + \frac{31}{18} C_F n_f
                \Bigg\} - 6
                C_F n_f  \Bigg]
                \intertext{and}
                \gamma_{JG} &=0\,.
\end{align}
As it happens, we note that $\gamma_{JJ}$'s are $\epsilon$-dependent
and in fact, this plays a crucial role in determining the other
quantities. Our results are in accordance with the existing ones,
$\gamma_{GG}$ and $\gamma_{GJ}$, which are available up to
${\cal O}(a_{s}^{2})$ \cite{Larin:1993tq} and ${\cal O}(a_{s}^{3})$
\cite{Zoller:2013ixa}, respectively. In addition to the existing ones,
here we compute the new result of $\gamma_{GG}$ at
${\cal O}(a_{s}^{3})$. It was observed through explicit computation in
the article \cite{Larin:1993tq} that
\begin{align}
  \label{eq:gammaGGbt}
  \gamma_{GG} = - \frac{\beta}{a_{s}} 
\end{align} 
holds true up to two loop level but there was no statement on the
validity of this relation beyond that order. In \cite{Zoller:2013ixa},
it was demonstrated in the operator product expansion that the
relation holds even at three loop. Here, through explicit calculation,
we arrive at the same conclusion that the relation is still valid at
three loop level which can be seen if we look at the $\gamma_{GG, 3}$
in Eq.~(\ref{eq:gammaGG}) which is equal to the $\beta_{2}$.

Before ending the discussion of $\gamma_{ij}$, we examine our results
against the axial anomaly relation. The renormalisation group
invariance of the anomaly equation (Eq.~(\ref{eq:Anomaly})), see
\cite{Larin:1993tq}, gives
\begin{align}
  \label{eq:AnomalyAlt}
  \gamma_{JJ} = \frac{\beta}{a_{s}} + \gamma_{G{G}} + a_{s}
  \frac{n_{f}}{2} \gamma_{GJ}\,.
\end{align}
Through our calculation up to three loop level we find that our
results are in complete agreement with the above anomaly equation
through
\begin{align}
  \label{eq:AnomalyHold}
  \gamma_{GG} &= - \frac{\beta}{a_{s}}
                \qquad \text{and} \qquad
                \gamma_{GJ} = \left( a_{s} \frac{n_{f}}{2} \right)^{-1} \gamma_{JJ}
\end{align}
in the limit of $\epsilon \rightarrow 0$. This serves as one of the
most crucial checks on our computation.

Additionally, if we conjecture the above relations to hold beyond
three loops (which could be doubted in light of recent
findings~\cite{Almelid:2015jia}), then we can even predict the
$\epsilon$-independent part of the $\gamma_{JJ}$ at
${\cal O}(a_{s}^{3})$:
\begin{align}
  \label{eq:gammaJJ4}
  \gamma_{JJ}|_{\epsilon \rightarrow 0} &=  a_{s}^{2} \Bigg[ - 6 C_{F} n_{f} \Bigg] + a_{s}^{3} \Bigg[
                                          - \frac{142}{3} C_{A} C_{F} n_{f}  + 18 C_{F}^2 n_{f} +
                                          \frac{4}{3} C_{F} n_{f}^{2} \Bigg]\,.
\end{align}

The results of $\gamma_{ij}$ uniquely specify $Z_{ij}$, through
Eq.~(\ref{eq:ZCoupSoln}).  We summarize the resulting expressions of
$Z_{ij}$ below:
\begin{align}
  \label{eq:ZGGtZGJ}
  Z_{GG} &= 1 +  a_s \Bigg[ \frac{22}{3\epsilon}
           C_{A}  -
           \frac{4}{3\epsilon} n_{f} \Bigg] 
           + 
           a_s^2 \Bigg[ \frac{1}{\epsilon^2}
           \Bigg\{ \frac{484}{9} C_{A}^2 - \frac{176}{9} C_{A}
           n_{f} + \frac{16}{9} n_{f}^2 \Bigg\}
           + \frac{1}{\epsilon} \Bigg\{ \frac{34}{3} C_{A}^2  
           \nonumber\\
         &-
           \frac{10}{3} C_{A} n_{f}  - 2 C_{F} n_{f} \Bigg\} \Bigg] 
           + 
           a_s^3 \Bigg[   \frac{1}{\epsilon^3} 
           \Bigg\{ \frac{10648}{27} C_{A}^3 - \frac{1936}{9}
           C_{A}^2 n_{f}  + \frac{352}{9} C_{A} n_{f}^2  -
           \frac{64}{27} n_{f}^3 \Bigg\}  
           \nonumber\\
         &+   \frac{1}{\epsilon^2}
           \Bigg\{ \frac{5236}{27} C_{A}^3 - \frac{2492}{27}
           C_{A}^2 n_{f}  - \frac{308}{9} C_{A} C_{F} n_{f}  + 
           \frac{280}{27} C_{A} n_{f}^2  + \frac{56}{9} C_{F}
           n_{f}^2 \Bigg\}
           \nonumber\\
         &  
           +  \frac{1}{\epsilon} \Bigg\{ \frac{2857}{81} C_{A}^3  -
           \frac{1415}{81} C_{A}^2 n_{f}  - \frac{205}{27} C_{A} C_{F} n_{f} + 
           \frac{2}{3} C_{F}^2 n_{f} + \frac{79}{81} C_{A}
           n_{f}^2  + \frac{22}{27} C_{F} n_{f}^2 \Bigg\}
           \Bigg]
           \nonumber \intertext{and} 
           Z_{GJ} &=  a_s \Bigg[ - \frac{24}{\epsilon} C_{F} \Bigg]
                    + 
                    a_s^2 \Bigg[ \frac{1}{\epsilon^2}
                    \Bigg\{ - 176 C_{A} C_{F} + 32 C_{F} n_{f} \Bigg\}
                    + \frac{1}{\epsilon} \Bigg\{ - \frac{284}{3} C_{A} C_{F} +
                    84 C_{F}^2 
                    \nonumber\\
         &+ \frac{8}{3} C_{F} n_{f} \Bigg\}  \Bigg]
           + a_{s}^{3} \Bigg[
           \frac{1}{\epsilon^3} \Bigg\{ - \frac{3872}{3} C_{A}^2 C_{F}  +
           \frac{1408}{3} C_{A} C_{F} n_{f}  - \frac{128}{3} C_{F}
           n_{f}^2 \Bigg\}  
           \nonumber\\
         &+ 
           \frac{1}{\epsilon^2} \Bigg\{ - \frac{9512}{9} C_{A}^2 C_{F}  +
           \frac{2200}{3} C_{A} C_{F}^2  + \frac{2272}{9} C_{A} C_{F}
           n_{f}  - 
           \frac{64}{3} C_{F}^2 n_{f} - \frac{32}{9} C_{F} n_{f}^2 \Bigg\}
           \nonumber\\
         &+ 
           \frac{1}{\epsilon} \Bigg\{ - \frac{3214}{9} C_{A}^2 C_{F}  +
           \frac{5894}{9} C_{A} C_{F}^2  - 356 C_{F}^3 - \frac{328}{9}
           C_{A} C_{F} n_{f}  + 
           \frac{1096}{9} C_{F}^2 n_{f} + \frac{104}{9} C_{F} n_{f}^2 
           \nonumber\\
         &+ 192
           C_{A} C_{F} n_{f} \zeta_3  - 
           192 C_{F}^2 n_{f} \zeta_3 \Bigg\} 
           \Bigg]\,.
\end{align}
$Z_{GG}$ and $Z_{GJ}$ are in agreement with the results already
available in the literature up to ${\cal{O}}(a^{2}_{s})$
\cite{Larin:1993tq} and ${\cal{O}}(a_{s}^{3})$ \cite{Zoller:2013ixa},
where a completely different approach and methodology was used.

\subsection{Results of UV Renormalised Form Factors}
\label{ss:Ren}

Using the renormalisation constants obtained in the previous section,
we get all the UV renormalised form factors
$\left[ {\cal F}^{\lambda}_{\beta} \right]_{R}$, defined in
Eq.~(\ref{eq:RenFFG}) and Eq.~(\ref{eq:RenFFJ}), up to three loops. In
this section we present the results for the choice of the scales
$\mu_R^2=\mu_F^2=q^2$.
\begin{align}
  \label{eq:FFRen1Ggg}
  \left[ {\cal F}^{G,(1)}_{g} \right]_{R} &=  {\dis{2 n_{f} T_{F}}} \Bigg\{ - \frac{4}{3
                                            \epsilon} \Bigg\}  + {\dis{C_{A}}} \Bigg\{ -
                                            \frac{8}{\epsilon^2}  +
                                            \frac{22}{3 \epsilon} + 4 + \zeta_2  + \epsilon \Bigg( - 6 -
                                            \frac{7}{3} \zeta_3 \Bigg) + 
                                            \epsilon^2 \Bigg( 7 -
                                            \frac{\zeta_2}{2}  
                                            \nonumber\\
                                          &+ \frac{47}{80} \zeta_2^2
                                            \Bigg) + 
                                            \epsilon^3 \Bigg( - \frac{15}{2}
                                            + \frac{3}{4} \zeta_2   +
                                            \frac{7}{6} \zeta_3  +
                                            \frac{7}{24} \zeta_2 \zeta_3  - 
                                            \frac{31}{20} \zeta_5 \Bigg) \Bigg\}\,,
  \\  \left[ {\cal F}^{G,(2)}_{g} \right]_{R} &=  {\dis{4 n_{f}^2 T_{F}^{2}}} \Bigg\{ \frac{16}{9
                                                \epsilon^2} \Bigg\}
                                                + 
                                                {\dis{C_{A}^2}} \Bigg\{
                                                \frac{32}{\epsilon^4}  -
                                                \frac{308}{3 \epsilon^3}  + 
                                                \Bigg( \frac{62}{9} - 4 \zeta_2
                                                \Bigg) \frac{1}{\epsilon^2}  + 
                                                \Bigg( \frac{2780}{27} +
                                                \frac{11}{3} \zeta_2  +
                                                \frac{50}{3} \zeta_3 \Bigg)
                                                \frac{1}{\epsilon} 
                                                \nonumber\\
                                          &-
                                            \frac{3293}{81} +
                                            \frac{115}{6} \zeta_2  -
                                            \frac{21}{5} \zeta_2^2  - 33 \zeta_3   + 
                                            \epsilon \Bigg( -
                                            \frac{114025}{972} -
                                            \frac{235}{18} \zeta_2  +
                                            \frac{1111}{120} \zeta_2^2  +
                                            \frac{1103}{54} \zeta_3  
                                            \nonumber\\
                                          &- 
                                            \frac{23}{6} \zeta_2 \zeta_3 -
                                            \frac{71}{10} \zeta_5 \Bigg) + 
                                            \epsilon^2 \Bigg(
                                            \frac{4819705}{11664} -
                                            \frac{694}{27} \zeta_2  -
                                            \frac{2183}{240} \zeta_2^2  + 
                                            \frac{2313}{280} \zeta_2^3 -
                                            \frac{7450}{81} \zeta_3  
                                            \nonumber\\
                                          &-
                                            \frac{11}{36} \zeta_2 \zeta_3  + 
                                            \frac{901}{36} \zeta_3^2 -
                                            \frac{341}{20} \zeta_5 \Bigg)  
                                            \Bigg\}  
                                            + 
                                            {\dis{2 C_{A} n_{f} T_{F}}} \Bigg\{ \frac{56}{3
                                            \epsilon^3}  - \frac{52}{3
                                            \epsilon^2}  + \Bigg( -
                                            \frac{272}{27} -  \frac{2}{3}
                                            \zeta_2 \Bigg) \frac{1}{\epsilon}
                                            \nonumber\\
                                          & - \frac{295}{81} 
                                           - 
                                            \frac{5}{3} \zeta_2 - 2 \zeta_3 +
                                            \epsilon \Bigg( \frac{15035}{486}
                                            + \frac{\zeta_2}{18} +
                                            \frac{59}{60} \zeta_2^2  + 
                                            \frac{383}{27} \zeta_3 \Bigg) +
                                            \epsilon^2 \Bigg( -
                                            \frac{116987}{1458}  +
                                            \frac{583}{108} \zeta_2  
                                            \nonumber\\
                                          &- 
                                            \frac{329}{72} \zeta_2^2 -
                                            \frac{1688}{81} \zeta_3  +
                                            \frac{61}{18} \zeta_2 \zeta_3  - 
                                            \frac{49}{10} \zeta_5 \Bigg)
                                            \Bigg\} 
                                            + 
                                            {\dis{2 C_{F} n_{f} T_{F}}} \Bigg\{  - \frac{2}{\epsilon} -
                                            \frac{71}{3} + 8
                                            \zeta_3 + 
                                            \epsilon \Bigg( \frac{2665}{36} 
                                            \nonumber\\
                                          &-
                                            \frac{19}{6} \zeta_2  -
                                            \frac{8}{3} \zeta_2^2  -
                                            \frac{64}{3} \zeta_3 \Bigg) + 
                                            \epsilon^2 \Bigg( -
                                            \frac{68309}{432} +
                                            \frac{505}{36} \zeta_2  +
                                            \frac{64}{9} \zeta_2^2 +
                                            \frac{455}{9} \zeta_3 -  
                                            \frac{10}{3} \zeta_2 \zeta_3  
                                            \nonumber\\
                                          &+  8 \zeta_5 \Bigg) \Bigg\}\,,
  \\  \left[ {\cal F}^{G,(3)}_{g} \right]_{R} &= {\dis{8 n_{f}^3 T_{F}^{3}}} \Bigg\{ -  
                                                \frac{64}{27 \epsilon^3} \Bigg\}  
                                                +
                                                {\dis{4 C_{F} n_{f}^2 T_{F}^{2}}} \Bigg\{ \frac{56}{9
                                                \epsilon^2} + \Bigg(
                                                \frac{874}{27}  - \frac{32}{3}
                                                \zeta_3 \Bigg) \frac{1}{\epsilon}
                                                -
                                                \frac{418}{27} + 2 \zeta_2  + 
                                                \frac{16}{5} \zeta_2^2 
                                                \nonumber\\
                                          &- \frac{80}{9} \zeta_3 \Bigg\} 
                                            + 
                                            {\dis{2 C_{F}^2 n_{f} T_{F}}} \Bigg\{
                                            \frac{2}{3
                                            \epsilon} + \frac{457}{6} + 104 \zeta_3  - 160
                                            \zeta_5 \Bigg\} 
                                            + 
                                            {\dis{2 C_{A}^2 n_{f} T_{F}}} \Bigg\{ -
                                            \frac{320}{3 \epsilon^5}  
\nonumber\\
&+
                                            \frac{28480}{81 \epsilon^4} 
                                           +
                                            \Bigg( - \frac{608}{243}  +
                                            \frac{56}{27} \zeta_2 \Bigg)
                                            \frac{1}{\epsilon^3}  + 
                                            \Bigg( - \frac{54088}{243} +
                                            \frac{676}{81} \zeta_2  +
                                            \frac{272}{27}  \zeta_3 \Bigg)
                                            \frac{1}{\epsilon^2} 
\nonumber\\
&+ 
                                            \Bigg( - \frac{623293}{2187} 
                                           -
                                            \frac{7072}{243} \zeta_2 -
                                            \frac{941}{90} \zeta_2^2  -
                                            \frac{7948}{81}  \zeta_3 \Bigg)
                                            \frac{1}{\epsilon} +
                                            \frac{6345979}{13122} - 
                                            \frac{42971}{729} \zeta_2 +
                                            \frac{687}{20} \zeta_2^2 
\nonumber\\
&+
                                            \frac{652}{3} \zeta_3  
                                           -
                                            \frac{301}{9} \zeta_2 \zeta_3   +
                                            \frac{4516}{45} \zeta_5 \Bigg\}  
                                            + 
                                            {\dis{4 C_{A} n_{f}^2 T_{F}^{2}}} \Bigg\{ -
                                            \frac{2720}{81 \epsilon^4}  +
                                            \frac{7984}{243 \epsilon^3}  + 
                                            \Bigg( \frac{560}{27} 
\nonumber\\
&+
                                            \frac{8}{27} \zeta_2 \Bigg)
                                            \frac{1}{\epsilon^2} 
                                           + \Bigg(
                                            \frac{10889}{2187}  +
                                            \frac{140}{81} \zeta_2  +
                                            \frac{328}{81} \zeta_3 \Bigg)
                                            \frac{1}{\epsilon} + 
                                            \frac{9515}{6561} +
                                            \frac{10}{27} \zeta_2  -
                                            \frac{157}{135} \zeta_2^2  - 
                                            \frac{20}{243} \zeta_3  \Bigg\}  
                                            \nonumber\\
                                          &+ 
                                            {\dis{2 C_{A} C_{F} n_{f} T_{F}}} \Bigg\{
                                            \frac{272}{9 \epsilon^3}   + 
                                            \Bigg( \frac{4408}{27} -
                                            \frac{640}{9} \zeta_3 \Bigg)
                                            \frac{1}{\epsilon^2}  + 
                                            \Bigg( - \frac{65110}{81} +
                                            \frac{74}{3} \zeta_2  +
                                            \frac{352}{15} \zeta_2^2 
                                            \nonumber\\
                                          &+
                                            \frac{6496}{27} \zeta_3 \Bigg)
                                            \frac{1}{\epsilon} + 
                                            \frac{1053625}{972} -
                                            \frac{311}{2} \zeta_2  -
                                            \frac{1168}{15} \zeta_2^2 -
                                            \frac{24874}{81} \zeta_3 +  48 \zeta_2 \zeta_3 + 
                                            \frac{32}{9} \zeta_5 \Bigg\}
                                            \nonumber\\
                                          &+ 
                                            {\dis{C_{A}^3}}  \Bigg\{  -
                                            \frac{256}{3 \epsilon^6}  + 
                                            \frac{1760}{3 \epsilon^5} -
                                            \frac{62264}{81 \epsilon^4}  + 
                                            \Bigg( - \frac{176036}{243} -
                                            \frac{308}{27} \zeta_2  -
                                            \frac{176}{3} \zeta_3 \Bigg)
                                            \frac{1}{\epsilon^3} + 
                                            \Bigg( \frac{207316}{243} 
                                            \nonumber\\
                                          &-
                                            \frac{8164}{81} \zeta_2  +
                                            \frac{494}{45} \zeta_2^2  +
                                            \frac{9064}{27} \zeta_3  \Bigg)
                                            \frac{1}{\epsilon^2} +
                                            \Bigg( \frac{2763800}{2187}  +
                                            \frac{36535}{243} \zeta_2  - 
                                            \frac{12881}{180} \zeta_2^2 -
                                            \frac{3988}{9} \zeta_3  
                                            \nonumber\\
                                          &+
                                            \frac{170}{9} \zeta_2 \zeta_3  + 
                                            \frac{1756}{15} \zeta_5 \Bigg)
                                            \frac{1}{\epsilon} -
                                            \frac{84406405}{26244} +
                                            \frac{617773}{1458} \zeta_2  + 
                                            \frac{144863}{1080} \zeta_2^2 -
                                            \frac{22523}{270} \zeta_2^3   
                                            \nonumber\\
                                          &+
                                            \frac{44765}{243} \zeta_3 -  
                                            \frac{1441}{18} \zeta_2 \zeta_3 -
                                            \frac{1766}{9} \zeta_3^2 +
                                            \frac{13882}{45} \zeta_5 \Bigg\}\,,
  \\  \left[ {\cal F}^{G,(1)}_{q} \right]_{R} &= {\dis{C_{F}}} \Bigg\{ -
                                                \frac{8}{\epsilon^2}  +
                                                \frac{6}{\epsilon}  -
                                                \frac{33}{4} + \zeta_2  + 
                                                \epsilon \Bigg( \frac{29}{16}
                                                + \frac{25}{48}
                                                \zeta_2  - \frac{7}{3}
                                                \zeta_3 \Bigg)  + 
                                                \epsilon^2 \Bigg( \frac{299}{192}
                                                - \frac{1327}{576} \zeta_2 
                                                \nonumber\\
                                          &+
                                            \frac{1387}{2880} \zeta_2^2  + 
                                            \frac{143}{48} \zeta_3 \Bigg) + \epsilon^3 \Bigg( -
                                            \frac{13763}{2304}  +
                                            \frac{32095}{6912} \zeta_2  - 
                                            \frac{1559}{3456} \zeta_2^2 + \frac{61}{6912} \zeta_2^3 -
                                            \frac{1625}{576} \zeta_3  
                                            \nonumber\\
                                          &+ 
                                            \frac{377}{864} \zeta_2 \zeta_3 - \frac{31}{20} \zeta_5 \Bigg)
                                            \Bigg\}  
                                            + 
                                            {2 \dis{n_{f} T_{F}}} \Bigg\{ -
                                            \frac{445}{162}  + \epsilon \Bigg( \frac{8231}{1944}  -
                                            \frac{239}{1944} \zeta_2  -
                                            \frac{2}{3} \zeta_3 \Bigg) 
                                            \nonumber\\
                                          &+ 
                                            \epsilon^2 \Bigg( -
                                            \frac{50533}{7776}  + \frac{1835}{7776} \zeta_2
                                            + \frac{22903}{116640} \zeta_2^2 + 
                                            \frac{9125}{5832} \zeta_3  +
                                            \frac{1}{18} \zeta_2 \zeta_3 \Bigg)  + 
                                            \epsilon^3 \Bigg(
                                            \frac{2754151}{279936}  
                                            \nonumber\\
                                          &- \frac{35083}{93312}
                                            \zeta_2  - \frac{316343}{699840} \zeta_2^2 - 
                                            \frac{22903}{1399680} \zeta_2^3
                                            - \frac{61121}{23328}  \zeta_3  + 
                                            \frac{2053}{34992} \zeta_2
                                            \zeta_3  - \frac{1}{216} \zeta_2^2
                                            \zeta_3  
                                            \nonumber\\
                                          &- \frac{7}{54} \zeta_3^2 - 
                                            \frac{7}{6} \zeta_5 \Bigg) \Bigg\}
                                            + 
                                            {\dis{C_{A}}} \Bigg\{ \frac{7115}{324}
                                            - \frac{2}{3} \zeta_2 - 2 \zeta_3 + 
                                            \epsilon \Bigg( -
                                            \frac{114241}{3888}  + \frac{7321}{3888} \zeta_2
                                            + \frac{53}{90} \zeta_2^2   
                                            \nonumber\\
                                          &+ \frac{13}{3} \zeta_3 + 
                                            \frac{1}{6} \zeta_2 \zeta_3 \Bigg)
                                            + \epsilon^2 \Bigg(
                                            \frac{692435}{15552}  -
                                            \frac{55117}{15552} \zeta_2  - 
                                            \frac{326369}{233280} \zeta_2^2
                                            - \frac{53}{1080} \zeta_2^3  -
                                            \frac{90235}{11664} \zeta_3  
                                            \nonumber\\
                                          &- 
                                            \frac{41}{108} \zeta_2 \zeta_3 -
                                            \frac{1}{72} \zeta_2^2 \zeta_3  -
                                            \frac{7}{18} \zeta_3^2  - 5 \zeta_5
                                            \Bigg)  + 
                                            \epsilon^3 \Bigg( - \frac{37171073}{559872} +
                                            \frac{1013165}{186624} \zeta_2  
                                            \nonumber\\
                                          &+ 
                                            \frac{3399073}{1399680} \zeta_2^2
                                            +  \frac{34037663}{19595520}
                                            \zeta_2^3  + 
                                            \frac{53}{12960} \zeta_2^4 +
                                            \frac{585439}{46656} \zeta_3   -
                                            \frac{56159}{69984} \zeta_2 \zeta_3
                                            + 
                                            \frac{3223}{12960} \zeta_2^2
                                            \zeta_3 
                                            \nonumber\\
                                          &+ \frac{1}{864} \zeta_2^3 \zeta_3 + \frac{8}{9}
                                            \zeta_3^2  + 
                                            \frac{7}{108} \zeta_2 \zeta_3^2 +
                                            8 \zeta_5   + \frac{5}{12} \zeta_2 \zeta_5 \Bigg) \Bigg\}\,,
  \\  \left[ {\cal F}^{G,(2)}_{q} \right]_{R} &= {\dis{4 n_{f}^2 T_{F}^{2}}}
                                                \Bigg\{
                                                \frac{9505}{1458}  +
                                                \epsilon \Bigg(  -
                                                \frac{146177}{5832}  +
                                                \frac{12419}{17496}
                                                \zeta_2 + \frac{38}{9}
                                                \zeta_3 \Bigg) \Bigg\}
                                                + 
                                                {\dis{2 C_{F} n_{f} T_{F}}} \Bigg\{
                                                \frac{8}{\epsilon^3}  +
                                                \frac{1636}{81 \epsilon^2}  
                                                \nonumber\\
                                          &+
                                            \Bigg( - \frac{12821}{243}  -
                                            \frac{247}{243} \zeta_2   +
                                            \frac{16}{3} \zeta_3 \Bigg)
                                            \frac{1}{\epsilon} + 
                                            \frac{20765}{324} + \frac{35}{486} \zeta_2 +
                                            \frac{85}{2916} \zeta_2^2  +
                                            \frac{6265}{729} \zeta_3  - 
                                            \frac{4}{9} \zeta_2 \zeta_3  
                                            \nonumber\\
                                          &+ 
                                            \epsilon \Bigg(-
                                            \frac{1457425}{34992}  -
                                            \frac{11146}{729} \zeta_2 -
                                            \frac{232457}{174960}  \zeta_2^2 - 
                                            \frac{85}{34992} \zeta_2^3
                                            + \frac{9907}{1458} \zeta_3 -
                                            \frac{7723}{4374} \zeta_2 \zeta_3
                                            \nonumber\\
                                          &+ 
                                            \frac{1}{27} \zeta_2^2
                                            \zeta_3  +
                                            \frac{28}{27} \zeta_3^2   - \frac{20}{9} \zeta_5
                                            \Bigg) \Bigg\} 
                                            + 
                                            {\dis{C_{A}^2}} \Bigg\{
                                            \frac{2796445}{5832}  -
                                            \frac{587}{18} \zeta_2  + 
                                            \frac{53}{30} \zeta_2^2 
                                            -
                                            \frac{185}{2} \zeta_3  -
                                            \frac{10}{3} \zeta_2 \zeta_3  
                                            \nonumber\\
                                          &+
                                            20 \zeta_5   + 
                                            \epsilon \Bigg( -
                                            \frac{34321157}{23328}  +
                                            \frac{10420379}{69984} \zeta_2  +
                                            \frac{589}{20} \zeta_2^2  + 
                                            \frac{7921}{2520} \zeta_2^3 
                                            +
                                            \frac{8411}{24} \zeta_3  -
                                            \frac{329}{72} \zeta_2 \zeta_3  
                                            \nonumber\\
                                          &+ 
                                            \frac{5}{18} \zeta_2^2 \zeta_3 +
                                            13 \zeta_3^2  - \frac{757}{18} \zeta_5 - 
                                            \frac{5}{3} \zeta_2 \zeta_5
                                            \Bigg)  \Bigg\}
                                            + 
                                            {\dis{2 C_{A} n_{f} T_{F}}} \Bigg\{ -
                                            \frac{178361}{1458}  + \frac{44}{9} \zeta_2 -
                                            \frac{76}{45} \zeta_2^2  
\nonumber\\
&-
                                            \frac{44}{9} \zeta_3  
                                           + 
                                            \epsilon \Bigg(
                                            \frac{2357551}{5832}  -
                                            \frac{478171}{17496} \zeta_2  -
                                            \frac{137}{135} \zeta_2^2  + 
                                            \frac{19}{135} \zeta_2^3 -
                                            \frac{1621}{27} \zeta_3  -
                                            \frac{40}{27} \zeta_2
  \zeta_3  
\nonumber\\
&+ 
                                            \frac{22}{3} \zeta_5 \Bigg)  
                                            \Bigg\} 
                                           + 
                                            {\dis{C_{A} C_{F}}} \Bigg\{  -
                                            \frac{44}{\epsilon^3}  
                                            + 
                                            \Bigg( - \frac{13654}{81} +
                                            \frac{28}{3} \zeta_2  + 16
                                            \zeta_3 \Bigg)
                                            \frac{1}{\epsilon^2} + 
                                            \Bigg( \frac{186925}{486} 
\nonumber\\
&-
                                            \frac{3919}{486} \zeta_2  -
                                            \frac{212}{45} \zeta_2^2  
                                           -
                                            \frac{218}{3} \zeta_3  
                                            - 
                                            \frac{4}{3} \zeta_2 \zeta_3
                                            \Bigg) \frac{1}{\epsilon} -
                                            \frac{61613}{81}  +
                                            \frac{59399}{972} \zeta_2  + 
                                            \frac{749513}{29160}
  \zeta_2^2 
\nonumber\\
&+
                                            \frac{53}{135} \zeta_2^3
                                            + \frac{213517}{1458} \zeta_3 + 
                                            \frac{91}{27} \zeta_2 \zeta_3 
                                           +
                                            \frac{1}{9} \zeta_2^2
                                            \zeta_3  + \frac{28}{9} \zeta_3^2  +
                                            \epsilon \Bigg(
                                            \frac{35327209}{34992}  -
                                            \frac{2158003}{23328}
                                            \zeta_2 
\nonumber\\
&-
                                            \frac{3532645}{69984}
                                            \zeta_2^2  - 
                                            \frac{11307767}{2449440}
                                            \zeta_2^3  -
                                            \frac{53}{1620} \zeta_2^4
                                           - \frac{1030169}{2916} \zeta_3 + 
                                            \frac{191915}{8748}
                                            \zeta_2 \zeta_3  -
                                            \frac{817}{405} \zeta_2^2
                                            \zeta_3  
\nonumber\\
&- \frac{1}{108} \zeta_2^3
                                            \zeta_3  - 
                                            \frac{121}{9} \zeta_3^2 -
                                            \frac{14}{27}  \zeta_2
                                            \zeta_3^2  - \frac{43}{6} \zeta_5
                                            \Bigg)  \Bigg\} 
                                           + 
                                            {\dis{C_{F}^2}} \Bigg\{
                                            \frac{32}{\epsilon^4}  -
                                            \frac{48}{\epsilon^3}   + 
                                            \Bigg( 84 - 8 \zeta_2 \Bigg)
                                            \frac{1}{\epsilon^2}  
                                            \nonumber\\
&+ \Bigg(
                                            - \frac{125}{2}  - \frac{61}{6}
                                            \zeta_2  + \frac{128}{3} \zeta_3
                                            \Bigg) \frac{1}{\epsilon} +
                                            \frac{6881}{216} +
                                            \frac{193}{12} \zeta_2  
                                           -
                                            \frac{281}{24} \zeta_2^2  - 
                                            \frac{1037}{18} \zeta_3  
\nonumber\\
&+ 
                                            \epsilon \Bigg(
                                            \frac{166499}{2592} -
                                            \frac{3761}{648} \zeta_2
                                            + \frac{3451}{480}
                                            \zeta_2^2 - \frac{31}{288}
                                            \zeta_2^3 + 
                                            \frac{10607}{108} \zeta_3
                                            - \frac{1081}{108}
                                            \zeta_2 \zeta_3  
                                           + \frac{328}{45} \zeta_5
                                            \Bigg)  
                                            \Bigg\}\,,
  \\  \left[ {\cal F}^{J,(1)}_{g} \right]_{R} &= {\dis{2 n_{f} T_{F}}}  \Bigg\{ - \frac{4}{3
                                                \epsilon} \Bigg\}
                                                + 
                                                {\dis{C_{A}}} \Bigg\{ -
                                                \frac{8}{\epsilon^2}  +
                                                \frac{22}{3 \epsilon} + 4 + \zeta_2  + 
                                                \epsilon \Bigg( - \frac{15}{2} +
                                                \zeta_2  - \frac{16}{3} \zeta_3
                                                \Bigg) 
\nonumber\\
&+ 
                                                \epsilon^2 \Bigg( \frac{287}{24}
                                           - 2 \zeta_2  + \frac{127}{80}
                                            \zeta_2^2 \Bigg)    + 
                                            \epsilon^3 \Bigg( -
                                            \frac{5239}{288}  +
                                            \frac{151}{48} \zeta_2  +
                                            \frac{19}{120} \zeta_2^2  +
                                            \frac{\zeta_3}{12}  + 
                                            \frac{7}{6} \zeta_2
  \zeta_3  
\nonumber\\
&-
                                            \frac{91}{20} \zeta_5 \Bigg)
                                            \Bigg\}  
                                           + 
                                            {\dis{C_{F}}} \Bigg\{  
                                            \epsilon \Bigg( - \frac{21}{2} +
                                            6 \zeta_3 \Bigg)  + \epsilon^2 \Bigg(
                                            \frac{155}{8} - \frac{5}{2}
                                            \zeta_2  - \frac{9}{5} \zeta_2^2
                                            - \frac{9}{2} \zeta_3
  \Bigg) 
\nonumber\\
&+ \epsilon^3
                                            \Bigg( - \frac{1025}{32}  +
                                            \frac{83}{16} \zeta_2  
                                           + 
                                            \frac{27}{20} \zeta_2^2 +
                                            \frac{20}{3} \zeta_3  -
                                            \frac{3}{4} \zeta_2 \zeta_3  + \frac{21}{2} \zeta_5 \Bigg) \Bigg\}\,,
  \\  \left[ {\cal F}^{J,(2)}_{g} \right]_{R} &= {\dis{4 n_{f}^2 T_{F}^{2}}} \Bigg\{ \frac{16}{9 \epsilon^2} \Bigg\}
                                                + 
                                                {\dis{C_{A} C_{F}}} \Bigg\{\Bigg( 84 - 48
                                                \zeta_3 \Bigg) \frac{1}{\epsilon}
                                                - 232 +
                                                20 \zeta_2  + \frac{72}{5}
                                                \zeta_2^2 + 80 \zeta_3 
\nonumber\\
&+ 
                                                \epsilon \Bigg( \frac{17545}{108}
                                           - 58 \zeta_2 - 24 \zeta_2^2  -
                                            \frac{38}{3} \zeta_3  + 10
                                            \zeta_2 \zeta_3  - 
                                            14 \zeta_5 \Bigg) + 
                                            \epsilon^2 \Bigg(
                                            \frac{402635}{1296} -
                                            \frac{233}{36} \zeta_2  
\nonumber\\
&+
                                            \frac{72}{5} \zeta_2^2  +
                                            \frac{17}{70} \zeta_2^3  
                                           + 
                                            \frac{535}{12} \zeta_3 - 2
                                            \zeta_2 \zeta_3 - 34 \zeta_3^2  -
                                            \frac{1355}{6} \zeta_5 \Bigg) \Bigg\}
                                            + {\dis{2 C_{A} n_{f} T_{F}}}
                                            \Bigg\{
                                            \frac{56}{3 \epsilon^3}  -
                                            \frac{52}{3 \epsilon^2}  
\nonumber\\
&+ 
                                            \Bigg( - \frac{272}{27} -
                                            \frac{2}{3} \zeta_2 \Bigg)
                                            \frac{1}{\epsilon}   
                                           - \frac{133}{81} - 3 \zeta_2 +
                                            2 \zeta_3  + 
                                            \epsilon \Bigg( \frac{7153}{243}
                                            - \frac{7}{18} \zeta_2 -
                                            \frac{13}{60} \zeta_2^2  +
                                            \frac{599}{27} \zeta_3
  \Bigg)  
\nonumber\\
&+ 
                                            \epsilon^2 \Bigg( -
                                            \frac{135239}{1458} 
                                           +
                                            \frac{1139}{108} \zeta_2  -
                                            \frac{167}{24} \zeta_2^2  - 
                                            \frac{3146}{81} \zeta_3 +
                                            \frac{73}{18} \zeta_2 \zeta_3  -
                                            \frac{137}{30}  \zeta_5 \Bigg)
                                            \Bigg\} 
                                            \nonumber\\
&+ 
                                            {\dis{2 C_{F} n_{f} T_{F}}} \Bigg\{
                                            - \frac{2}{\epsilon}  -
                                            \frac{29}{3} 
                                           +  \epsilon
                                            \Bigg( \frac{14989}{216} -
                                            \frac{25}{6} \zeta_2 -
                                            \frac{4}{15} \zeta_2^2  - 
                                            32 \zeta_3 \Bigg) + \epsilon^2
                                            \Bigg( -
  \frac{606661}{2592}  
\nonumber\\
&+
                                            \frac{2233}{72} \zeta_2  +
                                            \frac{158}{15} \zeta_2^2  
                                           + 
                                            \frac{1409}{18} \zeta_3 - 2
                                            \zeta_2 \zeta_3  + \frac{82}{3}
                                            \zeta_5 \Bigg) \Bigg\}  
                                            + 
                                            {\dis{C_{A}^2}} \Bigg\{
                                            + \frac{32}{\epsilon^4}  -
                                            \frac{308}{3 \epsilon^3}  
\nonumber\\
&+
                                            \Bigg( \frac{62}{9}  - 4 \zeta_2
                                            \Bigg) \frac{1}{\epsilon^2} + 
                                            \Bigg( \frac{3104}{27} 
                                           -
                                            \frac{13}{3} \zeta_2  +
                                            \frac{122}{3} \zeta_3  \Bigg)
                                            \frac{1}{\epsilon}  - \frac{7397}{81} + 
                                            \frac{77}{2} \zeta_2 -
                                            \frac{61}{5} \zeta_2^2 - 55
                                            \zeta_3   
\nonumber\\
&+ 
                                            \epsilon \Bigg( -
                                            \frac{32269}{972} -
                                            \frac{997}{36} \zeta_2  
                                           +
                                            \frac{1049}{120} \zeta_2^2  -
                                            \frac{2393}{108} \zeta_3  - 
                                            \frac{53}{6} \zeta_2 \zeta_3 +
                                            \frac{369}{10} \zeta_5
  \Bigg)  
\nonumber\\
&+ 
                                            \epsilon^2 \Bigg(
                                            \frac{4569955}{11664} -
                                            \frac{15323}{432} \zeta_2  +
                                            \frac{2129}{180} \zeta_2^2  
                                           - 
                                            \frac{7591}{840} \zeta_2^3 -
                                            \frac{4099}{1296} \zeta_3  -
                                            \frac{605}{36} \zeta_2 \zeta_3  + 
                                            \frac{775}{36} \zeta_3^2 
\nonumber\\
&+
                                            \frac{2011}{30} \zeta_5 \Bigg)
                                            \Bigg\}  
                                            + 
                                            {\dis{C_{F}^2}} \Bigg\{ \epsilon \Bigg(
                                            \frac{763}{12} +  17 \zeta_3 
                                           - 60
                                            \zeta_5 \Bigg)  + 
                                            \epsilon^2 \Bigg( -
                                            \frac{18857}{144} + \frac{31}{3}
                                            \zeta_2 -  \frac{76}{15}
                                            \zeta_2^2 
\nonumber\\
&+ \frac{120}{7}
                                            \zeta_2^3  - 
                                            145 \zeta_3 + 4 \zeta_2 \zeta_3 +
                                            30 \zeta_3^2 
                                           + \frac{470}{3} \zeta_5 \Bigg) \Bigg\}\,,
  \\  \left[ {\cal F}^{J,(1)}_{q} \right]_{R} &= {\dis{C_{F}}} \Bigg\{ -
                                                \frac{8}{\epsilon^2}  +
                                                \frac{6}{\epsilon}   - 6 + \zeta_2 +
                                                \epsilon \Bigg( - 1  -
                                                \frac{3}{4} \zeta_2   \frac{7}{3}
                                                \zeta_3 \Bigg)  + 
                                                \epsilon^2 \Bigg( \frac{5}{2} +
                                                \frac{\zeta_2}{4}  +
                                                \frac{47}{80} \zeta_2^2  +
                                                \frac{7}{4} \zeta_3 \Bigg)  
                                                \nonumber\\
                                          &+ 
                                            \epsilon^3 \Bigg( - \frac{13}{4}
                                            + \frac{\zeta_2}{8}  -
                                            \frac{141}{320} \zeta_2^2  -
                                            \frac{7}{12} \zeta_3  + 
                                            \frac{7}{24} \zeta_2 \zeta_3  -
                                            \frac{31}{20} \zeta_5  \Bigg) \Bigg\}\,,
  \\  \left[ {\cal F}^{J,(2)}_{q} \right]_{R} &= {\dis{2 C_{F} n_{f} T_{F}}} \Bigg\{ 
                                                \frac{8}{\epsilon^3}  -
                                                \frac{16}{9 \epsilon^2}  + \Bigg(
                                                - \frac{65}{27}  - 2 \zeta_2
                                                \Bigg) \frac{1}{\epsilon}  -
                                                \frac{3115}{324} + 
                                                \frac{23}{9} \zeta_2 +
                                                \frac{2}{9} \zeta_3  + \epsilon
                                                \Bigg( \frac{129577}{3888}  
                                                \nonumber\\
                                          &-
                                            \frac{731}{108} \zeta_2  - 
                                            \frac{\zeta_2^2}{10} +
                                            \frac{119}{27} \zeta_3 \Bigg)  + 
                                            \epsilon^2 \Bigg( -
                                            \frac{3054337}{46656} +
                                            \frac{20951}{1296} \zeta_2  -
                                            \frac{145}{144} \zeta_2^2  - 
                                            \frac{2303}{324} \zeta_3 
                                            \nonumber\\
                                          &-
                                            \frac{10}{9} \zeta_2 \zeta_3  -
                                            \frac{59}{30} \zeta_5 \Bigg)
                                            \Bigg\}  
                                            + 
                                            {\dis{C_{F}^2}} \Bigg\{
                                            \frac{32}{\epsilon^4}  -
                                            \frac{48}{\epsilon^3} + \Bigg( 66
                                            - 8 \zeta_2 \Bigg)
                                            \frac{1}{\epsilon^2}   +
                                            \Bigg( - \frac{53}{2}  +
                                            \frac{128}{3} \zeta_3  \Bigg)
                                            \frac{1}{\epsilon} 
                                            \nonumber\\
                                          &- \frac{121}{8} +
                                            \frac{\zeta_2}{2}  - 
                                            13 \zeta_2^2 - 58 \zeta_3 +
                                            \epsilon \Bigg( \frac{3403}{32}
                                            + \frac{27}{8} \zeta_2  +
                                            \frac{171}{10} \zeta_2^2  + 
                                            \frac{559}{6} \zeta_3 -
                                            \frac{56}{3} \zeta_2 \zeta_3  
                                            \nonumber\\
                                          &+
                                            \frac{92}{5} \zeta_5 \Bigg) + 
                                            \epsilon^2 \Bigg( -
                                            \frac{21537}{128} -
                                            \frac{825}{32} \zeta_2  -
                                            \frac{457}{16} \zeta_2^2  +
                                            \frac{223}{20} \zeta_2^3  - 
                                            \frac{4205}{24} \zeta_3 +
                                            \frac{27}{2} \zeta_2
                                            \zeta_3  
\nonumber\\
&+
                                            \frac{652}{9} \zeta_3^2  
                                           - 
                                            \frac{231}{10} \zeta_5 \Bigg) 
                                            \Bigg\}  
                                            + 
                                            {\dis{C_{A} C_{F}}} \Bigg\{ -
                                            \frac{44}{\epsilon^3} + 
                                            \Bigg( \frac{64}{9} + 4 \zeta_2
                                            \Bigg) \frac{1}{\epsilon^2}  +
                                            \Bigg( \frac{961}{54}  + 11
                                            \zeta_2 
\nonumber\\
&- 26 \zeta_3 \Bigg)
                                            \frac{1}{\epsilon}  
                                           -
                                            \frac{30493}{648} -
                                            \frac{193}{18} \zeta_2  +
                                            \frac{44}{5} \zeta_2^2  + 
                                            \frac{313}{9} \zeta_3 + \epsilon
                                            \Bigg( - \frac{79403}{7776}  +
                                            \frac{133}{216} \zeta_2  -
                                            \frac{229}{20} \zeta_2^2  
\nonumber\\
&- 
                                            \frac{4165}{54} \zeta_3 
                                           +
                                            \frac{89}{6} \zeta_2 \zeta_3  -
                                            \frac{51}{2} \zeta_5 \Bigg)  + 
                                            \epsilon^2 \Bigg(
                                            \frac{9732323}{93312} +
                                            \frac{41363}{2592} \zeta_2  +
                                            \frac{33151}{1440} \zeta_2^2  - 
                                            \frac{809}{280} \zeta_2^3 
\nonumber\\
&+
                                            \frac{89929}{648} \zeta_3  
                                            -
                                            \frac{80}{9} \zeta_2 \zeta_3  - 
                                            \frac{569}{12} \zeta_3^2 +
                                            \frac{2809}{60}  \zeta_5 \Bigg) \Bigg\}\,,
  \\  \left[ {\cal F}^{J,(3)}_{q} \right]_{R} &= Z^{s,(3)}_{5} 
                                                + {\dis{4 C_{F}
                                                n_{f}^2 T_{F}^{2}}} \Bigg\{ -
                                                \frac{704}{81 \epsilon^4}  +
                                                \frac{64}{243 \epsilon^3}  +
                                                \Bigg( \frac{184}{81}  +
                                                \frac{16}{9} \zeta_2 \Bigg)
                                                \frac{1}{\epsilon^2}    + 
                                                \Bigg( - \frac{4834}{2187} +
                                                \frac{40}{27} \zeta_2  
                                                \nonumber\\
                                          &+
                                            \frac{16}{81} \zeta_3 \Bigg)
                                            \frac{1}{\epsilon}  + 
                                            \frac{538231}{13122} - 
                                            \frac{680}{81} \zeta_2 -
                                            \frac{188}{135} \zeta_2^2-
                                            \frac{416}{243} \zeta_3 \Bigg\}  
                                            + 
                                            {\dis{C_{F}^3}} \Bigg\{ -
                                            \frac{256}{3 \epsilon^6}  +
                                            \frac{192}{\epsilon^5}    
                                            \nonumber\\
                                          &+
                                            \Bigg( - 336  + 32 \zeta_2 \Bigg)
                                            \frac{1}{\epsilon^4}  + 
                                            \Bigg( 280 + 24 \zeta_2 -
                                            \frac{800}{3}  \zeta_3 \Bigg)
                                            \frac{1}{\epsilon^3} + \Bigg(
                                            - 58 - 66 \zeta_2  +
                                            \frac{426}{5} \zeta_2^2  
                                            \nonumber\\
                                          &+ 552
                                            \zeta_3 \Bigg)
                                            \frac{1}{\epsilon^2}  + 
                                            \Bigg( - \frac{4193}{6} + 83
                                            \zeta_2 - \frac{1461}{10}
                                            \zeta_2^2  - \frac{3142}{3} \zeta_3 + 
                                            \frac{428}{3} \zeta_2 \zeta_3 -
                                            \frac{1288}{5}  \zeta_5 \Bigg)
                                            \frac{1}{\epsilon} 
                                            \nonumber\\
                                          &+ \frac{41395}{24} +
                                            \frac{1933}{12} \zeta_2  + 
                                            \frac{10739}{40} \zeta_2^2 -
                                            \frac{9095}{252} \zeta_2^3 + 1385
                                            \zeta_3 - 35 \zeta_2 \zeta_3  - 
                                            \frac{1826}{3} \zeta_3^2  -
                                            \frac{562}{5} \zeta_5 \Bigg\}  
                                            \nonumber\\
                                          &+ 
                                            {\dis{2 C_{F}^2 n_{f} T_{F}}} \Bigg\{ -
                                            \frac{64}{\epsilon^5}  +
                                            \frac{560}{9 \epsilon^4}  +
                                            \Bigg( - \frac{680}{27}  + 24
                                            \zeta_2 \Bigg)
                                            \frac{1}{\epsilon^3} + 
                                            \Bigg( \frac{5180}{81} -
                                            \frac{266}{9} \zeta_2  -
                                            \frac{440}{9} \zeta_3  \Bigg)
                                            \frac{1}{\epsilon^2}  
                                            \nonumber\\
                                          &+ 
                                            \Bigg( - \frac{78863}{243} +
                                            \frac{2381}{27} \zeta_2  +
                                            \frac{287}{18} \zeta_2^2  -
                                            \frac{938}{27} \zeta_3 \Bigg)
                                            \frac{1}{\epsilon}   + 
                                            \frac{1369027}{1458} -
                                            \frac{16610}{81} \zeta_2  - 
                                            \frac{8503}{1080} \zeta_2^2 
                                            \nonumber\\
                                          &+ 
                                            \frac{22601}{81} \zeta_3 +
                                            \frac{35}{3} \zeta_2 \zeta_3  -
                                            \frac{386}{9} \zeta_5 \Bigg\}  
                                            + 
                                            {\dis{C_{A}^2 C_{F}}} \Bigg\{ -
                                            \frac{21296}{81 \epsilon^4} +
                                            \Bigg( - \frac{22928}{243}  +
                                            \frac{880}{27} \zeta_2  \Bigg)
                                            \frac{1}{\epsilon^3}  
                                            \nonumber\\
                                          &+ \Bigg(
                                            \frac{23338}{243}  +
                                            \frac{6500}{81} \zeta_2  -
                                            \frac{352}{45} \zeta_2^2  - 
                                            \frac{3608}{27} \zeta_3 \Bigg)
                                            \frac{1}{\epsilon^2}    + 
                                            \Bigg( \frac{139345}{4374} +
                                            \frac{14326}{243} \zeta_2  +
                                            \frac{332}{15} \zeta_2^2 
                                            \nonumber\\
                                          & -
                                            \frac{7052}{27} \zeta_3  + 
                                            \frac{176}{9} \zeta_2 \zeta_3  +
                                            \frac{272}{3} \zeta_5  \Bigg)
                                            \frac{1}{\epsilon}  -
                                            \frac{10659797}{52488} -
                                            \frac{207547}{729} \zeta_2  + 
                                            \frac{19349}{270} \zeta_2^2 -
                                            \frac{6152}{189} \zeta_2^3 
                                            \nonumber\\
                                          &+
                                            \frac{361879}{486} \zeta_3  +
                                            \frac{344}{3} \zeta_2 \zeta_3  - 
                                            \frac{1136}{9} \zeta_3^2 -
                                            \frac{2594}{9} \zeta_5  \Bigg\}  
                                            + 
                                            {\dis{2 C_{A} C_{F} n_{f} T_{F}}} \Bigg\{ +
                                            \frac{7744}{81 \epsilon^4}  +
                                            \Bigg( \frac{6016}{243}  
                                            \nonumber\\
                                          &-
                                            \frac{160}{27} \zeta_2  \Bigg)
                                            \frac{1}{\epsilon^3} + \Bigg( -
                                            \frac{8272}{243}  -
                                            \frac{1904}{81} \zeta_2  +
                                            \frac{848}{27} \zeta_3 \Bigg)
                                            \frac{1}{\epsilon^2}   +
                                            \Bigg( \frac{17318}{2187}  -
                                            \frac{5188}{243} \zeta_2  -
                                            \frac{88}{15} \zeta_2^2  
                                            \nonumber\\
                                          &+ 
                                            \frac{1928}{81} \zeta_3 \Bigg)
                                            \frac{1}{\epsilon}    -
                                            \frac{4158659}{13122} +
                                            \frac{81778}{729} \zeta_2  -
                                            \frac{17}{135} \zeta_2^2  -
                                            \frac{5881}{27} \zeta_3  + 
                                            \frac{22}{3} \zeta_2 \zeta_3 +
                                            \frac{176}{3} \zeta_5 \Bigg\}  
                                            \nonumber\\
                                          &+
                                            {\dis{C_{A} C_{F}^2}} \Bigg\{ \frac{352}{\epsilon^5} + 
                                            \Bigg( - \frac{2888}{9} - 32
                                            \zeta_2 \Bigg)
                                            \frac{1}{\epsilon^4} + \Bigg(
                                            \frac{4436}{27}  - 108 \zeta_2 +
                                            208 \zeta_3 \Bigg)
                                            \frac{1}{\epsilon^3} 
                                            \nonumber\\
                                          &+
                                            \Bigg( \frac{39844}{81}  +
                                            \frac{983}{9} \zeta_2  -
                                            \frac{332}{5} \zeta_2^2  - 
                                            \frac{1928}{9} \zeta_3 \Bigg)
                                            \frac{1}{\epsilon^2}  + \Bigg( -
                                            \frac{97048}{243}  -
                                            \frac{12361}{54} \zeta_2  +
                                            \frac{2975}{36} \zeta_2^2  
                                            \nonumber\\
                                          &+ 
                                            \frac{3227}{3} \zeta_3 -
                                            \frac{430}{3} \zeta_2 \zeta_3  +
                                            284 \zeta_5 \Bigg)  \frac{1}{\epsilon}   -
                                            \frac{709847}{729} +
                                            \frac{36845}{324} \zeta_2  -
                                            \frac{536683}{2160} \zeta_2^2  - 
                                            \frac{18619}{1260} \zeta_2^3  
                                            \nonumber\\
                                          &-
                                            \frac{31537}{18} \zeta_3  -
                                            \frac{518}{3} \zeta_2 \zeta_3  + 
                                            \frac{1616}{3} \zeta_3^2  + 
                                            \frac{1750}{9} \zeta_5 \Bigg\}\,.
\end{align}

\subsection{Universal Behaviour of Leading Transcendentality
  Contribution}
\label{sec:SUYM}

In \cite{Gehrmann:2011xn}, the form factor of a scalar composite
operator belonging to the stress-energy tensor super-multiplet of
conserved currents of ${\cal N}=4$ super Yang-Mills (SYM) with gauge
group SU(N) was studied to three-loop level. Since the theory is UV
finite in $d=4$ space-time dimensions, it is an ideal framework to
study the IR structures of amplitudes in perturbation theory.  In this
theory, one observes that scattering amplitudes can be expressed as a
linear combinations of polylogarithmic functions of uniform degree
$2 l$, where $l$ is the order of the loop, with constant coefficients.
In other words, the scattering amplitudes in ${\cal N}=4$ SYM exhibit
uniform transcendentality, in contrast to QCD loop amplitudes, which
receive contributions from all degrees of transcendentality up to
$2l$.

The three-loop QCD quark and gluon form factors~\cite{Gehrmann:2010ue}
display an interesting relation to the SYM form factor. Upon
replacement~\cite{hep-th/0611204} of the color factors $C_A = C_F = N$
and $T_{f} n_{f}=N/2$, the
leading transcendental (LT) parts of the quark and gluon form factors
in QCD not only coincide with each other but also become identical, up
to a normalization factor of $2^{l}$, to the form factors of scalar
composite operator computed in ${\cal N}=4$ SYM
\cite{Gehrmann:2011xn}.

This correspondence between the QCD form factors and that of the
${\cal N}=4$ SYM can be motivated by the leading transcendentality
principle~\cite{hep-th/0611204, hep-th/0404092, Kotikov:2001sc} which
relates anomalous dimensions of the twist two operators in
${\cal N} =4$ SYM to the LT terms of such operators computed in QCD.
Examining the diagonal pseudo-scalar form factors ${\cal F}^G_g$ and
${\cal F}^J_q$, we find a similar behaviour: the LT terms of these
form factors with replacement $C_A = C_F = N$ and 
  $T_{f} n_{f}=N/2$ are not only identical to
each other but also coincide with the LT terms of the QCD form
factors~\cite{Gehrmann:2010ue} with the same replacement as well as
with the LT terms of the scalar form factors in ${\cal N}=4$ SYM
\cite{Gehrmann:2011xn}, up to a normalization factor of
  $2^{l}$. This observation holds true for the finite terms in
$\epsilon$, and could equally be validated for higher-order terms up
to transcendentality 8 (which is the highest order for which all
three-loop master integrals are available~\cite{Lee:2010ik}).  In
addition to checking the diagonal form factors, we also examined the
off-diagonal ones namely, ${\cal F}^{G}_{q}$, ${\cal F}^{J}_{g}$,
where we find that the LT terms these two form factors are identical
to each other after the replacement of colour factors. However, the LT
terms of these do not coincide with those of the diagonal ones.

\section{Gluon Form Factors for the Pseudo-scalar Higgs Boson Production}
\label{sec:pScalar-FullFF}

The complete form factor for the production of a pseudo-scalar Higgs boson through gluon
fusion, ${\hat \F}^{A,(n)}_{g}$, can be written in terms of the 
individual gluon form factors, Eq.~(\ref{eq:Mexp}), as follows:
\begin{align}
  \label{eq:FFDefMatEle}
  {\F}^{A}_{g} = {\F}^{G}_{g} + \Bigg(
  \frac{Z_{GJ}}{Z_{GG}} + \frac{4 C_{J}}{C_{G}} \frac{Z_{JJ}}{Z_{GG}} \Bigg)
  {\F}^{J}_{g} \frac{\langle{\hat{\cal
  M}}^{G,(0)}_{g}|{\hat{\cal M}}^{J,(1)}_{g}\rangle}{\langle{\hat{\cal
  M}}^{G,(0)}_{g}|{\hat{\cal M}}^{G,(0)}_{g}\rangle}\,.  
\end{align}
In the above expression, the quantities $Z_{ij} (i,j= G, J)$ are the
overall operator renormalization constants which are required to
introduce in the context of UV renormalization.  These are discussed in Sec.~\ref{ss:UV} in great
detail. The ingredients of the form factor ${\F}^{A}_{g}$, namely,
${\F}^{G}_{g}$ and ${\F}^{J}_{g}$ have been calculated up to three
loop level by us~~\cite{Ahmed:2015qpa} and are presented in the Appendix~\ref{App:pScalar-Results}. Using those results we obtain the three 
loop form factor for the pseudo-scalar Higgs boson production through gluon
fusion.
In this section, we present the unrenormalized form factors
${\hat \F}^{A,(n)}_{g}$ up to three loop where the components are
defined through the expansion
\begin{align}
  \label{eq:FF3}
  {\F}^{A}_{g} \equiv
  \sum_{n=0}^{\infty} \left[ {\hat a}_{s}^{n}
  \left( \frac{Q^{2}}{\mu^{2}} \right)^{n\frac{\epsilon}{2}}
  S_{\epsilon}^{n}  {\hat{\F}}^{A,(n)}_{g}\right] \, .
\end{align}
We present the unrenormalized results for the choice of the scale
$\mu_{R}^{2}=\mu_{F}^{2}=q^{2}$\, as follows:
\begin{align}
  \label{eq:FF}
  {\hat \F}^{A,(1)}_{g} &= {\dis{C_{A}}} \Bigg\{ - \frac{8}{\epsilon^2}
                          + 4 +
                          \zeta_2 
                          + 
                          \epsilon \Bigg( - 6 - \frac{7}{3} \zeta_3
                          \Bigg)
                          + \epsilon^2 \Bigg( 7  -
                          \frac{\zeta_2}{2} + \frac{47}{80}  \zeta_2^2
                          \Bigg) 
                          + 
                          \epsilon^3 \Bigg( - \frac{15}{2} 
                          \nonumber\\
                        &+
                          \frac{3}{4} \zeta_2 +  \frac{7}{6} \zeta_3 +
                          \frac{7}{24} \zeta_2 \zeta_3  - 
                          \frac{31}{20} \zeta_5 \Bigg)
                          + \epsilon^4 \Bigg(  \frac{31}{4} -
                          \frac{7}{8} \zeta_2 -  \frac{47}{160}
                          \zeta_2^2 +  
                          \frac{949}{4480} \zeta_2^3 - \frac{7}{4}
                          \zeta_3 -  \frac{49}{144} \zeta_3^2 \Bigg) 
                          \nonumber\\
                        &+  
                          \epsilon^5 \Bigg( - \frac{63}{8} +
                          \frac{15}{16} \zeta_2 +  \frac{141}{320}
                          \zeta_2^2 + \frac{49}{24} \zeta_3 -  
                          \frac{7}{48} \zeta_2 \zeta_3 +
                          \frac{329}{1920} \zeta_2^2 \zeta_3 +
                          \frac{31}{40} \zeta_5 +  
                          \frac{31}{160} \zeta_2 \zeta_5 
                          \nonumber\\
                        &-
                          \frac{127}{112} \zeta_7 \Bigg)
                          + \epsilon^6
                          \Bigg( \frac{127}{16}  - \frac{31}{32}
                          \zeta_2 - \frac{329}{640} \zeta_2^2  - 
                          \frac{949}{8960} \zeta_2^3 +
                          \frac{55779}{716800} \zeta_2^4  -
                          \frac{35}{16} \zeta_3 +  
                          \frac{7}{32} \zeta_2 \zeta_3 
                          \nonumber\\
                        &+ \frac{49}{288}
                          \zeta_3^2 +  \frac{49}{1152} \zeta_2 \zeta_3^2 - 
                          \frac{93}{80} \zeta_5 - \frac{217}{480}
                          \zeta_3 \zeta_5 \Bigg) 
                          +  
                          \epsilon^7 \Bigg( - \frac{255}{32} +
                          \frac{63}{64} \zeta_2 +  \frac{141}{256}
                          \zeta_2^2  + \frac{2847}{17920} \zeta_2^3 
                          \nonumber\\
                        &+ 
                          \frac{217}{96} \zeta_3 - \frac{49}{192}
                          \zeta_2 \zeta_3  - \frac{329}{3840} \zeta_2^2
                          \zeta_3 +  
                          \frac{949}{15360} \zeta_2^3 \zeta_3 -
                          \frac{49}{192} \zeta_3^2 -  \frac{343}{10368}
                          \zeta_3^3 +  
                          \frac{217}{160} \zeta_5 
                          \nonumber\\
                        &- \frac{31}{320}
                          \zeta_2 \zeta_5 +  \frac{1457}{12800}
                          \zeta_2^2 \zeta_5 +  
                          \frac{127}{224} \zeta_7 + \frac{127}{896}
                          \zeta_2 \zeta_7  - \frac{511}{576} \zeta_{9} \Bigg) \Bigg\}\,,
                          \nonumber\\
  {\hat \F}^{A,(2)}_{g} &= {\dis{C_{F} n_{f}}} \Bigg\{ - \frac{80}{3} +
                          6  \ln
                          \left(\frac{q^2}{m_t^2}\right)  + 8
                          \zeta_3 
                          +  \epsilon \Bigg(
                          \frac{2827}{36} - 9  \ln
                          \left(\frac{q^2}{m_t^2}\right)   -
                          \frac{19}{6} \zeta_2 -  
                          \frac{8}{3} \zeta_2^2 
                          \nonumber\\
                        &- \frac{64}{3} \zeta_3
                          \Bigg)  
                          +
                          \epsilon^2 \Bigg(  - \frac{70577}{432} +
                          \frac{21}{2}  \ln
                          \left(\frac{q^2}{m_t^2}\right)   + 
                          \frac{1037}{72} \zeta_2 - \frac{3}{4}
                          \ln \left(\frac{q^2}{m_t^2} \right)
                          \zeta_2  + \frac{64}{9} \zeta_2^2 +
                          \frac{455}{9} \zeta_3  
                          \nonumber\\
                        &- 
                          \frac{10}{3} \zeta_2 \zeta_3 + 8 \zeta_5
                          \Bigg) 
                          + 
                          \epsilon^3 \Bigg( \frac{1523629}{5184} -
                          \frac{45}{4}  \ln
                          \left(\frac{q^2}{m_t^2}\right)  -
                          \frac{14975}{432} \zeta_2  + 
                          \frac{9}{8}  \ln
                          \left(\frac{q^2}{m_t^2}\right) \zeta_2 
                          \nonumber\\
                        &- 
                          \frac{70997}{4320} \zeta_2^2 +  \frac{22}{35}
                          \zeta_2^3  - 
                          \frac{3292}{27} \zeta_3 + \frac{7}{4}
                          \ln \left(\frac{q^2}{m_t^2}\right)
                          \zeta_3  +  \frac{80}{9} \zeta_2 \zeta_3 + 
                          15 \zeta_3^2 - \frac{64}{3} \zeta_5 \Bigg)    
                          \nonumber\\
                        &+ \epsilon^4 \Bigg( -
                          \frac{30487661}{62208}  + \frac{93}{8}   \ln
                          \left( \frac{q^2}{m_t^2}\right) + 
                          \frac{43217}{648} \zeta_2 - \frac{21}{16}
                          \ln \left(\frac{q^2}{m_t^2}\right)
                          \zeta_2   + \frac{1991659}{51840} \zeta_2^2 
                          \nonumber\\
                        &- 
                          \frac{141}{320}  \ln \left(\frac{q^2}{m_t^2}\right) \zeta_2^2 -
                          \frac{176}{105} \zeta_2^3  +
                          \frac{694231}{2592} \zeta_3  - 
                          \frac{21}{8}  \ln \left(\frac{q^2}{m_t^2}\right) \zeta_3 -
                          \frac{9757}{432} \zeta_2 \zeta_3  - 
                          \frac{1681}{180} \zeta_2^2 \zeta_3 
                          \nonumber\\
                        &- 40
                          \zeta_3^2 + \frac{8851}{180} \zeta_5  - 
                          2 \zeta_2 \zeta_5 - \frac{127}{8} \zeta_7
                          \Bigg) \Bigg\} 
                          + 
                          {\dis{C_{A} n_{f}}} \Bigg\{  -
                          \frac{8}{3 \epsilon^3}  + \frac{20}{9
                          \epsilon^2} + 
                          \Bigg( \frac{106}{27} + 2 \zeta_2 \Bigg)
                          \frac{1}{\epsilon} 
                          \nonumber\\
                        &- \frac{1591}{81} -
                          \frac{5}{3} \zeta_2     - \frac{74}{9}
                          \zeta_3 
                          + 
                          \epsilon \Bigg( \frac{24107}{486} -
                          \frac{23}{18} \zeta_2  + \frac{51}{20}
                          \zeta_2^2  + \frac{383}{27} \zeta_3 \Bigg) 
                          + 
                          \epsilon^2 \Bigg( - \frac{146147}{1458} 
                          \nonumber\\
                        &+
                          \frac{799}{108} \zeta_2  - \frac{329}{72} \zeta_2^2 - 
                          \frac{1436}{81} \zeta_3 + \frac{25}{6}
                          \zeta_2 \zeta_3  - \frac{271}{30} \zeta_5
                          \Bigg) 
                          + 
                          \epsilon^3 \Bigg( \frac{6333061}{34992} -
                          \frac{11531}{648} \zeta_2  
                          \nonumber\\
                        &+ 
                        + \frac{1499}{240}
                          \zeta_2^2 
                          \frac{253}{1680} \zeta_2^3 +
                          \frac{19415}{972} \zeta_3  - \frac{235}{36}
                          \zeta_2 \zeta_3  - 
                          \frac{1153}{108} \zeta_3^2 + \frac{535}{36}
                          \zeta_5 \Bigg)  
                          + 
                          \epsilon^4 \Bigg( - \frac{128493871}{419904}
                          \nonumber\\
                        &+ \frac{133237}{3888} \zeta_2  -
                          \frac{21533}{2592} \zeta_2^2  + 
                          \frac{649}{1440} \zeta_2^3 -
                          \frac{156127}{5832} \zeta_3  + \frac{215}{27}
                          \zeta_2 \zeta_3  + 
                          \frac{517}{80} \zeta_2^2 \zeta_3 +
                          \frac{14675}{648} \zeta_3^2  
                          \nonumber\\
                        &-
                          \frac{2204}{135} \zeta_5  + 
                          \frac{171}{40} \zeta_2 \zeta_5 +
                          \frac{229}{336} \zeta_7 \Bigg) \Bigg\}  
                          + 
                          {\dis{C_{A}^2}} \Bigg\{ 
                          \frac{32}{\epsilon^4}  + \frac{44}{3
                          \epsilon^3} + \Bigg( - \frac{422}{9}  - 4
                          \zeta_2 \Bigg) \frac{1}{\epsilon^2}
                          +  
                          \Bigg( \frac{890}{27} 
                          \nonumber\\
                        &- 11 \zeta_2 +
                          \frac{50}{3} \zeta_3 \Bigg)
                          \frac{1}{\epsilon} + \frac{3835}{81} +  
                          \frac{115}{6} \zeta_2 - \frac{21}{5}
                          \zeta_2^2 + \frac{11}{9} \zeta_3  
                          + 
                          \epsilon \Bigg( - \frac{213817}{972} -
                          \frac{103}{18} \zeta_2  + \frac{77}{120}
                          \zeta_2^2 
                          \nonumber\\
                        &+ \frac{1103}{54} \zeta_3  - 
                          \frac{23}{6} \zeta_2 \zeta_3 - \frac{71}{10}
                          \zeta_5 \Bigg)  
                          + 
                          \epsilon^2 \Bigg( \frac{6102745}{11664} -
                          \frac{991}{27} \zeta_2  - \frac{2183}{240} \zeta_2^2 + 
                          \frac{2313}{280} \zeta_2^3 - \frac{8836}{81}
                          \zeta_3  
                          \nonumber\\
                        &- \frac{55}{12} \zeta_2 \zeta_3 + 
                          \frac{901}{36} \zeta_3^2 + \frac{341}{60}
                          \zeta_5 \Bigg)  
                          + 
                          \epsilon^3 \Bigg( - \frac{142142401}{139968}
                          + \frac{75881}{648} \zeta_2  +
                          \frac{79819}{2160} \zeta_2^2 -  
                          \frac{2057}{480} \zeta_2^3 
                          \nonumber\\
                        &+
                          \frac{606035}{1944} \zeta_3 -  \frac{251}{72}
                          \zeta_2 \zeta_3 -  
                          \frac{1291}{80} \zeta_2^2 \zeta_3 -
                          \frac{5137}{216} \zeta_3^2 +
                          \frac{14459}{360} \zeta_5 +  
                          \frac{313}{40} \zeta_2 \zeta_5 -
                          \frac{3169}{28} \zeta_7 \Bigg) 
                          \nonumber\\
                        &+  
                          \epsilon^4 \Bigg( \frac{2999987401}{1679616}
                          - \frac{1943429}{7776} \zeta_2  -
                          \frac{15707}{160} \zeta_2^2 -  
                          \frac{35177}{20160} \zeta_2^3 +
                          \frac{50419}{1600} \zeta_2^4  -
                          \frac{16593479}{23328} \zeta_3 
                          \nonumber\\
                        &+  
                          \frac{1169}{27} \zeta_2 \zeta_3 +
                          \frac{22781}{1440} \zeta_2^2 \zeta_3 +  
                          \frac{93731}{1296} \zeta_3^2 -
                          \frac{1547}{144} \zeta_2 \zeta_3^2 -
                          \frac{8137}{54} \zeta_5 -  
                          \frac{1001}{80} \zeta_2 \zeta_5 +
                          \frac{845}{24} \zeta_3 \zeta_5 
                          \nonumber\\
                        &-
                          \frac{33}{2} \zeta_{5,3} +  
                          \frac{56155}{672} \zeta_7 \Bigg) \Bigg\}\,,
                          \nonumber\\ 
  {\hat \F}^{A,(3)}_{g} &=  {\dis{n_{f} C^{(2)}_{J}}} \Bigg\{ - 2+3 \epsilon \Bigg\}
                          + 
                          {\dis{C_{F} n_{f}^2}} \Bigg\{ \Bigg( -
                          \frac{640}{9} + 16 \ln
                          \left(\frac{q^2}{m_t^2}\right) + \frac{64}{3}
                          \zeta_3  \Bigg) \frac{1}{\epsilon} +  
                          \frac{7901}{27}  
                          \nonumber\\
                        &- 24 \ln
                          \left(\frac{q^2}{m_t^2}\right) 
                          - \frac{32}{3}
                          \zeta_2  - \frac{112}{15} \zeta_2^2 - 
                          \frac{848}{9} \zeta_3 \Bigg\} 
                          + 
                          {\dis{C_{F}^2 n_{f}}} \Bigg\{ \frac{457}{6} + 104
                          \zeta_3  - 160 \zeta_5 \Bigg\} 
                          \nonumber\\
                        &+ 
                          {\dis{C_{A}^2 n_{f}}} \Bigg\{ 
                          \frac{64}{3 \epsilon^5} - \frac{32}{81
                          \epsilon^4}  
                          + 
                          \Bigg( - \frac{18752}{243}  - \frac{376}{27}
                          \zeta_2 \Bigg)  \frac{1}{\epsilon^3}  + 
                          \Bigg( \frac{36416}{243} - \frac{1700}{81}
                          \zeta_2  + \frac{2072}{27} \zeta_3 \Bigg)
                          \frac{1}{\epsilon^2} 
                          \nonumber\\
                        &+ \Bigg(
                          \frac{62642}{2187}  + \frac{22088}{243}
                          \zeta_2 
                          - \frac{2453}{90} \zeta_2^2  - 
                          \frac{3988}{81} \zeta_3 \Bigg)
                          \frac{1}{\epsilon} -
                          \frac{14655809}{13122} -
                          \frac{60548}{729} \zeta_2  + 
                          \frac{917}{60} \zeta_2^2 
                          \nonumber\\
                        &- \frac{772}{27} \zeta_3
                          - \frac{439}{9} \zeta_2 \zeta_3   
                          + \frac{3238}{45}
                          \zeta_5 \Bigg\}  
                          + 
                          {\dis{C_{A}  n_{f}^2}} \Bigg\{ - \frac{128}{81
                          \epsilon^4} + \frac{640}{243 \epsilon^3}  + \Bigg(
                          \frac{128}{27}  + \frac{80}{27} \zeta_2
                          \Bigg) \frac{1}{\epsilon^2}  
                          \nonumber\\
                        &+ 
                          \Bigg( - \frac{93088}{2187} 
                          - \frac{400}{81}
                          \zeta_2  
                          - \frac{1328}{81} \zeta_3 \Bigg)
                          \frac{1}{\epsilon} + 
                          \frac{1066349}{6561}  -
                          \frac{56}{27} \zeta_2  + 
                          \frac{797}{135} \zeta_2^2  + 
                          \frac{13768}{243} \zeta_3 \Bigg\} 
                          \nonumber\\
                        &+ 
                          {\dis{C_{A} C_{F} 
                          n_{f}}} \Bigg\{ -
                          \frac{16}{9 \epsilon^3} 
                          +  
                          \Bigg( \frac{5980}{27} - 48 \ln \left(\frac{q^2}{m_t^2}\right)  -
                          \frac{640}{9} \zeta_3 \Bigg)
                          \frac{1}{\epsilon^2} + \Bigg( -
                          \frac{20377}{81} 
                          \nonumber\\
                        &- 16 \ln \left(\frac{q^2}{m_t^2}\right)  + \frac{86}{3}
                          \zeta_2  + 
                          \frac{352}{15} \zeta_2^2 
                          + \frac{1744}{27}
                          \zeta_3 \Bigg) \frac{1}{\epsilon} + 72 \ln \left(\frac{q^2}{m_t^2}\right)   
                          - \frac{587705}{972} - 
                          \frac{551}{6} \zeta_2 
                          \nonumber\\
                        &+ 12 \ln \left(\frac{q^2}{m_t^2}\right) \zeta_2 -
                          \frac{96}{5} \zeta_2^2  + \frac{12386}{81}
                          \zeta_3  
                          + 
                          48 \zeta_2 \zeta_3  +
                          \frac{32}{9} \zeta_5 \Bigg\}   
                          + 
                          {\dis{C_{A}^3}} \Bigg\{ -
                          \frac{256}{3 \epsilon^6}  - \frac{352}{3
                          \epsilon^5}  
                          \nonumber\\
                        &+ \frac{16144}{81 \epsilon^4}  + 
                          \Bigg( \frac{22864}{243} + \frac{2068}{27}
                          \zeta_2  - \frac{176}{3} \zeta_3 \Bigg)
                          \frac{1}{\epsilon^3}  
                          + 
                          \Bigg( - \frac{172844}{243} - \frac{1630}{81}
                          \zeta_2  + \frac{494}{45} \zeta_2^2  
                          \nonumber\\
                        &-
                          \frac{836}{27} \zeta_3 \Bigg)
                          \frac{1}{\epsilon^2}  + \Bigg(
                          \frac{2327399}{2187}  - \frac{71438}{243}
                          \zeta_2 +  
                          \frac{3751}{180} \zeta_2^2 
                          - \frac{842}{9}
                          \zeta_3  + \frac{170}{9} \zeta_2 \zeta_3  + 
                          \frac{1756}{15} \zeta_5 \Bigg)
                          \frac{1}{\epsilon} 
                          \nonumber\\
                        &+ \frac{16531853}{26244} + 
                          \frac{918931}{1458} \zeta_2 +
                          \frac{27251}{1080} \zeta_2^2  -
                          \frac{22523}{270} \zeta_2^3   
                          - \frac{51580}{243}
                          \zeta_3  + \frac{77}{18} \zeta_2 \zeta_3 -
                          \frac{1766}{9} \zeta_3^2  
                          \nonumber\\
                        &+ 
                          \frac{20911}{45} \zeta_5 \Bigg\}\,.
\end{align}
The results up to two loop level is consistent with the existing
ones~\cite{Ravindran:2004mb} and the three loop result is the new
one. These are later used to compute the SV cross-section for the production
of a pseudo-scalar particle through gluon fusion at N$^{3}$LO QCD~\cite{Ahmed:2015pSSV}. 
This is an essential ingredient to compute all the other associated observables.

\section{Hard Matching Coefficients in SCET}
\label{sec:scet}

Soft-collinear effective theory (SCET, \cite{hep-ph/0005275,
  hep-ph/0011336, hep-ph/0107001, hep-ph/0109045, hep-ph/0206152,
  hep-ph/0211358, hep-ph/0202088}) is a systematic expansion of the
full QCD theory in terms of particle modes with different infrared
scaling behaviour. It provides a framework to perform threshold
resummation. In the effective theory, the infrared poles of the full
high energy QCD theory manifest themselves as ultraviolet
poles~\cite{Korchemsky:1985xj, Korchemsky:1987wg, Korchemsky:1988pn},
which then can be resummed by employing the renormalisation group
evolution from larger scales to the smaller ones. To ensure matching
of SCET and full QCD, one computes the matrix elements in both
theories and adjusts the Wilson coefficients of SCET accordingly. For
the on-shell matching of these two theories, the matching coefficients
relevant to pseudo-scalar production in gluon fusion can be obtained
directly from the gluon form factors.

The UV renormalised form factors in QCD contain IR
divergences. Since the IR poles in QCD turn into UV ones in SCET, we
can remove the IR divergences with the help of a renormalisation
constant $Z^{A,h}_{g}$, which essentially absorbs all residual IR poles
and produces finite results. The result is the matching coefficient
$C^{A, {\rm eff}}_{g}$, which is defined through the following
factorisation relation:
\begin{align}
  \label{eq:MatchCof}
  C^{A, {\rm eff}}_{g}\l(Q^{2},\mu_h^{2}\r) &\equiv \lim_{\epsilon
                                              \rightarrow 0} (Z_{g
                                              }^{A,h})^{-1}(\epsilon, Q^{2},\mu_h^{2})
                                              \left[ {\cal F}^{A}_{g} \right]_{R}(\epsilon,Q^{2})
\end{align}  
where, the UV renormalised form factor
$\left[ {\cal F}^{A}_{g} \right]_{R}$, is defined as
\begin{align}
  \label{eq:UVRenFF}
  \left[ {\cal F}^{A}_{g} \right]_{R} =   \left[ {\cal F}^{G}_{g}
  \right]_{R}  + \frac{4 C_J}{C_G}  \left[ {\cal F}^{J}_{g}
  \right]_{R} { \l( a_{s} \frac{S^{J,(1)}_g}{S^{G,(0)}_{g}}
  \r)} \,.
\end{align}
The parameter $\mu_h$ is the newly introduced mass scale at which the
above factorisation is carried out. For the UV renormalised form
factors $[{\cal F}^{A}_{g}]_{R}$ in Eq.~(\ref{eq:MatchCof}), we fixed
the other scales as $\mu_{R}^{2}=\mu_{F}^{2}=\mu_h^2$. Upon
expanding the $Z^{A,h}_{g}$ and $C^{A,{\rm eff}}_{g}$ in powers
of $a_{s}$ as
\begin{align}
  \label{eq:MatchCofExp}
  Z^{A,h}_{g}(\epsilon, Q^{2},\mu_h^{2}) &= 1 + \sum_{i=1}^{\infty}
                                                  a_{s}^{i}(\mu_h^{2})
                                                  Z^{A, h}_{g,i}(\epsilon,
                                                  Q^{2}, \mu_h^{2})\,,
                                                  \nonumber\\
  C^{A, {\rm eff}}_{g}\l(Q^{2},\mu_h^{2}\r) &= 1+\sum_{i=1}^{\infty}
                                              a_{s}^{i}(\mu_h^{2})
                                              C^{A, {\rm eff}}_{g,i}\l(Q^{2},\mu_h^{2}\r)
\end{align}
and utilising the above Eq.~(\ref{eq:MatchCof}), we compute the
$Z^{A,h}_{g,i}$ as well as $C^{A, {\rm eff}}_{g,i}$ up to three
loops ($i=3$). Demanding the cancellation of the residual IR poles of
$\left[ {\cal F}^{A}_{g} \right]_{R}$ against the poles of
$(Z^{A,h}_{g,i})^{-1}$, we compute $Z^{A,h}_{g,i}$ which
comes out to be
\begin{align}
  \label{eq:ZIR}
  Z^{A,h}_{g,1} &= {\dis{C_{A}}} \Bigg\{ - \frac{8}{\epsilon^2} +
                         \Bigg( - 4 L +
                         \frac{22}{3}   \Bigg) \frac{1}{\epsilon}
                         \Bigg\}  - {\dis{n_{f}}} \Bigg\{ \frac{4}{3 \epsilon} \Bigg\}\,,
                         \nonumber\\
  Z^{A,h}_{g,2} &= C_{F} n_{f} \Bigg\{ - \frac{2}{\epsilon} \Bigg\}
                         +  n_{f}^2 \Bigg\{ \frac{16}{9
                         \epsilon^2}  \Bigg\}
                         + 
                         C_{A} n_{f} \Bigg\{ \frac{56}{3 \epsilon^3} +
                         \Bigg( - \frac{52}{3} +  8 L \Bigg) \frac{1}{\epsilon^2} + 
                         \Bigg( - \frac{128}{27} + \frac{20}{9} L 
                         \nonumber\\
                       &+
                         \frac{2}{3} \zeta_2 \Bigg)
                         \frac{1}{\epsilon} \Bigg\} 
                         + 
                         C_{A}^2 \Bigg\{ \frac{32}{\epsilon^4} + \Bigg(
                         - \frac{308}{3} + 32 L \Bigg)  \frac{1}{\epsilon^3} + 
                         \Bigg( \frac{350}{9} - 44 L + 8 L^2 + 4
                         \zeta_2 \Bigg) \frac{1}{\epsilon^2}  + 
                         \Bigg( \frac{692}{27} 
                         \nonumber\\
                       &- \frac{134}{9} L -
                         \frac{11}{3} \zeta_2 +  4 L \zeta_2 - 2
                         \zeta_3 \Bigg)  \frac{1}{\epsilon} \Bigg\}\,,
                         \nonumber\\
  Z^{A,h}_{g,3} &= C_{F}^2 n_{f} \Bigg\{ \frac{2}{3 \epsilon}
                         \Bigg\} 
                         + C_{F} n_{f}^2
                         \Bigg\{ \frac{56}{9 \epsilon^2}  +
                         \frac{22}{27 \epsilon} \Bigg\}   
                         - 
                         n_{f}^3 \Bigg\{ \frac{64}{27 \epsilon^3}
                         \Bigg\} 
                         + C_{A}^2
                         n_{f} \Bigg\{  - \frac{320}{3 \epsilon^5} +
                         \Bigg( \frac{28480}{81} 
                         \nonumber\\
                       &-  96 L \Bigg) \frac{1}{\epsilon^4} + 
                         \Bigg( - \frac{18752}{243} + \frac{3152}{27}
                         L - \frac{64}{3} L^2  - \frac{448}{27}
                         \zeta_2 \Bigg) \frac{1}{\epsilon^3}  + 
                         \Bigg( - \frac{32656}{243} + \frac{7136}{81}
                         L 
                         \nonumber\\
                       &- \frac{80}{9} L^2  + \frac{1000}{81} \zeta_2 - 
                         \frac{104}{9} L \zeta_2 + \frac{344}{27}
                         \zeta_3 \Bigg) \frac{1}{\epsilon^2}  + 
                         \Bigg( - \frac{30715}{2187} + \frac{836}{81}
                         L + \frac{2396}{243} \zeta_2  -
                         \frac{160}{27} L \zeta_2 
                         \nonumber\\
                       &-  
                         \frac{328}{45} \zeta_2^2 - \frac{712}{81}
                         \zeta_3  + \frac{112}{9} L \zeta_3 \Bigg)
                         \frac{1}{\epsilon} \Bigg\} 
                         +  
                         C_{A} n_{f}^2 \Bigg\{ - \frac{2720}{81
                         \epsilon^4}  + \Bigg( \frac{7984}{243} -
                         \frac{352}{27} L \Bigg)  \frac{1}{\epsilon^3} + 
                         \Bigg( \frac{368}{27} 
                         \nonumber\\
                       &- \frac{400}{81} L -
                         \frac{40}{27} \zeta_2  \Bigg)
                         \frac{1}{\epsilon^2} +  
                         \Bigg( \frac{269}{2187} + \frac{16}{81} L -
                         \frac{40}{81} \zeta_2  + \frac{112}{81}
                         \zeta_3 \Bigg) \frac{1}{\epsilon}  \Bigg\} 
                         + C_{A} 
                         C_{F} n_{f} \Bigg\{ \frac{272}{9 \epsilon^3} 
                         \nonumber\\
                       &+
                         \Bigg( - \frac{704}{27}  + \frac{40}{3} L -
                         \frac{64}{9} \zeta_3 \Bigg)
                         \frac{1}{\epsilon^2} +  
                         \Bigg( - \frac{2434}{81} + \frac{110}{9} L +
                         \frac{4}{3} \zeta_2  + \frac{32}{15}
                         \zeta_2^2 +  
                         \frac{304}{27} \zeta_3 
                         \nonumber\\
                       &- \frac{32}{3} L
                         \zeta_3 \Bigg)  \frac{1}{\epsilon} \Bigg\}
                         +  
                         C_{A}^3 \Bigg\{ - \frac{256}{3 \epsilon^6} +
                         \Bigg( \frac{1760}{3}  - 128 L \Bigg)
                         \frac{1}{\epsilon^5} +  
                         \Bigg( - \frac{72632}{81} + 528 L - 64 L^2
                         \nonumber\\
                       &- 32 \zeta_2 \Bigg)  \frac{1}{\epsilon^4} + 
                         \Bigg( - \frac{29588}{243} - \frac{5824}{27}
                         L + \frac{352}{3} L^2  - \frac{32}{3} L^3 + 
                         \frac{2464}{27} \zeta_2 - 48 L \zeta_2 + 16
                         \zeta_3 \Bigg)  \frac{1}{\epsilon^3} 
                         \nonumber\\
                       &+ 
                         \Bigg( \frac{80764}{243} - \frac{25492}{81}
                         L + \frac{536}{9} L^2  - \frac{1486}{81}
                         \zeta_2 +  
                         \frac{572}{9} L \zeta_2 - 16 L^2 \zeta_2 -
                         \frac{352}{45} \zeta_2^2  - \frac{836}{27}
                         \zeta_3 
                         \nonumber\\
                       &+  
                         8 L \zeta_3 \Bigg) \frac{1}{\epsilon^2} +
                         \Bigg( \frac{194372}{2187}  - \frac{490}{9}
                         L - \frac{12218}{243} \zeta_2  + 
                         \frac{1072}{27} L \zeta_2 + \frac{1276}{45}
                         \zeta_2^2  - \frac{176}{15} L \zeta_2^2 - 
                         \frac{244}{9} \zeta_3 
                         \nonumber\\
                       &- \frac{88}{9} L
                         \zeta_3  + \frac{80}{9} \zeta_2 \zeta_3  +
                         \frac{32}{3} \zeta_5 \Bigg)  \frac{1}{\epsilon} \Bigg\}\,.
\end{align}
After cancellation of the IR poles, we are left with the following
finite matching coefficients:
\begin{align}
  \label{eq:SCETCof}
  C^{A, {\rm eff}}_{g,1} &= {\dis{C_{A}}} \Bigg\{ - L^2 + 4 + \zeta_2 \Bigg\}\,,
                           \nonumber\\
  C^{A, {\rm eff}}_{g,2} &=  C_{A}^2 \Bigg\{  \frac{1}{2} L^4 + 
                           \frac{11}{9} L^3  + L^2
                           \Bigg( - \frac{103}{9}  + \zeta_2 \Bigg) 
                           + L \Bigg( - \frac{10}{27}  - \frac{22}{3} 
                           \zeta_2 - 2 \zeta_3 \Bigg) + \frac{4807}{81} + 
                           \frac{91}{6} \zeta_2 
                           \nonumber\\
                         &+ \frac{1}{2}\zeta_2^2 -  
                           \frac{143}{9} \zeta_3 \Bigg\} 
                           + C_{A} n_{f}
                           \Bigg\{  - \frac{2}{9} L^3 +  \frac{10}{9}
                           L^2 + L \Bigg(
                           \frac{34}{27} + \frac{4}{3} \zeta_2 \Bigg)
                           - \frac{943}{81} -  
                           \frac{5}{3} \zeta_2 - \frac{46}{9} \zeta_3 \Bigg\}  
                           \nonumber\\
                         & + 
                           C_{F} n_{f} \Bigg\{ - \frac{80}{3} + 6
                           \ln \left(\frac{\mu_h^2}{m_t^2}\right) + 8 \zeta_3 \Bigg\}\,,
                           \nonumber\\
  C^{A, {\rm eff}}_{g,3} &= n_{f} C^{(2)}_{J} \Bigg\{ - {2}
                           \Bigg\} 
                           + C_{F} n_{f}^2 \Bigg\{ L \Bigg( -
                           \frac{320}{9} + 8 \ln \left(\frac{\mu_h^2}{m_t^2}\right) +  \frac{32}{3}
                           \zeta_3 \Bigg) +
                           \frac{749}{9} - \frac{20}{9} \zeta_2  -
                           \frac{16}{45} \zeta_2^2  
                           \nonumber\\
                         &- 
                           \frac{112}{3} \zeta_3  \Bigg\}  
                           + 
                           C_{F}^2 n_{f} \Bigg\{ \frac{457}{6} + 104
                           \zeta_3 - 160 \zeta_5 \Bigg\}  
                           + 
                           C_{A}^2 n_{f} \Bigg\{  
                           \frac{2}{9} L^5 - \frac{8}{27} L^4 + 
                           L^3 \Bigg( - \frac{752}{81} 
                           \nonumber\\
                         &- \frac{2}{3}
                           \zeta_2 \Bigg)  + 
                           L^2 \Bigg( \frac{512}{27} - \frac{103}{9}
                           \zeta_2  + \frac{118}{9} \zeta_3 \Bigg) +  
                           L \Bigg( \frac{129283}{729} +
                           \frac{4198}{81} \zeta_2 -  \frac{48}{5}
                           \zeta_2^2 +  \frac{28}{9} \zeta_3 \Bigg) 
                           \nonumber\\
                         &-
                           \frac{7946273}{13122} -  \frac{19292}{729} \zeta_2
                           + \frac{73}{45} \zeta_2^2  - 
                           \frac{2764}{81} \zeta_3 - \frac{82}{9}
                           \zeta_2 \zeta_3  +
                           \frac{428}{9} \zeta_5 \Bigg\} 
                           +  
                           C_{A}^3 \Bigg\{  - \frac{1}{6} L^6 -
                           \frac{11}{9} L^5   
                           \nonumber\\
                         &+ L^4
                           \Bigg( \frac{389}{54}  - \frac{3}{2}
                           \zeta_2 \Bigg) + 
                           L^3 \Bigg( \frac{2206}{81} + \frac{11}{3}
                           \zeta_2 + 2 \zeta_3 \Bigg) +  
                           L^2 \Bigg( - \frac{20833}{162} +
                           \frac{757}{18} \zeta_2  - \frac{73}{10}
                           \zeta_2^2  
                           \nonumber\\
                         &+ \frac{143}{9} \zeta_3 \Bigg) + 
                           \frac{2222}{9} \zeta_5 + L \Bigg( -
                           \frac{500011}{1458}  - \frac{16066}{81}
                           \zeta_2 + \frac{176}{5} \zeta_2^2  + 
                           \frac{1832}{27} \zeta_3 + \frac{34}{3}
                           \zeta_2 \zeta_3  
                           \nonumber\\
                         &+ 16 \zeta_5 \Bigg) + \frac{41091539}{26244} + 
                           \frac{316939}{1458} \zeta_2 -
                           \frac{1399}{270} \zeta_2^2 -
                           \frac{24389}{1890} \zeta_2^3 -  
                           \frac{176584}{243} \zeta_3 - \frac{605}{9}
                           \zeta_2 \zeta_3  
                           \nonumber\\
                         &- \frac{104}{9} \zeta_3^2  \Bigg\} 
                           + 
                           C_{A} n_{f}^2 \Bigg\{
                           - \frac{2}{27} L^4 + \frac{40}{81} L^3   + 
                           L^2 \Bigg( \frac{80}{81} + \frac{8}{9}
                           \zeta_2 \Bigg) + 
                           L \Bigg( - \frac{12248}{729} -
                           \frac{80}{27} \zeta_2 
                           \nonumber\\
                         &- \frac{128}{27}
                           \zeta_3 \Bigg) + \frac{280145}{6561} +
                           \frac{4}{9}  \zeta_2  + \frac{4}{27} \zeta_2^2  +  \frac{4576}{243} \zeta_3
                           \Bigg\}
                           + 
                           C_{A} C_{F} n_{f} \Bigg\{  -
                           \frac{2}{3} L^3  + L^2 \Bigg( \frac{215}{6}
                           \nonumber\\
                         &- 6 \ln \left(\frac{\mu_h^2}{m_t^2}\right)
                           - 16 \zeta_3 \Bigg) +
                           L \Bigg( \frac{9173}{54}  - 44 \ln
                           \left(\frac{\mu_h^2}{m_t^2}\right)  + 4 \zeta_2 + 
                           \frac{16}{5} \zeta_2^2 - \frac{376}{9}
                           \zeta_3 \Bigg) 
                           \nonumber\\
                         &+ 24 \ln
                           \left(\frac{\mu_h^2}{m_t^2}\right)  -
                           \frac{726935}{972} -
                           \frac{415}{18} \zeta_2  + 
                           6 \ln \left(\frac{\mu_h^2}{m_t^2}\right)
                           \zeta_2 - \frac{64}{45} \zeta_2^2    +  
                           \frac{20180}{81} \zeta_3 + \frac{64}{3}
                           \zeta_2 \zeta_3 
                           \nonumber\\
                         &+  \frac{608}{9} \zeta_5 \Bigg\}\,. 
\end{align}
In the above expressions,
$L=\ln \left( Q^{2}/\mu_h^{2} \right)=\ln \left( -q^{2}/\mu_h^{2}
\right)$.
These matching coefficients allow to perform the matching of the
SCET-based resummation onto the full QCD calculation up to three-loop
order.

Before ending the discussion of this section, we demonstrate the
universal factorisation property fulfilled by the anomalous dimension
of the $Z^{A,h}_{g}$ which is defined through the RG equation
\begin{align}
  \label{eq:RGEZIR}
  \mu_{h}^{2}\frac{d}{d\mu_{h}^{2}} \ln Z^{A,h}_{g}(\epsilon, Q^{2},\mu_{h}^{2}) \equiv
  \gamma^{A,h}_{g}(Q^{2},\mu_{h}^{2}) = \sum_{i=1}^{\infty} a_{s}^{i}(\mu_{h}^{2})
  \gamma^{A,h}_{g,i}(Q^{2},\mu_{h}^{2})\,.
\end{align}
The renormalisation group invariance of the UV renormalised
$[F^{A}_{g}]_{R}(\epsilon, Q^{2})$ with respect to the scale $\mu_{h}$
implies
\begin{align}
  \label{eq:RGEZIRC}
  \mu_{h}^{2}\frac{d}{d\mu_{h}^{2}} \ln Z^{A,h}_{g} +
  \mu_{h}^{2}\frac{d}{d\mu_{h}^{2}} \ln C^{A,{\rm eff}}_{g} = 0\,. 
\end{align}  
By explicitly evaluating the $\gamma^{A,h}_{g,i}$ using the
results of $Z^{A,h}_{g}$ (Eq.~(\ref{eq:ZIR})) up to three loops
($i=3$), we find that these satisfy the following decomposition in
terms of the universal factors $A_{g,i}, B_{g,i}$ and $f_{g.i}$:
\begin{align}
  \label{eq:gammaIRdecom}
  \gamma^{A,h}_{g,i} = - \frac{1}{2} A_{g,i} L + \l(B_{g,i} +
  \frac{1}{2} f_{g,i}\r)\,.
\end{align} 
This in turn implies the evolution equation of the matching
coefficients as
\begin{align}
  \label{eq:RGECeff}
  \mu_{h}^{2}\frac{d}{d\mu_{h}^{2}} \ln C^{A,{\rm eff}}_{g,i} = \frac{1}{2} A_{g,i} L - \l(B_{g,i} +
  \frac{1}{2} f_{g,i}\r)
\end{align}
which is in complete agreement with the existing
results~\cite{Becher:2006nr} upon identifying
\begin{align}
  \label{eq:gammaV}
  \gamma^{V} = B_{g.i} + \frac{1}{2} f_{g,i}\,.
\end{align}

\section{Summary}
\label{sec:conc}
In this part of the thesis, we derived the three-loop massless QCD corrections to
the quark and gluon form factors of pseudo-scalar operators.  Working
in dimensional regularisation, we used the 't~Hooft-Veltman
prescription for $\gamma_5$ and the Levi-Civita tensor, which requires
non-trivial finite renormalisation to maintain the symmetries of the
theory. By exploiting the universal behaviour of the infrared pole
structure at three loops in QCD, we were able to independently
determine the renormalisation constants and operator mixing, in
agreement with earlier results that were obtained in a completely
different approach~\cite{Larin:1993tq,Zoller:2013ixa}.

The three-loop corrections to the pseudo-scalar form factors are an
important ingredient to precision Higgs phenomenology. They will
ultimately allow to bring the gluon fusion cross section for
pseudo-scalar Higgs production to the same level of accuracy that has
been accomplished most recently for scalar Higgs production with fixed
order N$^3$LO~\cite{Anastasiou:2015ema} and soft-gluon resummation at
N$^3$LL~\cite{Ahrens:2008nc,Bonvini:2014joa,Catani:2014uta,Schmidt:2015cea}.

With our new results, the soft-gluon resummation for pseudo-scalar
Higgs production~\cite{deFlorian:2007sr,Schmidt:2015cea} can be
extended imminently to N$^3$LL
{accuracy~\cite{Ahmed:2015pSSV}}, given the established
formalisms at this order~\cite{Ahrens:2008nc,Catani:2014uta}. 
With the derivation of the three-loop pseudo-scalar form factors 
presented here, all ingredients to this calculation are now available. 
Another imminent application is
the threshold approximation to the N$^3$LO cross {section~\cite{Ahmed:2015pSSV}}.  By
exploiting the universal infrared structure~\cite{Catani:2014uta}, one
can use the result of an explicit computation of the threshold
contribution to the N$^3$LO cross section for scalar Higgs
production~\cite{Anastasiou:2014vaa} to derive threshold results for
other processes essentially through the ratios of the respective form
factors (which is no longer possible beyond
threshold~\cite{Anastasiou:2014lda,Anastasiou:2015ema}, where the
corrections become process-specific), as was done for the Drell-Yan
process~\cite{Ahmed:2014cla} and for Higgs production from bottom
quark annihilation~\cite{Ahmed:2014cha}. 

%


  


\chapter{\label{chap:ConclOutlook}Conclusions and Outlooks}

\begingroup
\hypersetup{linkcolor=blue}
\minitoc
\endgroup

No doubt, the whole particle physics community is standing on the
verge of a crucial era, where the main tasks can be largely
categorized into two parts: testing the SM with unprecedented accuracy
and searching for the physics beyond SM. In achieving these golden tasks,
precise theoretical predictions play a very crucial role. The field of 
precision studies at theoretical level is mostly controlled by the
higher order corrections to the scattering amplitudes, that are the
basic building blocks of constructing any observable in QFT. Among all the higher
order corrections, the QCD ones contribute substantially to any
typical observable. This thesis deals with this higher order QCD
radiative corrections to the observables associated with the
Drell-Yan, scalar and pseudo-scalar Higgs boson.

The Higgs boson is among the best candidates at hadron collider, and
hence it is of utmost importance to make the theoretical prediction as
precise as possible to the associated observables. In the first part
of the thesis, Chapter~\ref{chap:bBCS}, we have computed the N$^3$LO QCD
radiative corrections, arising 
from the soft gluons, to the
inclusive production cross section of the Higgs boson produced through
bottom quark annihilation~\cite{Ahmed:2014cha}. Of course, this is not the dominant
production channel of the scalar Higgs boson
in the SM,
nonetheless its contribution must also be taken into account in this spectacular
precision studies. In order to achieve this, we have systematically
employed an elegant
prescription~\cite{Ravindran:2005vv,Ravindran:2006cg}. The
factorisation of QCD 
amplitudes, gauge invariance, renormalisation group invariance and the
Sudakov resummation of soft gluons are at the heart of this
formalism. The recently available three loop $Hb{\bar b}$ QCD form
factors~\cite{Gehrmann:2014vha} and the soft gluon contributions
calculated~\cite{Ahmed:2014cla} 
from the threshold QCD corrections to the Higgs boson at
N$^3$LO~\cite{Anastasiou:2014vaa}, enable us to compute the full
N$^3$LO soft-virtual QCD corrections 
to the production cross section of the Higgs boson produced through
bottom quark annihilation. One of the most beautiful parts of this
calculation is that even without evaluating all the hundreds or
thousands of Feynman diagrams contributing to the real emissions, we
have obtained the required contribution arising from the soft gluons! The
universal nature of the soft gluons are the underlying reasons behind
this remarkable feature. We have also demonstrated the numerical
impact of this result at the LHC. This is the most accurate result for
this production channel
existing in the literature till date and it is expected to play an
important role in coming days.

In the second part of the thesis, Chapter~\ref{chap:Rap}, we have dealt with an another very
important observable, namely, the rapidity distributions of the Higgs
boson produced through gluon fusion and the leptonic pair in
Drell-Yan. The importance of these two processes are quite
self-evident! We have computed the threshold enhanced N$^3$LO QCD
corrections~\cite{Ahmed:2014uya} 
to these observables employing the formalism developed in the
article~\cite{Ravindran:2006bu}.
The skeleton of this elegant prescription which has been employed is also based on the
properties, like, the factorisation of QCD 
amplitudes, gauge invariance, renormalisation group invariance and the
Sudakov resummation of soft gluons. With the help of recently computed inclusive production
cross section of the Higgs boson~\cite{Anastasiou:2014vaa} and
Drell-Yan~\cite{Ahmed:2014cla} at 
threshold N$^3$LO QCD, we have computed the contributions arising from
the soft gluons to the processes under consideration. These were the
only missing ingredients to achieve our goal. Our newly calculated
part of this distribution is found to be the most dominant one
compared to the other contributions. We have demonstrated numerically
the impact of this result for the Higgs boson at the LHC. Indeed,
inclusion of this N$^3$LO contributions does reduce the dependence on
the unphysical renormalisation and factorisation scales. 
It is worth mentioning that, this beautiful formalism not only
helps us to compute the rapidity distribution at threshold, but also
enhance our understanding about the underlying structures of the QCD
amplitudes.

In the third part of the thesis, Chapter~\ref{chap:Multiloop}, we have
discussed the relatively modern techniques of the multiloop
computations which have been employed to get some of the results
calculated in this thesis. The backbone of this methodology is the
integration-by-parts~\cite{Tkachov:1981wb,
  Chetyrkin:1981qh} and Lorentz invariant~\cite{Gehrmann:1999as}
identities. The successful implementation of these in computer codes
revolutionizes the area of multiloop computations. 

The last part, Chapter~\ref{chap:pScalar}, is dealt with a particle,
pseudo-scalar, 
which is not included in particle spectrum of the SM, but is believed
to be present in the nature. Intensive search for this particle has
been going on for past several years, although nothing conclusive
evidence has been found. However, to make conclusive remark about the
existence of this particle, we need to revamp the understanding about
this particle and improve the precision of the theoretical
predictions. This work arises exactly at this context. In
these articles~\cite{Ahmed:2015qpa, Ahmed:2015pSSV}, we have computed
one of the important ingredients to 
calculate the inclusive production cross section or the differential
distributions for the pseudo-scalar at N$^3$LO QCD which is presently
the level of accuracy for the scalar Higgs boson, achieved very
recently~\cite{Anastasiou:2015ema}. In particular, we have derived 
the three loop massless QCD corrections to the quark and gluon form
factors of the pseudo-scalar. Unlike the scalar Higgs boson, this
problem involves the $\gamma_5$ which makes the life interesting as
well as challenging. We have handled them under the `t Hooft-Veltman
prescription for the $\gamma_5$ and Levi-Civita tensor in dimensional
regularisation. Employing this prescription, however, brings some
additional complication, namely, it violates the chiral Ward
identity. In order to rectify this, we need to perform an additional
and non-trivial finite renormalisation. By exploiting the universal
behaviour of the infrared pole structure at three loops in QCD, we
were able to independently determine the renormalisation constants and
operator mixing, in agreement with the earlier results that were
obtained in a completely different
approach~\cite{Larin:1993tq,Zoller:2013ixa}. We must emphasize the
approach which we have employed here for the first time is exactly
opposite to the usual one: the infrared pole
structures of the form factors have been taken to be universal that
dictates us to obtain the UV operator renormalisation constants upon
imposing the demand of UV finiteness.  With our new results, the
threshold approximation to the N$^3$LO inclusive production cross
section for the pseudo-scalar through gluon fusion are
obtained~\cite{Ahmed:2015pSSV} by us. This is also extended to the N$^3$LL resummed
accuracy in~\cite{Ahmed:2015pSSV}. We have also
computed the hard matching coefficients in the context of
soft-collinear effective theory which are later employed to obtain the
N$^3$LL' resummed cross section~\cite{Ahmed:2016otz}. We have also found some
interesting facts about the form factors in the context of Leading
Transcendentality principle~\cite{hep-th/0611204, hep-th/0404092,
  Kotikov:2001sc}: the LT terms of the diagonal 
form factors with replacement $C_A = C_F = N$ and 
  $T_{f} n_{f}=N/2$ are not only identical to
each other but also coincide with the LT terms of the QCD form
factors~\cite{Gehrmann:2010ue} with the same replacement as well as
with the LT terms of the scalar form factors in ${\cal N}=4$ SYM
\cite{Gehrmann:2011xn}, up to a normalization factor of
  $2^{l}$. This observation holds true for the finite terms in
$\epsilon$, and could equally be validated for higher-order terms up
to transcendentality 8 (which is the highest order for which all
three-loop master integrals are available~\cite{Lee:2010ik}). In
addition to checking the diagonal form factors, we also examined the
off-diagonal ones,
where we find that the LT terms these two form factors are identical
to each other after the replacement of colour factors. However, the LT
terms of these do not coincide with those of the diagonal ones.

The state-of-the-art techniques, which mostly use our in-house codes,
have been employed extensively to 
carry out all the computations presented in this thesis. The prescription of
computing the threshold correction is applicable for any colorless
final state particle. We are in the process of extending this
formalism to the case of threshold resummation of differential
rapidity distributions. The methodology of calculating the
pseudo-scalar form factors can be generalized to the cases involving any number
of operators which can mix among each others under UV renormalisation.

In conclusion, it has been a while the Higgs-like particle has been discovered at the LHC and finally, we are 
very close to having enough statistics for precision measurements of the Higgs quantum numbers and coupling 
constants to fermions and gauge bosons. This, along with the precise results from theoreticians like us, hopefully, 
would help to explore the underlying nature of the electroweak symmetry breaking and possibly open the door
of new physics.
\begin{appendices}

\chapter{Inclusive Production Cross Section}
\label{App:CX}

In QCD improved parton model, the inclusive cross-section for the
production of a colorless particle can be computed using 
\begin{align}
\label{eq:App-1}
\sigma^{I}(\tau,q^{2}) = \sum\limits_{a,b=q,{\bar q},g} \int\limits_0^1 dx_{1} \int\limits_0^1 dx_{2}
  f_{a}(x_{1},\mu_{F}^{2}) f_{b}(x_{2}, \mu_F^{2}) \sigma^{I}_{ab}
  \left( z, q^{2}, \mu_{R}^{2}, \mu_F^{2} \right)
\end{align}
where, $f$'s are the partonic distribution functions factorised at the
mass scale $\mu_{F}$. $\sigma^{I}_{ab}$ is the partonic cross section
for the production of colorless particle $I$ from the partons $a$ and
$b$. This is UV renormalised at renormalisation scale $\mu_{R}$ and
mass factorised at $\mu_{F}$. The other quantities are defined as
\begin{align}
\label{eq:App-2}
&q^{2} = m_{I}^{2}\,,
\nonumber\\
&\tau = \frac{q^2}{S}\,,
\nonumber\\
&z = \frac{q^2}{{\hat s}}\,.
\end{align}
In the above expression, $S$ and ${\hat s}$ are square of the hadronic
and partonic center of mass energies, respectively, and they are related
by
\begin{align}
\label{eq:App-3}
{\hat s} = x_{1} x_{2} S\,.
\end{align}
By introducing the identity
\begin{align}
\label{eq:App-4}
\int dz \delta(\tau - x_{1} x_{2} z) = \frac{1}{x_{1} x_{2}} =
  \frac{S}{{\hat s}}
\end{align}
in Eq.~(\ref{eq:App-1}), we can rewrite the Eq.~(\ref{eq:App-1}) as
\begin{align}
\label{eq:App-5}
\sigma^{I}(\tau, q^{2}) = \sigma^{I,(0)}(\mu_R^2) \sum\limits_{ab=q,{\bar q},g}
  \int\limits_{\tau}^{1} dx \;\Phi_{ab}(x,\mu_F^{2})\;
  \Delta^I_{ab}\left(\frac{\tau}{x}, q^{2}, \mu_R^2, \mu_F^2\right)\,.
\end{align}
The partonic flux $\Phi_{ab}$ is defined through
\begin{align}
\label{eq:App-6}
\Phi_{ab}(x, \mu_{F}^2) = \int\limits_x^1 \frac{dy}{y} f_a(y, \mu_F^2)
  \;f_b\left(\frac{x}{y}, \mu_F^2 \right)
\end{align}
and the dimensionless quantity $\Delta^{I}_{ab}$ is called the
coefficient function of the partonic level cross section. Upon
normalising the partonic level cross section by the born one, we
obtain $\Delta^I_{ab}$ i.e.
\begin{align}
\label{eq:App-7}
\Delta^I_{ab} \equiv \frac{\sigma^I_{ab}}{\sigma^{I,(0)}}\,.
\end{align}  


\chapter{Anomalous Dimensions}
\label{chpt:App-AnoDim}

Here we present $A$~\cite{Moch:2004pa, Moch:2005tm, Vogt:2004mw,
  Vogt:2000ci}, $f$~\cite{Ravindran:2004mb, Moch:2005tm},
and $B$~\cite{Vogt:2004mw, Moch:2005tm} up to three loop level. The
$A$'s are given by
\begin{align}
\label{eq:App-AnoDimA}
 A_{gg,1} &= {{C_A}} \Big\{4\Big\} \,, 
\nonumber\\
 A_{gg,2} &= {{C_A^{2}}} \left\{ \frac{268}{9} - 8 \zeta_2 \right\} + {{C_A n_f}} \left\{ -\frac{40}{9} \right\} \,,
\nonumber\\
 A_{gg,3} &= {{C_A^3}} \left\{ \frac{490}{3} 
- \frac{1072 \zeta_2 }{9} 
+ \frac{88 \zeta_3}{3} 
+ \frac{176 \zeta_2^2}{5} \right\}
+ {{C_{A} C_F n_f}} \left\{ - \frac{110}{3} + 32 \zeta_3 \right\}
\nonumber\\ \nonumber
& + {{C_A^{2} n_f}} \left\{ - \frac{836}{27} 
+ \frac{160 \zeta_2}{9} 
- \frac{112 \zeta_3}{3} \right\}
+ {{C_A n_f^2}} \left\{ - \frac{16}{27} \right\} 
\intertext{and}
&A_{q{\bar q},i} = A_{b{\bar b},i} = \frac{C_{F}}{C_{A}} A_{gg,i}\,.
\end{align}
The $f$'s are obtained as
\begin{align}
\label{eq:App-AnoDimf}
 f_{gg,1} &= 0 \,,
\nonumber \\
 f_{gg,2} &= {{C_A^{2}}} \left\{ -\frac{22}{3} {\zeta_2} - 28 {\zeta_3} + \frac{808}{27} \right\}
        + {{C_A n_f}} \left\{ \frac{4}{3} {\zeta_2} - \frac{112}{27} \right\} \,,
\nonumber \\
 f_{gg,3} &= {{{C_A}^3}} \left\{ \frac{352}{5} {\zeta_2}^2 + \frac{176}{3} {\zeta_2} {\zeta_3}
        - \frac{12650}{81} {\zeta_2} - \frac{1316}{3} {\zeta_3} + 192 {\zeta_5}
        + \frac{136781}{729}\right\}
\nonumber \\
&
        + {{{C_A^{2}} {n_f}}} \left\{ - \frac{96}{5} {\zeta_2}^2 
        + \frac{2828}{81} {\zeta_2}
        + \frac{728}{27} {\zeta_3} - \frac{11842}{729} \right\} 
\nonumber \\ \nonumber
&
        + {{C_{A} {C_F} {n_f}}} \left\{ \frac{32}{5} {\zeta_2}^2 + 4 {\zeta_2} 
        + \frac{304}{9} {\zeta_3} - \frac{1711}{27} \right\}
        + {{{C_A} {n_f}^2}} \left\{ - \frac{40}{27} {\zeta_2} + \frac{112}{27} {\zeta_3}
        - \frac{2080}{729} \right\} 
\intertext{and}
&f_{q{\bar q}, i} = f_{b{\bar b}, i} = \frac{C_{F}}{C_{A}} f_{gg, i}\, .
\end{align}
Similarly the $B$'s are given by
\begin{align}
\label{eq:App-AnoDimB}
B_{gg,1} &= {{C_A}} \left\{\frac{11}{3}\right\} - {{n_f}} \left\{\frac{2}{3}\right\}\, ,
\nonumber\\
B_{gg,2} &= {{C_A^2}} \left\{\frac{32}{3} + 12 \zeta_3\right\} - {{n_f C_A}} \left\{\frac{8}{3} \right\} - {{n_f C_F}} \Big\{2 \Big\}\, ,
\nonumber\\
B_{gg,3} &= {{C_A C_F n_f}} \left\{-\frac{241}{18}\right\}
        + {{C_A n_f^2}} \left\{\frac{29}{18}\right\}
        - {{C_A^2 n_f}} \left\{\frac{233}{18} + \frac{8}{3} \zeta_2+ \frac{4}{3} \zeta_2^2 + \frac{80}{3} \zeta_3\right\}
\nonumber\\       
&+ {{C_A^3}} \left\{\frac{79}{2} - 16 \zeta_2 \zeta_3 + \frac{8}{3} \zeta_2 + \frac{22}{3} \zeta_2^2
        + \frac{536}{3} \zeta_3 - 80 \zeta_5\right\}
        + {{C_F n_f^2}} \left\{\frac{11}{9}\right\}
        + {{C_F^2 n_f}} \Big\{{1}\Big\}\, ,
\nonumber\\
B_{q{\bar q},1} &= {{C_F}} \Big\{{3}\Big\}\,,
\nonumber \\
B_{q{\bar q},2} &=   {{C_F^2}} \Bigg\{ \frac{3}{2} - 12 \zeta_2 + 24 \zeta_3 \Bigg\}
              + {{C_A C_F}} \Bigg\{ \frac{17}{34} + \frac{88}{6} \zeta_2 - 12 \zeta_3 \Bigg\}
                + {{n_f C_F}}T_{F} \Bigg\{ - \frac{2}{3} - \frac{16}{3} \zeta_2 \Bigg\}\,,
\nonumber \\
 B_{q{\bar q},3} &=  {{{C_A}^2 {C_F}}} \Bigg\{ - 2 {\zeta_2}^2 + \frac{4496}{27} {\zeta_2}
          - \frac{1552}{9} {\zeta_3} + 40 {\zeta_5} - \frac{1657}{36} \Bigg\}
          + {{{C_A} {C_F}^2}} \Bigg\{ -\frac{988}{15} {\zeta_2}^2 
          \nonumber\\
&
          + 16 {\zeta_2} {\zeta_3} 
          - \frac{410}{3} {\zeta_2} +\frac{844}{3} {\zeta_3}
          + 120 {\zeta_5} + \frac{151}{4} \Bigg\}
+ {{{C_A} {C_F} {n_f}}} \Bigg\{ \frac{4}{5} {\zeta_2}^2 - \frac{1336}{27} {\zeta_2}
          + \frac{200}{9} {\zeta_3}         
\nonumber\\
&+ 20 \Bigg\}
+ {{{C_F}^3}} \Bigg\{ \frac{288}{5} {\zeta_2}^2 - 32 {\zeta_2} {\zeta_3}
          + 18 {\zeta_2} + 68 {\zeta_3} - 240 {\zeta_5} + \frac{29}{2} \Bigg\}
\nonumber\\
&
+ {{{C_F}^2 {n_f}}} \Bigg\{ \frac{232}{15} {\zeta_2}^2 + \frac{20}{3} {\zeta_2}
          -\frac{136}{3} {\zeta_3} - 23 \Bigg\}
+ {{{C_F} {n_f}^2}} \Bigg\{ \frac{80}{27} {\zeta_2} - \frac{16}{9}
  {\zeta_3} - \frac{17}{9} \Bigg\} \, , 
\nonumber
\intertext{and}
&B_{q{\bar q},i}=B_{b{\bar b}, i}\, .
\end{align}


\chapter{Solving Renormalisation Group Equation}
\label{chpt:App-SolRGEZas}

To demonstrate the methodoogy of solving RGE, let us consider a
general form of an RGE with respect to the renormalisation scale $\mu_{R}$:
\begin{align}
\label{eq:App-GenRGE}
\mu_R^2 \frac{d}{d\mu_R^2} \ln M = N\,
\end{align}
where, $M$ and $N$ are functions of $\mu_{R}$. We need to solve for
$M$ in terms of $N$. Our goal is to solve it 
order by order in perturbation theory. We start by expanding the
quantities in powers of $a_{s} \equiv a_{s}(\mu_R^2)$:
\begin{align}
\label{eq:App-ExpandMN}
M &= 1 + \sum\limits_{k=1}^{\infty} a_{s}^{k} M^{(k)}\,,
\nonumber\\
N &= \sum\limits_{k=1}^{\infty} a_{s}^{k} N^{(k)}\,.
\end{align}
The $\mu_{R}$ dependence of $M$ and $N$ on $\mu_R$ enters through
$a_s$. All the coefficients $M^{(k)}$ and $N^{(k)}$ are independent of
$\mu_{R}$. Hence, $\ln M$ can be written as
\begin{align}
\label{eq:App-lnM}
\ln M &= \sum\limits_{k=1}^{\infty} a_s^k M_{k}
\intertext{with}
M_{1} &= M^{(1)}\,,
\nonumber\\
M_{2} &= - \frac{1}{2} (M^{(1)})^2 + M^{(2)}\,,
\nonumber\\
M_{3} &= \frac{1}{3} (M^{(1)})^3 - M^{(1)} M^{(2)} + M^{(3)}\,,
\nonumber\\
M_{4} &= - \frac{1}{4} (M^{(1)})^4 + (M^{(1)})^2 M^{(2)} -
        \frac{1}{2} (M^{(2)})^2  - M^{(1)} M^{(3)} + M^{(4)}\,.
\end{align}
Using the above expansions in RGE.~(\ref{eq:App-GenRGE}) and using the
RGE of $a_{s}$
\begin{align}
\label{eq:App-RGEas}
\mu_R^2 \frac{d}{d\mu_R^2} a_s = \frac{\epsilon}{2} a_s -
  \sum\limits_{k=0}^{\infty} \beta_k \,a_s^{k+2}
\end{align}
we get $M_{k}$'s by comparing the coefficients of $a_s$ as
\begin{align}
\label{eq:App-SolMdk}
M_1 &= \frac{2}{\epsilon}  N^{(1)}\,,
\nonumber\\
M_2 &= \frac{2}{\epsilon^2}  \beta_0 N^{(1)} +
      \frac{1}{\epsilon} N^{(2)}\,,
\nonumber\\
M_3 &= \frac{8}{3 \epsilon^3}  \beta_0^2 N^{(1)} + \frac{1}{\epsilon^2}
      \Bigg\{ \frac{4}{3}  \beta_1 N^{(1)} + \frac{4}{3}  \beta_0 N^{(2)}
      \Bigg\} + \frac{2}{3 \epsilon}  N^{(3)}\,,
\nonumber\\
M_4 &= \frac{4}{\epsilon^4} \beta_0^3 N^{(1)} + \frac{1}{\epsilon^3}
      \Bigg\{ 4 \beta_0 \beta_1 N^{(1)} + 2 \beta_0^2 N^{(2)} \Bigg\} +
      \frac{1}{\epsilon^2} \Bigg\{ \beta_2 N^{(1)} + \beta_1 N^{(2)} +
      \beta_0 N^{(3)} \Bigg\} + \frac{1}{2 \epsilon} N^{(4)}\,.
\end{align}
By equating the Eq.~(\ref{eq:App-lnM}) and (\ref{eq:App-SolMdk}) we
obtain,
\begin{align}
\label{eq:App-SolMuk}
M^{(1)} &= \frac{2}{\epsilon} N^{(1)}\,,
\nonumber\\
M^{(2)} &= \frac{1}{\epsilon^2} \Bigg\{ 2 \beta_0 N^{(1)} + 2
          (N^{(1)})^{2} \Bigg\} + \frac{1}{\epsilon} N^{(2)}\,,
\nonumber\\
M^{(3)} &= \frac{1}{\epsilon^3} \Bigg\{ \frac{8}{3}  \beta_0^2 N^{(1)} +
          4 \beta_0 (N^{(1)})^2 + \frac{4}{3} (N^{(1)})^3 \Bigg\} +
          \frac{1}{\epsilon^{2}} \Bigg\{ \frac{4}{3}  \beta_1 N^{(1)} +
          \frac{4}{3}  \beta_0 N^{(2)} + 2 N^{(1)} N^{(2)} \Bigg\} 
\nonumber\\
&+
          \frac{2}{3 \epsilon}  N^{(3)}\,,
\nonumber\\
M^{(4)} &= \frac{1}{\epsilon^4} \Bigg\{ 4 \beta_0^3 N^{(1)} + \frac{22
          }{3}\beta_0^2 (N^{(1)})^2 + 4 \beta_0 (N^{(1)})^3 + \frac{2
          }{3} (N^{(1)})^4 \Bigg\} + \frac{1}{\epsilon^3} \Bigg\{ 4
          \beta_0 \beta_1 N^{(1)} + \frac{8}{3} \beta_1 (N^{(1)})^2 
\nonumber\\
&+
          2 \beta_0^2 N^{(2)} + \frac{14}{3}  \beta_0 N^{(1)} N^{(2)} +
          2 (N^{(1)})^2 N^{(2)} \Bigg\}
+ \frac{1}{\epsilon^2} \Bigg\{ \beta_2 N^{(1)} + \beta_1 N^{(2)} +
  \frac{1}{2}(N^{(2)})^2 + \beta_0 N^{(3)} 
\nonumber\\
&+ \frac{4}{3} N^{(1)}
  N^{(3)} \Bigg\} + \frac{1}{2 \epsilon} N^{(4)}\,.
\end{align}
We have presented the solution up to ${\cal O}(a_s^4)$. However, this procedure
can be easily generalised to all orders in $a_s$.

\begin{itemize}
\item \textbf{Example 1: RGE of $Z_{a_{s}}$} 
\begin{align}
\label{eq:App-RGEZ}
\mu_{R}^{2} \frac{d}{d\mu_{R}^{2}}\ln Z_{a_{s}} = \frac{1}{a_{s}}
  \sum_{k=0}^{\infty} a_{s}^{k+2}\beta_{k} 
\end{align}
Comparing this RGE of $Z_{a_{s}}$ with Eq.~(\ref{eq:App-GenRGE}) we
get
\begin{align}
\label{eq:App-N4Zas}
N^{(k)} &= \beta_{k-1} \qquad k \in [1, \infty)\,.
\end{align}
By putting the values of $N^{(k)}$ in the general
solution~(\ref{eq:App-SolMuk}), we get the corresponding solutions of
$Z_{a_s}$ as
\begin{align}
\label{eq:App-Zas}
Z_{a_{s}} &= 1+ \sum\limits_{k=1}^{\infty} a_s^k Z_{a_s}^{(k)}
\end{align}
where
\begin{align}
\label{eq:App-Zas-comp}
Z_{a_{s}}^{(1)} &=\frac{2}{\ep} \beta_0\,,
\nonumber\\
Z_{a_{s}}^{(2)} &= \frac{4}{\ep^2 } \beta_0^2  + \frac{1}{\ep} \beta_1\,,
\nonumber\\
Z_{a_{s}}^{(3)} &= \frac{8}{ \ep^3} \beta_0^3 +\frac{14}{3
            \ep^2} \beta_0 \beta_1 + \frac{2}{3 \ep}  \beta_2\,, 
\nonumber\\
Z_{a_{s}}^{(4)} &= \frac{16}{\ep^{4}}\beta_{0}^{4} +
            \frac{46}{3\ep^{3}}\beta_{0}^{2}\beta_{1} +
            \frac{1}{\ep^{2}}\l(\frac{3}{2}\beta_{1}^{2} +
            \frac{10}{3}\beta_{0}\beta_{2}\r) +
            \frac{1}{2\ep}\beta_{3}\,. 
\end{align}
The $Z_{a_s}$ can also be expressed in powers of ${\hat a}_{s}$ by
utilising the 
\begin{align}
\label{eq:App-ashatANDas}
a_s = {\hat a_s} S_{\epsilon}  \left( \frac{\mu_{R}^{2}}{\mu^{2}}
  \right)^{\epsilon/2}  Z_{a_{s}}^{-1} 
\end{align}
iteratively. We get
\begin{align}
\label{eq:App-Zashat}
Z_{a_{s}} &= 1+ \sum\limits_{k=1}^{\infty} {\hat a}_s^k S_{\epsilon}^k \left( \frac{\mu_R^2}{\mu^2}
            \right)^{k\frac{\epsilon}{2}} {\hat Z}_{a_s}^{(k)}
\end{align}
where
\begin{align}
\label{eq:App-Zashat-comp}
{\hat Z}_{a_{s}}^{(1)} &=\frac{2}{\ep} \beta_0\,,
\nonumber\\
{\hat Z}_{a_{s}}^{(2)} &= \frac{1}{\ep} \beta_1\,,
\nonumber\\
{\hat Z}_{a_{s}}^{(3)} &= - \frac{4}{3 \ep^2} \beta_0 \beta_1 + \frac{2}{3 \ep}  \beta_2\,, 
\nonumber\\
{\hat Z}_{a_{s}}^{(4)} &= \frac{2}{\ep^{3}}\beta_{0}^{2}\beta_{1} +
            \frac{1}{\ep^{2}}\left(-\frac{1}{2}\beta_{1}^{2} -
            2 \beta_{0}\beta_{2}\right) +
            \frac{1}{2\ep}\beta_{3}\,. 
\end{align}
To arrive at the above result, we need to use the $Z_{a_s}^{-1}$ in
powers of ${\hat a}_s$:
\begin{align}
\label{eq:App-ZashatInv}
Z_{a_{s}}^{-1} &= 1+ \sum\limits_{k=1}^{\infty} {\hat a}_s^k S_{\epsilon}^k \left( \frac{\mu_R^2}{\mu^2}
            \right)^{k\frac{\epsilon}{2}} {\hat Z}_{a_s}^{-1,(k)}
\end{align}
where
\begin{align}
\label{eq:App-ZashatInv-comp}
{\hat Z}_{a_{s}}^{-1, (1)} &= -\frac{2}{\ep} \beta_0\,,
\nonumber\\
{\hat Z}_{a_{s}}^{-1, (2)} &=  \frac{4}{\epsilon^2} \beta_0^2 - \frac{1}{\ep} \beta_1\,,
\nonumber\\
{\hat Z}_{a_{s}}^{-1, (3)} &= - \frac{8}{\epsilon^3} \beta_0^3 +
                             \frac{16}{3 \ep^2} \beta_0 \beta_1 -
                             \frac{2}{3 \ep}  \beta_2\,,  
\nonumber\\
{\hat Z}_{a_{s}}^{-1, (4)} &= \frac{16}{\epsilon^4} \beta_0^4 -
                             \frac{58}{3\ep^{3}}\beta_{0}^{2}\beta_{1}
                             + 
            \frac{1}{\ep^{2}}\left(\frac{3}{2}\beta_{1}^{2} +
            \frac{14}{3} \beta_{0}\beta_{2}\right) -
            \frac{1}{2\ep}\beta_{3}\,. 
\end{align}
\item \textbf{Example 2: Solution of the Mass Factorisation Kernel}

The mass factorisation kernel satisfies the RG equation~(\ref{eq:bBH-kernelRGE})
\begin{align}
  \label{eq:App-kernelRGE}
  \mu_F^2 \frac{d}{d\mu_F^2} \Gamma^I_{ij}(z,\mu_F^2,\epsilon) =
  \frac{1}{2} \sum\limits_{k} P^I_{ik} \left(z,\mu_F^2\right) \otimes \Gamma^I_{kj} \left(z,\mu_F^2,\epsilon \right)
\end{align}
where, $P^I\left(z,\mu_{F}^{2}\right)$ are Altarelli-Parisi splitting
functions (matrix valued). Expanding $P^{I}\left(z,\mu_{F}^{2}\right)$ and
$\Gamma^{I}(z,\mu_F^2,\epsilon)$ in powers of the strong coupling constant
we get
\begin{align}
  \label{eq:App-APexpand}
  &P^{I}(z,\mu_{F}^{2}) = \sum_{k=1}^{\infty} a_{s}^{k}(\mu_{F}^{2})P^{I,(k-1)}(z)\, 
    \intertext{and}
  &\Gamma^I(z,\mu_F^2,\epsilon) = \delta(1-z) + \sum_{k=1}^{\infty}
    {\hat a}_{s}^{k}  S_{\ep}^{k} \l(\frac{\mu_{F}^{2}}{\mu^{2}}\r)^{k
    \frac{\ep}{2}}  \Gamma^{I,(k)}(z,\ep)\, .
\end{align}
Following the techniques prescribed above, it can be solved. However,
unlike the previous cases here we have to take care of the fact
that $P^I$ and $\Gamma^I$ are matrix valued quantities i.e. they are
non-commutative. Upon solving we obtain the general solution as
\begin{align}
\label{eq:App-Gamma-GenSoln}
\Gamma^{I,(1)}(z,\epsilon) &= \frac{1}{\epsilon}   \Bigg\{
           P^{I,(0)}(z)
          \Bigg\} \,,
\nonumber\\
\Gamma^{I,(2)}(z,\epsilon) &= \frac{1}{\epsilon^2}   \Bigg\{
          - \beta_0 P^{I,(0)} (z) 
          + \frac{1}{2} P^{I,(0)} (z) \otimes P^{I,(0)} (z)
          \Bigg\}
       + \frac{1}{\epsilon}   \Bigg\{
           \frac{1}{2} P^{I,(1)}(z)
          \Bigg\}\,,
\nonumber\\
\Gamma^{I,(3)}(z,\epsilon) &= \frac{1}{\epsilon^3}   \Bigg\{
           \frac{4}{3} \beta_0^2 P^{I,(0)} 
          - \beta_0 P^{I,(0)} \otimes P^{I,(0)} 
          + \frac{1}{6} P^{I,(0)} \otimes P^{I,(0)} \otimes P^{I,(0)}
          \Bigg\}
\nonumber\\
&
       + \frac{1}{\epsilon^2}   \Bigg\{
          - \frac{1}{3} \beta_1 P^{I,(0)} 
          + \frac{1}{6} P^{I,(0)} \otimes P^{I,(1)}
          - \frac{4}{3} \beta_0 P^{I,(1)}
          + \frac{1}{3} P^{I,(1)} \otimes P^{I,(0)}
          \Bigg\}
\nonumber\\
&
       + \frac{1}{\epsilon}   \Bigg\{
           \frac{1}{3} P^{I,(2)}
          \Bigg\}\,,
\nonumber\\
\Gamma^{I,(4)}(z,\epsilon) &= \frac{1}{\epsilon^4}   \Bigg\{
          - 2\beta_0^3 P^{I,(0)} 
          + \frac{11}{6} \beta_0^2 P^{I,(0)} \otimes P^{I,(0)} 
          - \frac{1}{2} \beta_0 P^{I,(0)} \otimes P^{I,(0)} \otimes P^{I,(0)}
\nonumber\\
&          
+ \frac{1}{24} P^{I,(0)} \otimes P^{I,(0)} \otimes P^{I,(0)} \otimes P^{I,(0)}
          \Bigg\}
       + \frac{1}{\epsilon^3}   \Bigg\{
           \frac{4}{3}\beta_0 \beta_1 P^{I,(0)} 
          - \frac{1}{3} \beta_1 P^{I,(0)} \otimes P^{I,(0)}
\nonumber\\
&          
+ \frac{1}{24} P^{I,(0)} \otimes P^{I,(0)} \otimes P^{I,(1)}
          - \frac{7}{12} \beta_0 P^{I,(0)} \otimes P^{I,(1)}
          + \frac{1}{12} P^{I,(0)} \otimes P^{I,(1)} \otimes P^{I,(0)}
\nonumber\\
& 
         + 3 \beta_0^2 P^{I,(1)} 
          - \frac{5}{4}  \beta_0 P^{I,(1)} \otimes P^{I,(0)}
          + \frac{1}{8} P^{I,(1)} \otimes P^{I,(0)} \otimes P^{I,(0)}
          \Bigg\}
\nonumber\\
&
       + \frac{1}{\epsilon^2}   \Bigg\{
          - \frac{1}{6}\beta_2 P^{I,(0)} 
          + \frac{1}{12} P^{I,(0)} \otimes P^{I,(2)}
          - \frac{1}{2}  \beta_1 P^{I,(1)}
          + \frac{1}{8} P^{I,(1)} \otimes P^{I,(1)}
\nonumber\\
&
          - \frac{3}{2}  \beta_0 P^{I,(2)}
          + \frac{1}{4} P^{I,(2)} \otimes P^{I,(0)}
          \Bigg\}
       + \frac{1}{\epsilon}   \Bigg\{
           \frac{1}{4} P^{I,(3)}
          \Bigg\}\,.
\end{align}
In the soft-virtual limit, only the diagonal parts of the kernels
contribute. Our findings are consistent with the existing diagonal
solutions which can be found in the article~\cite{Ravindran:2005vv}.

\end{itemize}


\chapter{Solving KG Equation}
\label{chpt:App-KGSoln}

The form factor satisfies the KG differential equation (See Sec.~\ref{ss:bBH-FF}):
\begin{equation}
  \label{eq:App-KG}
  Q^2 \frac{d}{dQ^2} \ln {\cal F}^{I}_{ij} (\hat{a}_s, Q^2, \mu^2, \epsilon)
  = \frac{1}{2} \left[ K^{I}_{ij} \left(\hat{a}_s, \frac{\mu_R^2}{\mu^2}, \epsilon
    \right)  + G^{I}_{ij} \left(\hat{a}_s, \frac{Q^2}{\mu_R^2},
      \frac{\mu_R^2}{\mu^2}, \epsilon \right) \right]\,. 
\end{equation}
In this appendix we demonstrate the procedure to solve the KG
equation. RG invariance of the ${\cal F}$ with respect to the
renormalisation scale $\mu_{R}$ implies
\begin{align}
\label{eq:App-KG-muRInd}
\mu_R^2 \frac{d}{d\mu_R^2} K^I_{ij}\left(\hat{a}_s, \frac{\mu_R^2}{\mu^2}, \epsilon
    \right) = - \mu_R^2 \frac{d}{d\mu_R^2} G^{I}_{ij} \left(\hat{a}_s, \frac{Q^2}{\mu_R^2},
      \frac{\mu_R^2}{\mu^2}, \epsilon \right) = - A^I_{ij}\left(a_{s}(\mu_R^{2})\right) 
\end{align}
where, $A^I_{ij}$'s are the cusp anomalous dimensions. Unlike the
previous cases, we expand $K^I_{ij}$ in powers of unrenormalised
${\hat a}_{s}$ as
\begin{align}
\label{eq:App-Kexpand}
K^I_{ij}\left(\hat{a}_s, \frac{\mu_R^2}{\mu^2}, \epsilon
    \right) &= \sum\limits_{k=1}^{\infty} {\hat a}_s^k
              S_{\epsilon}^{k} \left(
              \frac{\mu_R^2}{\mu^2} \right)^{k \frac{\epsilon}{2}}
              {\hat K}^I_{ij,k}(\epsilon)
\end{align}
whereas we define the components $A^I_{ij, k}$ through 
\begin{align}
\label{eq:App-Aexpand}
A^I_{ij} = \sum\limits_{k=1}^{\infty} a_s^k\left( \mu_R^2 \right)
  A^I_{ij, k}\,.
\end{align}
Following the methodology discussed in
Appendix~\ref{chpt:App-SolRGEZas}, we can solve for ${\hat K}^I_{ij,k}(\epsilon)$
\begin{align}
\label{eq:App-SolnK}
{\hat K}^I_{ij,1}(\epsilon) &= \frac{1}{\epsilon} \Bigg\{ - 2 A^I_{ij,
                       1}\Bigg\}\,,
\nonumber\\
{\hat K}^I_{ij,2}(\epsilon) &= \frac{1}{\epsilon^2} \Bigg\{ 2 \beta_0
                       A^I_{ij,1} \Bigg\} + \frac{1}{\epsilon} \Bigg\{
                       - A^I_{ij, 2}\Bigg\}\,,
\nonumber\\
{\hat K}^I_{ij,3}(\epsilon) &= \frac{1}{\epsilon^3} \Bigg\{ - \frac{8
                       }{3} \beta_0^2 A^I_{ij,1} \Bigg\} +
                       \frac{1}{\epsilon^2} \Bigg\{ \frac{2}{3}  \beta_1
                       A^I_{ij,1} + \frac{8}{3}  \beta_0 A^I_{ij,2}
                       \Bigg\} + \frac{1}{\epsilon} \Bigg\{ - \frac{2
                       }{3} A^I_{ij, 3} \Bigg\}\,,
\nonumber\\
{\hat K}^I_{ij,4}(\epsilon) &= \frac{1}{\epsilon^4} \Bigg\{ 4 \beta_0^3
                       A^I_{ij,1} \Bigg\} + \frac{1}{\epsilon^3}
                       \Bigg\{ -\frac{8}{3} \beta_0 \beta_1 A^I_{ij,1}
                       - 6 \beta_0^2 A^I_{ij,2} \Bigg\} +
                       \frac{1}{\epsilon^{2}} \Bigg\{ \frac{1}{3} \beta_2
                       A^I_{ij,1} + \beta_1 A^I_{ij,2} + 3 \beta_0
                       A^I_{ij,3} \Bigg\} 
\nonumber\\
&+ \frac{1}{\epsilon} \Bigg\{
                       - \frac{1}{2} A^I_{ij,4} \Bigg\}
\end{align}
Due to dependence of $G^I_{ij}$ on $Q^2$, we need to handle it
differently. Integrating the RGE of $G^I_{ij}$,
(\ref{eq:App-KG-muRInd}), we get
\begin{align}
\label{eq:App-SolveG}
&G^I_{ij} \left( {\hat a}_s, \frac{Q^2}{\mu_R^2},
  \frac{\mu_R^2}{\mu^2}, \epsilon \right) - G^I_{ij} \left( {\hat
  a}_s, 1, \frac{Q^2}{\mu^2}, \epsilon \right) =
                                                 \int\limits_{Q^2}^{\mu_R^2}
                                                 \frac{d\mu_R^2}{\mu_R^2}
  A^I_{ij}
\nonumber\\
\Rightarrow \quad&G^I_{ij} \left( a_s(\mu_R^2), \frac{Q^2}{\mu_R^2}, \epsilon \right) =
  G^I_{ij} \left( a_s(Q^2), 1, \epsilon \right) + \int\limits_{Q^2}^{\mu_R^2}
                                                 \frac{d\mu_R^2}{\mu_R^2}
  A^I_{ij}
\end{align}
Consider the second part of the above Eq.~(\ref{eq:App-SolveG})
\begin{align}
\label{eq:App-SolveG-1}
\int\limits_{Q^2}^{\mu_R^2} \frac{d\mu_R^2}{\mu_R^2} A^I_{ij} 
&= \int\limits_{Q^2}^{\mu_R^2} \frac{d\mu_R^2}{\mu_R^2}
  \sum\limits_{k=1}^{\infty} a_s^k A^I_{ij,k}
\nonumber\\
&= \sum\limits_{k=1}^{\infty} \int\limits_{\frac{Q^2}{\mu^2}}^{\frac{\mu_R^2}{\mu^2}}
  \frac{dX^2}{X^2} {\hat a}_s^k S_{\epsilon}^k \left(
  X^2\right)^{k\frac{\epsilon}{2}} \left( Z_{a_s}^{-1} (X^2) \right)^k A^I_{ij,k} 
\end{align}
where we have made the change of integration variable from $\mu_R$ to
$X$ by $\mu_R^2 =
X^2 \mu^2$. By using the $Z_{a_s}^{-1} (X^2)$ from
Eq.~(\ref{eq:App-ZashatInv}) and evaluating the integral we obtain
\begin{align}
\label{eq:App-SolveG-2}
\int\limits_{Q^2}^{\mu_R^2} \frac{d\mu_R^2}{\mu_R^2} A^I_{ij} =
  \sum\limits_{k=1}^{\infty} {\hat a}_s^k S_{\epsilon}^k \left(
  \frac{\mu_R^2}{\mu^2}\right)^{k\frac{\epsilon}{2}} \left[ \left(
  \frac{Q^2}{\mu_R^2} \right)^{k \frac{\epsilon}{2}} -1 \right] {\hat K}^I_{ij,k}(\epsilon)\,.
\end{align}
The first part of $G^I_{ij}$ in Eq.~(\ref{eq:App-SolveG}) can be
expanded in powers of $a_s(Q^2)$ as
\begin{align}
\label{eq:App-SolveG-3}
G^I_{ij} \left( a_s(Q^2), 1, \epsilon \right) =
  \sum\limits_{k=1}^{\infty} a_s^k(Q^2) G^I_{ij}(\epsilon)\,.
\end{align}
By putting back the Eq.~(\ref{eq:App-Kexpand}),
(\ref{eq:App-SolveG-1}) and (\ref{eq:App-SolveG-2}) in the original KG
equation~(\ref{eq:App-KG}), we solve for $\ln {\cal F}^{I}_{ij}
(\hat{a}_s, Q^2, \mu^2, \epsilon)$:
\begin{align}
  \label{eq:App-lnFSoln}
  \ln {\cal F}^{I}_{ij}(\hat{a}_s, Q^2, \mu^2, \epsilon) =
  \sum_{k=1}^{\infty} {\hat a}_{s}^{k}S_{\epsilon}^{k} \left(\frac{Q^{2}}{\mu^{2}}\right)^{k
  \frac{\epsilon}{2}}  \hat {\cal L}_{ij, k}^{I}(\epsilon)
\end{align}
with
\begin{align}
  \label{eq:App-lnFitoCalLF}
  \hat {\cal L}_{ij,1}^{I}(\ep) &= { \frac{1}{\ep^2} } \Bigg\{-2 A^{I}_{{ij},1}\Bigg\}
                                  + { \frac{1}{\ep}
                                  }
                                  \Bigg\{G^{I}_{{ij},1}
                                  (\ep)\Bigg\}\, ,
                                  \nonumber\\
  \hat {\cal L}_{{ij},2}^{I}(\ep) &= { \frac{1}{\ep^3} } \Bigg\{\beta_0 A^{I}_{{ij},1}\Bigg\}
                                  + {
                                  \frac{1}{\ep^2} }
                                  \Bigg\{-  {
                                  \frac{1}{2} }  A^{I}_{{ij},2}
                                  - \beta_0   G^{I}_{{ij},1}(\ep)\Bigg\}
                                  + { \frac{1}{\ep}
                                  } \Bigg\{ {
                                  \frac{1}{2} }  G^{I}_{{ij},2}(\ep)\Bigg\}\, ,
                                  \nonumber\\
  \hat {\cal L}_{{ij},3}^{I}(\ep) &= { \frac{1}{\ep^4}
                                  } \Bigg\{- {
                                  \frac{8}{9} }  \beta_0^2 A^{I}_{{ij},1}\Bigg\}
                                  + {
                                  \frac{1}{\ep^3} }
                                  \Bigg\{ { \frac{2}{9} } \beta_1 A^{I}_{{ij},1}
                                  + { \frac{8}{9} }
                                  \beta_0 A^{I}_{{ij},2}  + { \frac{4}{3} }
                                  \beta_0^2 G^{I}_{{ij},1}(\ep)\Bigg\}
                                  \nonumber\\
                                &
                                  + { \frac{1}{\ep^2} } \Bigg\{- { \frac{2}{9} } A^{I}_{{ij},3}
                                  - { \frac{1}{3} } \beta_1 G^{I}_{{ij},1}(\ep)
                                  - { \frac{4}{3} } \beta_0 G^{I}_{{ij},2}(\ep)\Bigg\}
                                  + { \frac{1}{\ep}
                                  } \Bigg\{  { \frac{1}{3} }
                                  G^{I}_{ij,3}(\ep)\Bigg\}\,,
\nonumber\\
  \hat {\cal L}_{{ij},4}^{I}(\ep) &= \frac{1}{\epsilon^5} \Bigg\{ A^I_{ij,1} \beta_0^3 \Bigg\}
   + \frac{1}{\epsilon^4} \Bigg\{ - \frac{3}{2} A^I_{ij,2} \beta_0^2 - \frac{2}{3}
                                     A^I_{ij,1} \beta_0 \beta_1 - 2
                                     \beta_0^3 G^I_{ij,1}(\epsilon))
                                     \Bigg\} 
\nonumber\\
&   
+ \frac{1}{\epsilon^3} \Bigg\{ \frac{3}{4} A^I_{ij,3} \beta_0 +
                                     \frac{1}{4} A^I_{ij,2} \beta_1 +
                                     \frac{1}{12} A^I_{ij,1} \beta_2 + \frac{4}{3}
                                     \beta_0 \beta_1
                                     G^I_{ij,1}(\epsilon) + 
      3 \beta_0^2 G^I_{ij,2}(\epsilon)) \Bigg\}  
\nonumber\\
&   + \frac{1}{\epsilon^2} \Bigg\{ -\frac{1}{8} A^I_{ij,4} - \frac{1}{6} \beta_2
                                     G^I_{ij,1}(\epsilon) - \frac{1}{2}
                                     \beta_1 G^I_{ij,2}(\epsilon)
                                     - \frac{3}{2}\beta_0
                                     G^I_{ij,3}(\epsilon)  \Bigg\} 
 \nonumber\\
&  + \frac{1}{\epsilon}  \Bigg\{ \frac{1}{4} G^I_{ij,4}(\epsilon) \Bigg\}\,.
\end{align}
This methodology can easily be generalised to all orders in
perturbation theory.


\chapter{Soft-Collinear Distribution}
\label{chpt:App-Soft-Col-Dist}

In Sec.~\ref{ss:bBH-SCD}, we introduced the soft-collinear
distribution $\Phi^H_{b{\bar b}}$ in the context of computing SV
cross section of the Higgs boson production in $b{\bar b}$
annihilation. In this appendix, we intend to elaborate the methodology
of finding this distribution. For the sake of generalisation, we
use $I$ instead of $H$ and omit the partonic indices. To understand
the underlying logics behind finding $\Phi^{I}$, let us consider an
example at one loop level. The generalisation to higher loop is straightforward.

As discussed in the Sec.~\ref{sec:bBH-ThreResu}, the SV cross section in
$z$-space can be computed in $d=4+\epsilon$ 
dimensions using 
\begin{align}
\label{eq:App-sigma}
\Delta^{I, \sv} (z, q^2, \mu_{R}^{2}, \mu_F^2) = 
{\cal C} \exp \Big( \Psi^I \left(z, q^2, \mu_R^2, \mu_F^2, \epsilon
  \right)  \Big) \Big|_{\epsilon = 0}
\end{align}
where, $\Psi^I \left(z, q^2, \mu_R^2, \mu_F^2, \epsilon \right)$ is a
finite distribution and ${\cal C}$ is the convolution defined
through Eq.~(\ref{eq:bBH-conv}). The $\Psi^I$ is given by, Eq.~(\ref{eq:bBH-psi})
\begin{align}
\label{eq:App-psi}
\Psi^{I} \left(z, q^2, \mu_R^2, \mu_F^2, \epsilon \right) = &\left(
                                                              \ln
                                                              \Big[
                                                              Z^I
                                                              (\hat{a}_s,
                                                              \mu_R^2,
                                                              \mu^2,
                                                              \epsilon)
                                                              \Big]^2  
+ \ln \Big|  {\cal F}^I (\hat{a}_s, Q^2, \mu^2, \epsilon)  \Big|^2
                                                              \right)
                                                              \delta(1-z)  
\nonumber\\
& + 2 \Phi^I (\hat{a}_s, q^2, \mu^2, z, \epsilon) - 2 {\cal C} \ln
  \Gamma^{I} (\hat{a}_s, \mu^2, \mu_F^2, z, \epsilon) \, . 
\end{align}
For all the details about the notations, see
Sec.~\ref{sec:bBH-ThreResu}. Considering only the poles at ${\cal O}(a_{s})$ with $\mu_{R} =
\mu_{F}$ we obtain,
\begin{align}
\label{eq:App-Psi-Comp-1}
&\ln\l(Z^{I, (1)}\r)^{2} = a_{s}(\mu_{F}^{2}) \frac{4\gamma_{1}^{I}}{\epsilon}\,,
\nonumber\\
&\ln|{\cal F}^{I, (1)}|^{2} = a_{s}(\mu_{F}^{2}) \l(
  \frac{q^{2}}{\mu_{F}^{2}} \r)^{\frac{\epsilon}{2}} \l[
  -\frac{4A_{1}^{I}}{\epsilon^{2}} + \frac{1}{\epsilon} \l( 2f_{1}^{I}
  + 4B_{1}^{I} - 4\gamma_{1}^{I}\r)\r]\,,
\nonumber\\
&2{\cal C} \ln\Gamma^{I, (1)} = 2 a_{s}(\mu_{F}^{2}) \l[
  \frac{2B^{I}_{1}}{\epsilon} \delta(1-z) +
  \frac{2A_{1}^{I}}{\epsilon} {\cal D}_{0}\r]
\end{align}
where, the components are defined through the expansion of these
quantities in powers of $a_s(\mu_F^{2})$
\begin{align}
\label{eq:App-psi-comp}
&\Psi^{I} = \sum\limits_{k=1}^{\infty} a_s^k \left( \mu_F^2 \right) \Psi^{I,(k)}\,,
\nonumber\\
&\ln (Z^I)^2 = \sum\limits_{k=1}^{\infty} a_s^k \left( \mu_F^2 \right)
  Z^{I,(k)}\,,
\nonumber\\
&\ln |{\cal F}^{I}|^2 = \sum\limits_{k=1}^{\infty} a_s^k\left( \mu_F^2
  \right) \left( \frac{q^2}{\mu_F^2} \right)^{k \frac{\epsilon}{2}}
  \ln |{\cal F}^{I,(k)}|^2\,,
\nonumber\\
&\Phi^{I} = \sum\limits_{k=1}^{\infty} a_s^k(\mu_F^2) \Phi^I_k\,,
\nonumber\\
&\ln \Gamma^I = \sum\limits_{k=1}^{\infty} a_s^k(\mu_F^2) \ln
  \Gamma^{I,(k)}
\intertext{and}
&
{\cal{D}}_{i} \equiv \left[ \frac{\ln^{i}(1-z)}{1-z} \right]_{+}\, .
\end{align}
Collecting the coefficients of $a_s(\mu_F^2)$, we
get
\begin{align}
\label{eq:App-Psi-1}
\Psi^{I, (1)}|_{\rm poles}
&= \l[ {\l\{ - \frac{4A_{1}^{I}}{\epsilon^{2}} + \frac{ 2f_{1}^{I}}{\epsilon} \r\} \delta(1-z)} - {\frac{4A_{1}^{I}}{\epsilon} {\cal D}_{0}} \r] + {2\Phi^{I}_{1}}
\end{align}
where, we have not shown the $\ln (q^2/\mu_F^2)$ terms.
To cancel the remaining divergences appearing in the above
Eq.~(\ref{eq:App-Psi-1}) for obtaining a finite cross section, we must
demand that $\Phi^{I}_{1}$ have exactly the same poles with
opposite sign:
\begin{align}
\label{eq:App-Phi-1}
2 \Phi^{I}_{1}|_{\rm poles} = - \l[ {\l\{ - \frac{4A_{1}^{I}}{\epsilon^{2}} + \frac{ 2f_{1}^{I}}{\epsilon} \r\} \delta(1-z)} - {\frac{4A_{1}^{I}}{\epsilon} {\cal D}_{0}} \r]
\end{align} 
In addition, $\Phi^{I}$ also should be RG invariant with respect to
$\mu_R$:
\begin{align}
\label{eq:App-Phi-2}
\mu_R^2 \frac{d}{d\mu_R^2} \Phi^I = 0\,. 
\end{align}
We make an \textit{ansatz}, the above two demands,
Eq.~(\ref{eq:App-Phi-1}) and~(\ref{eq:App-Phi-2}) can be accomplished
if $\Phi^I$ satisfies the KG-type integro-differential equation which we call
$\overline{KG}$:
\begin{align}
  \label{eq:App-KGbarEqn}
  q^2 \frac{d}{dq^2} \Phi^I\left(\hat{a}_s, q^2, \mu^2, z,
    \ep\right)   = \frac{1}{2} \left[ \overline K^I
  \left(\hat{a}_s, \frac{\mu_R^2}{\mu^2}, z, 
      \ep \right)  + \overline G^I \left(\hat{a}_s,
      \frac{q^2}{\mu_R^2},  \frac{\mu_R^2}{\mu^2}, z, \ep \right) \right]\,.
\end{align}
$\overline{K}^I$ contains all the poles whereas $\overline{G}^I$ consists
of only the finite terms in $\epsilon$. RG
invariance~(\ref{eq:App-Phi-2}) of $\Phi^I$ dictates
\begin{align}
\label{eq:App-Phi-3}
\mu_R^2 \frac{d}{d\mu_R^2} \overline{K}^{I} = -\mu_R^2 \frac{d}{d\mu_R^2}
  \overline{G}^I \equiv Y^{I}
\end{align}
where, we introduce a quantity $Y^I$. Following the methodology of
solving the KG equation discussed in the
Appendix~\ref{chpt:App-KGSoln}, we can write the solution of $\Phi^I$
as
\begin{align}
  \label{eq:App-Phi-4}
  \Phi^I(\hat{a}_s, q^2, \mu^2, z, \epsilon) =
  \sum_{k=1}^{\infty} {\hat a}_{s}^{k}S_{\epsilon}^{k} \left(\frac{q^{2}}{\mu^{2}}\right)^{k
  \frac{\epsilon}{2}}  \hat {\Phi}^{I}_{k}(z,\epsilon)
\end{align}
with
\begin{align}
\label{eq:App-Phi-5}
\hat {\Phi}^{I}_{k}(z,\epsilon) = {\hat{\cal L}}^I_k \left( A^I_{i}
  \rightarrow Y^I_{i}, G^I_i \rightarrow \overline{G}^I_i(z,\epsilon) \right)\,.
\end{align}
where we define the components through the expansions
\begin{align}
\label{eq:App-YGexpans}
&Y^I  = \sum\limits_{k=1}^{\infty} a_s^k(\mu_F^2) Y^I_k\,,
\nonumber\\
&\overline{G}^I(z,\epsilon) = \sum\limits_{k=1}^{\infty}
  a_s^k(\mu_F^2) \overline{G}^I_{k}(z,\epsilon).
\end{align}
This solution directly follows from the Eq.~(\ref{eq:App-lnFSoln}). Hence we get
\begin{align}
  \label{eq:App-Phi-6}
  2 \hat {\Phi}^{I}_{1}(z,\ep) &= { \frac{1}{\ep^2} } \Bigg\{-4 Y^{I}_{1}\Bigg\}
                                  + { \frac{2}{\ep}
                                  }
                                  \Bigg\{\overline{G}^{I}_{1}
                                  (z,\ep)\Bigg\}\,.
\end{align}
By expressing the components of $\Phi^I$ in powers of $a_s(\mu_F^2)$,
we obtain
\begin{align}
\label{eq:App-Phi-7}
\Phi^I(\hat{a}_s, q^2, \mu^2, z, \epsilon) &=
  \sum_{k=1}^{\infty} {\hat a}_{s}^{k}S_{\epsilon}^{k} \left(\frac{q^{2}}{\mu^{2}}\right)^{k
  \frac{\epsilon}{2}}  \hat {\Phi}^{I}_{k}(z,\epsilon)
\nonumber\\
&= \sum_{k=1}^{\infty} a_s^k(\mu_F^2) \left( \frac{q^2}{\mu_F^2} \right)^{k
  \frac{\epsilon}{2}} Z_{a_s}^k \hat{\Phi}^I_k(z,\epsilon)
\nonumber\\
& \equiv \sum_{k=1}^{\infty} a_s^k(\mu_F^2) \left( \frac{q^2}{\mu_F^2} \right)^{k
  \frac{\epsilon}{2}} {\Phi}^I_{k}(z,\epsilon)
\end{align}
and at ${\cal O}(a_s(\mu_F^2)), \hat{\Phi}^I_1(z,\epsilon) =
\Phi^I_{1}(z,\epsilon)$ upon suppressing the terms like $\log (q^{2}/\mu_{F}^{2})$. Hence, by comparing the
Eq.~(\ref{eq:App-Phi-1}) and (\ref{eq:App-Phi-6}), we conclude
\begin{align}
\label{eq:App-Phi-8}
&Y^I_1 = -A^I_1 \delta(1-z)\,,
\nonumber\\
&\overline{G}^I_1(z,\epsilon) = -f^I_1 \delta(1-z) + 2 A^I_1 {\cal
  D}_0 + \sum\limits_{k=1}^{\infty} \epsilon^k \overline{g}^{I,k}_1(z)\,.
\end{align}
The coefficients of $\epsilon^k, \overline{g}^{I,k}_1(z)$ can only be
determined through explicit computations. These do not contribute to
the infrared poles associated with $\Phi^{I}$. This uniquely fixes the
unknown soft-collinear distribution $\Phi^I$ at one loop order. This
prescription can easily be generalised to higher orders in $a_s$. In
our calculation of the SV cross section, instead of solving in this way,
we follow a bit different methodology which is presented below.

Keeping the demands~(\ref{eq:App-Phi-1}) and~(\ref{eq:App-Phi-2}) in
mind, we propose the solution of the $\overline{KG}$ equation as (See
Eq.~(\ref{eq:App-Phi-4})) 
\begin{align}
\label{eq:App-Phi-9}
\hat{\Phi}^I_{k} (z,\epsilon) &\equiv \Bigg\{ k \epsilon \frac{1}{1-z}
                                \left[
  (1-z)^2 \right]^{k \frac{\epsilon}{2}}\Bigg\}
                                \hat{\Phi}^I_k(\epsilon)
\nonumber\\
&=\Bigg\{ \delta(1-z) + \sum\limits_{j=0}^{\infty} \frac{(k
  \epsilon)^{j+1}}{j!} {\cal D}_{j} \Bigg\} \hat{\Phi}^I_k(\epsilon)\,.
\end{align}
The RG invariance of $\Phi^{I}$, Eq.~(\ref{eq:App-Phi-2}), implies
\begin{align}
\label{eq:App-Phi-10}
\mu_R^2 \frac{d}{d\mu_R^2} \overline{K}^{I} = -\mu_R^2 \frac{d}{d\mu_R^2}
  \overline{G}^I \equiv Y'^{I}
\end{align}
where, we introduce a quantity $Y'$, analogous to $Y$. Hence, the
solution can be obtained as 
\begin{align}
\label{eq:App-Phi-11}
\hat {\Phi}^{I}_{k}(\epsilon) &= {\hat{\cal L}}^I_k \left( A^I_{i}
  \rightarrow Y'^I_{i}, G^I_i \rightarrow \overline{\cal G}^I_i(\epsilon) \right)\,.
\end{align}
Hence, according to the Eq.~(\ref{eq:App-lnFitoCalLF}), for $k=1$ we get
\begin{align}
\label{eq:App-Phi-12}
2\Phi^{I}_1 (z, \epsilon) &= \Bigg\{ \frac{1}{\epsilon^2}\l(
                               -4Y'^I_1\r) +
                               \frac{2}{\epsilon}{\overline{\cal
                               G}}^I_1 \l(\epsilon) \r) \Bigg\}
                               \delta(1-z) 
+ \Bigg\{ - \frac{4 Y'^I_1}{\epsilon^2} 
+ \frac{2}{\epsilon} {\overline{\cal G}}^I_1(\epsilon) \Bigg\}
                               \sum_{j=0}^{\infty}
                               \frac{\epsilon^{j+1}}{j!} {\cal D}_j
\end{align}
where, $Y'^{I}$ and $\overline{\cal G}^I$ are expanded similar to
Eq.~(\ref{eq:App-YGexpans}). Comparison between the two solutions
depicted in Eq.~(\ref{eq:App-Phi-1}) and (\ref{eq:App-Phi-12}), we 
can write
\begin{align}
\label{eq:App-Phi-13}
&Y'^I_1 = - A^I_1 
\nonumber\\
&{\overline{\cal G}}^I_1 \l(\epsilon\r) = - f^I_1 +
                                          \sum_{k=1}^{\infty}
                                          \epsilon^k \overline{\cal
                                          G}^{I, k}_1\,.
\end{align}
Explicit computation is required to determine the coefficients of
$\epsilon^k$, $\overline{\cal G}^{I,k}_{1}$. This solution is used in
Eq.~(\ref{eq:bBH-PhiSoln}) in
the context of SV cross section of Higgs boson production. The method
is generalised to higher orders in $a_s$ to obtain the results of the
soft-collinear distribution. In the next subsection, we present the
results of the soft-collinear distribution up to three loops.

\subsection{Results}
\label{app:ss-SCD-Res}

We define the renormalised components of the
$\Phi^I_{q{\bar q},k}$ through
\begin{align}
\label{eq:App-SCD-Re}
\Phi^I_{q{\bar q}}(\hat{a}_s, q^2, \mu^2, z \epsilon) &=
  \sum_{k=1}^{\infty} {\hat a}_{s}^{k}S_{\epsilon}^{k} \left(\frac{q^{2}}{\mu^{2}}\right)^{k
  \frac{\epsilon}{2}}  \hat {\Phi}^{I}_{q{\bar q},k}(z, \epsilon)
\nonumber\\
&= \sum\limits_{k=1}^{\infty} a_s^k\l( \mu_F^2 \r) \Phi^I_{q{\bar
  q},k} \left( z, \epsilon, q^2, \mu_F^2 \right)
\end{align} 
where, we make the choice of the renormalisation scale
$\mu_R=\mu_F$. The $\mu_R$ dependence can be easily restored by using
the evolution equation of strong coupling constant,
Eq.~(\ref{eq:bBH-asf2asr}). Below, we present the $\Phi^I_{i~{\bar
    i},k}$ for 
$i~{\bar i}=q{\bar q}$ up to three 
loops and the corresponding components for $i~{\bar
  i}=gg$ can be obtained using maximally non-Abelian property
fulfilled by this distribution:
\begin{align}
\label{eq:App-SCD-MaxNonAbe}
\Phi^I_{gg,k} = \frac{C_A}{C_F} \Phi^I_{q{\bar q},k}\,.
\end{align} 
The results are given by
\begin{align}
\label{eq:app-SCD-Res-1}
\Phi^I_{q{\bar q},1} &=
\delta(1-z) \Bigg[\frac{1}{\epsilon^2}  C_F   \Bigg\{ 8 \Bigg\}
       + \frac{1}{\epsilon}  \log\l(\frac{q^2}{\mu_F^2}\r) C_F   \Bigg\{ 4 \Bigg\}
       +  C_F   \Bigg\{  - 3 \zeta_2 \Bigg\}
       +  \log^2\l(\frac{q^2}{\mu_F^2}\r) C_F   \Bigg\{ 1 \Bigg\}
\Bigg]
\nonumber\\
&
       + {\cal D}_0 \Bigg[ \frac{1}{\epsilon} C_F   \Bigg\{ 8 \Bigg\}
       +  \log\l(\frac{q^2}{\mu_F^2}\r) C_F   \Bigg\{ 4 \Bigg\}
\Bigg]
       + {\cal D}_1 \Bigg[ C_F   \Bigg\{ 8 \Bigg\} \Bigg]\,,
\nonumber\\
\Phi^I_{q{\bar q},2} &=
\delta(1-z) \Bigg[ \frac{1}{\epsilon^3}  C_F C_A   \Bigg\{ 44 \Bigg\}
       + \frac{1}{\epsilon^3}  n_f C_F   \Bigg\{  - 8 \Bigg\}
       + \frac{1}{\epsilon^2}  C_F C_A   \Bigg\{ \frac{134}{9} - 4 \zeta_2 \Bigg\}
\nonumber\\
&       + \frac{1}{\epsilon^2}  n_f C_F   \Bigg\{  - \frac{20}{9} \Bigg\}
       + \frac{1}{\epsilon^2}  \log\l(\frac{q^2}{\mu_F^2}\r) C_F C_A
  \Bigg\{ \frac{44}{3} \Bigg\} 
       + \frac{1}{\epsilon^2}  \log\l(\frac{q^2}{\mu_F^2}\r) n_f C_F
  \Bigg\{  - \frac{8}{3} \Bigg\} 
\nonumber\\
&       + \frac{1}{\epsilon}  C_F C_A   \Bigg\{  - \frac{404}{27} + 14 \zeta_3 + \frac{11}{3}
          \zeta_2 \Bigg\}
       + \frac{1}{\epsilon}  n_f C_F   \Bigg\{ \frac{56}{27} - \frac{2}{3} \zeta_2 \Bigg\}
\nonumber\\
&       + \frac{1}{\epsilon}  \log\l(\frac{q^2}{\mu_F^2}\r) C_F C_A
  \Bigg\{ \frac{134}{9} - 4 
         \zeta_2 \Bigg\}
       + \frac{1}{\epsilon}  \log\l(\frac{q^2}{\mu_F^2}\r) n_f C_F
  \Bigg\{  - \frac{20}{9} \Bigg\}
       +  C_F C_A   \Bigg\{ \frac{1214}{81} 
\nonumber\\
&- \frac{187}{9} \zeta_3 - \frac{469}{18}
         \zeta_2 + 2 \zeta_2^2 \Bigg\}
       +  n_f C_F   \Bigg\{  - \frac{164}{81} + \frac{34}{9} \zeta_3 + \frac{35}{9}
         \zeta_2 \Bigg\}
\nonumber\\
&       +  \log\l(\frac{q^2}{\mu_F^2}\r) C_F C_A   \Bigg\{  - \frac{404}{27} + 14
         \zeta_3 + \frac{44}{3} \zeta_2 \Bigg\}
       +  \log\l(\frac{q^2}{\mu_F^2}\r) n_f C_F   \Bigg\{ \frac{56}{27} - \frac{8}{3} \zeta_2
          \Bigg\}
\nonumber\\
&       +  \log^2\l(\frac{q^2}{\mu_F^2}\r) C_F C_A   \Bigg\{ \frac{67}{9} - 2 \zeta_2
          \Bigg\}
       +  \log^2\l(\frac{q^2}{\mu_F^2}\r) n_f C_F   \Bigg\{  - \frac{10}{9} \Bigg\}
       +  \log^3\l(\frac{q^2}{\mu_F^2}\r) C_F C_A   \Bigg\{  - \frac{11}{9} \Bigg\}
\nonumber\\
&       +  \log^3\l(\frac{q^2}{\mu_F^2}\r) n_f C_F   \Bigg\{ \frac{2}{9} \Bigg\}
\Bigg]
       + {\cal D}_0 \Bigg[ \frac{1}{\epsilon^2} C_F C_A   \Bigg\{ \frac{88}{3} \Bigg\}
       +  \frac{1}{\epsilon^2} n_f C_F   \Bigg\{  - \frac{16}{3} \Bigg\}
       +  \frac{1}{\epsilon} C_F C_A   \Bigg\{ \frac{268}{9} 
\nonumber\\
&- 8 \zeta_2 \Bigg\}
       +  \frac{1}{\epsilon} n_f C_F   \Bigg\{  - \frac{40}{9} \Bigg\}
       +  C_F C_A   \Bigg\{  - \frac{808}{27} + 28 \zeta_3 + \frac{88}{3} \zeta_2 \Bigg\}
       +  n_f C_F   \Bigg\{ \frac{112}{27} - \frac{16}{3} \zeta_2 \Bigg\}
\nonumber\\
&       +  \log\l(\frac{q^2}{\mu_F^2}\r) C_F C_A   \Bigg\{ \frac{268}{9} - 8 \zeta_2 \Bigg\}
       +  \log\l(\frac{q^2}{\mu_F^2}\r) n_f C_F   \Bigg\{  - \frac{40}{9} \Bigg\}
       +  \log^2\l(\frac{q^2}{\mu_F^2}\r) C_F C_A   \Bigg\{  - \frac{22}{3} \Bigg\}
\nonumber\\
&       +  \log^2\l(\frac{q^2}{\mu_F^2}\r) n_f C_F   \Bigg\{ \frac{4}{3} \Bigg\}
\Bigg]
       + {\cal D}_1 \Bigg[ C_F C_A   \Bigg\{ \frac{536}{9} - 16 \zeta_2 \Bigg\}
       +  n_f C_F   \Bigg\{  - \frac{80}{9} \Bigg\}
\nonumber\\
&       +  \log\l(\frac{q^2}{\mu_F^2}\r) C_F C_A   \Bigg\{  - \frac{88}{3} \Bigg\}
       +  \log\l(\frac{q^2}{\mu_F^2}\r) n_f C_F   \Bigg\{ \frac{16}{3} \Bigg\}
\Bigg]
\nonumber\\
&       + {\cal D}_2 \Bigg[ C_F C_A   \Bigg\{  - \frac{88}{3} \Bigg\}
       + n_f C_F   \Bigg\{ \frac{16}{3} \Bigg\}
\Bigg]\,,
\nonumber\\
\Phi^I_{q{\bar q},3} &=
\delta(1-z) \Bigg[ \frac{1}{\epsilon^4} C_F C_A^2   \Bigg\{
           \frac{21296}{81}
          \Bigg\}
       + \frac{1}{\epsilon^4} n_f C_F C_A   \Bigg\{
          - \frac{7744}{81}
          \Bigg\}
       + \frac{1}{\epsilon^4} n_f^2 C_F   \Bigg\{
           \frac{704}{81}
          \Bigg\}
\nonumber\\
&       + \frac{1}{\epsilon^3} C_F C_A^2   \Bigg\{
           \frac{49064}{243}
          - \frac{880}{27} \zeta_2
          \Bigg\}
       + \frac{1}{\epsilon^3} n_f C_F C_A   \Bigg\{
          - \frac{15520}{243}
          + \frac{160}{27} \zeta_2
          \Bigg\}
\nonumber\\
&       + \frac{1}{\epsilon^3} n_f C_F^2   \Bigg\{
          - \frac{128}{9}
          \Bigg\}
       + \frac{1}{\epsilon^3} n_f^2 C_F   \Bigg\{
           \frac{800}{243}
          \Bigg\}
       + \frac{1}{\epsilon^3} \log\l(\frac{q^2}{\mu_F^2}\r) C_F C_A^2   \Bigg\{
           \frac{1936}{27}
          \Bigg\}
\nonumber\\
&       + \frac{1}{\epsilon^3} \log\l(\frac{q^2}{\mu_F^2}\r) n_f C_F C_A   \Bigg\{
          - \frac{704}{27}
          \Bigg\}
       + \frac{1}{\epsilon^3} \log\l(\frac{q^2}{\mu_F^2}\r) n_f^2 C_F   \Bigg\{
          + \frac{64}{27}
          \Bigg\}
\nonumber\\
&       + \frac{1}{\epsilon^2} C_F C_A^2   \Bigg\{
          - \frac{8956}{243}
          + \frac{2024}{27} \zeta_3
          - \frac{692}{81} \zeta_2
          + \frac{352}{45} \zeta_2^2
          \Bigg\}
       + \frac{1}{\epsilon^2} n_f C_F C_A   \Bigg\{
           \frac{4024}{243}
          - \frac{560}{27} \zeta_3
\nonumber\\
&          - \frac{208}{81} \zeta_2
          \Bigg\}
       + \frac{1}{\epsilon^2} n_f C_F^2   \Bigg\{
          - \frac{220}{27}
          + \frac{64}{9} \zeta_3
          \Bigg\}
       + \frac{1}{\epsilon^2} n_f^2 C_F   \Bigg\{
          - \frac{160}{81}
          + \frac{16}{27} \zeta_2
          \Bigg\}
 \nonumber\\
&      + \frac{1}{\epsilon^2} \log\l(\frac{q^2}{\mu_F^2}\r) C_F C_A^2   \Bigg\{
           \frac{8344}{81}
          - \frac{176}{9} \zeta_2
          \Bigg\}
       + \frac{1}{\epsilon^2} \log\l(\frac{q^2}{\mu_F^2}\r) n_f C_F C_A   \Bigg\{
          - \frac{2672}{81}
          + \frac{32}{9} \zeta_2
          \Bigg\}
\nonumber\\
&       + \frac{1}{\epsilon^2} \log\l(\frac{q^2}{\mu_F^2}\r) n_f C_F^2   \Bigg\{
          - \frac{16}{3}
          \Bigg\}
       + \frac{1}{\epsilon^2} \log\l(\frac{q^2}{\mu_F^2}\r) n_f^2 C_F   \Bigg\{
           \frac{160}{81}
          \Bigg\}
       + \frac{1}{\epsilon} C_F C_A^2   \Bigg\{
          - \frac{136781}{2187}
\nonumber\\
&          - 64 \zeta_5
          + \frac{1316}{9} \zeta_3
          + \frac{12650}{243} \zeta_2
          - \frac{176}{9} \zeta_2 \zeta_3
          - \frac{352}{15} \zeta_2^2
          \Bigg\}
       + \frac{1}{\epsilon} n_f C_F C_A   \Bigg\{
           \frac{11842}{2187}
          - \frac{728}{81} \zeta_3
\nonumber\\
&          - \frac{2828}{243} \zeta_2
          + \frac{32}{5} \zeta_2^2
          \Bigg\}
       + \frac{1}{\epsilon} n_f C_F^2   \Bigg\{
           \frac{1711}{81}
          - \frac{304}{27} \zeta_3
          - \frac{4}{3} \zeta_2
          - \frac{32}{15} \zeta_2^2
          \Bigg\}
       + \frac{1}{\epsilon} n_f^2 C_F   \Bigg\{
           \frac{2080}{2187}
 \nonumber\\
&         - \frac{112}{81} \zeta_3
          + \frac{40}{81} \zeta_2
          \Bigg\}
       + \frac{1}{\epsilon} \log\l(\frac{q^2}{\mu_F^2}\r) C_F C_A^2   \Bigg\{
           \frac{490}{9}
          + \frac{88}{9} \zeta_3
          - \frac{1072}{27} \zeta_2
          + \frac{176}{15} \zeta_2^2
          \Bigg\}
\nonumber\\
&       + \frac{1}{\epsilon} \log\l(\frac{q^2}{\mu_F^2}\r) n_f C_F C_A   \Bigg\{
          - \frac{836}{81}
          - \frac{112}{9} \zeta_3
          + \frac{160}{27} \zeta_2
          \Bigg\}
       + \frac{1}{\epsilon} \log\l(\frac{q^2}{\mu_F^2}\r) n_f C_F^2   \Bigg\{
          - \frac{110}{9}
\nonumber\\
&          + \frac{32}{3} \zeta_3
          \Bigg\}
       + \frac{1}{\epsilon} \log\l(\frac{q^2}{\mu_F^2}\r) n_f^2 C_F   \Bigg\{
          - \frac{16}{81}
          \Bigg\}
       + C_F C_A^2   \Bigg\{
           \frac{5211949}{26244}
          - \frac{484}{9} \zeta_5
          - \frac{128966}{243} \zeta_3
\nonumber\\
&          + \frac{536}{9} \zeta_3^2
          - \frac{578479}{1458} \zeta_2
          + 242 \zeta_2 \zeta_3
          + \frac{9457}{135} \zeta_2^2
          + \frac{152}{189} \zeta_2^3
          \Bigg\}
       + n_f C_F C_A   \Bigg\{
          - \frac{412765}{13122}
 \nonumber\\
&         - \frac{8}{3} \zeta_5
          + \frac{9856}{81} \zeta_3
          + \frac{75155}{729} \zeta_2
          - \frac{44}{3} \zeta_2 \zeta_3
          - \frac{2528}{135} \zeta_2^2
          \Bigg\}
       + n_f C_F^2   \Bigg\{
          - \frac{42727}{972}
          + \frac{112}{9} \zeta_5
 \nonumber\\
&         + \frac{2284}{81} \zeta_3
          + \frac{605}{18} \zeta_2
          - \frac{88}{3} \zeta_2 \zeta_3
          + \frac{152}{45} \zeta_2^2
          \Bigg\}
       + n_f^2 C_F   \Bigg\{
          - \frac{128}{6561}
          - \frac{1480}{243} \zeta_3
          - \frac{404}{81} \zeta_2
 \nonumber\\
&         + \frac{148}{135} \zeta_2^2
          \Bigg\}
       + \log\l(\frac{q^2}{\mu_F^2}\r) C_F C_A^2   \Bigg\{
          - \frac{297029}{1458}
          - 96 \zeta_5
          + \frac{10036}{27} \zeta_3
          + \frac{24556}{81} \zeta_2
          - \frac{88}{3} \zeta_2 \zeta_3
\nonumber\\
&          - \frac{748}{15} \zeta_2^2
          \Bigg\}
       + \log\l(\frac{q^2}{\mu_F^2}\r) n_f C_F C_A   \Bigg\{
           \frac{31313}{729}
          - \frac{620}{9} \zeta_3
          - \frac{7348}{81} \zeta_2
          + \frac{184}{15} \zeta_2^2
          \Bigg\}
\nonumber\\
&       + \log\l(\frac{q^2}{\mu_F^2}\r) n_f C_F^2   \Bigg\{
           \frac{1711}{54}
          - \frac{152}{9} \zeta_3
          - 8 \zeta_2
          - \frac{16}{5} \zeta_2^2
          \Bigg\}
       + \log\l(\frac{q^2}{\mu_F^2}\r) n_f^2 C_F   \Bigg\{
          - \frac{928}{729}
          + \frac{80}{27} \zeta_3
\nonumber\\
&          + \frac{160}{27} \zeta_2
          \Bigg\}
       + \log^2\l(\frac{q^2}{\mu_F^2}\r) C_F C_A^2   \Bigg\{
           \frac{15503}{162}
          - 44 \zeta_3
          - \frac{752}{9} \zeta_2
          + \frac{44}{5} \zeta_2^2
          \Bigg\}
\nonumber\\
&       + \log^2\l(\frac{q^2}{\mu_F^2}\r) n_f C_F C_A   \Bigg\{
          - \frac{2051}{81}
          + 24 \zeta_2
          \Bigg\}
       + \log^2\l(\frac{q^2}{\mu_F^2}\r) n_f C_F^2   \Bigg\{
          - \frac{55}{6}
          + 8 \zeta_3
          \Bigg\}
 \nonumber\\
&      + \log^2\l(\frac{q^2}{\mu_F^2}\r) n_f^2 C_F   \Bigg\{
           \frac{100}{81}
          - \frac{16}{9} \zeta_2
          \Bigg\}
       + \log^3\l(\frac{q^2}{\mu_F^2}\r) C_F C_A^2   \Bigg\{
          - \frac{1780}{81}
          + \frac{44}{9} \zeta_2
          \Bigg\}
 \nonumber\\
&      + \log^3\l(\frac{q^2}{\mu_F^2}\r) n_f C_F C_A   \Bigg\{
           \frac{578}{81}
          - \frac{8}{9} \zeta_2
          \Bigg\}
       + \log^3\l(\frac{q^2}{\mu_F^2}\r) n_f C_F^2   \Bigg\{
           \frac{2}{3}
          \Bigg\}
       + \log^3\l(\frac{q^2}{\mu_F^2}\r) n_f^2 C_F   \Bigg\{
          - \frac{40}{81}
          \Bigg\}
 \nonumber\\
&      + \log\l(\frac{q^2}{\mu_F^2}\r)^4 C_F C_A^2   \Bigg\{
           \frac{121}{54}
          \Bigg\}
       + \log\l(\frac{q^2}{\mu_F^2}\r)^4 n_f C_F C_A   \Bigg\{
          - \frac{22}{27}
          \Bigg\}
       + \log\l(\frac{q^2}{\mu_F^2}\r)^4 n_f^2 C_F   \Bigg\{
           \frac{2}{27}
          \Bigg\}
\Bigg]
 \nonumber\\
&      + {\cal D}_0 \Bigg[ \frac{1}{\epsilon^3} C_F C_A^2   \Bigg\{
           \frac{3872}{27}
          \Bigg\}
       +  \frac{1}{\epsilon^3} n_f C_F C_A   \Bigg\{
          - \frac{1408}{27}
          \Bigg\}
       +  \frac{1}{\epsilon^3} n_f^2 C_F   \Bigg\{
           \frac{128}{27}
          \Bigg\}
\nonumber\\
&       +  \frac{1}{\epsilon^2} C_F C_A^2   \Bigg\{
           \frac{16688}{81}
          - \frac{352}{9} \zeta_2
          \Bigg\}
       +  \frac{1}{\epsilon^2} n_f C_F C_A   \Bigg\{
          - \frac{5344}{81}
          + \frac{64}{9} \zeta_2
          \Bigg\}
       +  \frac{1}{\epsilon^2} n_f C_F^2   \Bigg\{
          - \frac{32}{3}
          \Bigg\}
 \nonumber\\
&      +  \frac{1}{\epsilon^2} n_f^2 C_F   \Bigg\{
           \frac{320}{81}
          \Bigg\}
       +  \frac{1}{\epsilon} C_F C_A^2   \Bigg\{
           \frac{980}{9}
          + \frac{176}{9} \zeta_3
          - \frac{2144}{27} \zeta_2
          + \frac{352}{15} \zeta_2^2
          \Bigg\}
 \nonumber\\
&      +  \frac{1}{\epsilon} n_f C_F C_A   \Bigg\{
          - \frac{1672}{81}
          - \frac{224}{9} \zeta_3
          + \frac{320}{27} \zeta_2
          \Bigg\}
       +  \frac{1}{\epsilon} n_f C_F^2   \Bigg\{
          - \frac{220}{9}
          + \frac{64}{3} \zeta_3
          \Bigg\}
 \nonumber\\
&      +  \frac{1}{\epsilon} n_f^2 C_F   \Bigg\{
          - \frac{32}{81}
          \Bigg\}
       +  C_F C_A^2   \Bigg\{
          - \frac{297029}{729}
          - 192 \zeta_5
          + \frac{20072}{27} \zeta_3
          + \frac{49112}{81} \zeta_2
          - \frac{176}{3} \zeta_2 \zeta_3
 \nonumber\\
&         - \frac{1496}{15} \zeta_2^2
          \Bigg\}
       +  n_f C_F C_A   \Bigg\{
           \frac{62626}{729}
          - \frac{1240}{9} \zeta_3
          - \frac{14696}{81} \zeta_2
          + \frac{368}{15} \zeta_2^2
          \Bigg\}
       +  n_f C_F^2   \Bigg\{
           \frac{1711}{27}
 \nonumber\\
&         - \frac{304}{9} \zeta_3
          - 16 \zeta_2
          - \frac{32}{5} \zeta_2^2
          \Bigg\}
       +  n_f^2 C_F   \Bigg\{
          - \frac{1856}{729}
          + \frac{160}{27} \zeta_3
          + \frac{320}{27} \zeta_2
          \Bigg\}
\nonumber\\
&       +  \log\l(\frac{q^2}{\mu_F^2}\r) C_F C_A^2   \Bigg\{
           \frac{31006}{81}
          - 176 \zeta_3
          - \frac{3008}{9} \zeta_2
          + \frac{176}{5} \zeta_2^2
          \Bigg\}
 \nonumber\\
&      +  \log\l(\frac{q^2}{\mu_F^2}\r) n_f C_F C_A   \Bigg\{
          - \frac{8204}{81}
          + 96 \zeta_2
          \Bigg\}
       +  \log\l(\frac{q^2}{\mu_F^2}\r) n_f C_F^2   \Bigg\{
          - \frac{110}{3}
          + 32 \zeta_3
          \Bigg\}
 \nonumber\\
&      +  \log\l(\frac{q^2}{\mu_F^2}\r) n_f^2 C_F   \Bigg\{
           \frac{400}{81}
          - \frac{64}{9} \zeta_2
          \Bigg\}
       +  \log^2\l(\frac{q^2}{\mu_F^2}\r) C_F C_A^2   \Bigg\{
          - \frac{3560}{27}
          + \frac{88}{3} \zeta_2
          \Bigg\}
 \nonumber\\
&      +  \log^2\l(\frac{q^2}{\mu_F^2}\r) n_f C_F C_A   \Bigg\{
           \frac{1156}{27}
          - \frac{16}{3} \zeta_2
          \Bigg\}
       +  \log^2\l(\frac{q^2}{\mu_F^2}\r) n_f C_F^2   \Bigg\{
           4
          \Bigg\}
 \nonumber\\
&      +  \log^2\l(\frac{q^2}{\mu_F^2}\r) n_f^2 C_F   \Bigg\{
          - \frac{80}{27}
          \Bigg\}
       +  \log^3\l(\frac{q^2}{\mu_F^2}\r) C_F C_A^2   \Bigg\{
           \frac{484}{27}
          \Bigg\}
       +  \log^3\l(\frac{q^2}{\mu_F^2}\r) n_f C_F C_A   \Bigg\{
          - \frac{176}{27}
          \Bigg\}
 \nonumber\\
&      +  \log^3\l(\frac{q^2}{\mu_F^2}\r) n_f^2 C_F   \Bigg\{
           \frac{16}{27}
          \Bigg\}
\Bigg]
       + {\cal D}_1 \Bigg[ C_F C_A^2   \Bigg\{
           \frac{62012}{81}
          - 352 \zeta_3
          - \frac{6016}{9} \zeta_2
          + \frac{352}{5} \zeta_2^2
          \Bigg\}
\nonumber\\
&       +  n_f C_F C_A   \Bigg\{
          - \frac{16408}{81}
          + 192 \zeta_2
          \Bigg\}
       +  n_f C_F^2   \Bigg\{
          - \frac{220}{3}
          + 64 \zeta_3
          \Bigg\}
       +  n_f^2 C_F   \Bigg\{
           \frac{800}{81}
          - \frac{128}{9} \zeta_2
          \Bigg\}
\nonumber\\
&       +  \log\l(\frac{q^2}{\mu_F^2}\r) C_F C_A^2   \Bigg\{
          - \frac{14240}{27}
          + \frac{352}{3} \zeta_2
          \Bigg\}
       +  \log\l(\frac{q^2}{\mu_F^2}\r) n_f C_F C_A   \Bigg\{
           \frac{4624}{27}
          - \frac{64}{3} \zeta_2
          \Bigg\}
 \nonumber\\
&      +  \log\l(\frac{q^2}{\mu_F^2}\r) n_f C_F^2   \Bigg\{
           16
          \Bigg\}
       +  \log\l(\frac{q^2}{\mu_F^2}\r) n_f^2 C_F   \Bigg\{
          - \frac{320}{27}
          \Bigg\}
       +  \log^2\l(\frac{q^2}{\mu_F^2}\r) C_F C_A^2   \Bigg\{
           \frac{968}{9}
          \Bigg\}
\nonumber\\
&       +  \log^2\l(\frac{q^2}{\mu_F^2}\r) n_f C_F C_A   \Bigg\{
          - \frac{352}{9}
          \Bigg\}
       +  \log^2\l(\frac{q^2}{\mu_F^2}\r) n_f^2 C_F   \Bigg\{
           \frac{32}{9}
          \Bigg\}
\Bigg]
       + {\cal D}_2 \Bigg[ C_F C_A^2   \Bigg\{
          - \frac{14240}{27}
\nonumber\\
&          + \frac{352}{3} \zeta_2
          \Bigg\}
       +  n_f C_F C_A   \Bigg\{
           \frac{4624}{27}
          - \frac{64}{3} \zeta_2
          \Bigg\}
       +  n_f C_F^2   \Bigg\{
           16
          \Bigg\}
       +  n_f^2 C_F   \Bigg\{
          - \frac{320}{27}
          \Bigg\}
\nonumber\\
&       +  \log\l(\frac{q^2}{\mu_F^2}\r) C_F C_A^2   \Bigg\{
          \frac{1936}{9}
          \Bigg\}
       +  \log\l(\frac{q^2}{\mu_F^2}\r) n_f C_F C_A   \Bigg\{
          - \frac{704}{9}
          \Bigg\}
       +  \log\l(\frac{q^2}{\mu_F^2}\r) n_f^2 C_F   \Bigg\{
           \frac{64}{9}
          \Bigg\}
\Bigg]
\nonumber\\
&       + {\cal D}_3 \Bigg[ C_F C_A^2   \Bigg\{
           \frac{3872}{27}
          \Bigg\}
       +  n_f C_F C_A   \Bigg\{
          - \frac{1408}{27}
          \Bigg\}
       +  n_f^2 C_F   \Bigg\{
           \frac{128}{27}
          \Bigg\}
\Bigg]\,.
\end{align}


\chapter{Soft-Collinear Distribution for Rapidity}
\label{chpt:App-Rap-Soft-Col-Dist}

In Sec.~\ref{ss:Rap-SCD}, we introduced the soft-collinear
distribution $\Phi^I_{i~{\bar i}}$ in the context of computing SV
correction to the differential rapidity distribution of a colorless
particle at Hadron collider. In this appendix, we intend to elaborate
the methodology 
of finding this distribution. For simplicity, we will omit the
partonic indices for our further calculation. To understand 
the underlying logics behind finding $\Phi^{I}$, let us consider an
example at one loop level. The generalisation to higher loop is
straightforward. The whole discussion of this appendix is closely
related to the Appendix~\ref{chpt:App-Soft-Col-Dist} where we
discussed the soft-collinear distribution for inclusive production
cross section for a colorless particle.

As discussed in the Sec.~\ref{sec:Rap-ThreResu}, the SV cross section in
$z$-space can be computed in $d=4+\epsilon$ 
dimensions using 
\begin{align}
\label{eq:App-Rap-sigma}
\Delta^{I, \sv}_{Y} (z_1, z_2, q^2, \mu_{R}^{2}, \mu_F^2) = 
{\cal C} \exp \Big( \Psi^I_{Y} \left(z_1, z_2, q^2, \mu_R^2, \mu_F^2, \epsilon
  \right)  \Big) \Big|_{\epsilon = 0} 
\end{align}
where, $\Psi^I_{Y} \left(z_1, z_2, q^2, \mu_R^2, \mu_F^2, \epsilon \right)$ is a
finite distribution and ${\cal C}$ is the double Mellin convolution defined
through Eq.~(\ref{eq:Rap-conv}). The $\Psi^I$ is given by, Eq.~(\ref{eq:Rap-psi})
\begin{align}
\label{eq:App-Rap-psi}
\Psi^{I}_{Y,ij} \left(z_1, z_2, q^2, \mu_R^2, \mu_F^2, \epsilon \right) 
= &\left( \ln \Big[ Z^I_{ij} (\hat{a}_s, \mu_R^2, \mu^2, \epsilon) \Big]^2 
+ \ln \Big|  {\cal F}^I_{ij} (\hat{a}_s, Q^2, \mu^2, \epsilon)
    \Big|^2 \right) \delta(1-z_1) \delta(1-z_2)     
\nonumber\\
& + 2 \Phi^I_{Y,ij} (\hat{a}_s, q^2, \mu^2, z_1, z_2, \epsilon) - {\cal C} \ln
  \Gamma^{I}_{ij} (\hat{a}_s, \mu^2, \mu_F^2, z_1,
  \epsilon)\delta(1-z_2) 
\nonumber\\
&
- {\cal C} \ln
  \Gamma^{I}_{ij} (\hat{a}_s, \mu^2, \mu_F^2, z_2,
  \epsilon)\delta(1-z_1)  \, .
\end{align}
For all the details about the notations, see
Sec.~\ref{sec:Rap-ThreResu}. Considering only the poles at ${\cal
  O}(a_{s})$ with $\mu_{R} = 
\mu_{F}$ we obtain,
\begin{align}
\label{eq:App-Rap-Psi-Comp-1}
&\ln\l(Z^{I, (1)}\r)^{2} = a_{s}(\mu_{F}^{2}) \frac{4\gamma_{1}^{I}}{\epsilon}\,,
\nonumber\\
&\ln|{\cal F}^{I, (1)}|^{2} = a_{s}(\mu_{F}^{2}) \l(
  \frac{q^{2}}{\mu_{F}^{2}} \r)^{\frac{\epsilon}{2}} \l[
  -\frac{4A_{1}^{I}}{\epsilon^{2}} + \frac{1}{\epsilon} \l( 2f_{1}^{I}
  + 4B_{1}^{I} - 4\gamma_{1}^{I}\r)\r]\,,
\nonumber\\
&{\cal C} \ln\Gamma^{I, (1)}(z_1) = a_{s}(\mu_{F}^{2}) \l[
  \frac{2B^{I}_{1}}{\epsilon} \delta(1-z_1) +
  \frac{2A_{1}^{I}}{\epsilon} {\cal D}_{0}\r]\,,
\nonumber\\
&{\cal C} \ln\Gamma^{I, (1)}(z_2) = a_{s}(\mu_{F}^{2}) \l[
  \frac{2B^{I}_{1}}{\epsilon} \delta(1-z_2) +
  \frac{2A_{1}^{I}}{\epsilon} \overline{\cal D}_{0}\r]
\end{align}
where, the components are defined through the expansion of these
quantities in powers of $a_s(\mu_F^{2})$
\begin{align}
\label{eq:App-Rap-psi-comp-2}
&\Psi^{I}_{Y} = \sum\limits_{k=1}^{\infty} a_s^k \left( \mu_F^2 \right) \Psi^{I,(k)}_{Y}\,,
\nonumber\\
&\ln (Z^I)^2 = \sum\limits_{k=1}^{\infty} a_s^k \left( \mu_F^2 \right)
  Z^{I,(k)}\,,
\nonumber\\
&\ln |{\cal F}^{I}|^2 = \sum\limits_{k=1}^{\infty} a_s^k\left( \mu_F^2
  \right) \left( \frac{q^2}{\mu_F^2} \right)^{k \frac{\epsilon}{2}}
  \ln |{\cal F}^{I,(k)}|^2\,,
\nonumber\\
&\Phi^{I}_{Y} = \sum\limits_{k=1}^{\infty} a_s^k(\mu_F^2) \Phi^I_{Y,k}\,,
\nonumber\\
&\ln \Gamma^I = \sum\limits_{k=1}^{\infty} a_s^k(\mu_F^2) \ln
  \Gamma^{I,(k)}
\intertext{and}
&{\cal{D}}_{i} \equiv \left[ \frac{\ln^{i}(1-z_1)}{1-z_1} \right]_{+}\,,
\nonumber\\
&\overline{\cal{D}}_{i} \equiv \left[ \frac{\ln^{i}(1-z_2)}{1-z_2} \right]_{+}\,.
\end{align}
Collecting the coefficients of $a_s(\mu_F^2)$, we
get
\begin{align}
\label{eq:App-Rap-Psi-1}
\Psi^{I, (1)}|_{Y,{\rm poles}}
&= \l[ \l\{ - \frac{4A_{1}^{I}}{\epsilon^{2}} + \frac{
  2f_{1}^{I}}{\epsilon} \r\} \delta(1-z_1) \delta(1-z_2) -
  \frac{2A_{1}^{I}}{\epsilon} \l\{ \delta(1-z_1) \overline{\cal
  D}_{0} + \delta(1-z_2) {\cal D}_0 \r\} \r]
\nonumber\\
&+ {2\Phi^{I}_{Y,1}} 
\end{align}
where, we have suppressed the $\ln (q^2/\mu_F^2)$ terms.
To cancel the remaining divergences appearing in the above
Eq.~(\ref{eq:App-Rap-Psi-1}) for obtaining a finite rapidity distribution, we must
demand that $\Phi^{I}_{Y,1}$ has exactly the same poles with
opposite sign:
\begin{align}
\label{eq:App-Rap-Phi-1}
2 \Phi^{I}_{Y,1}|_{\rm poles} = - \l[ \l\{ - \frac{4A_{1}^{I}}{\epsilon^{2}} + \frac{
  2f_{1}^{I}}{\epsilon} \r\} \delta(1-z_1) \delta(1-z_2) -
  \frac{2A_{1}^{I}}{\epsilon} \l\{ \delta(1-z_1) \overline{\cal
  D}_{0} + \delta(1-z_2) {\cal D}_0 \r\} \r]
\end{align} 
In addition, $\Phi^{I}_{Y}$ also should be RG invariant with respect to
$\mu_R$:
\begin{align}
\label{eq:App-Rap-Phi-2}
\mu_R^2 \frac{d}{d\mu_R^2} \Phi^I_{Y} = 0\,. 
\end{align}
We make an \textit{ansatz}, the above two demands,
Eq.~(\ref{eq:App-Phi-1}) and~(\ref{eq:App-Phi-2}) can be accomplished
if $\Phi^I_Y$ satisfies the KG-type integro-differential equation which we call
$\overline{KG}_{Y}$:
\begin{align}
  \label{eq:App-Rap-KGbarEqn}
  q^2 \frac{d}{dq^2} \Phi^I_Y\left(\hat{a}_s, q^2, \mu^2, z_1, z_2,
    \ep\right)   = \frac{1}{2} \left[ \overline K^I_Y
  \left(\hat{a}_s, \frac{\mu_R^2}{\mu^2}, z_1, z_2,
      \ep \right)  + \overline G^I_Y \left(\hat{a}_s,
      \frac{q^2}{\mu_R^2},  \frac{\mu_R^2}{\mu^2}, z_1, z_2, \ep \right) \right]\,.
\end{align}
$\overline{K}^I_Y$ contains all the poles whereas $\overline{G}^I_Y$ consists
of only the finite terms in $\epsilon$. RG
invariance~(\ref{eq:App-Rap-Phi-2}) of $\Phi^I_Y$ dictates
\begin{align}
\label{eq:App-Rap-Phi-3}
\mu_R^2 \frac{d}{d\mu_R^2} \overline{K}^{I}_Y = -\mu_R^2 \frac{d}{d\mu_R^2}
  \overline{G}^I_Y \equiv X^{I}_{Y}
\end{align}
where, we introduce a quantity $X^I_{Y}$. Following the methodology of
solving the KG equation discussed in the
Appendix~\ref{chpt:App-KGSoln}, we can write the solution of $\Phi^I_Y$
as
\begin{align}
  \label{eq:App-Rap-Phi-4}
  \Phi^I_Y(\hat{a}_s, q^2, \mu^2, z_1, z_2, \epsilon) =
  \sum_{k=1}^{\infty} {\hat a}_{s}^{k}S_{\epsilon}^{k} \left(\frac{q^{2}}{\mu^{2}}\right)^{k
  \frac{\epsilon}{2}}  \hat {\Phi}^{I}_{Y,k}(z_1, z_{2}, \epsilon)
\end{align}
with
\begin{align}
\label{eq:App-Rap-Phi-5}
\hat {\Phi}^{I}_{Y,k}(z_1, z_2, \epsilon) = {\hat{\cal L}}^I_k \left( A^I_{i}
  \rightarrow X^I_{Y,i}, G^I_i \rightarrow \overline{G}^I_{Y,i}(z,\epsilon) \right)\,.
\end{align}
where we define the components through the expansions
\begin{align}
\label{eq:App-Rap-YGexpans}
&X^I_{Y}  = \sum\limits_{k=1}^{\infty} a_s^k(\mu_F^2) X^I_{Y,k}\,,
\nonumber\\
&\overline{G}^I_Y(z_1, z_{2},\epsilon) = \sum\limits_{k=1}^{\infty}
  a_s^k(\mu_F^2) \overline{G}^I_{Y,k}(z_1, z_2, \epsilon).
\end{align}
This solution directly follows from the Eq.~(\ref{eq:App-lnFSoln}). Hence we get
\begin{align}
  \label{eq:App-Rap-Phi-6}
  2 \hat {\Phi}^{I}_{Y,1}(z,\ep) &= { \frac{1}{\ep^2} } \Bigg\{-4 X^{I}_{Y,1}\Bigg\}
                                  + { \frac{2}{\ep}
                                  }
                                  \Bigg\{\overline{G}^{I}_{Y,1}
                                  (z_1, z_2,\ep)\Bigg\}\,.
\end{align}
By expressing the components of $\Phi^I_Y$ in powers of $a_s(\mu_F^2)$,
we obtain
\begin{align}
\label{eq:App-Rap-Phi-7}
\Phi^I(\hat{a}_s, q^2, \mu^2, z_1, z_2, \epsilon) &=
  \sum_{k=1}^{\infty} {\hat a}_{s}^{k}S_{\epsilon}^{k} \left(\frac{q^{2}}{\mu^{2}}\right)^{k
  \frac{\epsilon}{2}}  \hat {\Phi}^{I}_{Y,k}(z_1, z_2, \epsilon)
\nonumber\\
&= \sum_{k=1}^{\infty} a_s^k(\mu_F^2) \left( \frac{q^2}{\mu_F^2} \right)^{k
  \frac{\epsilon}{2}} Z_{a_s}^k \hat{\Phi}^I_{Y,k}(z_1, z_2,\epsilon)
\nonumber\\
& \equiv \sum_{k=1}^{\infty} a_s^k(\mu_F^2) \left( \frac{q^2}{\mu_F^2} \right)^{k
  \frac{\epsilon}{2}} {\Phi}^I_{Y,k}(z,\epsilon)
\end{align}
and at ${\cal O}(a_s(\mu_F^2)), \hat{\Phi}^I_{Y,1}(z,\epsilon) =
\Phi^I_{Y,1}(z,\epsilon)$ upon suppressing the terms like $\log
(q^{2}/\mu_{F}^{2})$. Hence, by comparing the 
Eq.~(\ref{eq:App-Rap-Phi-1}) and (\ref{eq:App-Rap-Phi-6}), we conclude
\begin{align}
\label{eq:App-Rap-Phi-8}
&X^I_{Y,1} = -A^I_1 \delta(1-z_1) \delta(1-z_2)\,,
\nonumber\\
&\overline{G}^I_{Y,1}(z_1, z_2, \epsilon) = -f^I_1 \delta(1-z_1)
  \delta(1-z_2) +  A^I_1 \Bigg\{ \delta(1-z_1) \overline{\cal 
  D}_0 + \delta(1-z_2) {\cal D}_{0} \Bigg\}
\nonumber\\
&+ \sum\limits_{k=1}^{\infty}
  \epsilon^k \overline{g}^{I,k}_{Y,1}(z)\,. 
\end{align}
The coefficients of $\epsilon^k, \overline{g}^{I,k}_{Y,1}(z)$ can only be
determined through explicit computations. These do not contribute to
the infrared poles associated with $\Phi^{I}_{Y}$. This uniquely fixes the
unknown soft-collinear distribution $\Phi^I_{Y}$ at one loop order. This
prescription can easily be generalised to higher orders in $a_s$. In
our calculation of the SV correction to rapidity distribution, instead
of solving in this way, 
we follow a bit different methodology which is presented below.

Keeping the demands~(\ref{eq:App-Rap-Phi-1}) and~(\ref{eq:App-Rap-Phi-2}) in
mind, we propose the solution of the $\overline{KG}_{Y}$ equation as (See
Eq.~(\ref{eq:App-Rap-Phi-4})) (which is just the extension of the
Eq.~(\ref{eq:App-Phi-9}) from one variable $z$ to a case of two
variables $z_1$ and $z_2$)  
\begin{align}
\label{eq:App-Rap-Phi-9}
\hat{\Phi}^I_{Y,k} (z_1, z_2, \epsilon) &\equiv \Bigg\{ (k \epsilon)^2
                                          \frac{1}{4 (1-z_1) (1-z_2)}
                                \left[
  (1-z_1) (1-z_2) \right]^{k \frac{\epsilon}{2}}\Bigg\}
                                \hat{\Phi}^I_{Y,k}(\epsilon)
\nonumber\\
&= \Bigg\{ \frac{k\epsilon}{2} \frac{1}{(1-z_1)} \l[ (1-z_1)^2 \r]^{k
  \frac{\epsilon}{4}} \Bigg\} 
\Bigg\{ \frac{k\epsilon}{2} \frac{1}{(1-z_2)} \l[ (1-z_2)^2 \r]^{k
  \frac{\epsilon}{4}} \Bigg\}   \hat{\Phi}^I_{Y,k}(\epsilon)
\nonumber\\
&=\Bigg\{ \delta(1-z_1) + \sum\limits_{j=0}^{\infty} \frac{(k
  \epsilon/2)^{j+1}}{j!} {\cal D}_{j} \Bigg\} 
\Bigg\{ \delta(1-z_2) + \sum\limits_{l=0}^{\infty} \frac{(k
  \epsilon/2)^{l+1}}{l!} \overline{\cal D}_{l} \Bigg\} 
\hat{\Phi}^I_{Y,k}(\epsilon)\,.
\end{align}
The RG invariance of $\Phi^{I}_Y$, Eq.~(\ref{eq:App-Rap-Phi-2}), implies
\begin{align}
\label{eq:App-Rap-Phi-10}
\mu_R^2 \frac{d}{d\mu_R^2} \overline{K}^{I}_Y = -\mu_R^2 \frac{d}{d\mu_R^2}
  \overline{G}^I_Y \equiv X'^{I}_Y
\end{align}
where, we introduce a quantity $X'^I_Y$, analogous to $X^I_{Y}$. Hence, the
solution can be obtained as
\begin{align}
\label{eq:App-Rap-Phi-11}
\hat {\Phi}^{I}_{Y,k}(\epsilon) &= {\hat{\cal L}}^I_k \left( A^I_{i}
  \rightarrow X'^I_{Y,i}, G^I_{Y,i} \rightarrow \overline{\cal
                                  G}^I_{Y,i}(\epsilon) \right)\,. 
\end{align}
Hence, according to the Eq.~(\ref{eq:App-lnFitoCalLF}), for $k=1$ we get
\begin{align}
\label{eq:App-Rap-Phi-12}
2\Phi^{I}_{Y,1} (z_1, z_2, \epsilon) &= 2{\hat \Phi}^{I}_{Y,1} (z_1, z_2,
  \epsilon)
\nonumber\\
 &= 
\Bigg\{ \delta(1-z_1) + \sum\limits_{j=0}^{\infty} \frac{(k
  \epsilon/2)^{j+1}}{j!} {\cal D}_{j} \Bigg\} 
\Bigg\{ \delta(1-z_2) + \sum\limits_{l=0}^{\infty} \frac{(k
  \epsilon/2)^{l+1}}{l!} \overline{\cal D}_{l} \Bigg\} 
\nonumber\\
&~~~~\Bigg\{ \frac{1}{\epsilon^2}\l(
                               - 4X'^I_{Y,1} \r) +
                               \frac{2}{\epsilon}{\overline{\cal
                               G}}^I_{Y,1} \l(\epsilon) \r) \Bigg\}
\end{align}
where, $X'^{I}_{Y}$ and $\overline{\cal G}^I_{Y}$ are expanded similar to
Eq.~(\ref{eq:App-Rap-YGexpans}). Comparison between the two solutions
depicted in Eq.~(\ref{eq:App-Rap-Phi-1}) and (\ref{eq:App-Rap-Phi-12}), we 
can write
\begin{align}
\label{eq:App-Phi-13}
&X'^I_{Y,1} = - A^I_1 
\nonumber\\
&{\overline{\cal G}}^I_{Y,1} \l(\epsilon\r) = - f^I_1 +
                                          \sum_{k=1}^{\infty}
                                          \epsilon^k \overline{\cal
                                          G}^{I, k}_{Y,1}\,.
\end{align}
Explicit computation is required to determine the coefficients of
$\epsilon^k$, $\overline{\cal G}^{I,k}_{Y,1}$. This solution is used in
Eq.~(\ref{eq:Rap-PhiSoln}) in
the context of SV correction to differential rapidity distribution of
Higgs boson production or leptonic pair in DY production. The method
is generalised to higher orders in $a_s$ to obtain the results of the
soft-collinear distribution. Hence, the all order solution of
$\Phi^I_{Y}$ is
\begin{align}
\label{eq:App-Rap-Soln-PhiY}
&\Phi^I(\hat{a}_s, q^2, \mu^2, z_1, z_2, \epsilon) =
  \sum_{k=1}^{\infty} {\hat a}_{s}^{k}S_{\epsilon}^{k} \left(\frac{q^{2}}{\mu^{2}}\right)^{k
  \frac{\epsilon}{2}}  \hat {\Phi}^{I}_{Y,k}(z_1, z_2, \epsilon)
\intertext{with}
&\hat {\Phi}^{I}_{Y,k}(z_1, z_2, \epsilon) = \Bigg\{ (k \epsilon)^2
                                          \frac{1}{4 (1-z_1) (1-z_2)}
                                \left[
  (1-z_1) (1-z_2) \right]^{k \frac{\epsilon}{2}}\Bigg\}
                                \hat{\Phi}^I_{Y,k}(\epsilon)\,,
\nonumber\\
&\hat {\Phi}^{I}_{Y,k}(\epsilon) = {\hat{\cal L}}^I_k \left( A^I_{i}
  \rightarrow -A^I_{i}, G^I_{Y,i} \rightarrow \overline{\cal
                                  G}^I_{Y,i}(\epsilon) \right)\,.
\end{align}
Up to three loop, $\overline{\cal G}^I_{Y,i}(\epsilon)$ are found to be
\begin{align}
\label{eq:App-Rap-Sol-calGbar}
&\overline{\cal G}^I_{Y,i} (\epsilon) = -f^I_i + \overline{C}^I_{Y,i}
                                       + \sum\limits_{k=1}^{\infty}
                                       \epsilon^k \overline{\cal
                                       G}^{I,k}_{Y,i}
\intertext{where}
&\overline{C}^I_{Y,1} = 0\,,
\nonumber\\
&\overline{C}^I_{Y,2} = - 2 \beta_0 \overline{\cal G}^{I,1}_{Y,1}\,,
\nonumber\\
&\overline{C}^I_{Y,3} = - 2 \beta_1 \overline{\cal G}^{I,1}_{Y,1} - 2
  \beta_0 \l( \overline{\cal G}^{I,1}_{Y,2} + 2 \beta_0 \overline{\cal
  G}^{I,2}_{Y,1} \r)\,.
\end{align}
These are employed in the computation of rapidity distributions in
Chapter~\ref{chap:Rap}. In the next subsection, we present the results
of the soft-collinear distribution up to three loops.

\subsection{Results}
\label{app:ss-RapSCD-Res}

We define the renormalised components of the
$\Phi^I_{Y,i~{\bar i},k}$ through
\begin{align}
\label{eq:App-Rap-SCD-Re}
\Phi^I_{Y,i~{\bar i}}(\hat{a}_s, q^2, \mu^2, z_1, z_2, \epsilon) &=
  \sum_{k=1}^{\infty} {\hat a}_{s}^{k}S_{\epsilon}^{k} \left(\frac{q^{2}}{\mu^{2}}\right)^{k
  \frac{\epsilon}{2}}  \hat {\Phi}^{I}_{Y,i~{\bar i},k}(z_1, z_2, \epsilon)
\nonumber\\
&= \sum\limits_{k=1}^{\infty} a_s^k\l( \mu_F^2 \r) \Phi^I_{Y,i~{\bar
  i},k} \left( z_1, z_2, \epsilon, q^2, \mu_F^2 \right)
\end{align} 
where, we make the choice of the renormalisation scale
$\mu_R=\mu_F$. The $\mu_R$ dependence can be easily restored by using
the evolution equation of strong coupling constant,
Eq.~(\ref{eq:bBH-asf2asr}). Below, we present the $\Phi^I_{Y,i~{\bar
    i},k}$ for 
$I=H$ and $i~{\bar i}=gg$ up to three 
loops and the corresponding components for $I={\rm DY}$ and $i~{\bar
  i}=q{\bar q}$ can be obtained using maximally non-Abelian property
fulfilled by this distribution:
\begin{align}
\label{eq:App-Rap-SCD-MaxNonAbe}
\Phi^H_{Y,gg,k} = \frac{C_A}{C_F} \Phi^{\rm DY}_{Y,q~{\bar q},k}\,.
\end{align} 
The results are given by


\chapter{Results of the Unrenormalised Three Loop Form Factors for the Pseudo-Scalar}
\label{App:pScalar-Results}

In this appendix, we present the unrenormalised quark and gluon form
factors for the pseudo-scalar production up to three loops for the operators $\left[ O_{G} \right]_{B}$
and $\left[ O_{J} \right]_{B}$. Specifically, we present
${\hat{\cal F}}^{G,(n)}_{\beta}$ and ${\hat{\cal F}}^{J,(n)}_{\beta}$
for $\beta=q,g$ up to $n=3$ which are defined in
Sec.~\ref{sec:FF}. One and two loop results completely agree with the
existing literature~\cite{Ravindran:2004mb}. It should be noted
that the form factors at $n=2$ for ${\hat{\cal F}}^{G,(n)}_{q}$ and
${\hat{\cal F}}^{J,(n)}_{g}$ correspond to the contributions arising
from three loop diagrams since these processes start at one loop order.

%
\begin{align}
  \label{eq:FgG1}
  {\hat{\cal F}}^{G,(1)}_{g} &= {\dis{C_{A}}} \Bigg\{ - \frac{8}{\epsilon^2} + 4 + \zeta_2 + 
                               \epsilon \Bigg( - 6 - \frac{7}{3} \zeta_3 \Bigg) + \epsilon^2 \Bigg(
                               7 - \frac{\zeta_2}{2} + \frac{47}{80}
                               \zeta_2^2 \Bigg) + \epsilon^3 \Bigg( -
                               \frac{15}{2} + \frac{3}{4} \zeta_2 
                               \nonumber\\
                             &+ \frac{7}{6} \zeta_3 + 
                               \frac{7}{24} \zeta_2 \zeta_3 - \frac{31}{20} \zeta_5 \Bigg) \Bigg\}\,,
\end{align}

%
\begin{align}
  \label{eq:FgG2}
  {\hat{\cal F}}^{G,(2)}_{g} &= {\dis{2 C_{A} n_{f} T_{F}}} \Bigg\{ - \frac{8}{3 \epsilon^3} + \frac{20}{9 \epsilon^2} + 
                               \Bigg( \frac{106}{27} + 2 \zeta_2
                               \Bigg) \frac{1}{\epsilon}  -
                               \frac{1591}{81}  - \frac{5}{3} \zeta_2
                               - \frac{74}{9} \zeta_3 + 
                               \epsilon \Bigg( \frac{24107}{486}  
\nonumber\\
&-
                               \frac{23}{18} \zeta_2  
                               + \frac{51}{20} \zeta_2^2 + \frac{383}{27} \zeta_3 \Bigg) + 
                               \epsilon^2 \Bigg( - \frac{146147}{1458}
                               + \frac{799}{108} \zeta_2 - \frac{329}{72} \zeta_2^2 - 
                               \frac{1436}{81} \zeta_3  + \frac{25}{6}
                               \zeta_2 \zeta_3  
                               \nonumber\\
                             &- \frac{271}{30} \zeta_5 \Bigg) \Bigg\} 
                               + 
                               {\dis{C_{A}^2}} \Bigg\{ \frac{32}{\epsilon^4}
                               + \frac{44}{3 \epsilon^3}  + \Bigg( -
                               \frac{422}{9}  - 4 \zeta_2 \Bigg) \frac{1}{\epsilon^2} + 
                               \Bigg( \frac{890}{27} - 11 \zeta_2  +
                               \frac{50}{3} \zeta_3 \Bigg)
                               \frac{1}{\epsilon} 
                               \nonumber\\
                             &+ \frac{3835}{81} + 
                               \frac{115}{6} \zeta_2 - \frac{21}{5}
                               \zeta_2^2  + \frac{11}{9} \zeta_3 + 
                               \epsilon \Bigg( - \frac{213817}{972} -
                               \frac{103}{18} \zeta_2  +
                               \frac{77}{120} \zeta_2^2  +
                               \frac{1103}{54} \zeta_3  
                               \nonumber\\
                             &- 
                               \frac{23}{6} \zeta_2 \zeta_3  -
                               \frac{71}{10} \zeta_5 \Bigg)  + 
                               \epsilon^2 \Bigg( \frac{6102745}{11664}
                               - \frac{991}{27} \zeta_2  -
                               \frac{2183}{240} \zeta_2^2  + 
                               \frac{2313}{280} \zeta_2^3  -
                               \frac{8836}{81} \zeta_3   -
                               \frac{55}{12} \zeta_2 \zeta_3  
                               \nonumber\\
                             &+ 
                               \frac{901}{36} \zeta_3^2  + \frac{341}{60} \zeta_5 \Bigg) \Bigg\} 
                               + 
                               {\dis{2 C_{F} n_{f}  T_{F}}} \Bigg\{ \frac{12}{\epsilon}
                               - \frac{125}{3} + 8 \zeta_3  + \epsilon
                               \Bigg( \frac{3421}{36}  - \frac{14}{3}
                               \zeta_2  - \frac{8}{3} \zeta_2^2 
\nonumber\\
&- 
                               \frac{64}{3} \zeta_3 \Bigg) 
                              + \epsilon^2 \Bigg( - \frac{78029}{432}
                               + \frac{293}{18} \zeta_2 + 
                               \frac{64}{9} \zeta_2^2  +
                               \frac{973}{18} \zeta_3  - \frac{10}{3}
                               \zeta_2 \zeta_3  + 8 \zeta_5 \Bigg) \Bigg\}\,,
\end{align}
%


\begin{align}
  \label{eq:FgG3}
  {\hat{\cal F}}^{G,(3)}_{g} &= {\dis{4 C_{F} n_{f}^2 T_{F}^{2}}} \Bigg\{ 
                               \frac{16}{\epsilon^2} + \Bigg( -
                               \frac{796}{9} + \frac{64}{3} \zeta_3
                               \Bigg) \frac{1}{\epsilon} +
                               \frac{8387}{27} - \frac{38}{3} \zeta_2
                               - \frac{112}{15} \zeta_2^2 - 
                               \frac{848}{9} \zeta_3  \Bigg\} 
                               \nonumber\\
&+ {\dis{2 C_{F}^2 n_{f} T_{F}}} \Bigg\{ 
                               \frac{6}{\epsilon} 
                              - \frac{353}{6} + 176 \zeta_3 - 160 \zeta_5 \Bigg\} + 
                               {\dis{2 C_{A}^2 n_{f}  T_{F}}} \Bigg\{ \frac{64}{3
                               \epsilon^5} - \frac{32}{81 \epsilon^4} 
                               + \Bigg( - \frac{18752}{243}  
                               \nonumber\\
&- \frac{376}{27} \zeta_2 \Bigg) \frac{1}{\epsilon^3} + 
                               \Bigg( \frac{36416}{243} 
                              -
                               \frac{1700}{81} \zeta_2 +
                               \frac{2072}{27} \zeta_3 \Bigg)
                               \frac{1}{\epsilon^2} + \Bigg(
                               \frac{62642}{2187} + \frac{22088}{243} \zeta_2 - \frac{2453}{90}
                               \zeta_2^2 
\nonumber\\
&-
                               \frac{3988}{81} \zeta_3 \Bigg)
                               \frac{1}{\epsilon} -
                               \frac{14655809}{13122} 
                              - \frac{60548}{729} \zeta_2 + 
                               \frac{917}{60} \zeta_2^2  -
                               \frac{772}{27} \zeta_3 - \frac{439}{9}
                               \zeta_2 \zeta_3  
                               + \frac{3238}{45}
                               \zeta_5 \Bigg\}  
                               \nonumber\\
&+ 
                               {\dis{4 C_{A} n_{f}^2 T_{F}^{2}}} \Bigg\{ - \frac{128}{81
                               \epsilon^4} + \frac{640}{243
                               \epsilon^3} 
                              + \Bigg(
                               \frac{128}{27} + \frac{80}{27} \zeta_2
                               \Bigg) \frac{1}{\epsilon^2} + 
                               \Bigg( - \frac{93088}{2187} -
                               \frac{400}{81} \zeta_2 
\nonumber\\
&-
                               \frac{1328}{81} \zeta_3 \Bigg)
                               \frac{1}{\epsilon} +
                               \frac{1066349}{6561} - \frac{56}{27} \zeta_2 
                              + 
                               \frac{797}{135} \zeta_2^2  
                               +\frac{13768}{243} \zeta_3 \Bigg\} 
                               +
                               {\dis{2 C_{A} C_{F} n_{f} T_{F}}} \Bigg\{ -
                               \frac{880}{9 \epsilon^3} 
\nonumber\\
&+ \Bigg(
                               \frac{6844}{27} - \frac{640}{9} \zeta_3
                               \Bigg) \frac{1}{\epsilon^2} + \Bigg( -
                               \frac{16219}{81} 
                              + \frac{158}{3} \zeta_2 
                               + \frac{352}{15} \zeta_2^2 +
                               \frac{1744}{27} \zeta_3 \Bigg)
                               \frac{1}{\epsilon} - \frac{753917}{972}  
                               \nonumber\\
&- \frac{593}{6} \zeta_2 - \frac{96}{5} \zeta_2^2  + 
                               \frac{4934}{81} \zeta_3 + 48 \zeta_2
                               \zeta_3  
                              + \frac{32}{9} \zeta_5 \Bigg\} 
                               + 
                               {\dis{C_{A}^3}} \Bigg\{ - \frac{256}{3
                               \epsilon^6} - \frac{352}{3 \epsilon^5}
                               + \frac{16144}{81 \epsilon^4} 
\nonumber\\
&+ 
                               \Bigg( \frac{22864}{243} +
                               \frac{2068}{27} \zeta_2 - \frac{176}{3}
                               \zeta_3 \Bigg) \frac{1}{\epsilon^3} 
                              + 
                               \Bigg( - \frac{172844}{243} -
                               \frac{1630}{81} \zeta_2 +
                               \frac{494}{45} \zeta_2^2 -
                               \frac{836}{27} \zeta_3 \Bigg)
                               \frac{1}{\epsilon^2} 
                               \nonumber\\
&+ \Bigg( \frac{2327399}{2187} -
                               \frac{71438}{243} \zeta_2 
                              + 
                               \frac{3751}{180} \zeta_2^2 -
                               \frac{842}{9} \zeta_3 
                               + \frac{170}{9} \zeta_2 \zeta_3 + 
                               \frac{1756}{15} \zeta_5 \Bigg)
                               \frac{1}{\epsilon} 
                               + \frac{16531853}{26244} 
\nonumber\\
&+ 
                               \frac{918931}{1458} \zeta_2 +
                               \frac{27251}{1080} \zeta_2^2 
                              - \frac{22523}{270} \zeta_2^3   -
                               \frac{51580}{243} \zeta_3  +
                               \frac{77}{18} \zeta_2 \zeta_3  
                               - \frac{1766}{9} \zeta_3^2 + 
                               \frac{20911}{45} \zeta_5  \Bigg\} \,,
\end{align}


\begin{align}
  \label{eq:FgJ1}
  {\hat{\cal F}}^{J,(1)}_{g} &= {\dis{C_{A}}} \Bigg\{ - \frac{8}{\epsilon^2} +
                               4  + \zeta_2   + 
                               \epsilon \Bigg( - \frac{15}{2}  +
                               \zeta_2  - \frac{16}{3} \zeta_3 \Bigg)
                               + \epsilon^2 \Bigg( \frac{287}{24} - 2
                               \zeta_2  + \frac{127}{80} \zeta_2^2
                               \Bigg)  
                               \nonumber\\
                             &+ 
                               \epsilon^3 \Bigg( - \frac{5239}{288}  +
                               \frac{151}{48} \zeta_2  +
                               \frac{19}{120} \zeta_2^2  +
                               \frac{\zeta_3}{12}  + 
                               \frac{7}{6} \zeta_2 \zeta_3  -
                               \frac{91}{20} \zeta_5 \Bigg) \Bigg\}  + 
                               {\dis{C_{F}}} \Bigg\{ 4  + 
                               \epsilon \Bigg( - \frac{21}{2}  
                               \nonumber\\
                             &+ 6
                               \zeta_3 \Bigg)  + \epsilon^2 \Bigg(
                               \frac{155}{8}  - \frac{5}{2} \zeta_2  -
                               \frac{9}{5} \zeta_2^2  - \frac{9}{2}
                               \zeta_3 \Bigg)  + \epsilon^3 \Bigg( -
                               \frac{1025}{32}  + \frac{83}{16}
                               \zeta_2  + 
                               \frac{27}{20} \zeta_2^2  + \frac{20}{3}
                               \zeta_3  
                               \nonumber\\
                             &- \frac{3}{4} \zeta_2 \zeta_3
                               + \frac{21}{2} \zeta_5 \Bigg) \Bigg\}\,,
\end{align}


\begin{align}
  \label{eq:FgJ2}
  {\hat{\cal F}}^{J,(2)}_{g} &= {\dis{2 C_{A} n_{f} T_{F}}} \Bigg\{ - \frac{8}{3
                               \epsilon^3}  + \frac{20}{9 \epsilon^2} + 
                               \Bigg( \frac{106}{27} + 2 \zeta_2
                               \Bigg) \frac{1}{\epsilon}  -
                               \frac{1753}{81}   - \frac{\zeta_2}{3}
                               - \frac{110}{9} \zeta_3 + 
                               \epsilon \Bigg( \frac{14902}{243} 
\nonumber\\
&-
                               \frac{103}{18} \zeta_2 
                              + \frac{241}{60} \zeta_2^2 +
                               \frac{599}{27} \zeta_3 \Bigg)  + 
                               \epsilon^2 \Bigg( - \frac{411931}{2916}
                               + \frac{2045}{108} \zeta_2  -
                               \frac{2353}{360} \zeta_2^2  - 
                               \frac{3128}{81} \zeta_3 
\nonumber\\
&+ \frac{43}{6}
                               \zeta_2 \zeta_3 
                               - \frac{167}{10} \zeta_5 \Bigg) \Bigg\} + 
                               {\dis{C_{A}  C_{F}}} \Bigg\{ - \frac{32}{\epsilon^2} + 
                               \Bigg( \frac{208}{3} - 48 \zeta_3
                               \Bigg) \frac{1}{\epsilon} -
                               \frac{451}{9}  + 24 \zeta_2 +
                               \frac{72}{5} \zeta_2^2 
\nonumber\\
&- 8 \zeta_3 
                               + \epsilon \Bigg( - \frac{16385}{108} - \frac{52}{3} \zeta_2 + 
                               \frac{12}{5} \zeta_2^2 + 32 \zeta_3 +
                               10 \zeta_2 \zeta_3 - 14 \zeta_5 \Bigg) + \epsilon^2 \Bigg(
                               \frac{1073477}{1296} 
\nonumber\\
&- \frac{815}{9} \zeta_2 
                              + 
                               \frac{19}{20} \zeta_2^2  +
                               \frac{17}{70} \zeta_2^3  -
                               \frac{1915}{36} \zeta_3  + 9 \zeta_2
                               \zeta_3  - 
                               34 \zeta_3^2  - \frac{2279}{6} \zeta_5
                               \Bigg)  \Bigg\}  + 
                               {\dis{2 C_{F} n_{f} T_{F}}} \Bigg\{ \frac{26}{3
                               \epsilon} 
\nonumber\\
&- \frac{709}{18}  + 16 \zeta_3 
                              + \epsilon \Bigg( \frac{26149}{216} - \frac{65}{6} \zeta_2 - 
                               \frac{76}{15} \zeta_2^2 - 44 \zeta_3 \Bigg) + 
                               \epsilon^2 \Bigg( - \frac{828061}{2592}
                               + \frac{3229}{72} \zeta_2  
\nonumber\\
&+ \frac{212}{15} \zeta_2^2 + 
                               \frac{1729}{18} \zeta_3 
                              - 4 \zeta_2 \zeta_3 + \frac{166}{3} \zeta_5 \Bigg) \Bigg\} + 
                               {\dis{C_{A}^2}} \Bigg\{ \frac{32}{\epsilon^4}
                               + \frac{44}{3 \epsilon^3}  + \Bigg( -
                               \frac{422}{9}  - 4 \zeta_2 \Bigg)
                               \frac{1}{\epsilon^2}  
\nonumber\\
&+ 
                               \Bigg( \frac{1214}{27}  - 19 \zeta_2 
                              + \frac{122}{3} \zeta_3 \Bigg)
                               \frac{1}{\epsilon}  + \frac{1513}{81}  +
                               \frac{143}{6} \zeta_2 - \frac{61}{5}
                               \zeta_2^2  + \frac{209}{9} \zeta_3  + 
                               \epsilon \Bigg( - \frac{202747}{972}  
\nonumber\\
&+
                               \frac{59}{36} \zeta_2  - \frac{349}{24}
                               \zeta_2^2  
                              - \frac{2393}{108} \zeta_3  - 
                               \frac{53}{6} \zeta_2 \zeta_3  +
                               \frac{369}{10} \zeta_5 \Bigg)  + 
                               \epsilon^2 \Bigg( \frac{7681921}{11664}
                               - \frac{35255}{432} \zeta_2 +
                               \frac{1711}{180} \zeta_2^2  
\nonumber\\
&- 
                               \frac{7591}{840} \zeta_2^3 
                              -
                               \frac{5683}{1296} \zeta_3  -
                               \frac{407}{12} \zeta_2 \zeta_3  + 
                               \frac{775}{36} \zeta_3^2 +
                               \frac{4013}{30} \zeta_5 \Bigg) \Bigg\}
                               + 
                               {\dis{C_{F}^2}} \Bigg\{ - 6 + \epsilon \Bigg(
                               \frac{259}{12}  + 41 \zeta_3 
\nonumber\\
&- 60
                               \zeta_5 \Bigg)  
                              + 
                               \epsilon^2 \Bigg( - \frac{7697}{144}  +
                               \frac{\zeta_2}{3}  - \frac{184}{15}
                               \zeta_2^2  + \frac{120}{7} \zeta_2^3 - 
                               163 \zeta_3 + 4 \zeta_2 \zeta_3  + 30
                               \zeta_3^2  + \frac{470}{3} \zeta_5 \Bigg) \Bigg\}\,,
\end{align}


\begin{align}
  \label{eq:FqG1}
  {\hat{\cal F}}^{G,(1)}_{q} &= {\dis{2 n_{f} T_{F}}} \Bigg\{ \frac{4}{3 \epsilon}  -
                               \frac{19}{9}  + \epsilon \Bigg(
                               \frac{355}{108}  - \frac{\zeta_2}{6}
                               \Bigg)  + \epsilon^2 \Bigg( -
                               \frac{6523}{1296}  + \frac{19}{72}
                               \zeta_2  + \frac{25}{18} \zeta_3 \Bigg) + 
                               \epsilon^3 \Bigg( \frac{118675}{15552}
                               \nonumber\\
                             &- \frac{355}{864} \zeta_2  -
                               \frac{191}{480} \zeta_2^2  - 
                               \frac{475}{216} \zeta_3 \Bigg) \Bigg\} 
                               + 
                               {\dis{C_{F}}} \Bigg\{ - \frac{8}{\epsilon^2}  +
                               \frac{6}{\epsilon}  - \frac{11}{2}  +
                               \zeta_2  + 
                               \epsilon \Bigg( \frac{25}{8}  -
                               \frac{3}{4} \zeta_2  - \frac{7}{3}
                               \zeta_3 \Bigg)  
                               \nonumber\\
                             &+ 
                               \epsilon^2 \Bigg( - \frac{11}{32}  -
                               \frac{21}{16} \zeta_2  + \frac{47}{80}
                               \zeta_2^2  + \frac{7}{4} \zeta_3 \Bigg) + 
                               \epsilon^3 \Bigg( - \frac{415}{128}  +
                               \frac{223}{64} \zeta_2  -
                               \frac{141}{320} \zeta_2^2  -
                               \frac{155}{48} \zeta_3  
                               \nonumber\\
                             &+ 
                               \frac{7}{24} \zeta_2 \zeta_3  -
                               \frac{31}{20}  \zeta_5 \Bigg) \Bigg\} 
                               + 
                               {\dis{C_{A}}} \Bigg\{ - \frac{22}{3 \epsilon}
                               + \frac{269}{18}   + 
                               \epsilon \Bigg( - \frac{5045}{216}  +
                               \frac{23}{12} \zeta_2  + 3 \zeta_3
                               \Bigg)  + \epsilon^2 \Bigg(
                               \frac{90893}{2592}  
                               \nonumber\\
                             &- \frac{485}{144}
                               \zeta_2  - 
                               \frac{4}{5} \zeta_2^2 - \frac{275}{36}
                               \zeta_3 \Bigg)  + 
                               \epsilon^3 \Bigg( -
                               \frac{1620341}{31104}  +
                               \frac{8861}{1728} \zeta_2  +
                               \frac{751}{320} \zeta_2^2  + 
                               \frac{4961}{432} \zeta_3 +
                               \frac{\zeta_2 \zeta_3}{8}  
                               \nonumber\\
                             &+
                               \frac{15}{2} \zeta_5  \Bigg) \Bigg\}\,,
\end{align}


\begin{align}
  \label{eq:FqG2}
  {\hat{\cal F}}^{G,(2)}_{q} &= {\dis{4 n_{f}^2 T_{F}^{2}}} \Bigg\{ \frac{16}{9
                               \epsilon^2}  - \frac{152}{27 \epsilon}
                               + \frac{124}{9} - \frac{4}{9} \zeta_2  + 
                               \epsilon \Bigg( - \frac{7426}{243}  +
                               \frac{38}{27} \zeta_2  + \frac{136}{27}
                               \zeta_3 \Bigg)  + 
                               \epsilon^2 \Bigg( \frac{47108}{729}  
\nonumber\\
&-
                               \frac{31}{9} \zeta_2 
                               - \frac{43}{30}
                               \zeta_2^2  - \frac{1292}{81} \zeta_3 \Bigg)  \Bigg\} 
                               + 
                               {\dis{C_{A}^2}} \Bigg\{ \frac{484}{9
                               \epsilon^2}  - \frac{6122}{27 \epsilon}
                               + \frac{1865}{3} - \frac{319}{9}
                               \zeta_2  - 66 \zeta_3 
\nonumber\\
&+ 
                               \epsilon \Bigg( - \frac{702941}{486}  
                              +
                               \frac{14969}{108}  \zeta_2  +
                               \frac{299}{20} \zeta_2^2  + 
                               \frac{31441}{108} \zeta_3 +  5 \zeta_2
                               \zeta_3 - 30 \zeta_5 \Bigg)  + 
                               \epsilon^2 \Bigg( \frac{18199507}{5832}
                               \nonumber\\
&- \frac{5861}{16} \zeta_2 
                              -
                               \frac{63233}{720} \zeta_2^2  - 
                               \frac{691}{140} \zeta_2^3 -
                               \frac{995915}{1296} \zeta_3  +
                               \frac{52}{3} \zeta_2 \zeta_3  - 
                               \frac{39}{2} \zeta_3^2 -  \frac{1343}{12} \zeta_5 \Bigg)  \Bigg\} 
                               \nonumber\\
&+ 
                               {\dis{2 C_{F} n_{f} T_{F}}} \Bigg\{ - \frac{40}{3
                               \epsilon^3} 
                                + \frac{280}{9 \epsilon^2} + 
                               \Bigg( - \frac{1417}{27} +  2 \zeta_2
                               \Bigg) \frac{1}{\epsilon}  +
                               \frac{22021}{324} -  \frac{14}{3}
                               \zeta_2  -  \frac{82}{9} \zeta_3 
\nonumber\\
&+ 
                               \epsilon \Bigg( - \frac{238717}{3888} -
                               \frac{73}{12} \zeta_2 
                              + \frac{25}{12}
                               \zeta_2^2  + \frac{394}{27} \zeta_3 \Bigg) + 
                               \epsilon^2 \Bigg( -
                               \frac{290075}{46656}  +
                               \frac{6181}{144} \zeta_2 -
  \frac{499}{180} \zeta_2^2 
\nonumber\\
&- 
                               \frac{9751}{324} \zeta_3 + \frac{13}{6}
                               \zeta_2 \zeta_3  
                              - \frac{29}{6} \zeta_5 \Bigg) \Bigg\} 
                               + 
                               {\dis{C_{F}^2}} \Bigg\{ \frac{32}{\epsilon^4} -
                               \frac{48}{\epsilon^3} +  \Bigg( 62 -  8
                               \zeta_2 \Bigg)  \frac{1}{\epsilon^2} +
                               \Bigg( - \frac{113}{2}  
\nonumber\\
&+ \frac{128}{3}
                               \zeta_3 \Bigg) \frac{1}{\epsilon} +
                               \frac{581}{24} +  \frac{27}{2} \zeta_2 
                              - 
                               13 \zeta_2^2 - 58 \zeta_3 + 
                               \epsilon \Bigg( \frac{12275}{288} -
                               \frac{331}{24} \zeta_2  +
                               \frac{493}{30} \zeta_2^2  +
                               \frac{587}{6} \zeta_3  
\nonumber\\
&- 
                               \frac{56}{3} \zeta_2 \zeta_3 +
                               \frac{92}{5} \zeta_5 \Bigg) 
                               + 
                               \epsilon^2 \Bigg( - \frac{456779}{3456}
                               - \frac{2011}{96} \zeta_2 -
                               \frac{1279}{80} \zeta_2^2  + 
                               \frac{223}{20} \zeta_2^3 -
                               \frac{13363}{72} \zeta_3  
\nonumber\\
&- \frac{5}{2}
                               \zeta_2 \zeta_3  + 
                               \frac{652}{9} \zeta_3^2  
                              -
                               \frac{193}{30} \zeta_5  \Bigg)  \Bigg\} 
                               + 
                               {\dis{2 C_{A} n_{f} T_{F}}} \Bigg\{ - \frac{176}{9
                               \epsilon^2} +  \frac{1972}{27 \epsilon}
                               - \frac{1708}{9} +  \frac{80}{9} \zeta_2 + 
                               4 \zeta_3 
\nonumber\\
&+ 
                               \epsilon \Bigg( \frac{104858}{243}  -
                               \frac{853}{27} \zeta_2 
                               - \frac{2}{3}
                               \zeta_2^2  - \frac{1622}{27} \zeta_3
                               \Bigg) +  \epsilon^2 \Bigg( -
                               \frac{5369501}{5832} +  \frac{1447}{18}
                               \zeta_2 +  \frac{817}{45} \zeta_2^2 
\nonumber\\
&+ 
                               \frac{31499}{162} \zeta_3 + \frac{7}{3}
                               \zeta_2 \zeta_3  
                              + 19 \zeta_5 \Bigg) \Bigg\} 
                               + 
                               {\dis{C_{A} C_{F}}} \Bigg\{ \frac{220}{3 \epsilon^3}+ 
                               \Bigg( - \frac{1804}{9} + 4 \zeta_2
                               \Bigg) \frac{1}{\epsilon^2} +  \Bigg(
                               \frac{20777}{54} 
\nonumber\\
&-  19 \zeta_2 - 50
                               \zeta_3 \Bigg)  \frac{1}{\epsilon}  
                              -
                               \frac{397181}{648 } + \frac{161}{3}
                               \zeta_2 +  \frac{76}{5} \zeta_2^2 + 
                               \frac{1333}{9} \zeta_3 + \epsilon
                               \Bigg(  \frac{6604541}{7776} -
                               \frac{669}{8} \zeta_2  
\nonumber\\
&- 
                               \frac{5519}{120} \zeta_2^2 -
                               \frac{8398}{27} \zeta_3  
                              + \frac{89}{6}
                               \zeta_2 \zeta_3  - 
                               \frac{51}{2} \zeta_5 \Bigg) +
                               \epsilon^2 \Bigg(  -
                               \frac{93774821}{93312} +
                               \frac{20035}{288} \zeta_2 +  
                               \frac{33377}{360} \zeta_2^2 
\nonumber\\
&+
                               \frac{1793}{840} \zeta_2^3  +
                               \frac{390731}{648} \zeta_3 
                              -  
                               \frac{445}{12} \zeta_2 \zeta_3 -
                               \frac{425}{12} \zeta_3^2  +
                               \frac{641}{12} \zeta_5 \Bigg) \Bigg\}\,,
\end{align}


\begin{align}
  \label{eq:FqJ1}
  {\hat{\cal F}}^{J,(1)}_{q} &= {\dis{C_{F}}} \Bigg\{ - \frac{8}{\epsilon^2}
                               + \frac{6}{\epsilon}  - 2 + \zeta_2  +
                               \epsilon \Bigg( - 1  - \frac{3}{4}
                               \zeta_2  - \frac{7}{3} \zeta_3 \Bigg)  + 
                               \epsilon^2 \Bigg( \frac{5}{2} +
                               \frac{\zeta_2}{4}  + \frac{47}{80}
                               \zeta_2^2  + \frac{7}{4} \zeta_3 \Bigg) 
                               \nonumber\\
                             &+ 
                               \epsilon^3 \Bigg( - \frac{13}{4}  +
                               \frac{\zeta_2}{8}  - \frac{141}{320}
                               \zeta_2^2  - \frac{7}{12} \zeta_3 + 
                               \frac{7}{24} \zeta_2 \zeta_3  -
                               \frac{31}{20} \zeta_5  \Bigg) \Bigg\}\,,
\end{align}


\begin{align}
  \label{eq:FqJ2}
  {\hat{\cal F}}^{J,(2)}_{q} &={\dis{2 C_{F} n_{f} T_{F}}} \Bigg\{ - \frac{8}{3
                               \epsilon^3}  + \frac{56}{9 \epsilon^2}
                               + \Bigg( - \frac{47}{27} - \frac{2}{3}
                               \zeta_2 \Bigg)  \frac{1}{\epsilon} -
                               \frac{4105}{324}  + 
                               \frac{14}{9} \zeta_2  - \frac{26}{9}
                               \zeta_3  + \epsilon \Bigg(
                               \frac{142537}{3888}  
                               \nonumber\\
                             &- \frac{695}{108}
                               \zeta_2  + 
                               \frac{41}{60} \zeta_2^2 +
                               \frac{182}{27} \zeta_3 \Bigg)  + 
                               \epsilon^2 \Bigg( -
                               \frac{3256513}{46656}  +
                               \frac{21167}{1296} \zeta_2 
                               - \frac{287}{180} \zeta_2^2 - 
                               \frac{2555}{324} \zeta_3  
                               \nonumber\\
                             &-
                               \frac{13}{18} \zeta_2 \zeta_3  -
                               \frac{121}{30} \zeta_5  \Bigg) \Bigg\} 
                               + 
                               {\dis{C_{F}^2}} \Bigg\{ \frac{32}{\epsilon^4}
                               - \frac{48}{\epsilon^3}  + \Bigg( 34 -
                               8 \zeta_2 \Bigg)  \frac{1}{\epsilon^2}
                               + \Bigg( - \frac{5}{2} + \frac{128}{3}
                               \zeta_3 \Bigg)  \frac{1}{\epsilon} -
                               \frac{361}{8}  
                               \nonumber\\
                             &+ \frac{9}{2} \zeta_2 - 
                               13 \zeta_2^2 - 58 \zeta_3  + 
                               \epsilon \Bigg( \frac{3275}{32}  +
                               \frac{3}{8} \zeta_2  + \frac{171}{10}
                               \zeta_2^2  + \frac{503}{6} \zeta_3 - 
                               \frac{56}{3} \zeta_2 \zeta_3  +
                               \frac{92}{5} \zeta_5 \Bigg)  
                               \nonumber\\
                             &+ 
                               \epsilon^2 \Bigg( - \frac{20257}{128}
                               - \frac{793}{32} \zeta_2  -
                               \frac{2097}{80} \zeta_2^2  + 
                               \frac{223}{20} \zeta_2^3 -
                               \frac{4037}{24} \zeta_3  + \frac{27}{2}
                               \zeta_2 \zeta_3  + 
                               \frac{652}{9} \zeta_3^2 -
                               \frac{231}{10}  \zeta_5 \Bigg) \Bigg\} 
                               \nonumber\\
                             &+ 
                               {\dis{C_{A} C_{F}}} \Bigg\{ \frac{44}{3
                               \epsilon^3}  + 
                               \Bigg( - \frac{332}{9} + 4 \zeta_2
                               \Bigg) \frac{1}{\epsilon^2}  + \Bigg(
                               \frac{2545}{54}  + \frac{11}{3} \zeta_2
                               - 26 \zeta_3 \Bigg) \frac{1}{\epsilon}
                               - \frac{18037}{648} - \frac{47}{9}
                               \zeta_2  
                               \nonumber\\
                             &+ \frac{44}{5} \zeta_2^2 + 
                               \frac{467}{9} \zeta_3 + \epsilon \Bigg(
                               - \frac{221963}{7776}  -
                               \frac{263}{216} \zeta_2  -
                               \frac{1891}{120} \zeta_2^2  - 
                               \frac{2429}{27} \zeta_3 + \frac{89}{6}
                               \zeta_2 \zeta_3  - \frac{51}{2} \zeta_5
                               \Bigg)  
                               \nonumber\\
                             &+ 
                               \epsilon^2 \Bigg(
                               \frac{11956259}{93312}  +
                               \frac{38987}{2592} \zeta_2  +
                               \frac{9451}{360} \zeta_2^2  - 
                               \frac{809}{280} \zeta_2^3 +
                               \frac{92701}{648} \zeta_3  -
                               \frac{397}{36} \zeta_2 \zeta_3  - 
                               \frac{569}{12} \zeta_3^2  
                               \nonumber\\
                             &+ \frac{3491}{60} \zeta_5 \Bigg) \Bigg\}\,,
\end{align}


\begin{align}
  \label{eq:FqJ3}
  {\hat{\cal F}}^{J,(3)}_{q} &= {\dis{4 C_{F} n_{f}^2 T_{F}^{2}}} \Bigg\{ - \frac{128}{81
                               \epsilon^4}  + \frac{1504}{243
                               \epsilon^3}  + 
                               \Bigg( - \frac{16}{9}  - \frac{16}{9}
                               \zeta_2 \Bigg) \frac{1}{\epsilon^2}  + 
                               \Bigg( - \frac{73432}{2187}  +
                               \frac{188}{27} \zeta_2  
\nonumber\\
&-
                               \frac{272}{81} \zeta_3 \Bigg)
                               \frac{1}{\epsilon}  
                              +
                               \frac{881372}{6561}  - 26 \zeta_2  -
                               \frac{83}{135} \zeta_2^2   + \frac{3196}{243} \zeta_3 \Bigg\} 
                               + 
                               {\dis{C_{F}^3}} \Bigg\{ - \frac{256}{3
                               \epsilon^6}  + \frac{192}{\epsilon^5} +
                               \Bigg( - 208  
\nonumber\\
&+ 32 \zeta_2 \Bigg)
                               \frac{1}{\epsilon^4}  
                              + 
                               \Bigg( 88 + 24 \zeta_2  - \frac{800}{3}
                               \zeta_3 \Bigg)  \frac{1}{\epsilon^3}  +
                               \Bigg( 254  - 98 \zeta_2 +
                               \frac{426}{5} \zeta_2^2  + 552 \zeta_3
                               \Bigg) \frac{1}{\epsilon^2}  
\nonumber\\
&+ 
                               \Bigg( - \frac{5045}{6}  + 83 \zeta_2
                              - \frac{1461}{10} \zeta_2^2  -
                               \frac{2630}{3} \zeta_3  + 
                               \frac{428}{3} \zeta_2 \zeta_3  -
                               \frac{1288}{5} \zeta_5 \Bigg)
                               \frac{1}{\epsilon} + \frac{38119}{24}
                               + \frac{1885}{12} \zeta_2  
\nonumber\\
&+ 
                               \frac{8659}{40} \zeta_2^2  
                              -
                               \frac{9095}{252} \zeta_2^3   + 1153
                               \zeta_3 - 35 \zeta_2 \zeta_3  - 
                               \frac{1826}{3} \zeta_3^2  - \frac{562}{5} \zeta_5 \Bigg\} 
                               + 
                               {\dis{2 C_{F}^2 n_{f} T_{F}}} \Bigg\{ \frac{64}{3
                               \epsilon^5}  
\nonumber\\
&- \frac{592}{9 \epsilon^4}
+ \Bigg( \frac{1480}{27}  
                              + \frac{8}{3}
                               \zeta_2 \Bigg)  \frac{1}{\epsilon^3} + 
                               \Bigg( \frac{7772}{81} - \frac{266}{9}
                               \zeta_2  + \frac{584}{9} \zeta_3 \Bigg)
                               \frac{1}{\epsilon^2} + 
                               \Bigg( - \frac{116735}{243}  +
                               \frac{2633}{27} \zeta_2  
\nonumber\\
&-
                               \frac{337}{18} \zeta_2^2  
                              -
                               \frac{5114}{27} \zeta_3 \Bigg)
                               \frac{1}{\epsilon}  +
                               \frac{3396143}{2916}  -
                               \frac{32329}{162} \zeta_2  + 
                               \frac{8149}{216} \zeta_2^2 + 
                               \frac{39773}{81} \zeta_3  -
                               \frac{343}{9} \zeta_2 \zeta_3  
\nonumber\\
&+ \frac{278}{45} \zeta_5 \Bigg\} 
                              + 
                               {\dis{C_{A}^2 C_{F}}} \Bigg\{ - \frac{3872}{81
                               \epsilon^4}  + \Bigg( \frac{52168}{243}
                               - \frac{704}{27} \zeta_2 \Bigg)
                               \frac{1}{\epsilon^3}  + 
                               \Bigg( - \frac{117596}{243} -
                               \frac{2212}{81} \zeta_2  
\nonumber\\
&-
                               \frac{352}{45} \zeta_2^2  
                              +
                               \frac{6688}{27} \zeta_3 \Bigg)
                               \frac{1}{\epsilon^2}  + \Bigg(
                               \frac{1322900}{2187}  +
                               \frac{39985}{243} \zeta_2  - 
                               \frac{1604}{15} \zeta_2^2 -
                               \frac{24212}{27} \zeta_3  +
                               \frac{176}{9} \zeta_2 \zeta_3  
\nonumber\\
&+ 
                               \frac{272}{3} \zeta_5 \Bigg)
                               \frac{1}{\epsilon}  
                              +
                               \frac{1213171}{13122} -
                               \frac{198133}{729} \zeta_2  +
                               \frac{146443}{540} \zeta_2^2  - \frac{6152}{189} \zeta_2^3 + 
                               \frac{970249}{486} \zeta_3 -
                               \frac{926}{9} \zeta_2 \zeta_3  
\nonumber\\
&-
                               \frac{1136}{9} \zeta_3^2  
                              + \frac{772}{9} \zeta_5  \Bigg\} 
                               + 
                               {\dis{2 C_{A} C_{F} n_{f} T_{F}}} \Bigg\{
                               \frac{1408}{81 \epsilon^4}  + \Bigg( -
                               \frac{18032}{243}  + \frac{128}{27}
                               \zeta_2 \Bigg)  \frac{1}{\epsilon^3} + 
                               \Bigg( \frac{24620}{243} 
\nonumber\\
&+
                               \frac{1264}{81} \zeta_2  
                              -
                               \frac{1024}{27} \zeta_3 \Bigg)
                               \frac{1}{\epsilon^2}  + 
                               \Bigg( \frac{212078}{2187} -
                               \frac{16870}{243} \zeta_2  +
                               \frac{88}{5} \zeta_2^2  +
                               \frac{12872}{81} \zeta_3 \Bigg)
                               \frac{1}{\epsilon}  -
                               \frac{5807647}{6561}  
                               \nonumber\\
                             &+
                               \frac{299915}{1458} \zeta_2  - 
                               \frac{5492}{135} \zeta_2^2 - 
                               \frac{42941}{81} \zeta_3 +
                               \frac{422}{9} \zeta_2 \zeta_3  - \frac{28}{3} \zeta_5 \Bigg\} 
                               + 
                               {\dis{C_{A} C_{F}^2}} \Bigg\{ - \frac{352}{3 \epsilon^5} + 
                               \Bigg( \frac{3448}{9}  
                               \nonumber\\
                             &- 32 \zeta_2
                               \Bigg) \frac{1}{\epsilon^4}  + \Bigg( -
                               \frac{16948}{27}  + \frac{28}{3}
                               \zeta_2 + 208 \zeta_3 \Bigg)
                               \frac{1}{\epsilon^3}  + \Bigg(
                               \frac{44542}{81}  + \frac{1127}{9}
                               \zeta_2  - \frac{332}{5} \zeta_2^2 
                               \nonumber\\
                             &- 
                               840 \zeta_3 \Bigg) \frac{1}{\epsilon^2}
                               + \Bigg( \frac{149299}{486}  -
                               \frac{12757}{54} \zeta_2  +
                               \frac{9839}{36} \zeta_2^2  + 
                               \frac{5467}{3} \zeta_3 - \frac{430}{3}
                               \zeta_2 \zeta_3  + 284 \zeta_5 \Bigg)
                               \frac{1}{\epsilon}  
                               \nonumber\\
                             &-
                               \frac{15477463}{5832}  +
                               \frac{21455}{324} \zeta_2  -
                               \frac{1002379}{2160} \zeta_2^2  - 
                               \frac{18619}{1260} \zeta_2^3 -
                               \frac{51781}{18} \zeta_3  +
                               \frac{910}{9} \zeta_2 \zeta_3  + 
                               \frac{1616}{3} \zeta_3^2 
                               \nonumber\\
                             &- 
                               \frac{3394}{45} \zeta_5 \Bigg\}\,. 
\end{align}

\chapter{Harmonic Polylogarithms}
\label{chpt:App-HPL}

The logarithms, polylogarithms (Li$_n(x)$) and Nielsen's polylogarithm
(S$_{n,p}(x)$)  appear naturally in the analytical expressions of
radiative corrections in pQCD which are defined through
\begin{align}
\label{eq:App-HPL-1}
&\text{ln}(x) = \int_{1}^{x} \frac{dt}{t}\,,
\nonumber\\
&\text{Li}_n(x) \equiv \sum_{k=1}^{\infty} \frac{x^k}{k^n}  =
  \int_{0}^{x} \frac{dt}{t} \text{Li}_{n-1}(t)\,, \quad\quad
  \text{e.g.} \quad \text{Li}_1(x) = - \text{ln}(1-x)\,,
\nonumber\\
&S_{n,p}(x) \equiv \frac{(-1)^{n+p-1}}{(n-1)! p!} \int_{0}^{1}
  \frac{dt}{t} [\text{ln} (t)]^{n-1} [\text{ln}(1-x t)]^p\,,
\nonumber\\
&\quad\text{e.g.} \quad S_{n-1,1}(x) = \text{Li}_n(x)\,.
\end{align}
However, for higher order radiative corrections (2-loops and beyond), these functions are not
sufficient to evaluate  all the loop integrals appearing in
the Feynman graphs. This is overcome by introducing a new set of
functions which are called \textbf{Harmonic Polylogarithms
  (HPLs)}. These are essentially a generalisation of Nielsen's
polylogarithms. In this appendix, we briefly describe the definition
and properties of HPL~\cite{Remiddi:1999ew} and 2dHPL. HPL is
represented by $H(\vec{m}_w;y)$  
with a $w$-dimensional vector $\vec{m}_w$ of parameters and its argument $y$. $w$ is called the weight of the HPL. The elements of $\vec{m}_w$ belong to $\{ 1, 0, -1 \}$ through which 
the following rational functions are represented  
\begin{equation}
 f(1;y) \equiv \frac{1}{1-y}, \qquad f(0;y) \equiv \frac{1}{y},  \qquad f(-1;y) \equiv \frac{1}{1+y} \, .
\end{equation}
The weight 1 $(w = 1)$ HPLs are defined by
\begin{equation}
 H(1, y) \equiv - \ln (1 - y), \qquad  H(0, y) \equiv \ln y, \qquad  H(-1, y) \equiv \ln (1 + y) \, .
\end{equation}
For $w > 1$, $H(m, \vec{m}_{w};y)$ is defined by 
\begin{equation}\label{1dhpl}
 H(m, \vec{m}_w;y) \equiv \int_0^y dx ~ f(m, x) ~ H(\vec{m}_w;x),  \qquad \qquad  m \in 0, \pm 1  \, .
\end{equation}
The 2dHPLs are defined in the same way as Eq.~(\ref{1dhpl}) with the new elements $\{ 2, 3 \}$ in $\vec{m}_w$ representing a new 
class of rational functions
\begin{equation}
 f(2;y) \equiv f(1-z;y) \equiv \frac{1}{1-y-z}, \qquad f(3;y) \equiv f(z;y) \equiv \frac{1}{y+z} 
\end{equation}
and correspondingly with the weight 1 $(w = 1)$ 2dHPLs
\begin{equation}
 H(2, y) \equiv - \ln \Big(1 - \frac{y}{1-z} \Big), \qquad  H(3, y) \equiv \ln \Big( \frac{y+z}{z} \Big) \, .
\end{equation}

\subsection{Properties}

\underline{Shuffle algebra} : A product of two HPL with weights $w_1$ and $w_2$ of the same argument $y$ is a combination of HPLs with weight
$(w_1 + w_2)$ and argument $y$, such that all possible permutations of the elements of $\vec{m}_{w_1}$ and $\vec{m}_{w_2}$ are considered preserving the 
relative orders of the elements of $\vec{m}_{w_1}$ and $\vec{m}_{w_2}$,
\begin{equation}
 H(\vec{m}_{w_1};y) H(\vec{m}_{w_2};y) = \sum_{\text{\tiny $\vec{m}_{w} = \vec{m}_{w_1} \uplus \vec{m}_{w_2}$}}  H(\vec{m}_{w};y).
\end{equation}
\underline{Integration-by-parts identities} : The ordering of the elements of $\vec{m}_{w}$ in an HPL with weight $w$ and argument $y$ can be reversed using 
integration-by-parts and in the process, some products of two HPLs are generated in the following way
\begin{eqnarray}
 H(\vec{m}_{w};y) \equiv H(m_1, m_2, ... , m_w; y ) &=& H(m_1, y) H(m_2, ... , m_w; y )
\nonumber\\
 &-& H(m_2, m_1, y) H(m_3, ... , m_w; y )
\nonumber\\
 &+& ... + (-1)^{w+1} H ( m_w, ... , m_2, m_1 ; y ) \, .
\end{eqnarray}

\end{appendices}

\newpage
\thispagestyle{empty}
~
\vskip 1.0cm
\centerline{{\bf{\large ACKNOWLEDGMENTS}}}
\vskip 0.5cm
%

\textit{Finally, we are very close to the moment! The moment, every graduate
student aspires for! The moment, I have been craving for past several
years! Yes, the ship was sailed almost six years back and here, I am
about to write the closing chapter with an intention of revealing the
marvelous experiences of that long journey! The journey which has
played an important role to enrich me not only academically but also 
non-academically. Before joining to
Harish-Chandra Research Institute (HRI), situated at the bank of Ganges and
Yamuna in Allahabad and later to The Institute of Mathematical Sciences
(IMSc), Chennai, I could hardly imagine the experiences which I would
gather in those upcoming days. And none of these would have been possible
without the people whom I have come across during my PhD and
before. Let me begin by expressing my deepest gratitude to my
supervisor Prof. V. Ravindran for his unwavering support, collegiality
and extraordinary mentorship throughout my PhD life! His deep insight about the
subject has helped me to enhance my understanding about the world of QCD. Not only that, 
to me, he truly
symbolises the say, `A Teacher is a Friend, a Philosopher and a
Guide'! His unconditional support from the happiest to the darkest moment of my life is what 
makes all the differences. I would like to extend my gratitude to all those people I had
the opportunity to work with. To
my friend and colleague Narayan Rana for being a wonderful
collaborator. Working with you has made many things possible over a
short span of time. To Prof. Prakash Mathews for his guidance, collaboration and
support throughout my PhD life. To Prof. Thomas Gehrmann for his wonderful
collaboration and several helps about the master integrals. To Dr. M.C. Kumar for fruitful
collaborations, discussions and many helps for academic as well as non-academic
purposes. To my colleagues Dr. Manoj K.
Mandal, Dr. Maguni Mahakhud and Goutam Das for fruitful collaborations. 
To Prof. Roman
N. Lee for his insightful help regarding the reduction of Feynman
integrals through a 
wonderful 
package LiteRed. To Dr. Lorenzo Tancredi for
sharing his deep understanding about the computation of multiloop amplitude
through a set of fantastic lectures and many helpful discussions. To two new youths in our group, 
Pulak Banerjee and Prasanna K. Dhani for fantastic collaborations  and many valuable discussions. 
To Dr. Marco Bonvini and Luca Rottoli for fruitful collaborations and sharing their nice ideas about 
resummation with us. The format of this thesis is inspired by looking at your thesis, Marco, thank you.
To Prof. Ashoke Sen for his extraordinary course, advanced QFT at HRI. To all the professors
of HRI for some wonderful courses during my PhD course work. To all the professors and members IMSc. To Prof. N. D. Hari Dass for his 
constant encouragement starting from my B.Sc and indeed, you are the one who initiated the initial ignition of research into my mind at a very early stage of my life, my deepest gratitude to you Professor!}

\textit{Special thanks to my friend Jahanur. You were the very first inspiration for choosing Physics as 
my career! Our outstanding discussions, academic as well as non-academic, starting from the BSc really have helped us to explore and understand many things in a different way. I would like to express my deepest
appreciation to Satya da for your continuous motivations, help and valuable advice in choosing PhD supervisor.
I am thankful to 
my fellow HRI mates for their wonderful company during my stay at HRI for almost 5 years, in particular, Gupta, Joshi, Shankha,
Avinanda, Roji, Ushoshi, Utkarsh, Swapnomoy, Abhishek, Masud, Avijit, Shrobona, Dibya, Titas, Arijit,
Nabarun, Atri and `hopeless' Anirban da. My gratitude goes out to all of my friends at IMSc, specifically, Rusa, Pinaki, Shilpa,
Biswas, Trisha, Upayan, Anish, Arghya and my multitalented officemates Atanu, Arnab, Rupam. 
I convey my heartiest thanks to my wonderful friends Annwesha, Chitrak, Atreyee, Sundaram, Madhurima, Kawsar, Gazi,
Shahnawaj, Poulami, Saheli, Khadiza and Rudra. I would like to extend my special thanks to my beloved
friend Farha for her trust, love and constant encouragement. I thank Kamrin, Moonfared and Abid for their trust and supports.}

\textit{Finally, I am thankful to my beloved parents, sister, brother-in-law and whole family
for their unconditional love, supports and sacrifices, without those it would have been impossible to achieve 
my PhD.}
\thispagestyle{empty}

\vskip 3cm
\begin{flushright}
\textit{Taushif Ahmed}
\end{flushright}

\bibliography{main} 
\bibliographystyle{utphysM}

\end{document}